\documentclass[a4paper,fleqn,usenatbib]{mnras}

\usepackage{newtxtext,newtxmath}
\usepackage[T1]{fontenc}
\usepackage{ae,aecompl}

\usepackage{graphicx}	% Including figure files
\usepackage{amsmath}	% Advanced maths commands
\usepackage{amssymb}	% Extra maths symbols
\usepackage{booktabs}
\usepackage{relsize}
\usepackage[dvipsnames]{xcolor}
\usepackage[none]{hyphenat}
\newcommand{\HII}{H~\textsc{ii}}

\title[Metallicity calibrations for DIG/LIERs]{Metallicity calibrations for diffuse ionised gas and low ionisation emission regions}

\author[Nimisha Kumari]{
Nimisha Kumari,$^{1,2,3}$\thanks{E-mail: nkumari@ast.cam.ac.uk (NK)}
Roberto Maiolino,$^{1,2}$
Francesco Belfiore$^{4}$
and Mirko Curti$^{1,2}$
\\
$^{1}$Kavli Institute for Cosmology,  University of Cambridge CB3 0HA, UK\\
$^{2}$Cavendish Laboratory, University of Cambridge CB3 0HE, UK\\
$^{3}$Institute of Astronomy, University of Cambridge CB3 0HA, UK\\
$^{4}$European Southern Observatory, Karl-Schwarzschild-Str. 2, D-85748 Garching, Germany
}

\date{Accepted XXX. Received YYY; in original form ZZZ}

\pubyear{2018}

\begin{document}
%	\tableofcontents
\label{firstpage}
\pagerange{\pageref{firstpage}--\pageref{lastpage}}
\maketitle

\begin{abstract}
Using integral field spectroscopic data of 24 nearby spiral galaxies obtained with the Multi-Unit Spectroscopic Explorer (MUSE), we derive empirical calibrations to determine the metallicity of the diffuse ionized gas (DIG) and/or of the low-ionisation emission region (LI(N)ER) in passive regions of galaxies. To do so, we identify a large number of \HII--DIG/LIER pairs that are close enough to be chemically homogeneous and we measure the metallicity difference of each DIG/LIER region relative to its \HII~ region companion when applying the same strong line calibrations. The O3N2 diagnostic ($=$log [([O ~\textsc{iii}]/H$\beta$)/([N \textsc{ii}]/H$\alpha$)]) shows a minimal offset (0.01--0.04 dex)
between DIG/LIER and \HII~regions and little dispersion of the metallicity differences (0.05 dex),
suggesting that the O3N2 metallicity calibration for \HII~ regions can be applied to DIG/LIER regions
and that, when used on poorly resolved galaxies, this diagnostic provides reliable results by suffering
little from DIG contamination. We also derive second-order corrections which further reduce the scatter (0.03--0.04 dex) in the differential metallicity of \HII-DIG/LIER pairs. Similarly, we explore other metallicity diagnostics such as O3S2 ($=$log([O~\textsc{iii}]/H$\beta$+[S~\textsc{ii}]/H$\alpha$)) and N2S2H$\alpha$ ($=$ log([N~\textsc{ii}]/[S~\textsc{ii}]) + 0.264log([N~\textsc{ii}]/H$\alpha$)) and provide corrections for O3S2 to measure the metallicity of DIG/LIER regions. We propose that the corrected O3N2 and O3S2 diagnostics are used to measure the gas-phase metallicity in quiescent galaxies or in quiescent regions of star-forming galaxies.
\end{abstract}


%%%%%%%%%%%%%%%%%%%%%%%%%%%%%%%%%%%%%%%%%%%%%%%%%%

%%%%%%%%%%%%%%%%% BODY OF PAPER %%%%%%%%%%%%%%%%%%

\section{Introduction}
\label{section:introduction}

\indent Knowledge of the gas-phase metallicity of galaxies is essential to study various aspects of the
overall evolution of galaxies. In particular, a robust metallicity measurement within galaxies is
imperative for studying the physical correlations between the gas-phase metallicity of galaxies and physical
properties associated with the history of chemical enrichment of galaxies such as stellar mass (i.e.
mass-metallicity relation, MZR, see, e.g. \citealt[][]{Tremonti2004}) and star-formation rate (i.e.
fundamental metallicity relation, FMR, see, e.g. \citealt{Mannucci2010}), or to explore the abundance
gradients within galaxies \citep[see, e.g.][]{Sanchez2013, Belfiore2017}. The most reliable
method for measuring gas-phase metallicity is the so-called direct T$_e$-method. This method requires the
detection of auroral emission lines (e.g. [O \textsc{iii}] $\lambda$4363, [N \textsc{ii}]
$\lambda$5755), which are generally weak and difficult to detect. Accurate metallicity estimates using the
direct T$_e$-method become even more difficult in high metallicity environments as the auroral
lines are even weaker in these environments \citep{Garnett2004, Kewley2008}. Moreover, the auroral lines may also underestimate the abundances of high-metallicity (super-solar) \HII ~regions with temperature fluctuations or gradients \citep{Stasinska2005}.

\indent To circumvent such problems, almost a dozen indirect methods have been devised
which involve the use of relatively strong emission lines \citep[see, e.g.][]{Aller1942, Alloin1979,
Storchi-Bergmann1994, Oey2000, Denicolo2002,  Pettini2004, Nagao2006, Stasinska2006, Perez-Montero2007a,
Kewley2008, Maiolino2008, Marino2013, Dopita2016, Curti2017, Maiolino2018}. These indirect (``strong line'') metallicity
diagnostics have either been calibrated empirically against direct method metallicity estimates of H
$\textsc{ii}$ regions or theoretically by using photoionisation models of H $\textsc{ii}$ regions such as
\textsc{mappings} \citep{Sutherland1993} and  \textsc{cloudy} \citep{Ferland2013} in codes such as
\textsc{hii-chi-mistry} \citep{Perez-Montero2014}, \textsc{IZI} \citep{Blanc2015} and \textsc{bond}
\citep{ValeAsari2016}, which utilise grids of photoionisation models. These theoretical models suffer from
problems such as limitations on the geometry of H \textsc{ii} regions, poorly constrained dust-depletion and no implementation of the
clumpiness of the density distribution of gas and dust \citep{Kewley2008}. However, another generally overlooked problem with using the
indirect metallicity calibrators is that both empirical and theoretical indirect diagnostics are calibrated for H \textsc{ii} regions
and do not consider the investigation of the more diffuse component of the ionised gas, the so-called diffuse ionised gas (DIG), which dominates the observed nebular emission in several classes of galaxies, or may contaminate the emission from \HII~ regions.     

\indent The DIG, also known as the warm ionised medium (WIM), is an important ionised gas component of the interstellar medium (ISM) in
galaxies, along with the denser H \textsc{ii} regions. It is not only found in the Milky Way \citep{Reynolds1984, Reynolds1990}, but
also in the discs and halos of the other spiral galaxies, as well as in many quiescent galaxies \citep[see,
e.g.][]{RossaDettmar2003a,RossaDettmar2003b,Oey2007, Belfiore2016}. Various classification methods have been devised
and used to distinguish between \HII~ regions and DIG, or to distinguish between normal star-forming galaxies and passive quiescent
galaxies. These methods are based on the classical emission line ratio diagnostic diagrams \citep[the so-called BPT
diagrams,][]{Baldwin1981} involving [S \textsc{ii}]/H$\alpha$ and [N \textsc{ii}]/H$\alpha$ line ratios \citep[see e.g.][]{Kewley2001,
Kauffmann2003, Kewley2006, Belfiore2016},  equivalent width of H$\alpha$ (EW$_{\rm{H\alpha}}$) \citep[see e.g.][]{CidFernandes2011,
Belfiore2016}, a combination of EW$_{\rm{H\alpha}}$ and  [N \textsc{ii}]/H$\alpha$ line ratio \citep{CidFernandes2011} or surface
brightness of H$\alpha$ emission \citep[see e.g.][]{Oey2007, Zhang2017, Sanders2017}. The fraction of DIG increases with a decrease in
the H$\alpha$ surface brightness of galaxies \citep{Oey2007, Zhang2017}. As mentioned
above, the emission line ratios of DIG and H \textsc{ii} regions are
remarkably different. For example, [O \textsc{i}] $\lambda$6300/H$\alpha$,   [O \textsc{ii}] $\lambda$3727/H$\beta$,  [N \textsc{ii}]
$\lambda$6584/H$\alpha$,  [S \textsc{ii}] $\lambda\lambda$6717,6731/H$\alpha$ are known to increase in DIG regions compared to the H
\textsc{ii} regions \citep[see, e.g.][]{Haffner1999, Voges2006, Madsen2006, Blanc2009, Zhang2017}. As a consequence,
in terms of classification
on the BPT diagrams, the emission line ratios in DIG are
nearly always consistent with low-ionisation nuclear emission regions (LINER) or
low-ionisation emission regions \citep[LIER, when the low-ionisation is not concentrated in the nucleus of galaxies ,][]{Belfiore2016},
but also spill into the Seyfert-like region of the BPT diagrams. Given the nearly univocal DIG-LIER association many papers
simply identify DIG through the LIER BPT classification.
For this reason, in the following we will often refer interchangeably to DIG and LIER, except in a few sections where
we will identify the DIG emission through the EW$_{\rm_{H\alpha}}$ or H$\alpha$ surface brightness.

DIG and LIER-like line emission in red sequence galaxies is associated with regions where no star-formation is taking place despite the
available reservoir of gas, and are devoid of young stellar population \citep{Belfiore2016}. However, LIER may also appear in the
central region of blue sequence or green valley galaxies where star-formation is mainly taking place in their spiral discs,
and is also seen in the inter-arm regions \citep{Singh2013, Gomes2016, Belfiore2017}.

\indent The ionisation source of DIG/LIER is not very well understood.
The radiation from O and B stars which is filtered
and hardened by photoelectric absorption may result in DIG \citep{Belfiore2016}. However, in early-type, elliptical and lenticular
galaxies, and also in the old component of stellar discs (e.g. in the interarm regions or the innermost
parts of some discs), the hot evolved post-asymptotic giant branch (pAGB) stars and white dwarfs may emit hard ionising radiation
required to produce the ionised gas  \citep{Binette1994, Stasinska2008, Singh2013}. Weak active galactic nuclei (AGNs), low-mass X-ray
binaries and extreme horizontal branch stars have also been suggested to be the plausible sources of radiation resulting in DIG, though
their contribution might be almost negligible \citep{Sarzi2010, Yan2012, Belfiore2017}.
We shall also mention that LIER emission may also be associated with shocked regions, but this applies only
to very active galaxies with strong winds and/or in strongly interacting systems.

\indent Despite the ubiquity of DIG in active and quiescent galaxies, no metallicity calibrator exists for
estimating the abundances in LIERs or DIG dominated regions. However, the impact of DIG contamination on metallicity measurements has
been recently studied at global and spatially-resolved scales. For example, the work of \citet{Sanders2017} shows that
contamination of global galaxy spectrum by DIG results in an increase in the strength of low-ionisation lines and a decrease in the
temperature of the low-ionisation zone, hence leading to an overestimate of metallicity. Their study also shows that
combining DIG contamination with flux weighting effects may bias metallicity estimates by more than 0.3 dex.
Spatially-resolved observations may mitigate the effects of flux weighing. \citet{Zhang2017} used integral field
spectroscopic (IFS) data from Mapping Nearby Galaxies at APO (MaNGA) survey to study the effect of DIG on metallicity measurements at
kiloparsec scales. However, the resolution of MaNGA or comparable IFS surveys is still much lower than the physical
scale of \HII~regions, therefore providing only a partial solution to the problem of DIG contamination.

\cite{Yan2018} has attempted to directly derive the metallicity in galaxies dominated by DIG/LIER emission by
stacking large number of galaxy spectra from the Sloan Digital Sky Survey (SDSS) and measuring the auroral lines in these systems, hence the
gas temperature and, therefore, the gas metallicity. However, they point out that the results are somewhat puzzling
as the inferred metallicity is significantly lower than inferred from other diagnostics, such as the [N \textsc{ii}]/[O \textsc{ii}] ratio, sensitive
to the metallicity through the secondary nitrogen enrichment. The issue seems similar to the ``temperature problem'' found in AGNs,
where the use of the auroral lines gives metallicities that are up to 2~dex lower than the extrapolation of the metallicity gradient
inferred from \HII ~regions \citep{Dors2015}. These results seem to suggest that the `direct' T$_e$ method has some intrinsic problems that prevent
its application to regions ionized by hard radiation.

%Regardless of the
%contamination issue, determining the metallicity of the DIG/LIER emission would enable to investigate the gas
%metallicity over a much wider range of galaxy types and galactic regions. For instance, while most quiescent, passive galaxies
%do not show traces of HII regions with active star formation, a large fraction of them do have nebular emission
%but with DIG/LIER-like properties; hence, extending the metallicity calibrations to the DIG/LIER-like regime would, for instance,
%enable us to investigate the M--Z relation and Fundamental Metallicity Relation to passive galaxies. The central
%region of most green-valley galaxies are, as mentioned, characterized by DIG/LIER-like emission, hence determining the metallicity
%in these regions would enable us to determine the radial metallicity gradients down to the central region also for this
%class of galaxies, hence enabling a proper comparison with Main Sequence, star forming galaxies.
	
	\indent In this paper, we aim to study the impact of DIG contamination on metallicity estimates, and
	also to devise metallicity calibrators applicable to DIG and to a wide variety of galaxies including quiescent galaxies, which will
	allow us to explore their properties in depth. We utilise publicly available IFS data of nearby spiral galaxies obtained from
	the Multi Unit Spectroscopic Explorer (MUSE) instrument on the Very Large Telescope to separate \HII~regions from the DIG-dominated
	regions at scales of $\sim$ 50--100 pc. These scales are comparable to the size of \HII~region, which may vary between a few and a few hundred
	parsecs \citep{Kennicutt1984}. We exploit the unprecedented spatial resolution of MUSE to select DIG and \HII~
	regions which are sufficiently close to share the same gas-phase metallicity. In particular, we define a set of closeby
	\HII-DIG regions pairs, which are expected to share the same chemical abundances, and test the application of typical strong line metallicity
	calibrators on the DIG regions. The true metallicity of these regions is assumed to be that of the nearby \HII~ region (and measured
	using the standard metallicity diagnostics). In this framework we can therefore study the biases introduced by DIG contamination on metallicities estimated with strong line indicators.

%\indent We utilise the publicly available IFS data of nearby spiral galaxies obtained from the Multi Unit Spectroscopic Explorer (MUSE) instrument on the Very Large Telescope to separate \HII~regions from the DIG-dominated regions at scales of $\sim$ 100 pc, comparable to the size of \HII~region. \textcolor{magenta}{Such spatial resolution allow us to identify the closeby \HII-DIG regions, which must have the same gas-phase abundance}. The goal of the paper is to verify whether the existing metallicity calibrators apply to DIG regions as well \textcolor{magenta}{or some corrective factors are needed}, with the ultimate aim to devise \textcolor{magenta}{global} strong line metallicity calibrations for DIG\textcolor{magenta}{, LIERs and Seyferts}. This is achieved by comparing the metallicity of chemically-homogeneous \HII-DIG pairs within the sample galaxies.

\indent The paper is organised as follows. Section \ref{section:data} describes the MUSE dataset and spectral fitting technique used in estimating fluxes in emission lines. Section \ref{section:method} presents the methodology to identify the chemically-homogeneous \HII-DIG pairs, and metallicity calibrations used in this work.  Section \ref{section:results} presents the main results of the analysis, and a discussion of the biases involved in using existing metallicity calibrations, new methods to mitigate such biases and a discussion on the applicability of the new calibrations. In Section \ref{section:summary}, we summarise our main results.
%Throughout this study, we use Z for 12 + log(O/H) and the following shorthand notation for the strong line ratios for a compact presentation in equations and figures:
%\setlength{\belowdisplayskip}{-1.5pt} \setlength{\belowdisplayshortskip}{-1.5pt}
%\setlength{\abovedisplayskip}{-1.5pt} \setlength{\abovedisplayshortskip}{-1.5pt}

\iffalse
\begin{equation}
\centering
\rm O3N2 = log(([O \textsc{iii}] \lambda5007/H\beta) / ([N \textsc{ii}] \lambda6584/H\alpha))
\end{equation} 
\begin{equation}
\centering
\rm O3S2 = log(([O \textsc{iii}] \lambda5007/H\beta) + ([S \textsc{ii}] \lambda6717,6731/H\alpha))
\end{equation}

\fi

\begin{figure*}
	\centering
	\includegraphics[width=0.30\textwidth]{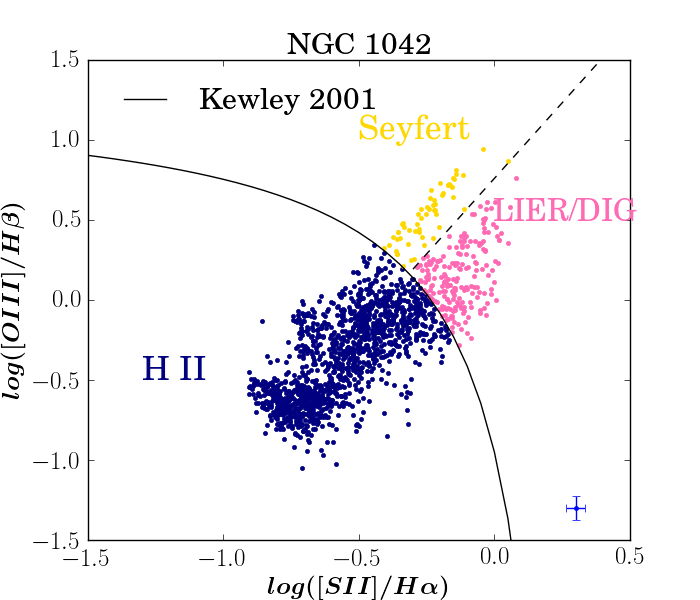}
	\includegraphics[width=0.30\textwidth]{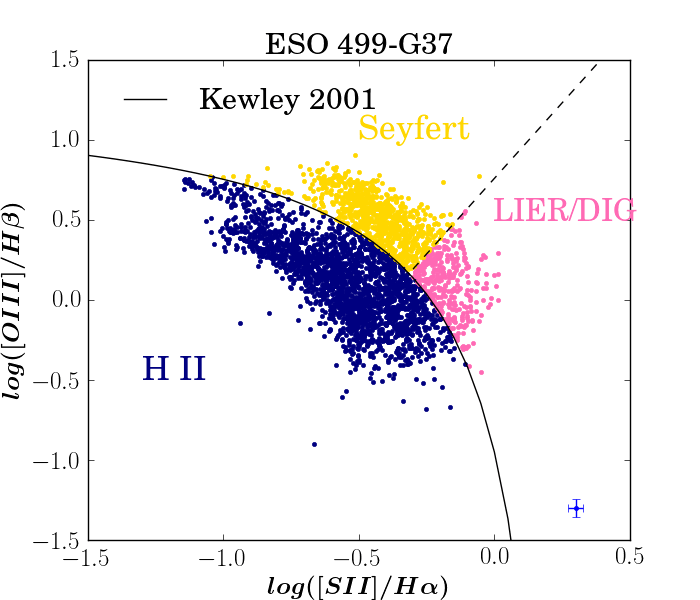}
	\includegraphics[width=0.30\textwidth]{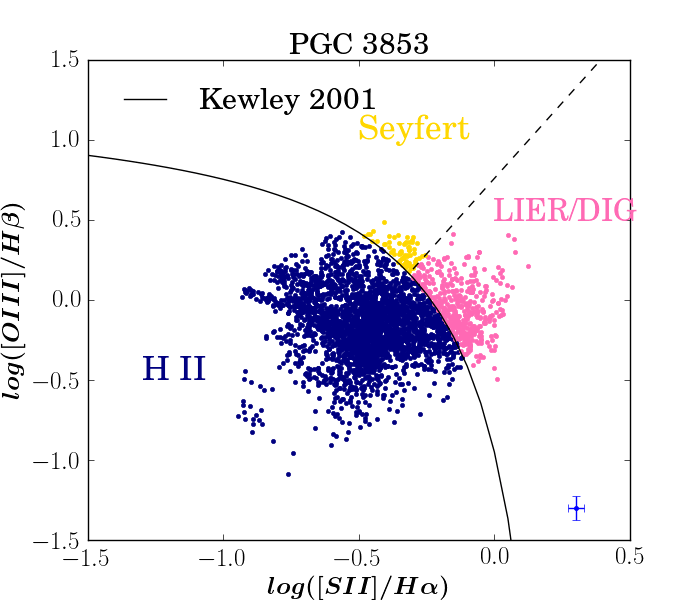}
	\includegraphics[width=0.30\textwidth, trim={2.2cm 1.2cm 2.2cm 0}, clip]{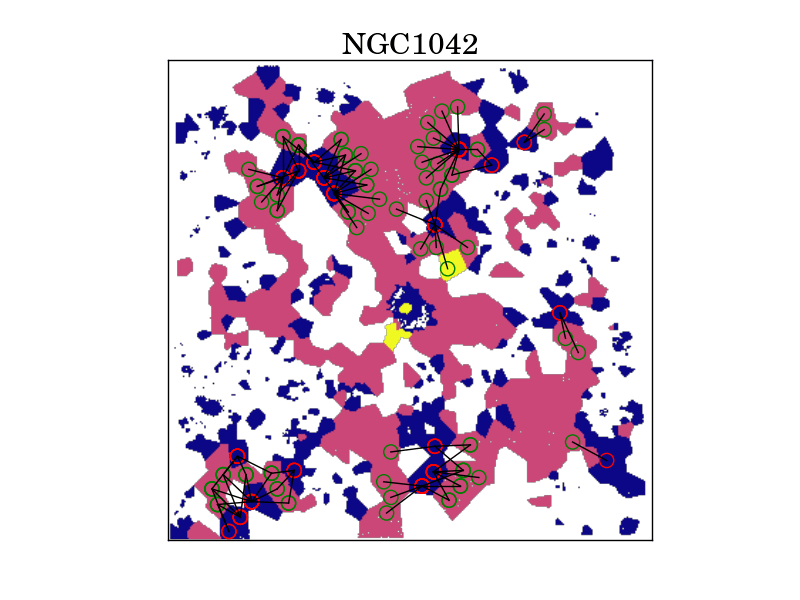}
	\includegraphics[width=0.30\textwidth, trim={2.2cm 1.2cm 2.2cm 0}, clip]{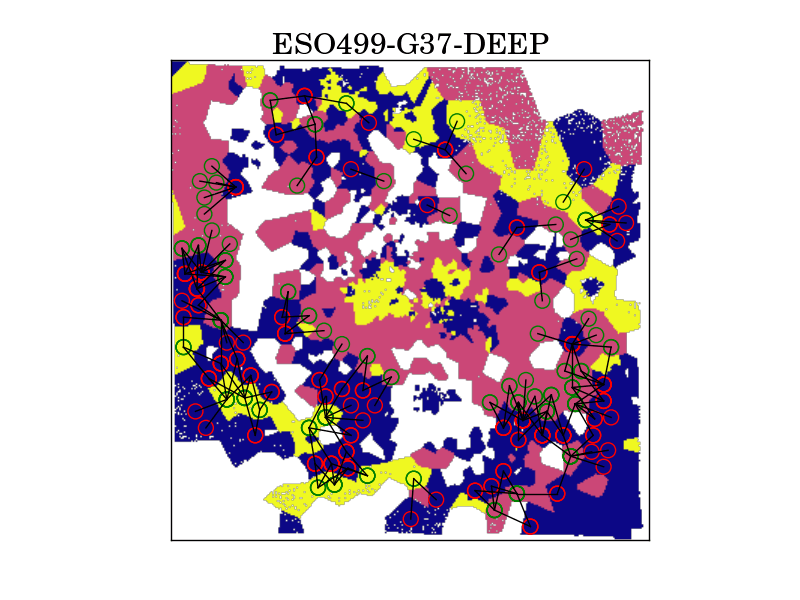}
	\includegraphics[width=0.30\textwidth, trim={2.2cm 1.2cm 2.2cm 0}, clip]{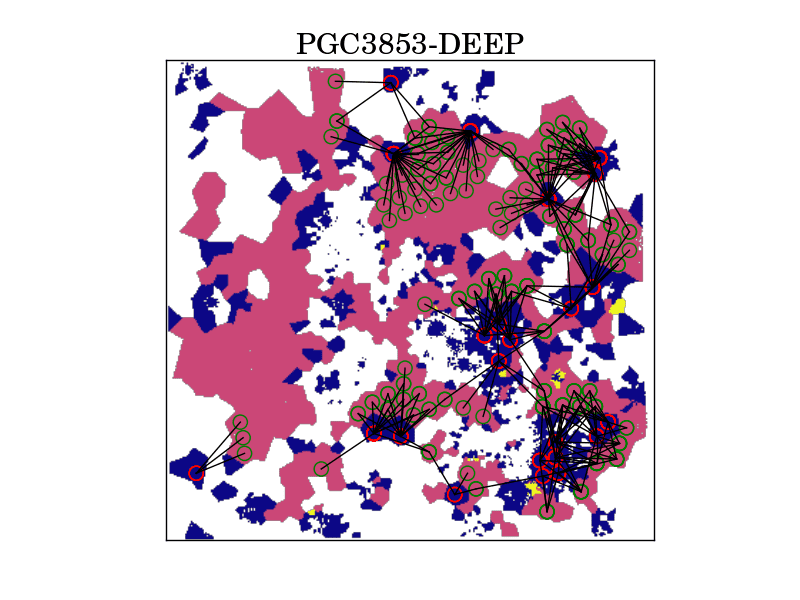}
	\caption{Emission line ratio diagnostic [S \textsc{ii}]-BPT diagrams (upper panel) and spatially-resolved [S \textsc{ii}]-BPT maps (lower panel) of NGC 1042 (left panel), ESO
	499-G37 (middle panel) and PGC 3853 (right panel). In all panels, blue, pink and yellow data points denote the data points with emission line ratios
	corresponding to \HII, LIER/DIG and Seyfert, respectively. In the upper panel, each point corresponds to line-ratios in a Voronoi bin, the black solid curve
	represents the theoretical maximum starburst line from \citet{Kewley2001}, providing a demarcation between the bins with H \textsc{ii} regions and
	DIG/LIER-dominated regions. In the DIG/LIER-dominated regime on the BPT diagnostic diagram, the dashed line shows the demarcation between the DIG/LIER and Seyfert regions. The median error bars (in blue) on the emission line ratios are shown in the bottom right-hand corner. In the lower panel, green circles denote the DIG/LIER/Seyfert circles connected by black lines to all red \HII~circles within a distance of $\approx$ 500 pc. White spaxels denote the bins where emission lines have S/N$<$5.}
	\label{fig:BPT}
\end{figure*}

\begin{figure*}
	\centering
	\includegraphics[width=0.30\textwidth]{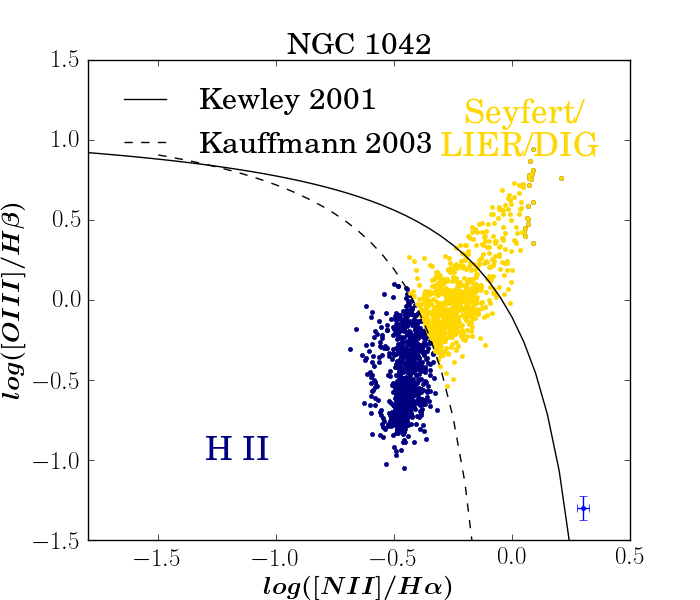}
	\includegraphics[width=0.30\textwidth]{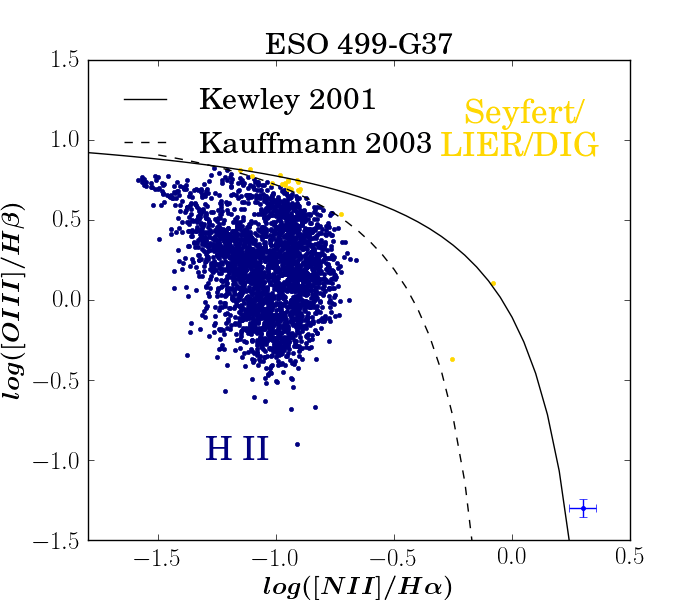}
	\includegraphics[width=0.30\textwidth]{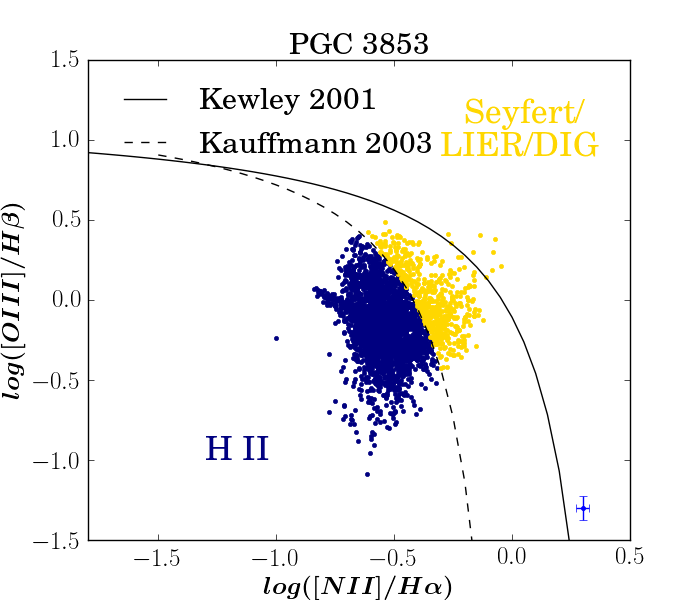}
	\includegraphics[width=0.30\textwidth, trim={2.2cm 1.2cm 2.2cm 0}, clip]{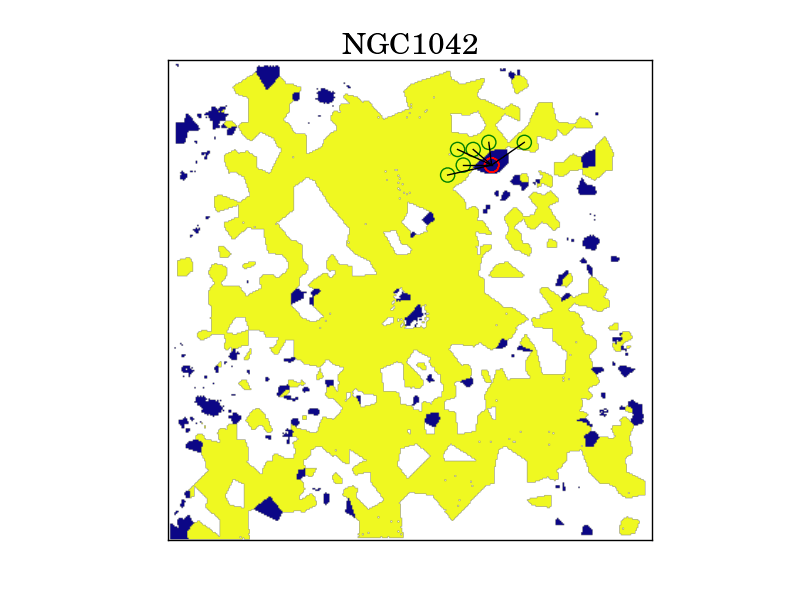}
	\includegraphics[width=0.30\textwidth, trim={2.2cm 1.2cm 2.2cm 0}, clip]{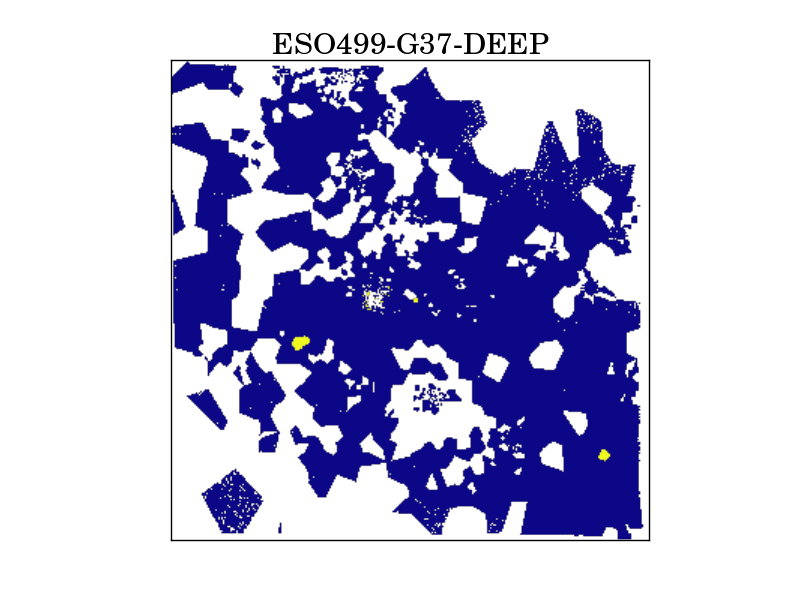}
	\includegraphics[width=0.30\textwidth, trim={2.2cm 1.2cm 2.2cm 0}, clip]{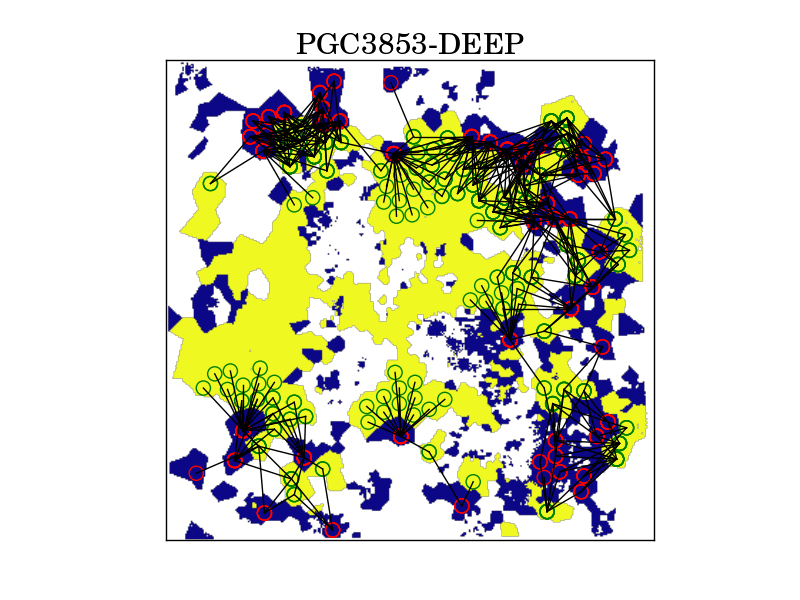}
	\caption{Emission line ratio diagnostic [N \textsc{ii}]-BPT diagrams (upper panel) and spatially-resolved [N \textsc{ii}]-BPT maps (lower panel) of NGC 1042 (left panel), ESO
	499-G37 (middle panel) and PGC 3853 (right panel). In all panels, blue and yellow data points denote the data points with emission line ratios corresponding to
	\HII and DIG/LIER/Seyfert, respectively. In the upper panel, each point corresponds to line-ratios in a Voronoi bin, the black solid curve represents the
	theoretical maximum starburst line from \citet{Kewley2001}, while dashed black curve represents the empirical Kauffmann line \citep{Kauffmann2003} used here to distinguish between the bins with H \textsc{ii} regions and DIG-dominated regions. The median error bars (in blue) on the emission line ratios are shown in the bottom right-hand corner. In the lower panel, green circles denote the DIG/LIER/Seyfert circles connected by black lines to all \HII~circles within a distance of $\approx$ 500 pc. A comparison of resolved BPT maps of ESO 499-G37 (bottom middle panels in Figures \ref{fig:BPT} and \ref{fig:BPT-NII}) points out that [S \textsc{ii}]-BPT classifies this galaxy as dominated by DIG/LIER while [N \textsc{ii}]-BPT classifies it as dominated by red \HII~regions. White spaxels denote the bins where emission lines have S/N$<$5.}
	\label{fig:BPT-NII}
\end{figure*}

\section{The Data}
\label{section:data}
\subsection{Galaxy sample and data}
\label{section:sample}

We have made use of archival MUSE data for 24 nearby spiral galaxies observed as part of the MUSE Atlas of Discs guaranteed time
observations program (program IDs 095.B-0532(A), 096.B-0309(A), 097.B-0165(A)). The observations of these galaxies were taken between
June -- August 2014, and October 2014 -- September 2015 (i.e. ESO P95, P96 and P97) in the wide field mode of the MUSE instrument. These
observations were performed in variable seeing conditions resulting in a seeing FWHM = 0.5 \arcsec-- 1.2\arcsec (corrected by airmass).  

Fully reduced datacubes were downloaded from the Phase 3 ESO archive. These have been reduced using the MUSE pipeline version muse-1.4 and higher \citep{Freudling2013}, which includes removal of instrumental artefacts, astrometric calibration, sky-subtraction, wavelength calibration and flux calibration. Each datacube covers a field of view of $\sim$1\arcmin $\times$ 1\arcmin, sampled by 0.2 arcsec per spatial pixel, leading to a total of $\sim$ 100,000 individual spectra per galaxy. The wavelength range covered for each datacube is $\sim$ 4800--9300 \AA ~with a spectral resolution of 1750 at 4650 \AA~ to 3750 at 9300 \AA.

The observed galaxies are at an average distance of $\sim$24 Mpc, corresponding to a spatial sampling of 23 pc per spatial pixel
($\sim$0.2\arcsec) and a typical spatial resolution of $\sim$50--100~pc. The galaxies in the sample are mostly massive Sa and Sb galaxies, with three Sc and one Sd galaxy.

\subsection{Spectral fitting}
\label{section:fluxes}

We measure emission line fluxes after subtracting the stellar continuum  using a customised spectral fitting routine described in \cite{Belfiore2016}. An initial Voronoi binning \citep{CappellariCopin2003} is first performed on the datacubes to optimise the S/N in the continuum. A set of simple stellar population templates from the MIUSCAT library \citep{Vazdekis2012, Ricciardelli2012} is then used to fit the stellar continuum within each Voronoi bin using penalised pixel fitting \citep{CappellariEmsellem2004, Cappellari2017}. The best-fit continuum within each bin is scaled to the continuum level (6000-6200 \AA) in the individual spaxels making up the bin and is subtracted from the observed spectrum in each spaxel to obtain a continuum-subtracted datacube containing only the emission lines. A second Voronoi binning is then performed to optimise the S/N on the H$\alpha$ emission line, hence the size of Voronoi bins varies (between a single pixel to several pixels) depending on the S/N of H$\alpha$ flux in
	a given pixel. The emission line flux maps are created by fitting a Gaussian profile to each emission line of interest, [O \textsc{iii}] $\lambda$5007, H$\beta$, H$\alpha$, [N \textsc{ii}] $\lambda$6584, [S \textsc{ii}] $\lambda\lambda$6717, 6731. A S/N cut of 5 is performed on all emission line flux maps used before carrying out further analysis. Note that the Voronoi-binned emission line flux maps are used as such a binning scheme allows us to preserve the initial spatial resolution of high S/N regions and also allows us to analyse the regions which had low S/N at the original sampling.

\section{Method}
\label{section:method}
\subsection{Selecting HII--DIG/LIER pairs}
\label{section:pairs}

\indent  The aim of the work is to compare the nominal gas-phase metallicity of closeby \HII~regions and LIER/DIG regions inferred from the classical strong-line metallicity diagnostics, in order to verify whether these diagnostics for determining metallicity in \HII~regions hold also for the DIG/LIER or, on the contrary require some correction factor, which can therefore be calculated. For this purpose, we define a set of closeby \HII-DIG/LIER pairs, which therefore are expected to be characterised by the same level of chemical enrichment in the gas-phase of the interstellar medium. % We  \textcolor{magenta}{choose closeby \HII-DIG pairs which therefore must be chemically homogeneous,} so that we can compare their \textcolor{magenta}{nominal} metallicities inferred from the metallicity diagnostics for H \textsc{ii} region, and verify whether these diagnostics hold also for the DIG. 
 Indeed, the ISM comprising of \HII~regions and DIG/LIER is chemically homogeneous at sufficiently small spatial scales
 because the dynamical time scale internal to these small regions is shorter than the supernovae burst time scale responsible for
 ejecting metals into the ISM, or the cooling time scales.  A few observational studies have explored
 azimuthal variations of the metallicity in galactic discs both by using the direct $T_e$ method \citep{Li2013, Berg2015} and the strong
 line method \citep{Zinchenko2016, Ho2017, Ho2018}. Generally the  azimuthal metallicity variations are found to be very small.
 The strongest detected azimuthal variations are about 0.1~dex, but these are inferred through the strong line method and
 residual variation due to the ionization parameter may be present
(despite efforts to correct for this effect).
% With the exception of a few individual H \textsc{ii} regions, these studies did not find evidence of significant azimuthal variation of the metallicity in galactic discs.}
We determined the spatial scale at which chemical homogeneity start to become important, by using the catalogue of \HII~regions in the
nearby spiral galaxy NGC 0628 from \citet{Berg2013}, which provides coordinates and metallicities of \HII~regions robustly
derived from the direct T$_e$ method. We estimated the metallicity difference and distances between all possible pairs of \HII~regions
in the catalogue and found that the metallicity difference between two \HII~regions is less than 0.1 dex when those regions were closer
than $\approx$ 500 parsecs apart. We use the distance of $\approx$ 500 parsecs or less 
for selecting chemically-homogeneous \HII-DIG/LIER pairs as described later in this section.

\indent Selection of \HII-DIG/LIER pairs requires an appropriate criterion to separate H \textsc{ii} regions from the DIG/LIER. We experimented with four different separation criteria, which include the use of two BPT diagrams ([S \textsc{ii}]-BPT and [N \textsc{ii}]-BPT), surface brightness of H$\alpha$ ($\Sigma_{\rm{H\alpha}}$) and equivalent width of H$\alpha$ (EW$_{\rm{H\alpha}}$). Among these methods, using [S \textsc{ii}]-BPT diagnostic diagram for separating \HII~ region from LIER/DIG/AGN ensures cleaner separation compared to that from H$\alpha$ surface brightness, and is less affected by nitrogen abundance variation than using the [N~\textsc{ii}]-BPT diagnostic diagram for separating \HII~ and DIG/LIER/AGN. Moreover, subsequent analysis shows that the separation based on [S \textsc{ii}]-BPT diagram proves superior to all other criteria. Hence, in the following, we describe the method involving  [S \textsc{ii}]-BPT in detail while other methods only briefly. 

\subsubsection{[S \textsc{ii}]-BPT}
\label{section:sii}
\indent  We use the classical BPT diagram of [S \textsc{ii}]/H$\alpha$ versus [O~\textsc{iii}]/H$\beta$ to classify spaxels within
galaxies as star-forming, LIER/DIG-dominated or exhibiting line-ratios of AGN, and create spatially-resolved [S
\textsc{ii}]-BPT maps of all galaxies. Figure \ref{fig:BPT} (upper panel) shows the spatially-resolved [S \textsc{ii}]-BPT diagram of
three example galaxies (NGC 1042, ESO499-G37 and PGC 3853), where blue, pink and yellow coloured data points represent the spaxels with
line-ratios corresponding to H \textsc{ii}, LIER/DIG and Seyfert regions, respectively. The solid black curve is
the theoretical maximum starburst line from \citet{Kewley2001} which is used to separate \HII~regions from
LIER/DIG/AGN when calibrating metallicity diagnostics, while the dashed black line from
\citet{Kewley2006} separates DIG from AGN. Figure \ref{fig:BPT} (lower panel) shows the corresponding spatially-resolved [S
\textsc{ii}]-BPT map of the same three sample galaxies (NGC 1042, ESO499-G37 and PGC 3853), where spaxels are colour-coded following the
same scheme as in the BPT diagram in the upper panel. We use these maps as reference for pair selection. We have
also included Seyfert-like regions in pair selection as these regions are seen well outside the nuclear regions and therefore
are unlikely associated with an AGN, but simply LIER-DIG emission whose diagnostics are formally spilling into
the Seyfert region. Moreover, we will argue that potentially the revised calibration obtained by us
can potentially be extended to the Seyfert region in Section \ref{section:applications}.

\indent The procedure for selecting \HII-DIG pairs and the subsequent method to create the final dataset for analysis is described as
follows. On the resolved [S \textsc{ii}]-BPT map of each galaxy, star-forming, Seyfert and LIER/DIG regions were
populated with non-overlapping circles of 1 arcsec radius. This radius was chosen on the basis of being larger than the seeing so that chosen circles within each map are independent of each other. Next, all star-forming (SF) circles around a LIER/DIG/Seyfert circle within a distance of $\sim$ 500 pc were identified, such that the distance between the centres of SF circle and LIER/DIG/Seyfert circle $<$ 500 pc. Figure \ref{fig:BPT} (lower panel) shows examples of pair-selection in three sample galaxies (NGC 1042, ESO499-G37 and PGC 3853) where each green circle denotes LIER/DIG/Seyfert circle connected by black lines to red circles representing all \HII~regions within a distance of $\sim$ 500 pc. The number of \HII-DIG pairs is, in fact, the number of LIER/DIG/Seyfert circles connected to surrounding star-forming circles. The representative value of any quantity (metallicity or emission line ratios) for any circle is the average of the unique values within that circle. Within a \HII-DIG pair, the average of representative values of all connected star-forming circles is the representative value of \HII~region. The whole procedure was performed automatically for each galaxy. No pairs could be identified on four galaxies in the sample (NGC 4030, NGC 4603, NGC4980 and NGC 5334), mainly because of the irregular Voronoi bins which could not accommodate circles. 

\subsubsection{[N \textsc{ii}]-BPT}
\label{section:nii}
\indent As in the case of [S \textsc{ii}]-BPT, we use the classical BPT diagram of [N~\textsc{ii}]/H$\alpha$ versus [O \textsc{iii}]/H$\beta$ to distinguish between star-forming spaxels and those spaxels exhibiting DIG/LIER/Seyfert like line ratios. However, we use the empirical Kauffmann line \citep{Kauffmann2003} rather than Kewley line for such a distinction as the metallicity diagnostics (O3N2 and O3S2, see Section \ref{section:maps}) which we test here, were initially calibrated for galaxies lying on the left-hand sequence defined by Kauffmann line \citep{Curti2017}. Figure \ref{fig:BPT-NII} (upper panel) shows the spatially-resolved [N \textsc{ii}]-BPT diagram of three sample galaxies (NGC 1042, ESO499-G37 and PGC 3853), where blue and yellow coloured data points represent the spaxels with line-ratios corresponding to H \textsc{ii} and LIER/DIG/Seyfert regions, respectively. The solid black line corresponds to the theoretical maximum star-burst line from \citet{Kewley2001} and the dashed black line corresponds to the empirical Kauffmann line.  We create the spatially-resolved [N \textsc{ii}]-BPT maps for each galaxy using the corresponding BPT diagrams, which we then use to select pairs of \HII~ regions and DIG region following exactly the same procedure as for [S \textsc{ii}]-BPT separation and pair selection. Figure \ref{fig:BPT-NII} (lower panel) shows examples of resolved [N \textsc{ii}]-BPT for three galaxies where spaxels are color-coded following the same scheme as in the corresponding BPT diagram (upper panel), and \HII-DIG pairs are also marked as red and green circles, respectively connected by black lines.  

\subsubsection{Surface brightness of H$\alpha$}
\label{section:Sigma}
\indent We use a threshold log $\Sigma_{\rm{H\alpha}}$ of 39 erg s$^{-1}$ kpc$^{-2}$ adopted by \citet{Zhang2017} to distinguish between a \HII~region spaxel and a spaxel dominated by DIG. A spaxel with $\Sigma_{\rm{H\alpha}}$ above this threshold is designated as \HII ~region provided that it also lies on the left-hand sequence defined by the Kauffmann line on the [N \textsc{ii}]-BPT diagram. This additional criterion of [N \textsc{ii}]-BPT was imposed as the metallicity diagnostics to be tested here were calibrated on star-forming galaxies defined on the basis of Kauffmann line. A spaxel with $\Sigma_{\rm{H\alpha}}$ $<$ 39 erg s$^{-1}$ kpc$^{-2}$ is designated as DIG region. Hence, we obtain spatially-resolved \HII-DIG map which are then used to select \HII-DIG pairs as described in Section \ref{section:sii}. Note that we have not applied the foreground Galactic extinction
	correction to the surface brightness to be consistent with \citet{Zhang2017}, who defined the threshold without applying
	foreground extinction. Moreover, Galactic extinction is very small for
	our sample galaxies, with E(B-V) varying between 0.006 and 0.145
	\citep{Schlafly2011}.       

\subsubsection{Equivalent Width of H$\alpha$}
\label{section:EW}
\indent LIER emission in extended-LIER (eLIER) and central-LIER (cLIER) are found to be associated with low EW$_{\rm{H\alpha}}$ ($<$
3\AA) \citep{Belfiore2016}. Early-type galaxies are also known to exhibit such low EW$_{\rm{H\alpha}}$ associated with ionising
radiation reprocessed from hot evolved stars \citep{Binette1994, Stasinska2008, CidFernandes2011}. Hence, we use a threshold
EW$_{\rm{H\alpha}}$ of 3 \AA~ to create spatially-resolved \HII-DIG map where every spaxel with EW$_{\rm{H\alpha}}$ $<$ 3 \AA~ is
classified as LIER/DIG and \HII~ region otherwise. We then select the \HII-DIG pairs using these resolved maps and create dataset in the
same manner as described for  [S \textsc{ii}]-BPT selection (see Section \ref{section:sii}). However, we find that an \HII~region
selected on the basis of EW$_{\rm{H\alpha}}$ often does not have line ratios corresponding to those of \HII ~region sequence on BPT
diagrams. As a consequence, this selection criterion results in \HII-DIG pairs both of which have line ratios corresponding to
DIG/LIER/Seyfert, i.e. the region where the metallicity diagnostics are not calibrated, and therefore this
prevents a proper comparison of their nominal metallicities. Hence, we do not undertake the analysis of the DIG regions
based  on the EW analysis any further.

%\indent \textcolor{teal}{As in the case of [S \textsc{ii}]-BPT selection method, no pairs could be selected for certain galaxies in each selection method.}

\begin{figure*}
	\centering
	\includegraphics[width=0.28\textwidth, trim={0 1.2cm 0 5.5cm}, clip]{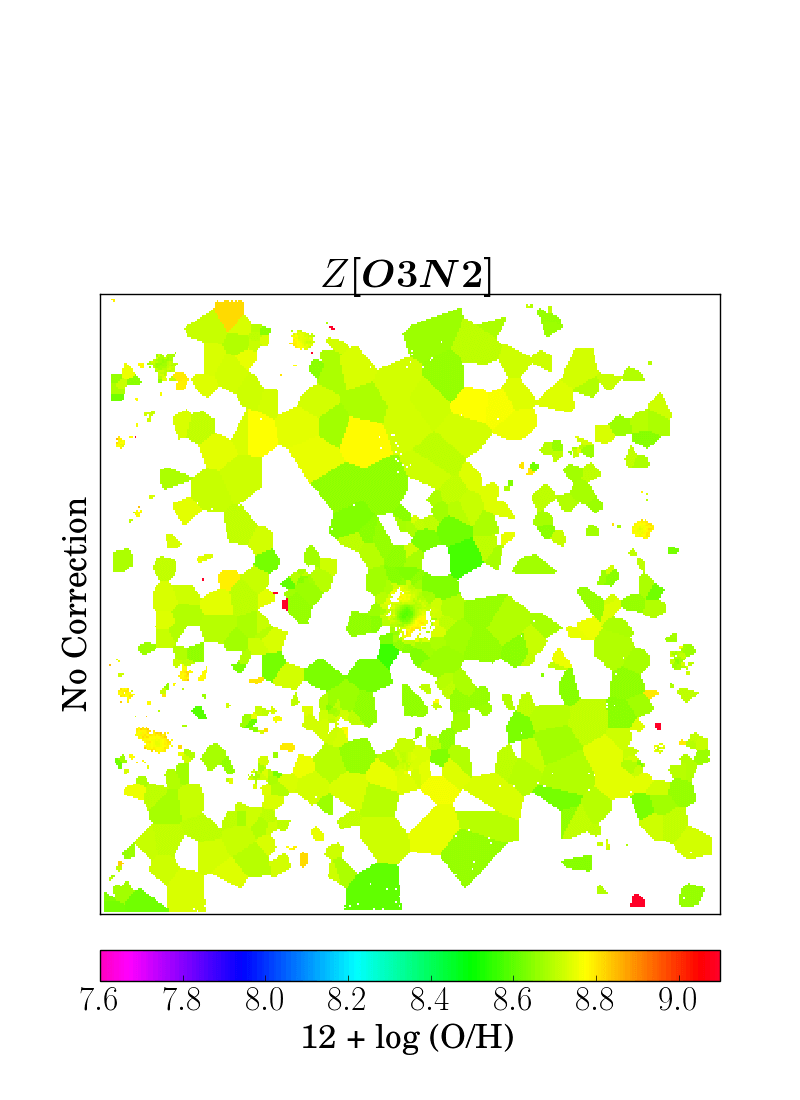}
	\includegraphics[width=0.28\textwidth, trim={0 1.2cm 0 5.5cm}, clip]{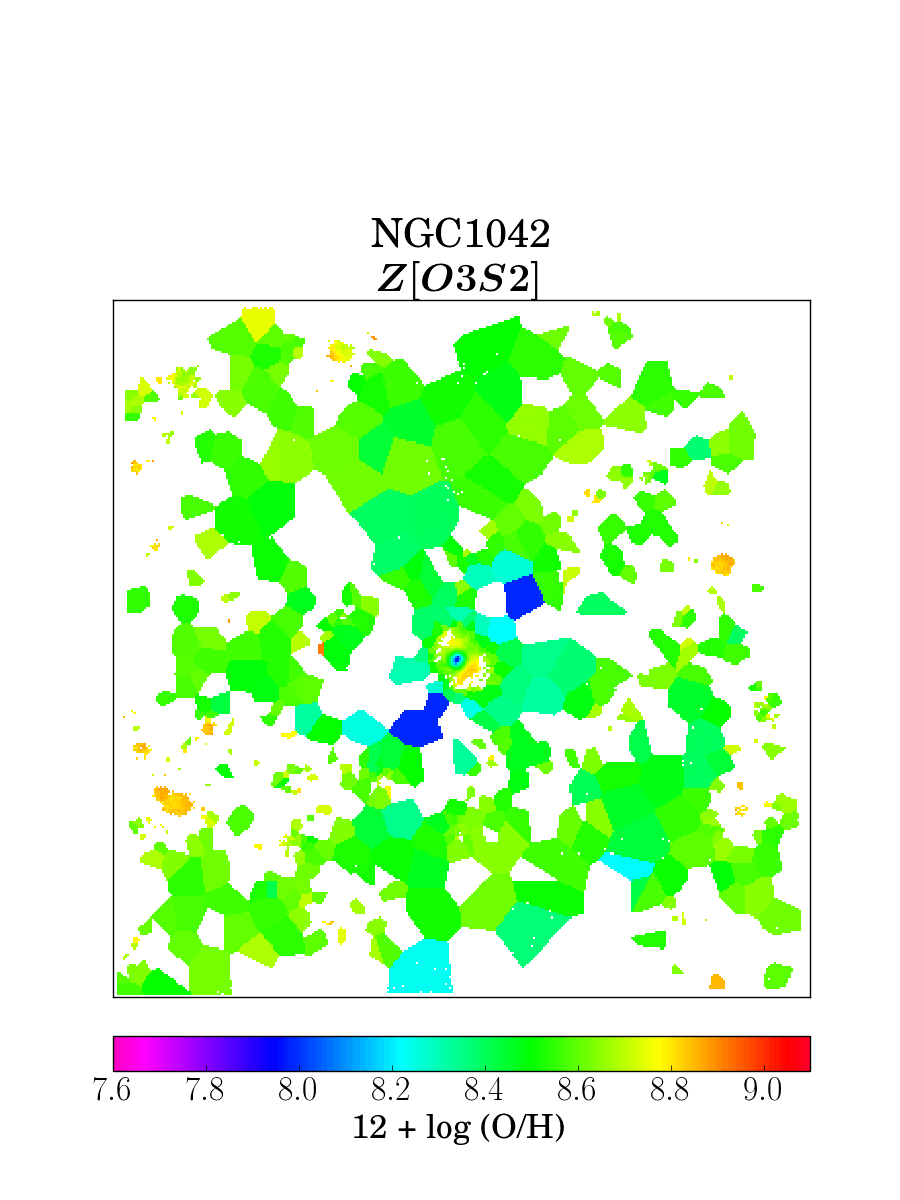}
	\includegraphics[width=0.28\textwidth, trim={0 1.2cm 0 5.5cm}, clip]{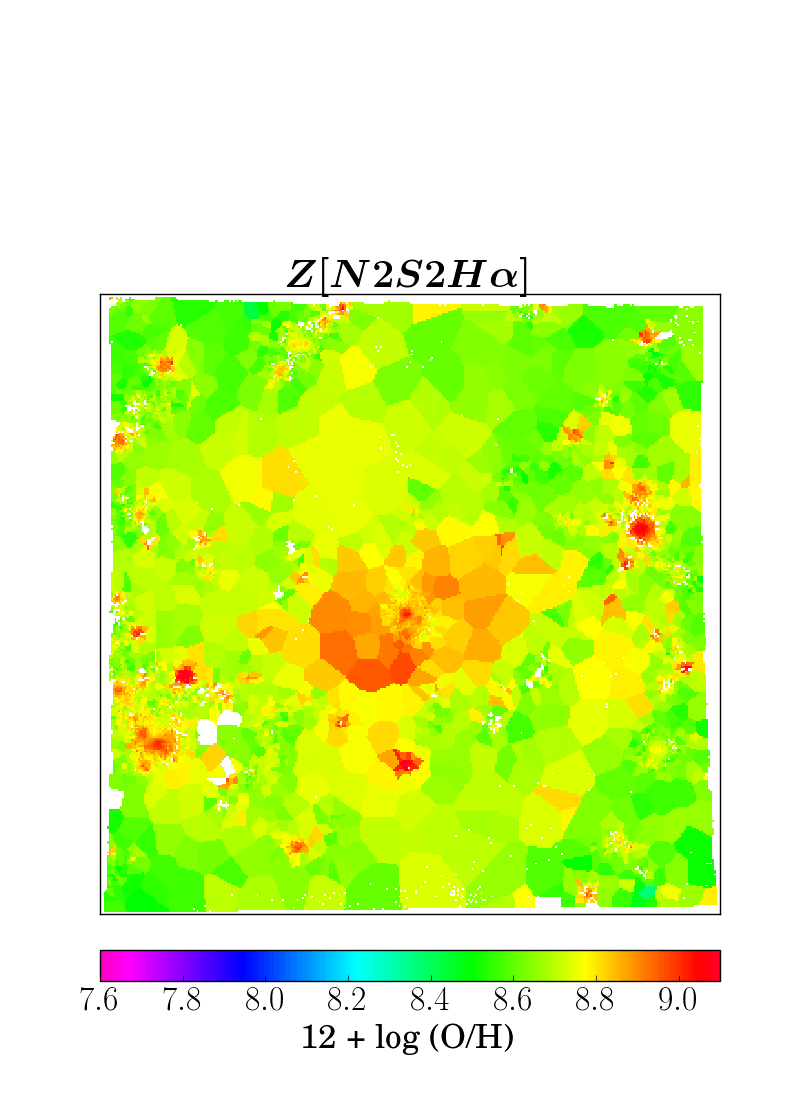}
	\includegraphics[width=0.28\textwidth, trim={0 1.2cm 0 5.5cm}, clip]{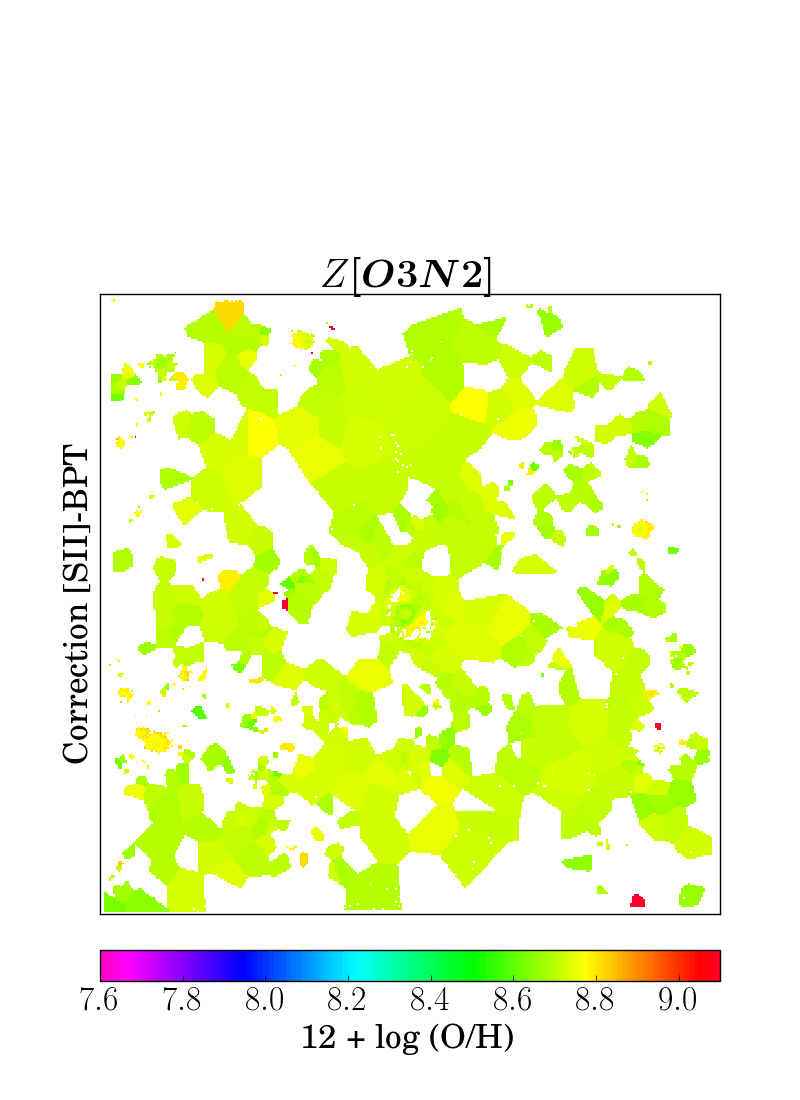}
	\includegraphics[width=0.28\textwidth, trim={0 1.2cm 0 5.5cm}, clip]{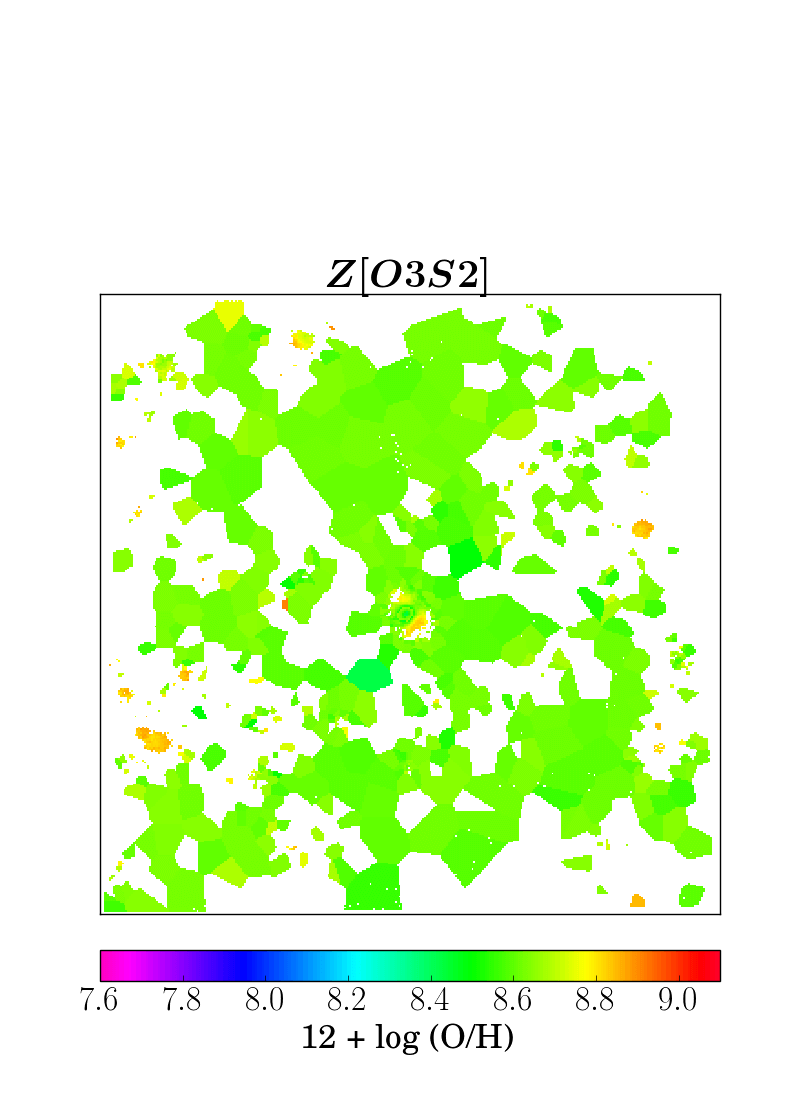}
	\includegraphics[width=0.28\textwidth, trim={2.8cm 0 2.8cm 0}, clip ]{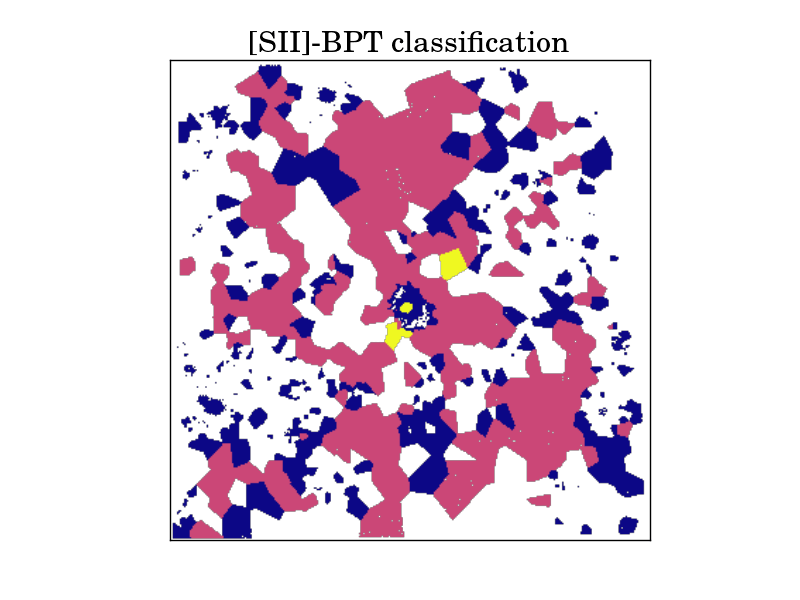}
	\includegraphics[width=0.28\textwidth, trim={0 1.2cm 0 5.5cm}, clip]{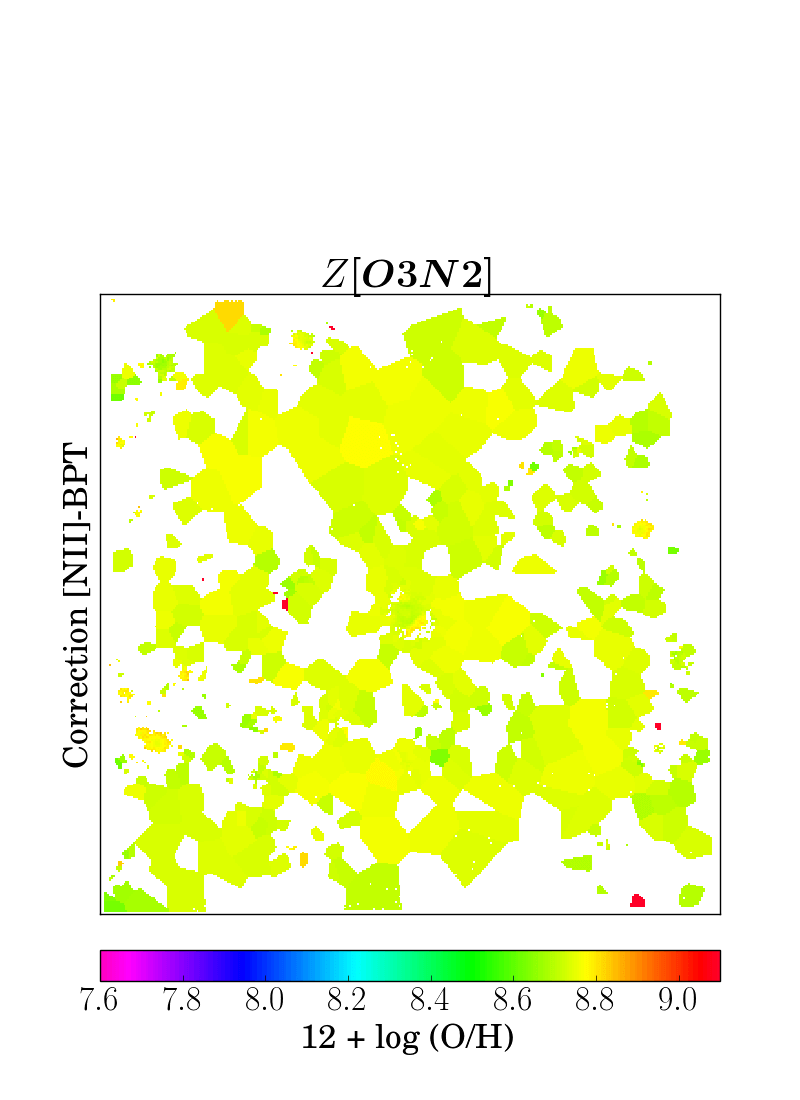}
	\includegraphics[width=0.28\textwidth, trim={0 1.2cm 0 5.5cm}, clip]{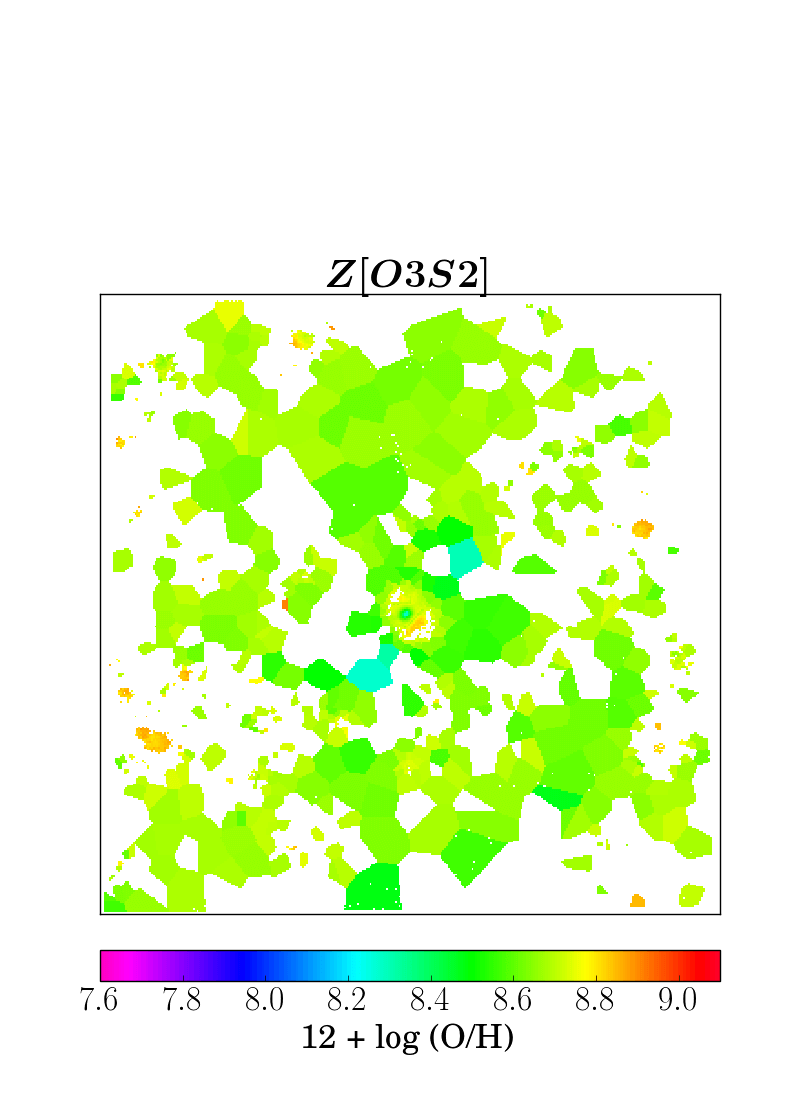}
	\includegraphics[width=0.28\textwidth, trim={2.8cm 0 2.8cm 0}, clip ]{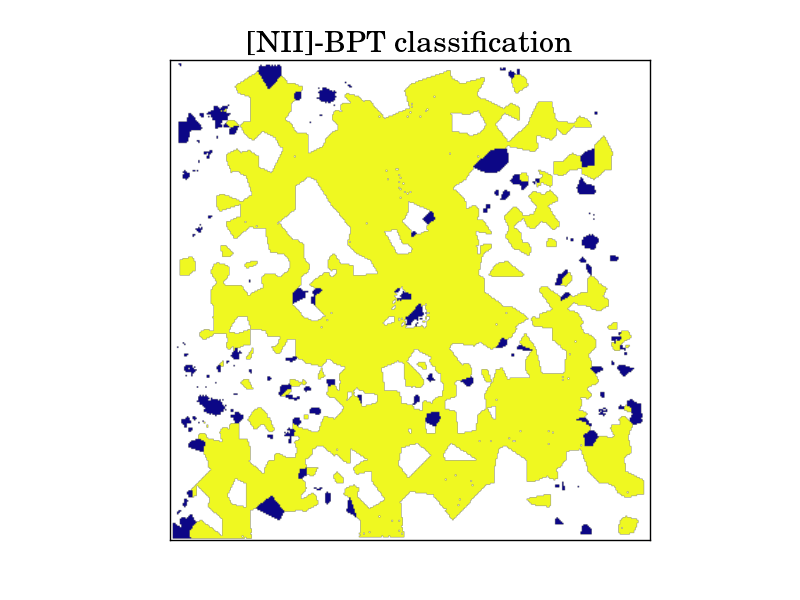}
	\caption{Upper row: Metallicity maps of  NGC 1042 using original metallicity calibrators, O3N2 (left panel), O3S2 (middle panel), N2S2H$\alpha$ (right panel).
	Middle-row, left and central panels: Metallicity maps using O3N2 (left panel) and O3S2 (middle panel), after applying the corrections obtained in Figure
	\ref{fig:correction sii}. We also present the spatially-resolved [S \textsc{ii}]-BPT map in the middle-right panel to show the spaxels where metallicity
	corrections are applied, i.e those spaxels with emission line ratios corresponding to DIG/LIER (pink spaxels) or Seyfert (yellow spaxels) rather than
	\HII~~regions (blue spaxels). Bottom row, left and central panels: Metallicity maps using O3N2 (left panel) and O3S2 (middle panel), after applying the corrections obtained in Figure \ref{fig:correction nii}. Bottom-right panel: Spatially-resolved [N \textsc{ii}]-BPT to show spaxels where metallicity corrections are applied, i.e. spaxels having line ratios of Seyfert (yellow spaxels) rather than \HII~ regions (blue spaxels).  White spaxels denote the bins with S/N of emission lines $<$ 5 or spaxels whose emission line ratios do not permit the use of a particular metallicity calibrator.}
	\label{fig:NGC1042}
\end{figure*}

\begin{figure*}
	\centering
	\includegraphics[width=0.45\textwidth]{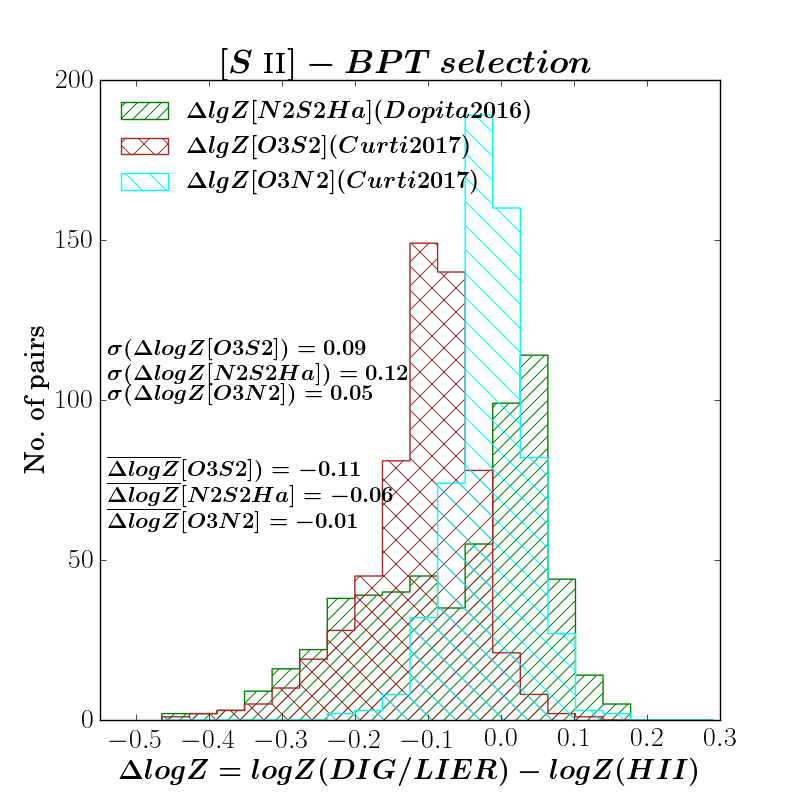}
	\includegraphics[width=0.45\textwidth]{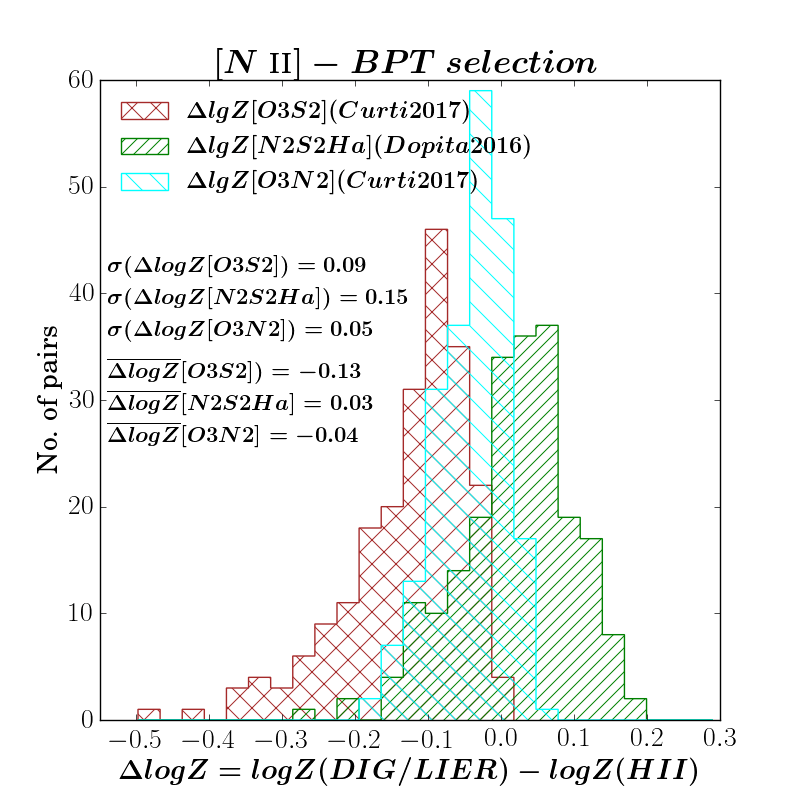}
	\includegraphics[width=0.45\textwidth]{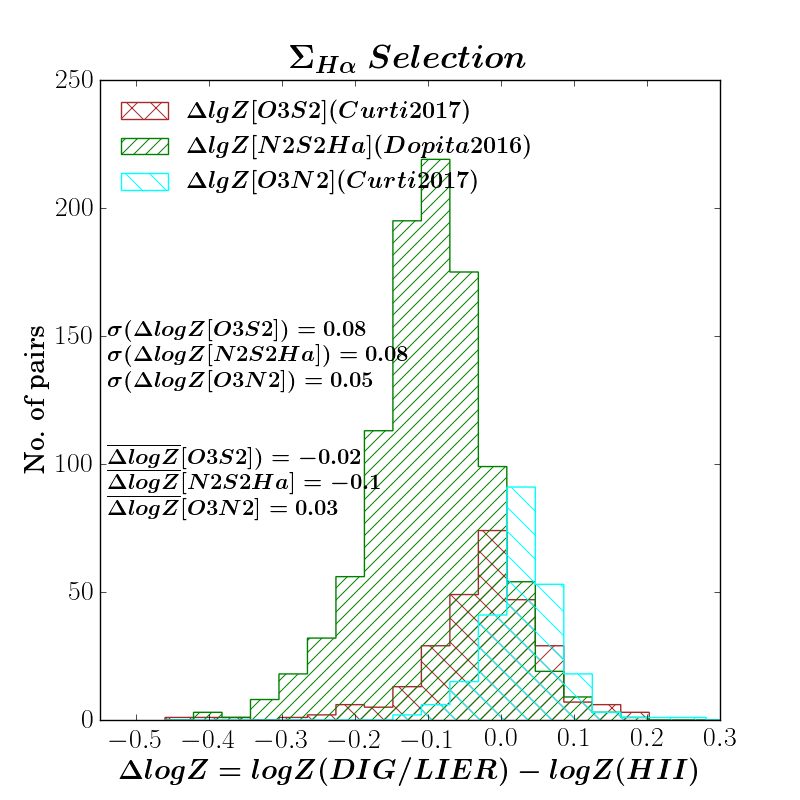}
	\caption{Distribution of differential metallicity ($\Delta$ log Z)  of the \HII--DIG/LIER pairs selected on the basis of the [S~\textsc{ii}]-BPT classification
	(upper-left panel), [N~\textsc{ii}]-BPT classification (upper-right panel) and surface brightness of H$\alpha$ (bottom panel). In each panel, the
	distribution of the differential metallicities obtained from N2S2H$\alpha$, O3S2 and O3N2 are shown in green, brown and cyan colours, respectively. In each
	panel $\overline{\Delta log(Z)}$ and $\sigma(\Delta log(Z))$ are the mean and standard deviation of the distributions, respectively.}
	\label{fig:differential}
\end{figure*}

\iffalse

\begin{table}
	\centering
	\caption{Metallicity calibration of the form R = $\mathlarger{\sum} c_n x^n$, where R is either O3N2 or O3S2, x is the oxygen abundance normalised to the solar value (12 + log(O/H)$_{\odot}$ = 8.69), and coefficients ``cn" are tabulated below. }
	\label{table:O3N2O3S2}
	\begin{tabular}{@{}lllll@{}}
		\toprule
		Diagnostic & c0     & c1     & c2     & c3    \\ \midrule
		O3N2       & 0.281  & -4.765 & -2.268 & --    \\
		O3S2       & -0.046 & -2.223 & -1.073 & 0.533 \\ \bottomrule
	\end{tabular}
\end{table}

\fi

\begin{table*}
	\centering
	\caption{Summary table for statistics on differential metallicities of \HII-DIG/LIER pairs from different selection criteria before and after applying corrections.}
	\label{tab:stats}
		\scalebox{0.9}{
	\begin{tabular}{ccclcclcclcclcc}
		\toprule
		Calibrator & \multicolumn{5}{c}{{[}SII{]}-BPT}                         &  & \multicolumn{5}{c}{{[}NII{]}-BPT}                                                      &  & \multicolumn{2}{c}{$\Sigma_{\rm{H\alpha}}$} \\
		\cline{2-6}\cline{8-12}\cline{14-15}\\
		& \multicolumn{2}{c}{Before} &  & \multicolumn{2}{c}{After} &  & \multicolumn{2}{c}{Before} &  & \multicolumn{2}{c}{After}                              &  & \multicolumn{2}{c}{Before}       \\ 
		%& Mean         & Std         &  & Mean        & Std         &  & Mean         & Std         &  & Mean                        & Std                      &  & Mean            & Std            \\
		\cline{2-3}\cline{5-6}\cline{8-9}\cline{11-12}\cline{14-15}\\
		& $\overline{\Delta log Z}$      & $\sigma(\Delta log Z)$    &  & $\overline{\Delta log Z}$      & $\sigma(\Delta log Z)$   &  & $\overline{\Delta log Z}$      & $\sigma(\Delta log Z)$   &  & $\overline{\Delta log Z}$      & $\sigma(\Delta log Z)$ &  & $\overline{\Delta log Z}$      & $\sigma(\Delta log Z)$  \\
		\midrule
		N2S2H$\alpha$ (linear)    & -0.06        & 0.12        &  & --          & --          &  & 0.03         & 0.15        &  & --                          & --                       &  & -0.1            & 0.08           \\
		N2S2H$\alpha$ (non-linear)     & -0.07        & 0.15        &  & --          & --          &  & 0.03         & 0.16        &  & --                          & --                       &  & -0.1            & 0.09           \\
		O3N2 \citep{Curti2017}     & -0.01        & 0.05        &  & -0.0002     & 0.04        &  & -0.04        & 0.05        &  & \multicolumn{1}{c}{-0.0004} & \multicolumn{1}{c}{0.03} &  & 0.03            & 0.05           \\
		O3N2  \citep{Marino2013}     & -0.02        & 0.05        &  & --     & --        &  & -0.04        & 0.05        &  & \multicolumn{1}{c}{--} & \multicolumn{1}{c}{--} &  & 0.03            & 0.05           \\
		O3S2       & -0.11        & 0.09        &  & 0.006       & 0.05        &  & -0.13        & 0.09        &  & \multicolumn{1}{c}{-0.003}  & \multicolumn{1}{c}{0.07} &  & -0.02           & 0.08  \\        
		\bottomrule  
	\end{tabular}%
%}
}
\\
Notes: $\overline{\Delta log Z}$  and $\sigma(\Delta log Z)$ indicate the mean and standard deviation of differential metallicities. 
\end{table*}

\begin{figure*}
	\includegraphics[width=0.30\textwidth, trim={0.2cm 2.0cm 0.2cm 2.0cm}, clip]{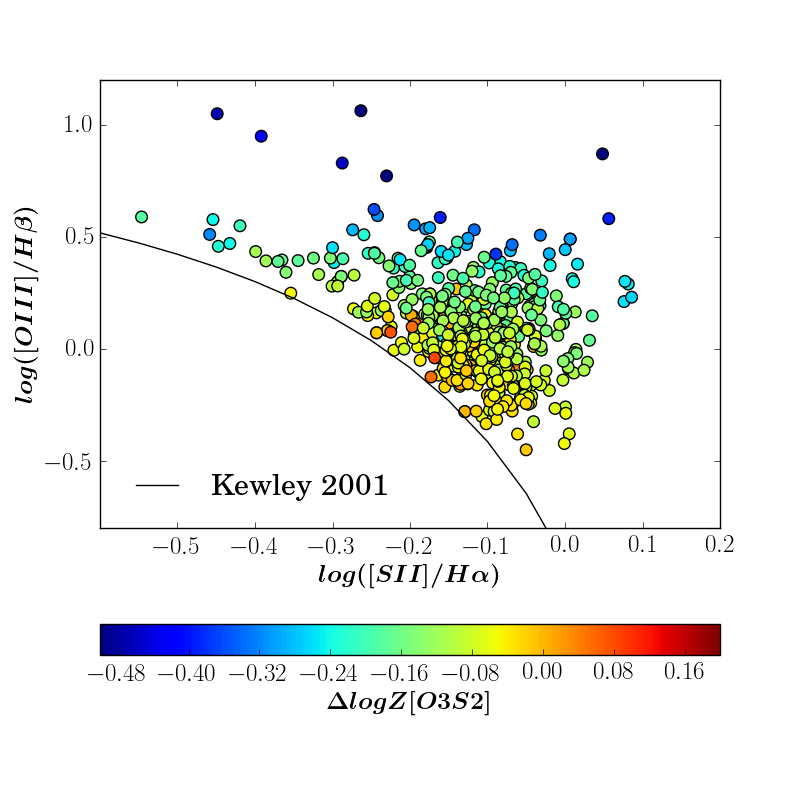}
	\includegraphics[width=0.30\textwidth,  trim={0.2cm 2.0cm 0.2cm 2.0cm}, clip]{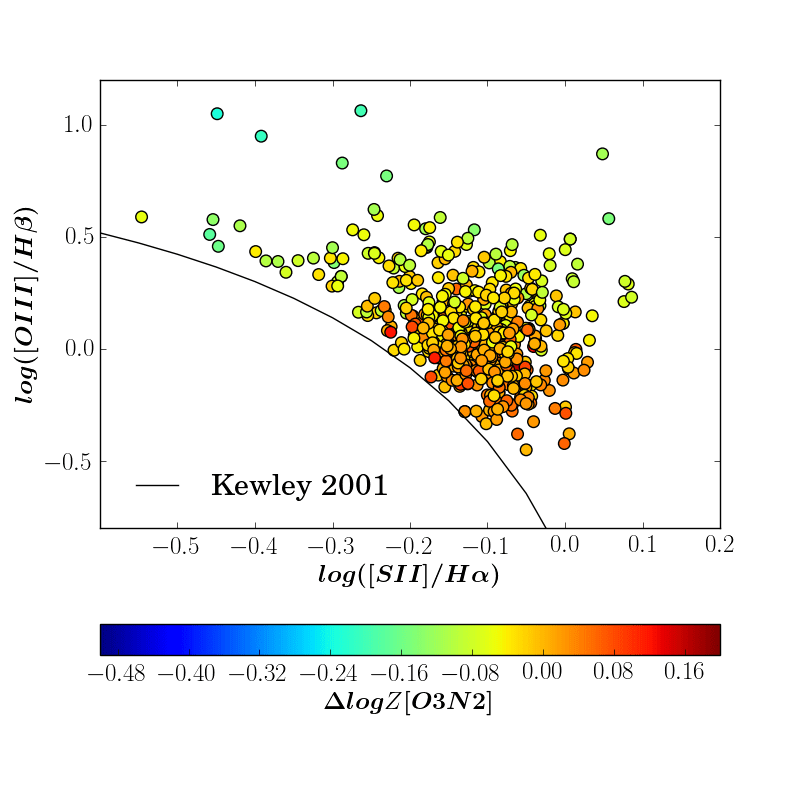}
	\includegraphics[width=0.30\textwidth,  trim={0.2cm 2.0cm 0.2cm 2.0cm}, clip]{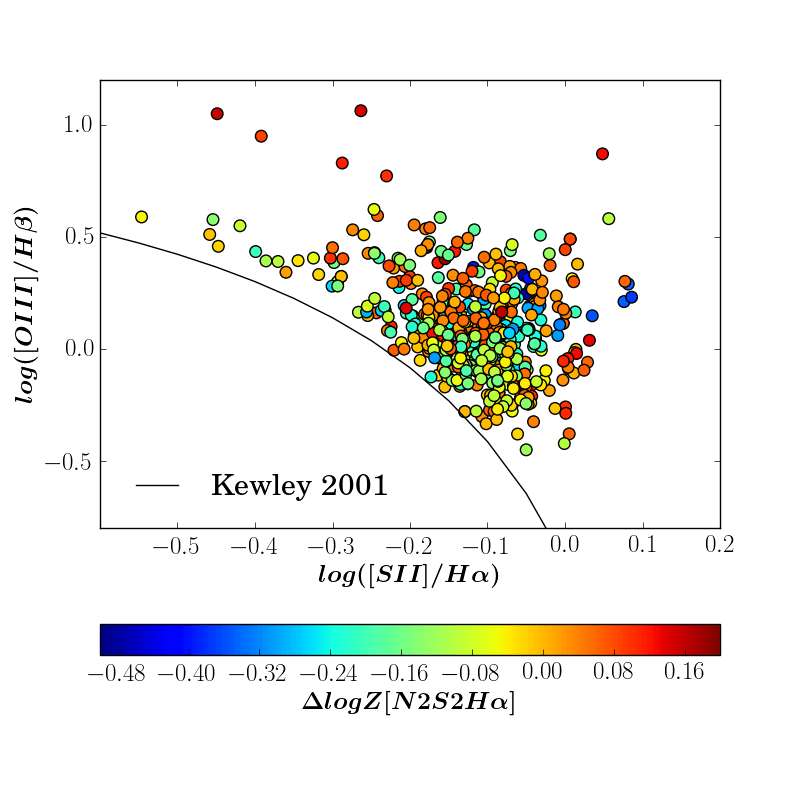}
	\includegraphics[width=0.30\textwidth,  trim={0.2cm 2.2cm 0.2cm 1.4cm}, clip]{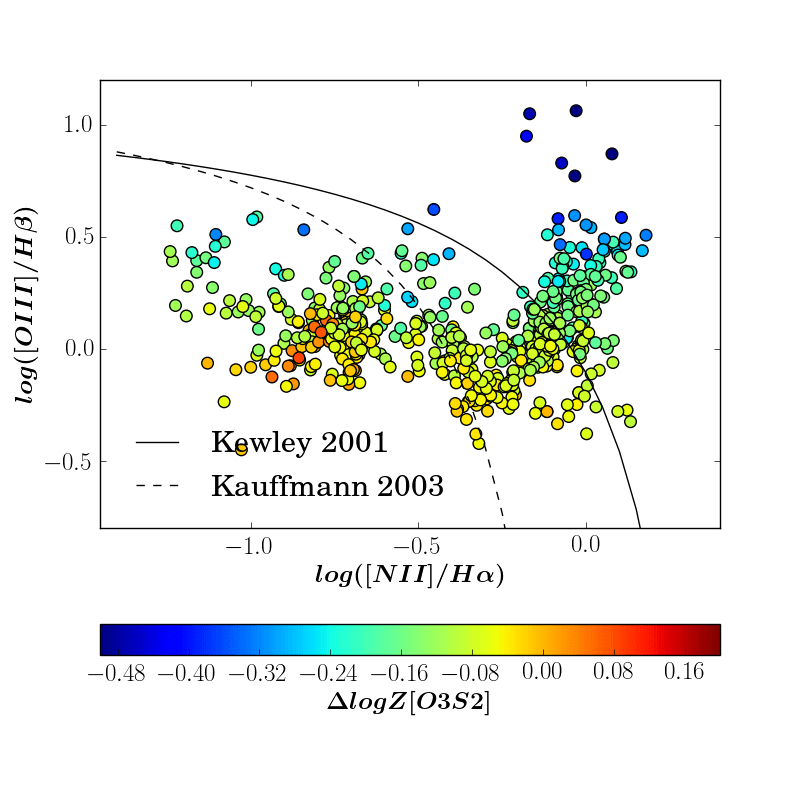}
	\includegraphics[width=0.30\textwidth,  trim={0.2cm 2.2cm 0.2cm 1.4cm}, clip]{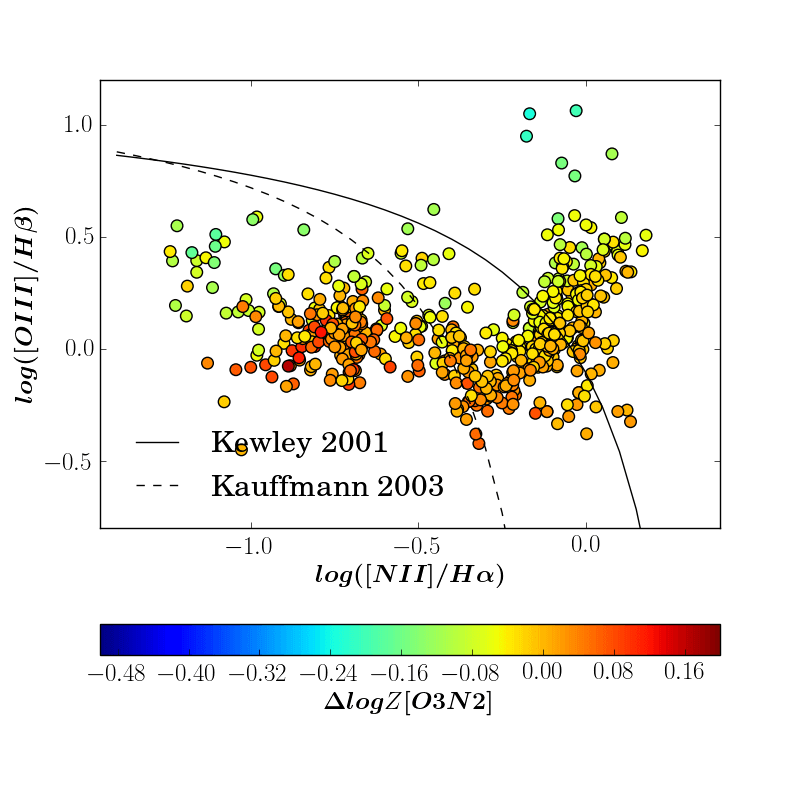}
	\includegraphics[width=0.30\textwidth,  trim={0.2cm 2.2cm 0.2cm 1.4cm}, clip]{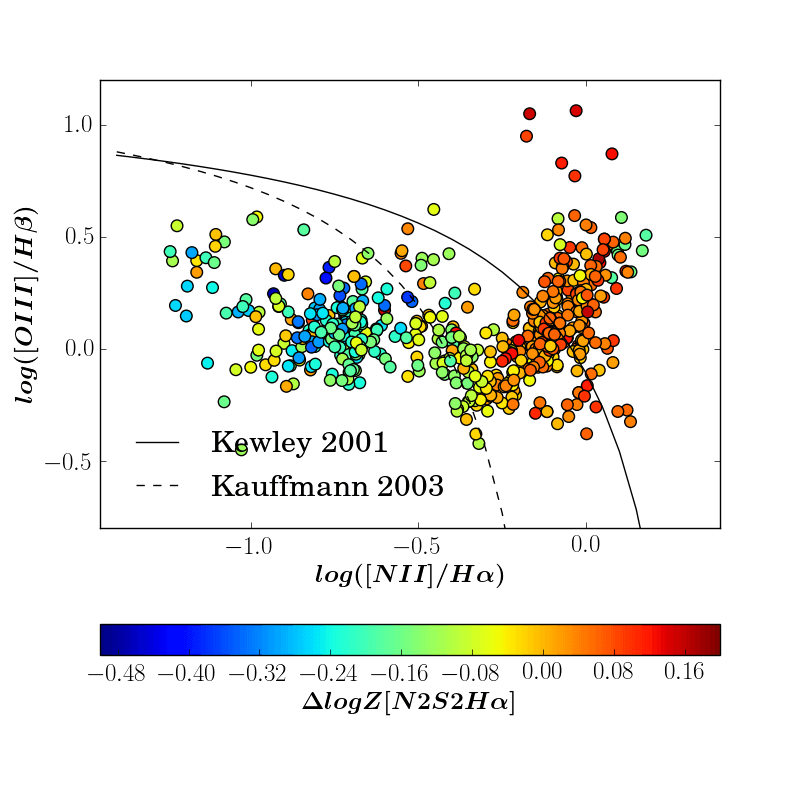}
	\caption{BPT diagnostic diagrams of the DIG/LIER counterpart in all selected pairs  from [S \textsc{ii}]-BPT selection method.
	The upper panel shows the 
 [O \textsc{iii}]/H$\beta$ versus [S \textsc{ii}]/H$\alpha$ diagram and the lower panel shows the [O \textsc{iii}]/H$\beta$ versus [N \textsc{ii}]/H$\alpha$ 
 diagram. Here each data point is colour-coded by the differential metallicity of the DIG/LIER-component relative to the \HII-region counterpart.
 The colour-bar and scale for all panels are fixed to be the same for all panels to give a better visual comparison. Metallicities have been determined using
 O3S2 (left panel), O3N2 (middle panel) and N2S2H$\alpha$ (right panel) diagnostics, respectively. On all panels, the solid black line indicates the maximum starburst line from \citet{Kewley2001}. On lower panel, the dashed black line indicates the demarcation line from \citet{Kauffmann2003}.}
	\label{fig:BPT sii pairs}
\end{figure*}

\subsection{\HII-regions metallicity diagnostics}

%\subsection{Metallicity Maps}
\label{section:maps}

\indent The wavelength range of the MUSE data enables us
to use three indirect metallicity calibrators, N2S2H$\alpha$,
O3N2 and O3S2, which will be described in the
following.
We anticipate that these diagnostics rely on the ratio of emission lines which are sufficiently
close in wavelength space not to require any significant reddening correction. 

In the following we define and briefly discuss these calibrators:

%\begin{itemize}

%\item {\bf N2S2H$\alpha$}

\subsubsection{N2S2H$\alpha$}
This diagnostic is defined as

\begin{equation}
\rm N2S2H\alpha = log([N\textsc{ii}]/[S\textsc{ii}]) + 0.264log([N \textsc{ii}]/H\alpha)
\label{eq:n2s2ha_def}
\end{equation}

\noindent where $[N\textsc{ii}] = [N \textsc{ii}]  \lambda 6584$ and $[S \textsc{ii}] \lambda\lambda 6717,6731$. This diagnostics
has been theoretically calibrated by
\citet{Dopita2016} by using a grid of photoionisation models implemented in \textsc{mappings}
\citep{Sutherland1993}.
According to their models the gas metallicity follow a linear relation with this diagnostic in the form

\begin{equation}
\rm 12+log(O/H) = 8.77 + N2S2H\alpha
\label{eq:n2s2ha_c1}
\end{equation}

\noindent for metallicities up to  $12+log(O/H)<9.05$. At higher metallicity their calibration deviates from such linear
relation and they derive the following polynomial relation that matches the theoretical models out to high metallicities:

\begin{equation}
\rm 12+log(O/H) = 8.77 + N2S2H\alpha +0.45(N2S2H\alpha +0.3)^5
\label{eq:n2s2ha_c2}
\end{equation}

This diagnostic is sensitive to metallicity primarily through the $[N\textsc{ii}]/[S\textsc{ii}]$ and through
the fact that $N/S$ (or $N/O$) is observed, on average, to increase linearly with metallicity at 12+log(O/H)$>$8.
This relationship is a crucial assumption of the model, as it provides the main sensitivity to metallicity.
However, for galaxies which do not follow the assumed N/S-O/H (or equivalently N/O-O/H relation which may be an issue specifically at high redshift, see also \citealt{Kumari2018}),
this diagnostic may provide results that are significantly offset. The $[N \textsc{ii}]/H\alpha$ term
provides a correction for secondary dependences on ionization parameter and gas pressure. \citet{Dopita2016} claim that, as a consequence,
this diagnostic is a fairly reliable metallicity tracer with only small residual dependence on other physical parameters,
hence a diagnostic that can be used over a wide range of environments.

The small wavelength range required to measure all three nebular lines involved in this diagnostic makes it particularly
attractive at high redshift, where optical nebular lines are shifted into the near-infrared and where instrumental limitations
and atmospheric absorption make it difficult to observe broad wavelength ranges.

%\item {\bf O3N2}
\subsubsection{O3N2}
This diagnostic is defined as

\begin{equation}
\rm O3N2 = log([O \textsc{iii}]/H\beta) - log([N \textsc{ii}]/H\alpha)
\label{eq:o3n2_def}
\end{equation}
\noindent where [O \textsc{iii}] = [O \textsc{iii}] $\lambda$5007 and [N \textsc{ii}] = [N \textsc{ii}] $\lambda$6584. This calibration was originally proposed by \cite{Pettini2004}, who calibrated it empirically mostly through direct T$_e$ measurements
of \HII-regions. We adopt the more recent calibration obtained by \citet{Curti2017} by using stacked spectra of several thousands
local galaxies in the SDSS-DR7 survey. The latter calibration provides the following relation:

\begin{equation}
%\rm 12+log(O/H) = 8.971-4.765~O3N2-2.268~O3N2^2
\rm O3N2 = 0.281-4.765~x_{O3N2}-2.268~x_{O3N2}^2
\label{eq:o3n2_c}
\end{equation}
\noindent where x$\rm_{O3N2}$ is oxygen abundance normalised to the solar value in the form 12 + log(O/H)$_{\odot}$ = 8.69. This calibration has been obtained in the range  7.6 $<$ 12 + log(O/H) $<$ 8.85. Of course, also this parameter is expected to be sensitive to the nitrogen abundance (N/O) and also to the ionization parameter. 

\indent 
Note that the calibrations of Curti et al 2017 are based on
	stacked spectra of the central regions of galaxies (from SDSS) with
	relatively large projected aperture (often a few kpc) which may
	include both \HII~regions and DIG rather than individual \HII~regions. For robustness of our analysis, we also test the following O3N2 relation from \citet{Marino2013} which is calibrated using 3423 \HII~ regions from the Calar Alto Legacy Integral Field Area project: 
	12 + log(O/H) = 8.505 $-$0.221 O3N2.
%\item {\bf O3S2}

\subsubsection{O3S2}

This diagnostic is defined as

\begin{equation}
\rm O3S2 = log([O \textsc{iii}]/H\beta + [S \textsc{ii}]/H\alpha)
\label{eq:o3s2_def}
\end{equation}

\noindent where [O \textsc{iii}] = [O \textsc{iii}] $\lambda$5007 and [S \textsc{ii}] = [S \textsc{ii}] $\lambda\lambda$6717,6731. This diagnostic is very similar to the more commonly used $\rm R_{23} = ([O \textsc{iii}]+[O \textsc{ii}])/H\beta$ \citep{Pagel1979}, but where the term
[O \textsc{ii}]/H$\beta$ is replaced by [S \textsc{ii}]/H$\alpha$, and essentially exploits the fact that the [S \textsc{ii}] flux is tightly correlated
with the [O \textsc{ii}] flux and that sulphur and oxygen are mostly produced by the same population of stars and on similar timescales.
MUSE's wavelength range does not allow us to observe [O \textsc{ii}] in local galaxies, while both [O \textsc{iii}] and [S \textsc{ii}] are observable.
The additional advantage of O3S2 (relative to $\rm R_{23}$) is that it is unaffected by dust reddening.
The additional advantage, relative to the other two diagnostics discussed above, is that O3S2 is not affected by variations
of the N/O relative abundance.

We adopt the calibration of this diagnostic derived by Curti et al. (in prep) and is based on same data and methodology presented in \cite{Curti2017}:

\begin{equation}
%\rm 12+log(O/H) = 8.644-2.223~O3S2-1.073~O3S2^2+0.533~03S2^3
\rm O3S2 = -0.046-2.223~x_{O3S2}-1.073~x_{O3S2}^2+0.533~x_{O3S2}^3
\label{eq:o3s2_c}
\end{equation}
\noindent where x$\rm_{O3S2}$ is oxygen abundance normalised to the solar value in the form 12 + log(O/H)$_{\odot}$ = 8.69. The disadvantage of this diagnostic is that, like $\rm R_{23}$ it is double-valued, hence it often requires another
diagnostic to identify whether the metallicity is in the low- or high-metallicity branch. In the specific
case of this paper this is not an issue, as for the central regions
of these local relatively massive galaxies the gas is certainly in the high metallicity branch of the relation.

%\item {\bf Other calibrators}
\subsubsection{Other calibrators}
Unfortunately we cannot investigate and calibrate other commonly used diagnostics. We have already discussed
that we cannot measure $\rm R_{23}$ because of MUSE's spectral coverage.
For the same reason we cannot use the N2O2 diagnostic ($\rm = [N \textsc{ii}]/[O \textsc{ii}]$, \citealt{Kewley2002}), again because $\rm [O \textsc{ii}]$ is out of the spectral range.
N2 ($\rm = [N \textsc{ii}]/H\alpha $, \citealt{Pettini2004}) has been widely adopted especially at high redshift,
exploiting the narrow wavelength range needed for this diagnostic. However, it cannot
be used in the DIG/LIER range, as the N2--metallicity calibration saturates 
and flattens at the BPT boundary between \HII-regions and LIERs \citep[see Fig 9 of][]{Curti2017}, as such it cannot be extrapolated into the LIER region.

%\end{itemize}
%\indent The other two metallicity diagnostics, O3N2 and O3S2, are fully empirical polynomial relations between line ratios and \textcolor{magenta}{direct T$_e$-metallicity} (Table \ref{table:O3N2O3S2}) derived by stacking spectra of more than 110 000 galaxies from Sloan Digital Sky Survey. The O3N2 calibration used here was presented in \citet{Curti2017}, and O3S2 calibration will be described in an upcoming paper (Curti et al. in prep.). Both of these calibrations \textcolor{magenta}{have been calibrated} in the range of 7.6 $<$ 12 + log(O/H) $<$ 8.85. The uncertainties on O3N2 and O3S2 diagnostics are at least 0.09 and 0.12 dex, respectively.  

%\indent \textcolor{magenta}{The O3S2 calibration has the advantage of being independent of nitrogen abundance in comparison to O3N2 or N2S2H$\alpha$ calibrations. It is equivalent to the metallicity diagnostic R$_{23}$ (= ([O \textsc{ii}] $\lambda\lambda3727,3729$ + [O \textsc{iii}] $\lambda\lambda4959,5007$)/H$\beta$), where [O \textsc{ii}] $\lambda\lambda3727,3729$/H$\beta$ is replaced with [S \textsc{ii}] $\lambda\lambda6717, 6731$/H$\alpha$ because of close values of the ionisation potentials (IP) of O+ (IP: 13.6 eV) and S+ (IP: 10.3 eV). We also find a linear relation between the emission line fluxes of  [O \textsc{ii}] $\lambda\lambda3727,3729$ and [S \textsc{ii}] $\lambda\lambda6717, 6731$/H$\alpha$ of SDSS galaxies. However, O3S2 does not require reddening correction unlike R$_{23}$.}

\subsection{Metallicity maps}

\indent We aim to verify if \HII~metallicity calibrations hold for the DIG/LIER,
both within the region where these relations have been calibrated and also on their extrapolation. Hence, we use the above metallicity calibrations even beyond
their range of calibration. The upper panels in Figures \ref{fig:NGC1042}, %\ref{fig:NGC289}--\ref{fig:ESO499-G37-DEEP}
and in figures in the supplementary online material (Appendix \textcolor{blue}{A})
show metallicity maps obtained using O3N2 (left-panel), O3S2 (middle panel) and N2S2H$\alpha$ (linear version, right panel) of NGC 1042 and other galaxies in
the sample, respectively. Metallicity maps of the same galaxy obtained from different metallicity calibrators show no apparent
correlation, which points towards the inherent discrepancies between the strong line metallicity calibrators. %Galaxiessuch as NGC 1483 (Figure \ref{fig:NGC1483}), NGC 4592 (Figure \ref{fig:NGC4592}), NGC4980 (Figure \ref{fig:NGC4980}) and ESO 499-G37 (Figure \ref{fig:ESO499-G37-DEEP}) show a much lower metallicity when estimated from N2S2H$\alpha$ with respect to O3S2 or O3N2. 
Galaxies such as NGC 1483, NGC 4592, NGC4980 and ESO 499-G37 (see supplementary online material Appendix \textcolor{blue}{A}) show a much lower metallicity when estimated from N2S2H$\alpha$ with respect to O3S2 or O3N2. The theoretical metallicity diagnostics are known to predict higher metallicities compared to those derived from the direct-T$_e$ method \citep[see e.g.][]{Kewley2008, Kumari2017, Maiolino2018}. Given that O3N2 and O3S2 diagnostics here are calibrated for a sample whose metallicities are determined from direct-T$_e$ methods, we expect that the metallicities estimated from the theoretical N2S2H$\alpha$ calibration are higher than the O3N2 or O3S2 diagnostics. The unexpected behaviour seen in these galaxies may be due to the following reasons: Firstly, these galaxies may have relatively lower nitrogen-content than sulphur which results in an underestimate of metallicity due to the use of [N \textsc{ii}]/[S \textsc{ii}] line ratio in N2S2H$\alpha$ diagnostic. Secondly, it is also possible that the assumption of an inherent relation between N/O and O/H in N2S2H$\alpha$ calibration is not valid at spatially-resolved scale within these particular galaxies. For example, \citet{Kumari2018} found an unusual negative N/O vs O/H relation on a spaxel-by-spaxel basis in a blue compact dwarf galaxy NGC 4670. In addition, we should take into account the well-known discrepancies in metallicities obtained from different diagnostics \citep[see e.g.][]{Kewley2008}, the reasons for which are yet to be understood. See \citet{Maiolino2018} for a review.

%\indent  We could not employ the frequently used R$_{23}$ parameter or N2O2 (= [N \textsc{ii}]/[O \textsc{ii}]) for mapping metallicity because the spectral range of MUSE does not cover the [O \textsc{ii}] $\lambda\lambda3727,3729$ doublet. We decide not to use N2 (=  [N \textsc{ii}]/H$\alpha$) which flattens at higher metallicity, and hence can not be extrapolated to the LIER region \citep[see Fig 9 of][]{Curti2017}.

\begin{figure*}
	\includegraphics[width=0.30\textwidth, trim={0.2cm 2.0cm 0.2cm 2.0cm}, clip]{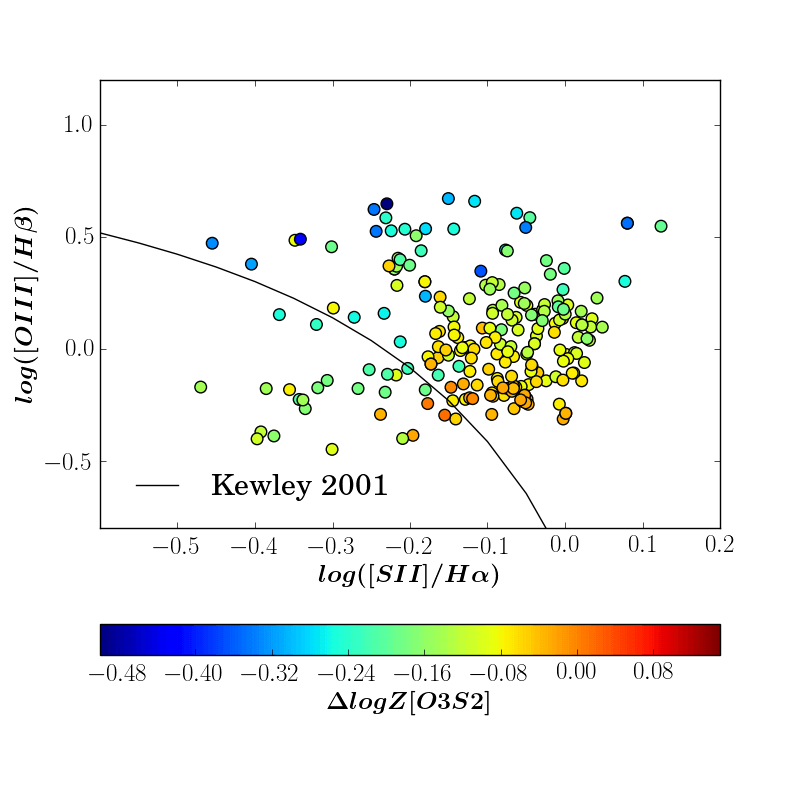}
	\includegraphics[width=0.30\textwidth, trim={0.2cm 2.0cm 0.2cm 2.0cm}, clip]{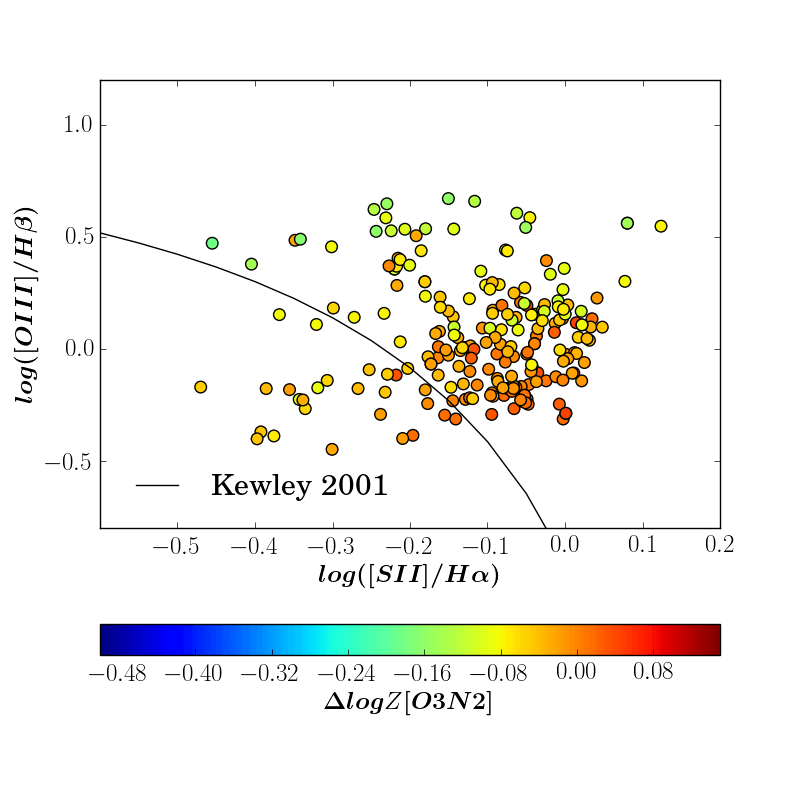}
	\includegraphics[width=0.30\textwidth, trim={0.2cm 2.0cm 0.2cm 2.0cm}, clip]{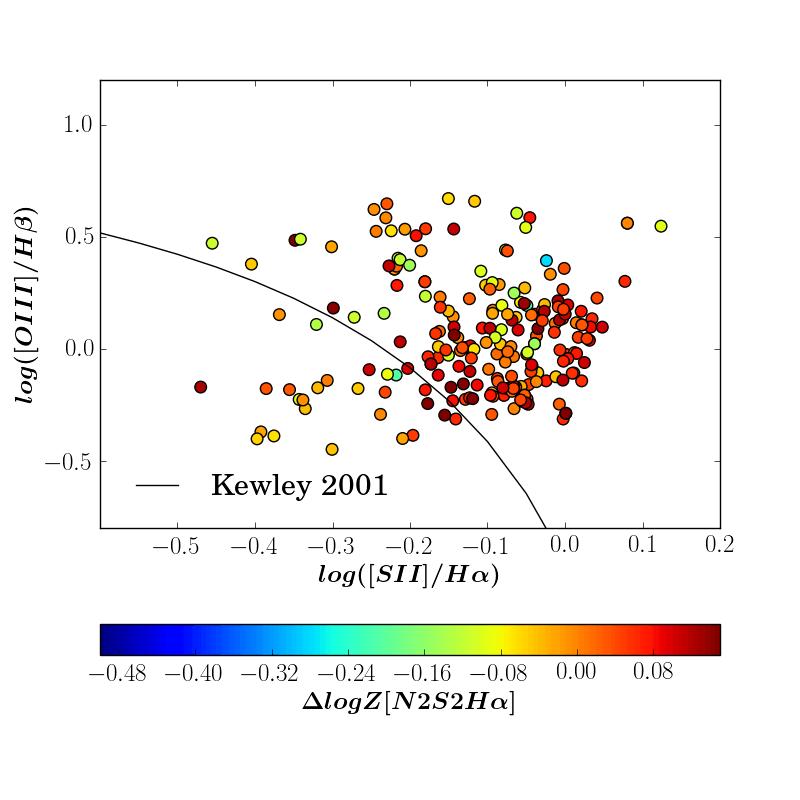}
	\includegraphics[width=0.30\textwidth, trim={0.2cm 2.0cm 0.2cm 2.0cm}, clip]{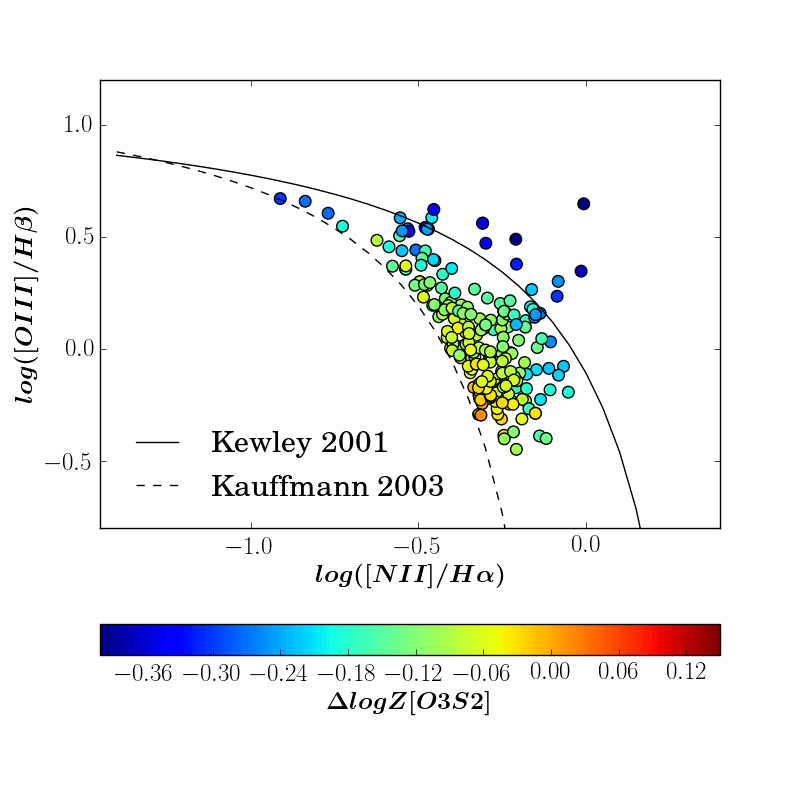}
	\includegraphics[width=0.30\textwidth, trim={0.2cm 2.0cm 0.2cm 2.0cm}, clip]{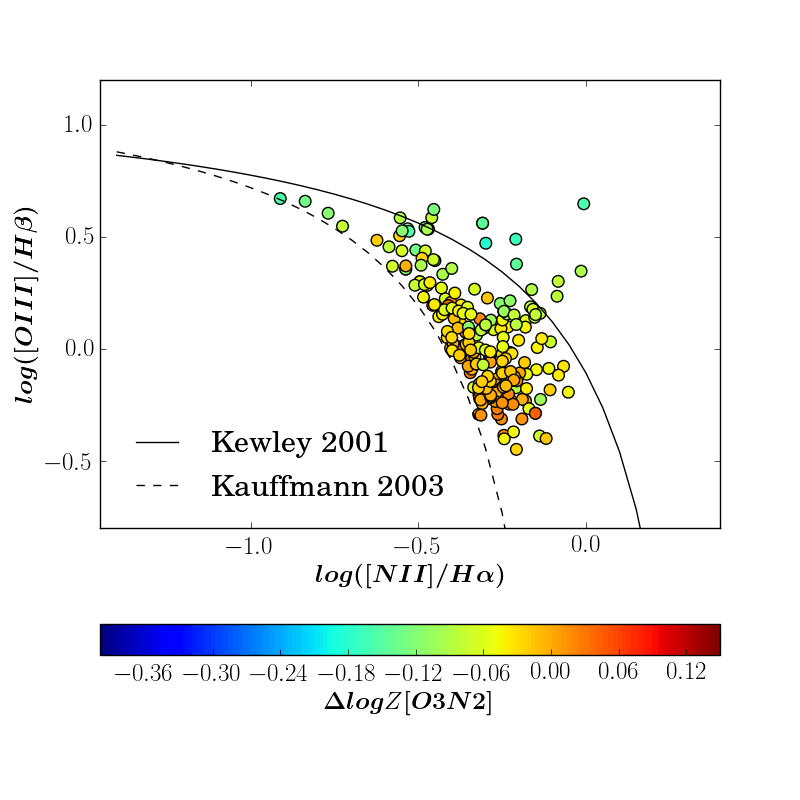}
	\includegraphics[width=0.30\textwidth, trim={0.2cm 2.0cm 0.2cm 2.0cm}, clip]{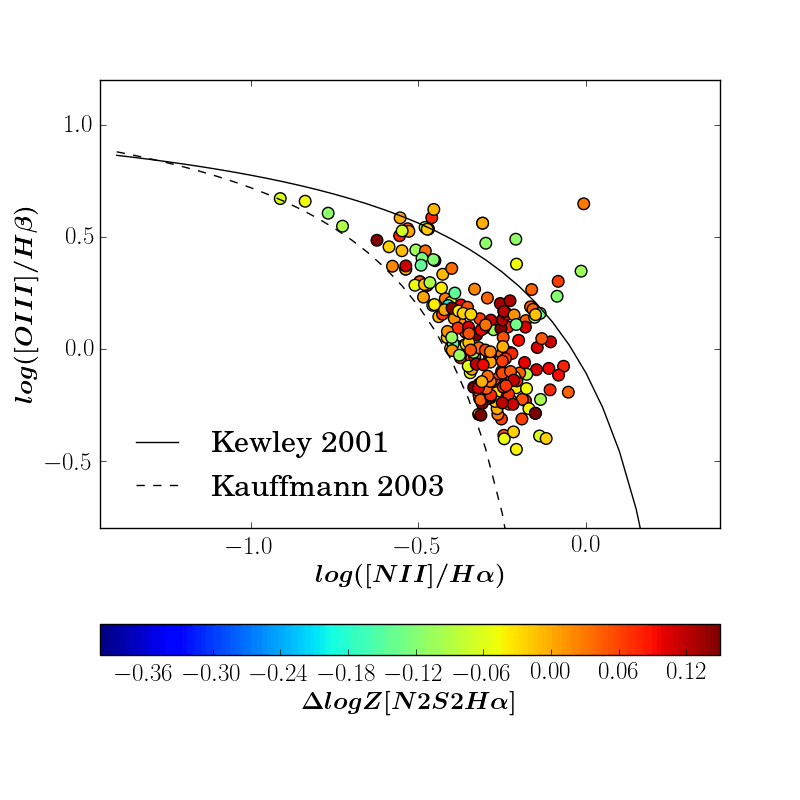}
	\caption{BPT diagnostic diagrams of the DIG/LIER counterpart in all selected pairs  from [N \textsc{ii}]-BPT selection method.
	The upper panel shows the 
 [O \textsc{iii}]/H$\beta$ versus [S \textsc{ii}]/H$\alpha$ diagram and the lower panel shows the [O \textsc{iii}]/H$\beta$ versus [N \textsc{ii}]/H$\alpha$ 
 diagram. Here each data point is colour-coded by the differential metallicity of the DIG/LIER-component relative to the HII-region counterpart.
 The colour-bar and scale for all panels are fixed to be the same for all panels to give a better visual comparison. Metallicities have been determined using
 O3S2 (left panel), O3N2 (middle panel) and N2S2H$\alpha$ (right panel) diagnostics, respectively. On all panels, the solid black line indicates the maximum starburst line from \citet{Kewley2001}. On lower panel, the dashed black line indicates the demarcation line from \citet{Kauffmann2003}.}

	\label{fig:BPT nii pairs}
\end{figure*}

\begin{figure*}
	\includegraphics[width=0.4\textwidth]{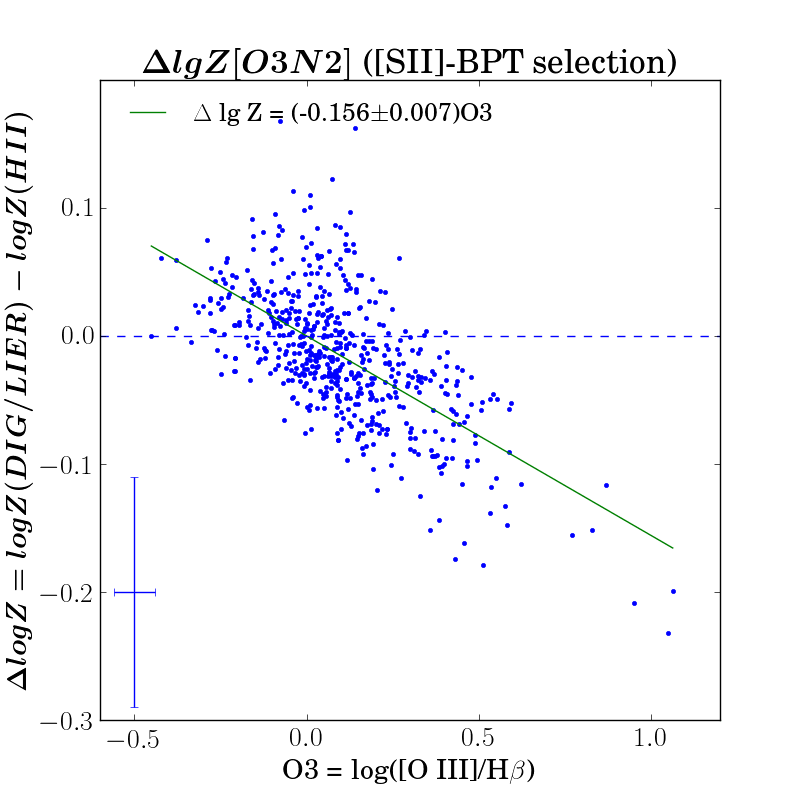}
	\includegraphics[width=0.4\textwidth]{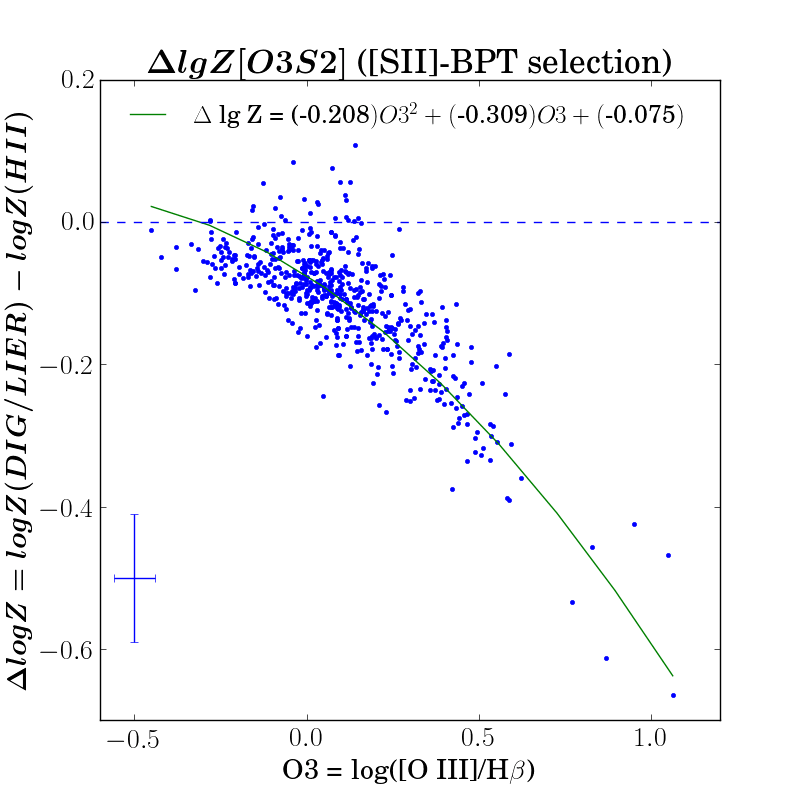}
	\caption{Least-square best-fit to differential metallicity between DIG/LIER regions and their \HII-counterparts, inferred from  the
	O3N2 (left-panel) and O3S2 (right panel) diagnostics, versus log ([O \textsc{iii}]/H$\beta$). On both panels, blue data points correspond to \HII-DIG/LIER pairs from the [S \textsc{ii}]-BPT selection, and the green line is the best-fit. The blue error bars in the left-hand corner are the median uncertainties on log ([O \textsc{iii}]/H$\beta$) and differential metallicities. The uncertainties on log ([O \textsc{iii}]/H$\beta$) were obtained by propagating the uncertainties on flux measurements, whereas the uncertainties on differential metallicities included a systematic calibration uncertainty along with the measurement uncertainties. The best-fit coefficients are shown at the top of each panel.}
	\label{fig:correction sii}
\end{figure*}

\begin{figure*}
	\includegraphics[width=0.4\textwidth]{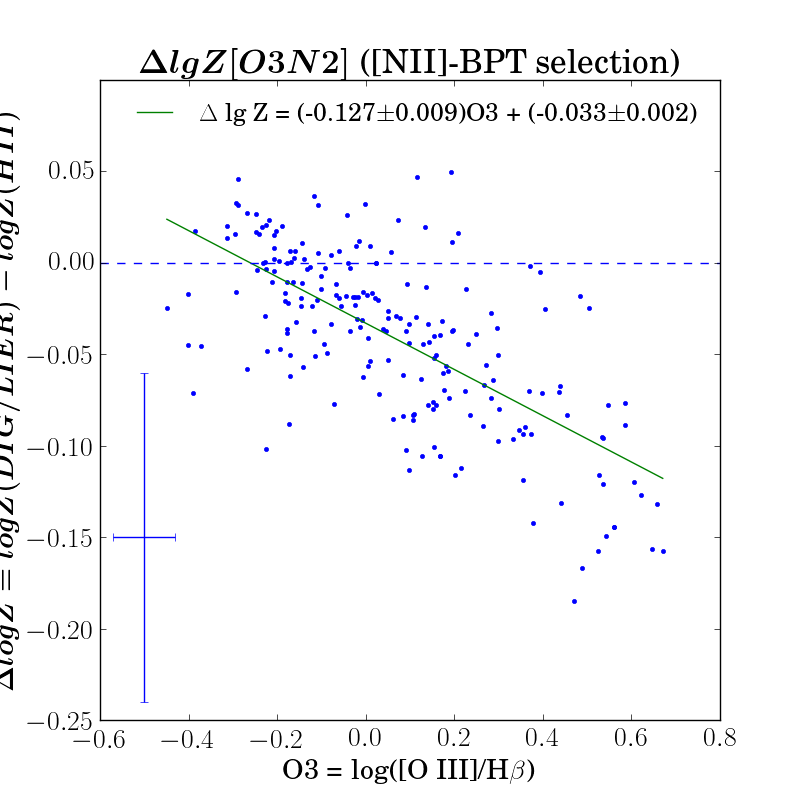}
	\includegraphics[width=0.4\textwidth]{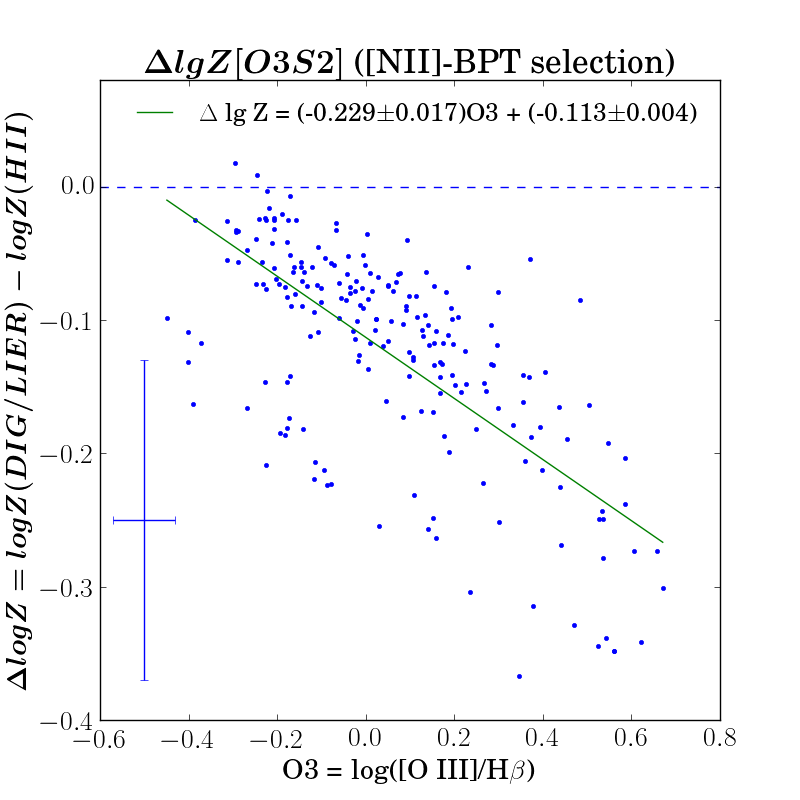}
	\caption{Least-square best-fit to differential metallicity between DIG/LIER regions and their \HII-counterparts, inferred from  the
	O3N2 (left-panel) and O3S2 (right panel) diagnostics, versus log ([O \textsc{iii}]/H$\beta$). On both panels, blue data points correspond to \HII-DIG/LIER pairs from the
	[N \textsc{ii}]-BPT selection, and the green line is the best-fit. The blue error bars in the left-hand corner are the median uncertainties on log ([O \textsc{iii}]/H$\beta$) and differential metallicities. The uncertainties on log ([O \textsc{iii}]/H$\beta$) were obtained by propagating the uncertainties on flux measurements, whereas the uncertainties on differential metallicities included a systematic calibration uncertainty along with the measurement uncertainties. The best-fit coefficients are shown at the top of each panel.
}
	\label{fig:correction nii}
\end{figure*}

\section{Results}
\label{section:results}
\subsection{Biases in metallicity estimates of DIG/LIER-dominated regions}
\label{section:impact}
\indent Figure \ref{fig:differential} shows the distribution of the differential metallicities ($\Delta$ log Z) of the pairs of H \textsc{ii}-DIG/LIER regions selected
on the basis of [S \textsc{ii}]-BPT (upper-left panel), [N \textsc{ii}]-BPT (upper-right panel) and $\Sigma_{\rm{H\alpha}}$ (lower panel), where metallicities
are  determined from the three metallicity calibrators described in Section \ref{section:maps}. On each panel, mean offset and standard deviation of differential metallicities are denoted
by $\overline{\Delta \log{Z}}$ and $\sigma(\Delta \log{Z})$, respectively (which are also reported in Table \ref{tab:stats}). 

\indent Considering [S \textsc{ii}]-BPT selected \HII-DIG/LIER pairs (Figure \ref{fig:differential}, upper-left panel), we find that O3N2 shows a negligible average
offset
and the least dispersion (0.05 dex), demonstrating that it is equally applicable to both \HII~regions and DIG/LIER-dominated regions. O3S2 underestimates the
metallicity of the DIG/LIER by 0.11 dex with a dispersion of 0.09 dex. The distribution of $\Delta$ log Z[N2S2H$\alpha$] appears to be bimodal, with an average offset of
$-$0.06 dex and a large dispersion of 0.12 dex. The latter refers to the linear calibration of the N2S2H$\alpha$ diagnostic; if using the
non-linear calibration the dispersion is even larger (see table \ref{tab:stats}).

\indent Pairs selected from [N \textsc{ii}]-BPT  (Figure \ref{fig:differential}, upper-right panel) appear to have comparable nominal metallicities when O3N2
diagnostic is used, like the case of [S \textsc{ii}]-BPT selected \HII-DIG/LIER pairs. The distribution of differential metallicities obtained from O3S2 shows a
large mean offset ($-$0.13) as well as a large dispersion (0.09 dex). N2S2H$\alpha$, on the contrary shows negligible average offset but a large dispersion (0.15 dex).

\indent Even for  $\Sigma_{\rm{H\alpha}}$-selected \HII-DIG/LIER pairs, O3N2 diagnostic appears to predict comparable metallicities (negligible mean offset) with
the least dispersion. O3S2 diagnostic also shows negligible offset unlike the previous two cases, although the dispersion is still large. N2S2H$\alpha$ (linear
calibration) produces a large mean offset accompanied by a large dispersion. The dispersion increases even further if adopting
the non-linear calibration of N2S2H$\alpha$.
 
\indent Above analysis shows that except for O3N2, the other two diagnostics (N2S2H$\alpha$ and O3S2) will significantly bias metallicity maps of galaxies
containing DIG/LIER irrespective of the adopted criterion to classify the DIG/LIER regions, hence they can introduce biases if DIG/LIER emission
is mixed with the emission in unresolved/poor-resolution observations. O3N2 shows negligible average offset, indicating that this diagnostic
is the least affected by DIG/LIER contamination, and can potentially be exploited to trace the metallicity in these regions. Results remain practically unchanged if we were to use O3N2 relation from \citet{Marino2013} calibrated on \HII~ regions rather than the relation
	from \citet{Curti2017} (see Table \ref{tab:stats}). Our further analysis is hence
	restricted to the more recent calibrations of \citet{Curti2017}, though we also report the results
	for the \citet{Marino2013} calibration in Table 1.

\indent  However, these calibrators are also characterized by some secondary trends, as discussed in the following,
which can be used to correct for the offsets and further
reduce the dispersion.
Figures \ref{fig:BPT sii pairs} and \ref{fig:BPT nii pairs} show the emission line ratio diagnostic diagrams  ([O \textsc{iii}]/H$\beta$ versus [S
\textsc{ii}]/H$\alpha$ (upper panel) and [O \textsc{iii}]/H$\beta$ versus [N \textsc{ii}]/H$\alpha$ (lower panel)) of the DIG/LIER counterpart in \HII-DIG/LIER
pairs selected from [S \textsc{ii}]-BPT and [N \textsc{ii}]-BPT criteria, respectively. It is interesting to note that the [S \textsc{ii}]-BPT DIG/LIER selection
(Fig.\ref{fig:BPT sii pairs}) results into a significant fraction of
regions that are not consistently classified as LIER in the [N \textsc{ii}]-BPT diagram. The inconsistency between the two BPT diagrams
is an issue that has already been highlighted
in the past \citep{Perez-Montero2009}. The [N \textsc{ii}]-BPT selection is better on this regard, by selecting in DIG/LIER regions most of which
are classified as such also in the [SII]-BPT diagram.
In both figures, data points in each panel are colour coded with respect to the differential metallicity determined from O3S2 (left panel), O3N2 (middle panel) and N2S2H$\alpha$ (right panel). The colour-bar and scale of differential metallicities are fixed to be the same for all calibrators for a comparative view of biases in metallicities introduced by different calibrators. These BPT diagnostics clearly demonstrate that differential metallicities determined from O3S2 (left panels in Figures \ref{fig:BPT sii pairs} and \ref{fig:BPT nii pairs}) and O3N2 (middle panels in Figures \ref{fig:BPT sii pairs} and \ref{fig:BPT nii pairs}) vary with respect to log~([O~\textsc{iii}]/H$\beta$). Instead, no clear trend is observed for the N2S2H$\alpha$ diagnostic (right panels in Figure \ref{fig:BPT sii pairs} and \ref{fig:BPT nii pairs}). 

\indent The origin of the trends of the DIG/LIER metallicity offsets with log~([O~\textsc{iii}]/H$\beta$), for the O3N2 and O3S2 diagrams, is likely associated with variation of the ionization
parameter. However, it is beyond the scope of this paper to investigate the physical origin of these correlations. Here we mostly exploit such correlations
to empirically
identify calibration correction factors
for the O3N2 and O3S2 diagnostics when used in the DIG/LIER region.
We quantify and calibrate these trends in Figure \ref{fig:correction sii} and \ref{fig:correction nii}.
We use a least square method which minimises the scatter between the data-points and the best-fit line to fit differential metallicities from O3N2 and O3S2 with respect to log~([O~\textsc{iii}]/H$\beta$) ratio. Figure \ref{fig:correction sii} shows the best-fit for O3N2 (left-panel) and O3S2 (right panel) for [S~\textsc{ii}]-BPT selected DIG/LIER pairs while Figure \ref{fig:correction nii} shows the best-fit for O3N2 (left-panel) and O3S2 (right panel) for [N \textsc{ii}]-BPT selected DIG/LIER pairs. In Figure \ref{fig:correction nii} (right panel), we find that the distribution of points on $\Delta$ log Z-log~([O~\textsc{iii}]/H$\beta$) plane is bimodal, hence coefficients of the best-fit line would be driven/biased by the over-density points.

\indent For \HII-DIG/LIER pairs selected on the basis of $\Sigma_{\rm{H\alpha}}$, we studied the emission line ratio diagnostic diagrams, though we did not find any trend which could be used to remedy the observed offset in the differential metallicities. As such, \HII-DIG/LIER pairs selected from $\Sigma_{\rm{H\alpha}}$ are not investigated any further.

\begin{figure*}
	\centering
	\includegraphics[width=0.4\textwidth]{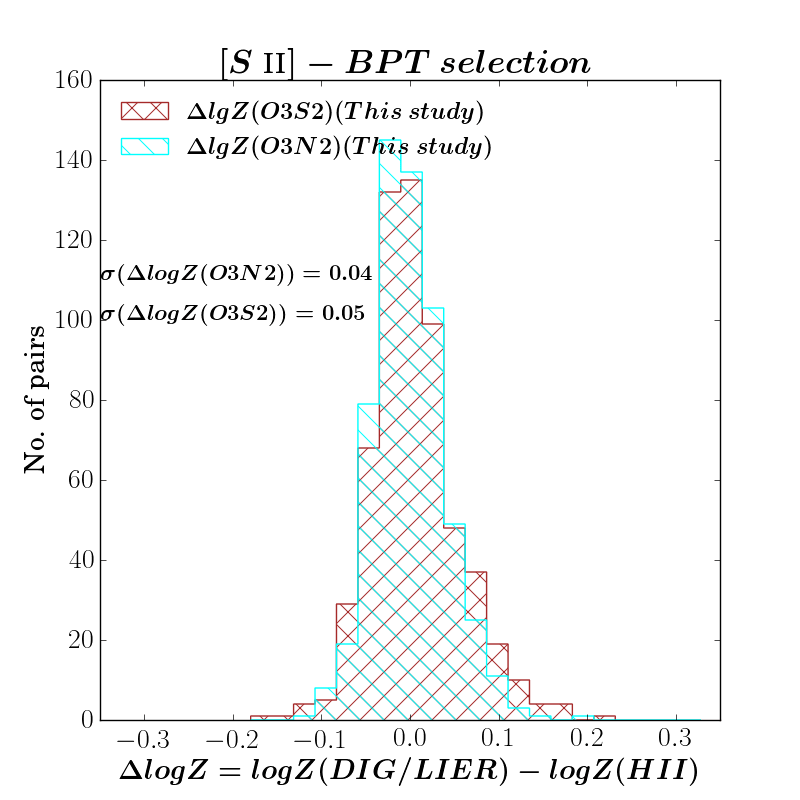}
	\includegraphics[width=0.4\textwidth]{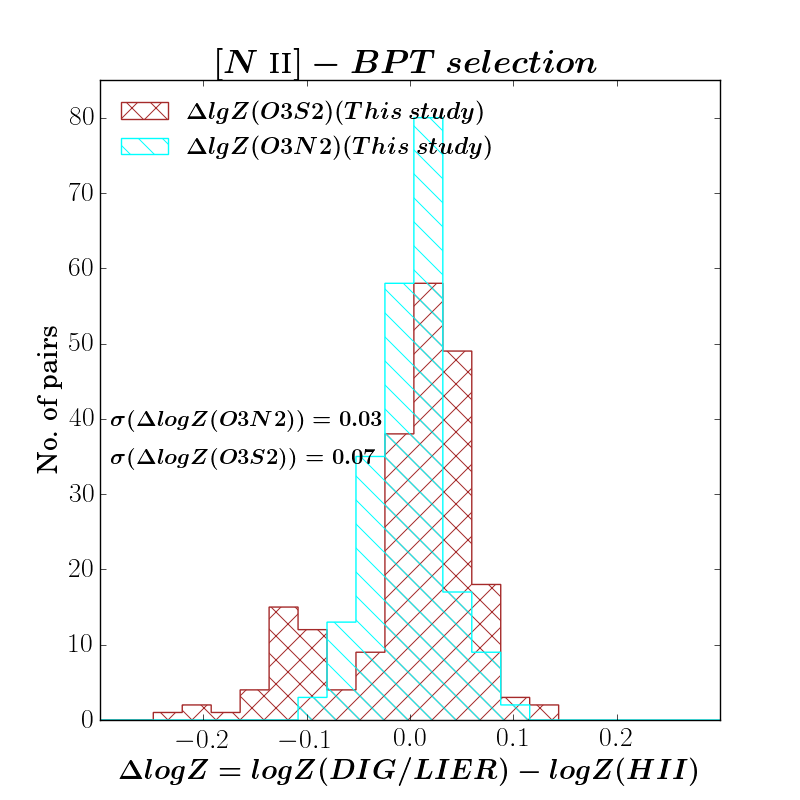}
	\caption{Distribution of corrected differential metallicity ($\Delta$ log Z) of \HII--DIG/LIER pairs selected on the basis of the [S \textsc{ii}]-BPT (left panel)
	and [N \textsc{ii}]-BPT (right panel) diagrams. In each panel brown and cyan histograms represent the differential metallicity distribution from O3S2, and O3N2,
	respectively  after applying the correction to the DIG/LIER/Seyfert spaxels presented in section \ref{section:new calibrations} (equations \ref{eq:cal o3n2 sii}, \ref{eq:cal o3s2 sii}, \ref{eq:cal o3n2 nii}, \ref{eq:cal o3s2 nii}).
	$\sigma(\Delta log(Z))$ denotes the standard deviation of the differential metallicity distribution. In each case the average offsets of the distribution are zero by construction.}
	\label{fig:differential correction}
\end{figure*}

\subsection{New calibrations}
\label{section:new calibrations}
\indent Based on the above corrections for the O3N2 and O3S2 metallicity calibrations,
we propose metallicity calibrations for the DIG/LIER regions
of the form, log[Z(DIG/LIER)$_{true}$] = log[Z(DIG/LIER)$_{orig}$] $-$ $\Delta$ log(Z),
on the basis of the best-fits to differential metallicity versus log~([O~\textsc{iii}]/H$\beta$). In the proposed form, log Z(DIG/LIER)$_{true}$ and
log Z(DIG/LIER)$_{orig}$ are the true metallicity and the
metallicity obtained after simply applying the \HII-calibrated diagnostics
\citep{Curti2017}, respectively, and $\Delta$~log Z is the correction term obtained
in Figures \ref{fig:correction sii} and \ref{fig:correction nii}. Hence, we
suggest the following two sets of corrections and calibrations, where O3 = log([O \textsc{iii}]/H$\beta$:

\subsubsection{Calibration for [S \textsc{ii}]-BPT selected DIG/LIER regions}
	\indent For data points whose emission line ratios lie beyond the theoretical Kewley line \citep{Kewley2001} on the [S \textsc{ii}]-BPT diagram, we first need to determine metallicity from the \HII~calibration given in Section \ref{section:maps} and correct by subtracting the following terms depending on the calibration used:
	\setlength{\belowdisplayskip}{-1.5pt} \setlength{\belowdisplayshortskip}{-1.5pt}
	\setlength{\abovedisplayskip}{-1.5pt} \setlength{\abovedisplayshortskip}{-1.5pt}

	\begin{equation}
	\rm \Delta log(Z)_{O3N2} = -0.156\times O3,   
	\label{eq:corr o3n2 sii}
	\end{equation}	
	\begin{equation}
	%\rm \Delta Z(O3S2) = -0.316\times O3 - 0.081,  
	\rm \Delta log(Z)_{O3S2} = -0.208\times O3^2 - 0.309\times O3 -0.075,
	\label{eq:corr o3s2 sii}
	\end{equation}

	\indent 
	
\indent	Combining these with equations \ref{eq:o3n2_c} and \ref{eq:o3s2_c}, they result into
	 the following final calibrations

\iffalse
\begin{equation}
\rm 12+log(O/H)_{DIG/LIER} = 8.971-4.765~O3N2-2.268~O3N2^2-0.156\times O3
\end{equation}

\begin{equation}
%\rm 12+log(O/H) = 8.644-2.223~O3S2-1.073~O3S2^2+0.533~03S2^3
\rm 12+log(O/H)_{DIG/LIER} = 8.569-2.223~O3S2-1.073~O3S2^2+0.533~03S2^3-0.309~O3-0.208~O3^2
\end{equation}

\begin{equation}
\begin{split}
\rm 12+log(O/H)_{DIG/LIER} = 8.971-4.765~O3N2 -2.268~O3N2^2\\
\rm -0.156\times O3
\end{split}
\label{eq:cal o3n2 sii}
\end{equation}

\begin{equation}
\begin{split}
%\rm 12+log(O/H) = 8.644-2.223~O3S2-1.073~O3S2^2+0.533~03S2^3
\rm 12+log(O/H)_{DIG/LIER} = 8.569-2.223~O3S2 -1.073~O3S2^2\\
\rm +0.533~03S2^3-0.309~O3-0.208~O3^2
\end{split}
\label{eq:cal o3s2 sii}
\end{equation}

\fi

\begin{equation}
%\begin{split}
\rm 12+log(O/H)_{DIG/LIER} = x_{O3N2} + 8.69 +0.156~O3
%\end{split}
\label{eq:cal o3n2 sii}
\end{equation}

\begin{equation}
%\begin{split}
%\rm 12+log(O/H) = 8.644-2.223~O3S2-1.073~O3S2^2+0.533~03S2^3
\rm 12+log(O/H)_{DIG/LIER} = x_{O3S2} + 8.765 +0.309~O3+0.208~O3^2
%\end{split}
\label{eq:cal o3s2 sii}
\end{equation}

\noindent where x$\rm_{O3N2}$ and x$\rm_{O3S2}$ are determined from equations \ref{eq:o3n2_c} and \ref{eq:o3s2_c}, respectively. Figure \ref{fig:NGC1042} (middle row) shows metallicity maps of NGC 1042 obtained
	from O3N2 (left-panel) and O3S2 (middle-panel) diagnostics after applying the above
	corrections to the DIG/LIER-dominated spaxels (i.e. those with emission line ratios above the
	Kewley line) shown by yellow/pink spaxels in spatially-resolved [S \textsc{ii}]-BPT (right
	panel). %Corresponding maps the other galaxies in the sample are shown in Figures \ref{fig:NGC289}--\ref{fig:ESO499-G37-DEEP} (middle rows). 
	Corresponding maps of other galaxies in the sample are shown in figures in the supplementary online material (Appendix \textcolor{blue}{A}, middle rows).
	The corrected metallicity maps are more uniform, especially azimuthally,
	than those obtained from original metallicity calibration, even for those galaxies where we could not identify any \HII-DIG/LIER pairs (which include NGC 4030, NGC 4603, NGC 4980 and NGC 5334), showing the robustness of our calibration.  
	
\subsubsection{Calibration for [N \textsc{ii}]-BPT selected DIG/LIER regions}
	 \indent For data points whose emission line ratios lie beyond the empirical Kauffmann line \citep{Kauffmann2003} on the [N \textsc{ii}]-BPT diagram, we propose that the following correction terms are subtracted from the metallicity determined from \HII~calibration given in Section \ref{section:maps}:
	\setlength{\belowdisplayskip}{-1.5pt} \setlength{\belowdisplayshortskip}{-1.5pt}
	\setlength{\abovedisplayskip}{-1.5pt} \setlength{\abovedisplayshortskip}{-1.5pt}
	
	\begin{equation}
	\rm \Delta log(Z)_{O3N2} = -0.127\times O3 - 0.033,   
	\label{eq:corr o3n2 nii}
	\end{equation}

	\begin{equation}
	\rm \Delta log(Z)_{O3S2} = -0.229\times O3 - 0.113,  
	\label{eq:corr o3s2 nii}
	\end{equation}
	
\indent	
		
\indent	Combining these with equations \ref{eq:o3n2_c} and \ref{eq:o3s2_c}, they result into
	 the following final calibrations
\iffalse
\begin{equation}
\rm 12+log(O/H)_{DIG/LIER} = 8.938-4.765~O3N2-2.268~O3N2^2-0.127\times O3
\end{equation}

\begin{equation}
%\rm 12+log(O/H) = 8.644-2.223~O3S2-1.073~O3S2^2+0.533~03S2^3
\rm 12+log(O/H)_{DIG/LIER} = 8.531-2.223~O3S2-1.073~O3S2^2+0.533~03S2^3-0.229~O3
\end{equation}

\begin{equation}
\begin{split}
\rm 12+log(O/H)_{DIG/LIER} = 8.938-4.765~O3N2 -2.268~O3N2^2\\
\rm -0.127\times O3
\end{split}
\label{eq:cal o3n2 nii}
\end{equation}

\begin{equation}
\begin{split}
%\rm 12+log(O/H) = 8.644-2.223~O3S2-1.073~O3S2^2+0.533~03S2^3
\rm 12+log(O/H)_{DIG/LIER} = 8.531-2.223~O3S2 -1.073~O3S2^2\\
\rm+0.533~O3S2^3-0.229~O3
\end{split}
\label{eq:cal o3s2 nii}
\end{equation}

\fi

\begin{equation}
%\begin{split}
\rm 12+log(O/H)_{DIG/LIER} = x_{O3N2} + 8.723 + 0.127~O3
%\end{split}
\label{eq:cal o3n2 nii}
\end{equation}

\begin{equation}
%\begin{split}
%
\rm 12+log(O/H)_{DIG/LIER} = x_{O3S2} + 8.803 + 0.229~O3
%\end{split}
\label{eq:cal o3s2 nii}
\end{equation}

\indent 

\noindent where x$\rm_{O3N2}$ and x$\rm_{O3S2}$ are determined from equations \ref{eq:o3n2_c} and \ref{eq:o3s2_c}, respectively. We apply the above corrections to the DIG/LIER-dominated spaxels (i.e. those with emission line ratios beyond the Kauffmann line) to all galaxies in sample. Figure \ref{fig:NGC1042} (bottom row) show metallicity maps of NGC 1042 obtained from O3N2 (left-panel) and O3S2 (middle-panel) after correction, %while the corresponding maps for all other galaxies are shown in Figures \ref{fig:NGC289}--\ref{fig:ESO499-G37-DEEP} (bottom rows).  
while the corresponding maps for all other galaxies are shown in figures (bottom rows) in the supplementary online material (Appendix \textcolor{blue}{A}).
\subsection{Residual differential metallicity after correction}
\label{section:differential correction}
The metallicity maps obtained by using these new calibrations for the DIG/LIER regions
already show an improvement, especially in terms of much reduced metallicity
contrast between \HII-regions and nearby DIG/LIER regions. In this subsection we quantify
such improvement.

\indent  From the corrected O3N2 and O3S2 metallicity maps of each galaxy,  we estimate
representative values of metallicities for \HII-DIG/LIER pairs using the same methodology described in
Section \ref{section:sii}, and study the distribution of differential metallicities in Figure
\ref{fig:differential correction}. Left panel shows the distribution for [S \textsc{ii}]-BPT
selected \HII-DIG/LIER pairs while right panel corresponds to  [N \textsc{ii}]-BPT selected \HII-DIG/LIER
pairs. By construction, mean differential metallicity is zero for each of the two
metallicity diagnostic and for both samples. A comparison of dispersions before and after applying corrections can be found in Table \ref{tab:stats}.

\indent O3N2 shows comparable metallicity dispersions for \HII-DIG/LIER samples obtained from both
selection criterion, i.e. 0.04 dex for [S \textsc{ii}]-BPT selection and 0.03 dex for [N
\textsc{ii}]-BPT selection. The dispersions are smaller than the scatter in differential
metallicities before applying correction (i.e. 0.05 dex).

\indent Like O3N2, O3S2 also shows comparable metallicity dispersion for \HII-DIG/LIER samples
obtained from both selection criterion, i.e. 0.06 dex for [S \textsc{ii}]-BPT selection and 0.07
dex for [N \textsc{ii}]-BPT selection. Although the dispersions after correction are still larger
than those obtained from O3N2, the improvement is significant with respect to the dispersions
before the correction (0.09~dex).
However, the [N \textsc{ii}]-BPT selected pairs shows a clear bimodal distribution of differential metallicities obtained from O3S2
after correction, pointing towards the systematic biases in the correction resulting from the bimodal distribution described in
Section \ref{section:impact}. Hence, we recommend not to use O3S2 calibration and associated correction when DIG/LIER/Seyferts are
classified using Kauffmann line. Instead, if using [N \textsc{ii}]-BPT we recommend using O3N2 calibration along with the associated correction. 

\indent We also point out that [S \textsc{ii}]-BPT selected sample indicates a Gaussian distribution of differential metallicities after correction for both diagnostics (O3N2 and O3S2), which emphasises that [S \textsc{ii}]-BPT is a better discriminator of \HII~regions and DIG/LIER/Seyfert in comparison to [N \textsc{ii}]-BPT. A comparison of O3N2 and O3S2 calibrations after corrections are presented in the supplementary material online (Appendix \textcolor{blue}{B}).

\subsection{Applicability of new calibrations}
\label{section:applications}
\indent The calibrations derived here have a wide range of applicability. Firstly, since these calibrations have been derived using
spatially-resolved data with line ratios corresponding to DIG/LIER/Seyfert, these diagnostics will now allow us to reliably
estimate the spatially-resolved abundances of the ionised gas which are not directly associated with ongoing star-formation but
probably to the older stellar population. Until now, studies of metallicity gradients within galaxies have not taken into account
the \HII~ and DIG components of the ISM, even though spatially resolved studies have revealed that both star-forming galaxies and
quiescent galaxies host DIG/LIER emission extending at kpc scales \citep{Belfiore2017}. \citet{Zhang2017} show that biases
introduced by DIG may be as large as the gradient in radial metallicity itself. Such biases can now be dealt with by using the
corrections derived in this work: either by directly using the O3N2 diagnostic (which we have shown to be the least deviating
in the DIG/LIER regions) in the case of poorly resolved galaxies, or,
if the DIG/LIER regions are properly resolved from the \HII-regions, by applying
our new calibrations. By doing so, spatial correlations of metallicity and other physical properties (e.g. stellar mass surface
density, star-formation rate surface density)
may also be studied more reliably, allowing for a deeper insight on the physics of the ISM.

\indent Secondly, the new calibrations particularly derived from [S~\textsc{ii}]-BPT selection (Equations
\ref{eq:corr o3n2 sii} and \ref{eq:corr o3s2 sii}) may be applied to a wide variety of late-type and early-type
galaxies as well. Using spatially-resolved  [S \textsc{ii}]-BPT maps of MaNGA galaxies, \citet{Belfiore2017}
established that  cLIERs (characterized by LIER-like emission in the central region)
are late-type galaxies lying on the transition green valley region between the main star-forming sequence and
quiescent galaxies (generally with more massive bulges), while eLIERs (i.e. those with LIER-like emission
extending across the entire galaxy) 
are similar to early-type galaxies and the passive galaxies devoid of line emission. Our new calibrations will enable to extend
the metallicity scaling relation studies to these class of galaxies, whose metallicity scaling relation have been so far
limited to the stellar component \citep[e.g.][]{Peng2015}.

\indent Although the ionisation source of DIG/LIERs is not completely understood, the most favoured scenario (at least in local normal galaxies)
is that it is associated with the hard photoionizing radiation produced by evolved stellar populations (post-AGB stars), filtered and hardened UV
radiation from star forming regions, and weak radiatively inefficient AGNs, combined with low ionization parameter. Even LIER-like emission
associated with shocks can be included in the same category, as eventually shock ionization and excitation results from the hard emission produced
by the hot gas in the post-shock gas \citep{Allen2008}. Line ratios over the LIER--Seyfert regions of the BPT diagram can therefore be generally seen
as overall associated with hard ionizing radiation, with the variations of the ionization parameter being primarily responsible for the
variations of the nebular line ratios in this region, together with secondary effect such as detailed shape of the ionizing radiation field,
pressure, density, etc. The fact that we find that the O3N2 and O3S2 diagnostics in the DIG/LIER region show a smooth regular trend with O3,
spilling well into the Seyfert regions, suggest that these diagnostics and their correction can likely be extended to the Seyfert region.
Of course, this should be tested further by extending our investigation to a sample of nearby Seyfert galaxies,  with ionization cones and Narrow Line Regions (NLR) distributed
on the galactic disc (not out of the galactic plane) so that similar \HII-regions/Seyfert-NLR pairs can be identified to further test and/or
correct/expand these metallicity calibrations.

\section{Summary \& Conclusion}
\label{section:summary}

\indent In this paper, we utilise the integral field spectroscopic data of 24 nearby spiral galaxies from MUSE to study the biases in the metallicities of  DIG/LIER-dominated regions when strong line calibrators N2S2H$\alpha$ \citep{Dopita2016}, O3N2 \citep{Curti2017} and O3S2 (Curti et al in prep) are used. We define close by \HII-DIG/LIER pairs which are expected to have the same metallicity, and compare their nominal metallicites using above three classical strong line metallicity diagnostics devised for \HII ~regions. We also present suitable methods to correct the observed biases hence providing calibrations for DIG/LIERs. Our main findings are summarised below:

\begin{enumerate}
	
	\item We separate the H \textsc{ii} regions and DIG/LIER/Seyfert region using four different criteria: [S \textsc{ii}]-BPT \citep{Kewley2001}, [N~\textsc{ii}]-BPT \citep{Kauffmann2003}, $\Sigma_{\rm{H\alpha}}$ and EW$_{\rm{H\alpha}}$. We find that BPT diagrams provide cleaner separation than threshold cuts on $\Sigma_{\rm{H\alpha}}$ or on EW$_{\rm{H\alpha}}$. [S \textsc{ii}]-BPT proves to be a better discriminator than [N \textsc{ii}]-BPT. 
	
	%\item We define close by \HII-DIG/LIER pairs with similar gas-phase abundance, and compare their nominal metallicites using three classical strong line metallicity diagnostics (O3N2, O3S2 and N2S2H$\alpha$) devised for \HII ~regions.
	
	\item The original O3N2 calibration derived for \HII~regions appears to be the best metallicity diagnostic among the three diagnostics tested here, as the average differential metallicity predicted by O3N2 diagnostic vary by only 0.01--0.04 dex for the close by \HII~regions-DIG/Seyfert/LIER regions. Moreover, the dispersion of the differential metallicites for \HII-DIG/LIER pairs is small (0.05 dex) irrespective of the method for selecting \HII-DIG/LIER pairs.
	
	\item The original O3S2 diagnostic is very similar to R$_{23}$ but less affected by redenning. This calibration shows large mean offsets (0.11--0.13 dex) in differential metallicities along with large dispersion (0.09 dex), for both samples of \HII-DIG/LIER pairs selected from either [S \textsc{ii}]-BPT or [N \textsc{ii}]-BPT. However, pairs selected on the basis of $\Sigma_{\rm{H\alpha}}$ show negligible mean offset but considerable scatter (0.08 dex) in the distribution of differential metallicities from O3S2. 
	
	\item N2S2H$\alpha$ diagnostic underestimates the metallicities of DIG/LIER/Seyfert by 0.06--0.1 dex compared to the close by \HII~region counterpart when \HII-DIG/LIER pairs are selected using [S \textsc{ii}]-BPT or $\Sigma_{\rm{H\alpha}}$. No significant offset (0.03 dex) is observed in the differential metallicites of \HII-DIG/LIER pairs selected from [N~\textsc{ii}]-BPT. However, dispersion is always found to be large (0.08--0.16 dex) irrespective of the selection criterion used. Unfortunately, the observed dispersion shown by N2S2H$\alpha$ can not be corrected through any obvious trend.

	\item We estimate the correction offsets in O3N2 and O3S2 metallicity diagnostics as a function of [O \textsc{iii}]/H$\beta$ line ratio, i.e. $\Delta log Z = aO3 + b$, where a and b are constants. For [S \textsc{ii}]-BPT selected DIG/LIER/Seyferts, corrections are given by equations \ref{eq:corr o3n2 sii} and \ref{eq:corr o3s2 sii}, while for [N \textsc{ii}]-BPT selected DIG/LIER/Seyferts, corrections are given by equations \ref{eq:corr o3n2 nii} and \ref{eq:corr o3s2 nii}.
	 
	 \item After correcting the O3N2 and O3S2 derived metallicities using equations \ref{eq:cal o3n2 sii}, \ref{eq:cal o3s2 sii}, \ref{eq:cal o3n2 nii} and \ref{eq:cal o3s2 nii}, no offset is found in differential metallicity distributions and scatters are considerably reduced. Corrected O3N2 shows dispersions of 0.04 dex and 0.03 dex for [S~\textsc{ii}]-BPT and [N \textsc{ii}]-BPT selected \HII-DIG/LIER pairs, respectively. Corrected O3S2 shows higher dispersions (than corrected O3N2 diagnostic), i.e. 0.05  and 0.07 dex for [S~\textsc{ii}]-BPT and [N \textsc{ii}]-BPT selected \HII-DIG/LIER pairs, respectively. 
	 
	 \item We recommend using the corrected O3N2 diagnostic (Equations \ref{eq:cal o3n2 sii} and \ref{eq:cal o3n2 nii}) for estimating metallicities of DIG/LIER/Seyfert like regions, irrespective of the BPT digram used for identifying such regions. The corrected O3S2 diagnostic is also a viable option though it should be used only when DIG/LIER/Seyfert like regions is identified using [S~\textsc{ii}]-BPT. We need to perform further experiments to confirm if these calibrations can be used for estimating metallicities of Seyfert galaxies.
	
\end{enumerate}

\indent The new calibrations derived in this work will allow future studies to reliably measure the spatially-resolved gas-phase abundances of DIG/LIER/Seyfert like regions in nearby star-forming galaxies but also gas-phase abundances of a wide variety of galaxies including early-type and passive quiescent galaxies (with nebular emission). This will enable extending the studies of chemical composition of the ISM in various types of galaxies, including the interlink between chemical abundance, star-formation and gas flows, which will further our understanding of galaxy formation and evolution. 

\section*{Acknowledgements}
\indent  We thank the anonymous referee for a useful and supportive report.  It is a pleasure to thank Mike Irwin for discussions on multi-regression analysis. NK acknowledges the financial support from the Institute of Astronomy, Cambridge, the Nehru Trust for Cambridge University during the PhD and the Schlumberger Foundation for the post-doctoral research. R.M. and M.C. acknowledge ERC Advanced Grant 695671 ``QUENCH" and support by the Science and Technology Facilities Council (STFC). This research made use of the NASA/IPAC Extragalactic Database (NED) which is operated by the Jet Propulsion Laboratory, California Institute of Technology, under contract with the National Aeronautics and Space Administration; SAOImage DS9, developed by Smithsonian Astrophysical Observatory"; Astropy, a community-developed core Python package for Astronomy \citep{Astropy2013}. Based on data products from observations made with ESO Telescopes at the La Silla Paranal Observatory under programme ID 095.B-0532(A), 096.B-0309(A), 097.B-0165(A).

%%%%%%%%%%%%%%%%%%%%%%%%%%%%%%%%%%%%%%%%%%%%%%%%%%



\bibliographystyle{mnras}
\bibliography{biblio} % if your bibtex file is called example.bib

\begin{thebibliography}{}
\makeatletter
\relax
\def\mn@urlcharsother{\let\do\@makeother \do\$\do\&\do\#\do\^\do\_\do\%\do\~}
\def\mn@doi{\begingroup\mn@urlcharsother \@ifnextchar [ {\mn@doi@}
  {\mn@doi@[]}}
\def\mn@doi@[#1]#2{\def\@tempa{#1}\ifx\@tempa\@empty \href
  {http://dx.doi.org/#2} {doi:#2}\else \href {http://dx.doi.org/#2} {#1}\fi
  \endgroup}
\def\mn@eprint#1#2{\mn@eprint@#1:#2::\@nil}
\def\mn@eprint@arXiv#1{\href {http://arxiv.org/abs/#1} {{\tt arXiv:#1}}}
\def\mn@eprint@dblp#1{\href {http://dblp.uni-trier.de/rec/bibtex/#1.xml}
  {dblp:#1}}
\def\mn@eprint@#1:#2:#3:#4\@nil{\def\@tempa {#1}\def\@tempb {#2}\def\@tempc
  {#3}\ifx \@tempc \@empty \let \@tempc \@tempb \let \@tempb \@tempa \fi \ifx
  \@tempb \@empty \def\@tempb {arXiv}\fi \@ifundefined
  {mn@eprint@\@tempb}{\@tempb:\@tempc}{\expandafter \expandafter \csname
  mn@eprint@\@tempb\endcsname \expandafter{\@tempc}}}

\bibitem[\protect\citeauthoryear{{Allen}, {Groves}, {Dopita}, {Sutherland}  \&
  {Kewley}}{{Allen} et~al.}{2008}]{Allen2008}
{Allen} M.~G.,  {Groves} B.~A.,  {Dopita} M.~A.,  {Sutherland} R.~S.,
  {Kewley} L.~J.,  2008, \mn@doi [\apjs] {10.1086/589652}, \href
  {http://adsabs.harvard.edu/abs/2008ApJS..178...20A} {178, 20}

\bibitem[\protect\citeauthoryear{{Aller}}{{Aller}}{1942}]{Aller1942}
{Aller} L.~H.,  1942, \mn@doi [\apj] {10.1086/144372}, \href
  {http://adsabs.harvard.edu/abs/1942ApJ....95...52A} {95, 52}

\bibitem[\protect\citeauthoryear{{Alloin}, {Collin-Souffrin}, {Joly}  \&
  {Vigroux}}{{Alloin} et~al.}{1979}]{Alloin1979}
{Alloin} D.,  {Collin-Souffrin} S.,  {Joly} M.,   {Vigroux} L.,  1979, \aap,
  \href {http://adsabs.harvard.edu/abs/1979A%26A....78..200A} {78, 200}

\bibitem[\protect\citeauthoryear{{Astropy Collaboration} et~al.,}{{Astropy
  Collaboration} et~al.}{2013}]{Astropy2013}
{Astropy Collaboration} et~al., 2013, \mn@doi [\aap]
  {10.1051/0004-6361/201322068}, \href
  {http://adsabs.harvard.edu/abs/2013A%26A...558A..33A} {558, A33}

\bibitem[\protect\citeauthoryear{{Baldwin}, {Phillips}  \&
  {Terlevich}}{{Baldwin} et~al.}{1981}]{Baldwin1981}
{Baldwin} J.~A.,  {Phillips} M.~M.,   {Terlevich} R.,  1981, \mn@doi [\pasp]
  {10.1086/130766}, \href {http://adsabs.harvard.edu/abs/1981PASP...93....5B}
  {93, 5}

\bibitem[\protect\citeauthoryear{{Belfiore} et~al.,}{{Belfiore}
  et~al.}{2016}]{Belfiore2016}
{Belfiore} F.,  et~al., 2016, \mn@doi [\mnras] {10.1093/mnras/stw1234}, \href
  {http://adsabs.harvard.edu/abs/2016MNRAS.461.3111B} {461, 3111}

\bibitem[\protect\citeauthoryear{{Belfiore} et~al.,}{{Belfiore}
  et~al.}{2017}]{Belfiore2017}
{Belfiore} F.,  et~al., 2017, \mn@doi [\mnras] {10.1093/mnras/stw3211}, \href
  {http://adsabs.harvard.edu/abs/2017MNRAS.466.2570B} {466, 2570}

\bibitem[\protect\citeauthoryear{{Berg}, {Skillman}, {Garnett}, {Croxall},
  {Marble}, {Smith}, {Gordon}  \& {Kennicutt}}{{Berg} et~al.}{2013}]{Berg2013}
{Berg} D.~A.,  {Skillman} E.~D.,  {Garnett} D.~R.,  {Croxall} K.~V.,  {Marble}
  A.~R.,  {Smith} J.~D.,  {Gordon} K.,   {Kennicutt} Jr. R.~C.,  2013, \mn@doi
  [\apj] {10.1088/0004-637X/775/2/128}, \href
  {http://adsabs.harvard.edu/abs/2013ApJ...775..128B} {775, 128}

\bibitem[\protect\citeauthoryear{{Berg}, {Skillman}, {Croxall}, {Pogge},
  {Moustakas}  \& {Johnson-Groh}}{{Berg} et~al.}{2015}]{Berg2015}
{Berg} D.~A.,  {Skillman} E.~D.,  {Croxall} K.~V.,  {Pogge} R.~W.,  {Moustakas}
  J.,   {Johnson-Groh} M.,  2015, \mn@doi [\apj] {10.1088/0004-637X/806/1/16},
  \href {http://adsabs.harvard.edu/abs/2015ApJ...806...16B} {806, 16}

\bibitem[\protect\citeauthoryear{{Binette}, {Magris}, {Stasi{\'n}ska}  \&
  {Bruzual}}{{Binette} et~al.}{1994}]{Binette1994}
{Binette} L.,  {Magris} C.~G.,  {Stasi{\'n}ska} G.,   {Bruzual} A.~G.,  1994,
  \aap, \href {http://adsabs.harvard.edu/abs/1994A%26A...292...13B} {292, 13}

\bibitem[\protect\citeauthoryear{{Blanc}, {Heiderman}, {Gebhardt}, {Evans}  \&
  {Adams}}{{Blanc} et~al.}{2009}]{Blanc2009}
{Blanc} G.~A.,  {Heiderman} A.,  {Gebhardt} K.,  {Evans} II N.~J.,   {Adams}
  J.,  2009, \mn@doi [\apj] {10.1088/0004-637X/704/1/842}, \href
  {http://adsabs.harvard.edu/abs/2009ApJ...704..842B} {704, 842}

\bibitem[\protect\citeauthoryear{{Blanc}, {Kewley}, {Vogt}  \&
  {Dopita}}{{Blanc} et~al.}{2015}]{Blanc2015}
{Blanc} G.~A.,  {Kewley} L.,  {Vogt} F.~P.~A.,   {Dopita} M.~A.,  2015, \mn@doi
  [\apj] {10.1088/0004-637X/798/2/99}, \href
  {http://adsabs.harvard.edu/abs/2015ApJ...798...99B} {798, 99}

\bibitem[\protect\citeauthoryear{{Cappellari}}{{Cappellari}}{2017}]{Cappellari2017}
{Cappellari} M.,  2017, \mn@doi [\mnras] {10.1093/mnras/stw3020}, \href
  {http://adsabs.harvard.edu/abs/2017MNRAS.466..798C} {466, 798}

\bibitem[\protect\citeauthoryear{{Cappellari} \& {Copin}}{{Cappellari} \&
  {Copin}}{2003}]{CappellariCopin2003}
{Cappellari} M.,  {Copin} Y.,  2003, \mn@doi [\mnras]
  {10.1046/j.1365-8711.2003.06541.x}, \href
  {http://adsabs.harvard.edu/abs/2003MNRAS.342..345C} {342, 345}

\bibitem[\protect\citeauthoryear{{Cappellari} \& {Emsellem}}{{Cappellari} \&
  {Emsellem}}{2004}]{CappellariEmsellem2004}
{Cappellari} M.,  {Emsellem} E.,  2004, \mn@doi [\pasp] {10.1086/381875}, \href
  {http://adsabs.harvard.edu/abs/2004PASP..116..138C} {116, 138}

\bibitem[\protect\citeauthoryear{{Cid Fernandes}, {Stasi{\'n}ska}, {Mateus}  \&
  {Vale Asari}}{{Cid Fernandes} et~al.}{2011}]{CidFernandes2011}
{Cid Fernandes} R.,  {Stasi{\'n}ska} G.,  {Mateus} A.,   {Vale Asari} N.,
  2011, \mn@doi [\mnras] {10.1111/j.1365-2966.2011.18244.x}, \href
  {http://adsabs.harvard.edu/abs/2011MNRAS.413.1687C} {413, 1687}

\bibitem[\protect\citeauthoryear{{Curti}, {Cresci}, {Mannucci}, {Marconi},
  {Maiolino}  \& {Esposito}}{{Curti} et~al.}{2017}]{Curti2017}
{Curti} M.,  {Cresci} G.,  {Mannucci} F.,  {Marconi} A.,  {Maiolino} R.,
  {Esposito} S.,  2017, \mn@doi [\mnras] {10.1093/mnras/stw2766}, \href
  {http://adsabs.harvard.edu/abs/2017MNRAS.465.1384C} {465, 1384}

\bibitem[\protect\citeauthoryear{{Denicol{\'o}}, {Terlevich}  \&
  {Terlevich}}{{Denicol{\'o}} et~al.}{2002}]{Denicolo2002}
{Denicol{\'o}} G.,  {Terlevich} R.,   {Terlevich} E.,  2002, \mn@doi [\mnras]
  {10.1046/j.1365-8711.2002.05041.x}, \href
  {http://adsabs.harvard.edu/abs/2002MNRAS.330...69D} {330, 69}

\bibitem[\protect\citeauthoryear{{Dopita}, {Kewley}, {Sutherland}  \&
  {Nicholls}}{{Dopita} et~al.}{2016}]{Dopita2016}
{Dopita} M.~A.,  {Kewley} L.~J.,  {Sutherland} R.~S.,   {Nicholls} D.~C.,
  2016, \mn@doi [\apss] {10.1007/s10509-016-2657-8}, \href
  {http://adsabs.harvard.edu/abs/2016Ap%26SS.361...61D} {361, 61}

\bibitem[\protect\citeauthoryear{{Dors}, {Cardaci}, {H{\"a}gele}, {Rodrigues},
  {Grebel}, {Pilyugin}, {Freitas-Lemes}  \& {Krabbe}}{{Dors}
  et~al.}{2015}]{Dors2015}
{Dors} O.~L.,  {Cardaci} M.~V.,  {H{\"a}gele} G.~F.,  {Rodrigues} I.,  {Grebel}
  E.~K.,  {Pilyugin} L.~S.,  {Freitas-Lemes} P.,   {Krabbe} A.~C.,  2015,
  \mn@doi [\mnras] {10.1093/mnras/stv1916}, \href
  {http://ukads.nottingham.ac.uk/abs/2015MNRAS.453.4102D} {453, 4102}

\bibitem[\protect\citeauthoryear{{Ferland} et~al.,}{{Ferland}
  et~al.}{2013}]{Ferland2013}
{Ferland} G.~J.,  et~al., 2013, \rmxaa, \href
  {http://adsabs.harvard.edu/abs/2013RMxAA..49..137F} {49, 137}

\bibitem[\protect\citeauthoryear{{Freudling}, {Romaniello}, {Bramich},
  {Ballester}, {Forchi}, {Garc{\'{\i}}a-Dabl{\'o}}, {Moehler}  \&
  {Neeser}}{{Freudling} et~al.}{2013}]{Freudling2013}
{Freudling} W.,  {Romaniello} M.,  {Bramich} D.~M.,  {Ballester} P.,  {Forchi}
  V.,  {Garc{\'{\i}}a-Dabl{\'o}} C.~E.,  {Moehler} S.,   {Neeser} M.~J.,  2013,
  \mn@doi [\aap] {10.1051/0004-6361/201322494}, \href
  {http://adsabs.harvard.edu/abs/2013A%26A...559A..96F} {559, A96}

\bibitem[\protect\citeauthoryear{{Garnett}, {Kennicutt}  \&
  {Bresolin}}{{Garnett} et~al.}{2004}]{Garnett2004}
{Garnett} D.~R.,  {Kennicutt} Jr. R.~C.,   {Bresolin} F.,  2004, \mn@doi
  [\apjl] {10.1086/421489}, \href
  {http://adsabs.harvard.edu/abs/2004ApJ...607L..21G} {607, L21}

\bibitem[\protect\citeauthoryear{{Gomes} et~al.,}{{Gomes}
  et~al.}{2016}]{Gomes2016}
{Gomes} J.~M.,  et~al., 2016, \mn@doi [\aap] {10.1051/0004-6361/201525974},
  \href {http://adsabs.harvard.edu/abs/2016A%26A...585A..92G} {585, A92}

\bibitem[\protect\citeauthoryear{{Haffner}, {Reynolds}  \& {Tufte}}{{Haffner}
  et~al.}{1999}]{Haffner1999}
{Haffner} L.~M.,  {Reynolds} R.~J.,   {Tufte} S.~L.,  1999, \mn@doi [\apj]
  {10.1086/307734}, \href {http://adsabs.harvard.edu/abs/1999ApJ...523..223H}
  {523, 223}

\bibitem[\protect\citeauthoryear{{Ho} et~al.,}{{Ho} et~al.}{2017}]{Ho2017}
{Ho} I.-T.,  et~al., 2017, \mn@doi [\apj] {10.3847/1538-4357/aa8460}, \href
  {http://adsabs.harvard.edu/abs/2017ApJ...846...39H} {846, 39}

\bibitem[\protect\citeauthoryear{{Ho} et~al.,}{{Ho} et~al.}{2018}]{Ho2018}
{Ho} I.,  et~al., 2018, preprint, \href
  {http://adsabs.harvard.edu/abs/2018arXiv180702043H} {} (\mn@eprint {arXiv}
  {1807.02043})

\bibitem[\protect\citeauthoryear{{Kauffmann} et~al.,}{{Kauffmann}
  et~al.}{2003}]{Kauffmann2003}
{Kauffmann} G.,  et~al., 2003, \mn@doi [\mnras]
  {10.1111/j.1365-2966.2003.07154.x}, \href
  {http://adsabs.harvard.edu/abs/2003MNRAS.346.1055K} {346, 1055}

\bibitem[\protect\citeauthoryear{{Kennicutt}}{{Kennicutt}}{1984}]{Kennicutt1984}
{Kennicutt} Jr. R.~C.,  1984, \mn@doi [\apj] {10.1086/162669}, \href
  {http://adsabs.harvard.edu/abs/1984ApJ...287..116K} {287, 116}

\bibitem[\protect\citeauthoryear{{Kewley} \& {Dopita}}{{Kewley} \&
  {Dopita}}{2002}]{Kewley2002}
{Kewley} L.~J.,  {Dopita} M.~A.,  2002, \mn@doi [\apjs] {10.1086/341326}, \href
  {http://adsabs.harvard.edu/abs/2002ApJS..142...35K} {142, 35}

\bibitem[\protect\citeauthoryear{{Kewley} \& {Ellison}}{{Kewley} \&
  {Ellison}}{2008}]{Kewley2008}
{Kewley} L.~J.,  {Ellison} S.~L.,  2008, \mn@doi [\apj] {10.1086/587500}, \href
  {http://adsabs.harvard.edu/abs/2008ApJ...681.1183K} {681, 1183}

\bibitem[\protect\citeauthoryear{{Kewley}, {Dopita}, {Sutherland}, {Heisler}
  \& {Trevena}}{{Kewley} et~al.}{2001}]{Kewley2001}
{Kewley} L.~J.,  {Dopita} M.~A.,  {Sutherland} R.~S.,  {Heisler} C.~A.,
  {Trevena} J.,  2001, \mn@doi [\apj] {10.1086/321545}, \href
  {http://adsabs.harvard.edu/abs/2001ApJ...556..121K} {556, 121}

\bibitem[\protect\citeauthoryear{{Kewley}, {Groves}, {Kauffmann}  \&
  {Heckman}}{{Kewley} et~al.}{2006}]{Kewley2006}
{Kewley} L.~J.,  {Groves} B.,  {Kauffmann} G.,   {Heckman} T.,  2006, \mn@doi
  [\mnras] {10.1111/j.1365-2966.2006.10859.x}, \href
  {http://adsabs.harvard.edu/abs/2006MNRAS.372..961K} {372, 961}

\bibitem[\protect\citeauthoryear{{Kumari}, {James}  \& {Irwin}}{{Kumari}
  et~al.}{2017}]{Kumari2017}
{Kumari} N.,  {James} B.~L.,   {Irwin} M.~J.,  2017, \mn@doi [\mnras]
  {10.1093/mnras/stx1414}, \href
  {http://adsabs.harvard.edu/abs/2017MNRAS.470.4618K} {470, 4618}

\bibitem[\protect\citeauthoryear{{Kumari}, {James}, {Irwin}, {Amor{\'{\i}}n}
  \& {P{\'e}rez-Montero}}{{Kumari} et~al.}{2018}]{Kumari2018}
{Kumari} N.,  {James} B.~L.,  {Irwin} M.~J.,  {Amor{\'{\i}}n} R.,
  {P{\'e}rez-Montero} E.,  2018, \mn@doi [\mnras] {10.1093/mnras/sty402}, \href
  {http://ukads.nottingham.ac.uk/abs/2018MNRAS.476.3793K} {476, 3793}

\bibitem[\protect\citeauthoryear{{Li}, {Bresolin}  \& {Kennicutt}}{{Li}
  et~al.}{2013}]{Li2013}
{Li} Y.,  {Bresolin} F.,   {Kennicutt} Jr. R.~C.,  2013, \mn@doi [\apj]
  {10.1088/0004-637X/766/1/17}, \href
  {http://adsabs.harvard.edu/abs/2013ApJ...766...17L} {766, 17}

\bibitem[\protect\citeauthoryear{{Madsen}, {Reynolds}  \& {Haffner}}{{Madsen}
  et~al.}{2006}]{Madsen2006}
{Madsen} G.~J.,  {Reynolds} R.~J.,   {Haffner} L.~M.,  2006, \mn@doi [\apj]
  {10.1086/508441}, \href {http://adsabs.harvard.edu/abs/2006ApJ...652..401M}
  {652, 401}

\bibitem[\protect\citeauthoryear{{Maiolino} \& {Mannucci}}{{Maiolino} \&
  {Mannucci}}{2018}]{Maiolino2018}
{Maiolino} R.,  {Mannucci} F.,  2018, arXiv e-prints, \href
  {http://ukads.nottingham.ac.uk/abs/2018arXiv181109642M} {}

\bibitem[\protect\citeauthoryear{{Maiolino} et~al.,}{{Maiolino}
  et~al.}{2008}]{Maiolino2008}
{Maiolino} R.,  et~al., 2008, \mn@doi [\aap] {10.1051/0004-6361:200809678},
  \href {http://adsabs.harvard.edu/abs/2008A%26A...488..463M} {488, 463}

\bibitem[\protect\citeauthoryear{{Mannucci}, {Cresci}, {Maiolino}, {Marconi}
  \& {Gnerucci}}{{Mannucci} et~al.}{2010}]{Mannucci2010}
{Mannucci} F.,  {Cresci} G.,  {Maiolino} R.,  {Marconi} A.,   {Gnerucci} A.,
  2010, \mn@doi [\mnras] {10.1111/j.1365-2966.2010.17291.x}, \href
  {http://adsabs.harvard.edu/abs/2010MNRAS.408.2115M} {408, 2115}

\bibitem[\protect\citeauthoryear{{Marino} et~al.,}{{Marino}
  et~al.}{2013}]{Marino2013}
{Marino} R.~A.,  et~al., 2013, \mn@doi [\aap] {10.1051/0004-6361/201321956},
  \href {http://adsabs.harvard.edu/abs/2013A%26A...559A.114M} {559, A114}

\bibitem[\protect\citeauthoryear{{Nagao}, {Maiolino}  \& {Marconi}}{{Nagao}
  et~al.}{2006}]{Nagao2006}
{Nagao} T.,  {Maiolino} R.,   {Marconi} A.,  2006, \mn@doi [\aap]
  {10.1051/0004-6361:20065216}, \href
  {http://adsabs.harvard.edu/abs/2006A%26A...459...85N} {459, 85}

\bibitem[\protect\citeauthoryear{{Oey} \& {Shields}}{{Oey} \&
  {Shields}}{2000}]{Oey2000}
{Oey} M.~S.,  {Shields} J.~C.,  2000, \mn@doi [\apj] {10.1086/309276}, \href
  {http://adsabs.harvard.edu/abs/2000ApJ...539..687O} {539, 687}

\bibitem[\protect\citeauthoryear{{Oey} et~al.,}{{Oey} et~al.}{2007}]{Oey2007}
{Oey} M.~S.,  et~al., 2007, \mn@doi [\apj] {10.1086/517867}, \href
  {http://adsabs.harvard.edu/abs/2007ApJ...661..801O} {661, 801}

\bibitem[\protect\citeauthoryear{{Pagel}, {Edmunds}, {Blackwell}, {Chun}  \&
  {Smith}}{{Pagel} et~al.}{1979}]{Pagel1979}
{Pagel} B.~E.~J.,  {Edmunds} M.~G.,  {Blackwell} D.~E.,  {Chun} M.~S.,
  {Smith} G.,  1979, \mn@doi [\mnras] {10.1093/mnras/189.1.95}, \href
  {http://adsabs.harvard.edu/abs/1979MNRAS.189...95P} {189, 95}

\bibitem[\protect\citeauthoryear{{Peng}, {Maiolino}  \& {Cochrane}}{{Peng}
  et~al.}{2015}]{Peng2015}
{Peng} Y.,  {Maiolino} R.,   {Cochrane} R.,  2015, \mn@doi [\nat]
  {10.1038/nature14439}, \href
  {http://adsabs.harvard.edu/abs/2015Natur.521..192P} {521, 192}

\bibitem[\protect\citeauthoryear{{P{\'e}rez-Montero}}{{P{\'e}rez-Montero}}{2014}]{Perez-Montero2014}
{P{\'e}rez-Montero} E.,  2014, \mn@doi [\mnras] {10.1093/mnras/stu753}, \href
  {http://cdsads.u-strasbg.fr/abs/2014MNRAS.441.2663P} {441, 2663}

\bibitem[\protect\citeauthoryear{{P{\'e}rez-Montero} \&
  {Contini}}{{P{\'e}rez-Montero} \& {Contini}}{2009}]{Perez-Montero2009}
{P{\'e}rez-Montero} E.,  {Contini} T.,  2009, \mn@doi [\mnras]
  {10.1111/j.1365-2966.2009.15145.x}, \href
  {http://adsabs.harvard.edu/abs/2009MNRAS.398..949P} {398, 949}

\bibitem[\protect\citeauthoryear{{P{\'e}rez-Montero}, {H{\"a}gele}, {Contini}
  \& {D{\'{\i}}az}}{{P{\'e}rez-Montero} et~al.}{2007}]{Perez-Montero2007a}
{P{\'e}rez-Montero} E.,  {H{\"a}gele} G.~F.,  {Contini} T.,   {D{\'{\i}}az}
  {\'A}.~I.,  2007, \mn@doi [\mnras] {10.1111/j.1365-2966.2007.12213.x}, \href
  {http://adsabs.harvard.edu/abs/2007MNRAS.381..125P} {381, 125}

\bibitem[\protect\citeauthoryear{{Pettini} \& {Pagel}}{{Pettini} \&
  {Pagel}}{2004}]{Pettini2004}
{Pettini} M.,  {Pagel} B.~E.~J.,  2004, \mn@doi [\mnras]
  {10.1111/j.1365-2966.2004.07591.x}, \href
  {http://adsabs.harvard.edu/abs/2004MNRAS.348L..59P} {348, L59}

\bibitem[\protect\citeauthoryear{{Reynolds}}{{Reynolds}}{1984}]{Reynolds1984}
{Reynolds} R.~J.,  1984, \mn@doi [\apj] {10.1086/162190}, \href
  {http://adsabs.harvard.edu/abs/1984ApJ...282..191R} {282, 191}

\bibitem[\protect\citeauthoryear{{Reynolds}}{{Reynolds}}{1990}]{Reynolds1990}
{Reynolds} R.~J.,  1990, \mn@doi [\apjl] {10.1086/185640}, \href
  {http://adsabs.harvard.edu/abs/1990ApJ...349L..17R} {349, L17}

\bibitem[\protect\citeauthoryear{{Ricciardelli}, {Vazdekis}, {Cenarro}  \&
  {Falc{\'o}n-Barroso}}{{Ricciardelli} et~al.}{2012}]{Ricciardelli2012}
{Ricciardelli} E.,  {Vazdekis} A.,  {Cenarro} A.~J.,   {Falc{\'o}n-Barroso} J.,
   2012, \mn@doi [\mnras] {10.1111/j.1365-2966.2012.21178.x}, \href
  {http://adsabs.harvard.edu/abs/2012MNRAS.424..172R} {424, 172}

\bibitem[\protect\citeauthoryear{{Rossa} \& {Dettmar}}{{Rossa} \&
  {Dettmar}}{2003a}]{RossaDettmar2003a}
{Rossa} J.,  {Dettmar} R.-J.,  2003a, \mn@doi [\aap]
  {10.1051/0004-6361:20030615}, \href
  {http://adsabs.harvard.edu/abs/2003A%26A...406..493R} {406, 493}

\bibitem[\protect\citeauthoryear{{Rossa} \& {Dettmar}}{{Rossa} \&
  {Dettmar}}{2003b}]{RossaDettmar2003b}
{Rossa} J.,  {Dettmar} R.-J.,  2003b, \mn@doi [\aap]
  {10.1051/0004-6361:20030698}, \href
  {http://adsabs.harvard.edu/abs/2003A%26A...406..505R} {406, 505}

\bibitem[\protect\citeauthoryear{{S{\'a}nchez Almeida},
  {Mu{\~n}oz-Tu{\~n}{\'o}n}, {Elmegreen}, {Elmegreen}  \&
  {M{\'e}ndez-Abreu}}{{S{\'a}nchez Almeida} et~al.}{2013}]{Sanchez2013}
{S{\'a}nchez Almeida} J.,  {Mu{\~n}oz-Tu{\~n}{\'o}n} C.,  {Elmegreen} D.~M.,
  {Elmegreen} B.~G.,   {M{\'e}ndez-Abreu} J.,  2013, \mn@doi [\apj]
  {10.1088/0004-637X/767/1/74}, \href
  {http://adsabs.harvard.edu/abs/2013ApJ...767...74S} {767, 74}

\bibitem[\protect\citeauthoryear{{Sanders}, {Shapley}, {Zhang}  \&
  {Yan}}{{Sanders} et~al.}{2017}]{Sanders2017}
{Sanders} R.~L.,  {Shapley} A.~E.,  {Zhang} K.,   {Yan} R.,  2017, \mn@doi
  [\apj] {10.3847/1538-4357/aa93e4}, \href
  {http://adsabs.harvard.edu/abs/2017ApJ...850..136S} {850, 136}

\bibitem[\protect\citeauthoryear{{Sarzi} et~al.,}{{Sarzi}
  et~al.}{2010}]{Sarzi2010}
{Sarzi} M.,  et~al., 2010, \mn@doi [\mnras] {10.1111/j.1365-2966.2009.16039.x},
  \href {http://adsabs.harvard.edu/abs/2010MNRAS.402.2187S} {402, 2187}

\bibitem[\protect\citeauthoryear{{Schlafly} \& {Finkbeiner}}{{Schlafly} \&
  {Finkbeiner}}{2011}]{Schlafly2011}
{Schlafly} E.~F.,  {Finkbeiner} D.~P.,  2011, \mn@doi [\apj]
  {10.1088/0004-637X/737/2/103}, \href
  {http://adsabs.harvard.edu/abs/2011ApJ...737..103S} {737, 103}

\bibitem[\protect\citeauthoryear{{Singh} et~al.,}{{Singh}
  et~al.}{2013}]{Singh2013}
{Singh} R.,  et~al., 2013, \mn@doi [\aap] {10.1051/0004-6361/201322062}, \href
  {http://adsabs.harvard.edu/abs/2013A%26A...558A..43S} {558, A43}

\bibitem[\protect\citeauthoryear{{Stasi{\'n}ska}}{{Stasi{\'n}ska}}{2005}]{Stasinska2005}
{Stasi{\'n}ska} G.,  2005, \mn@doi [\aap] {10.1051/0004-6361:20042216}, \href
  {http://adsabs.harvard.edu/abs/2005A%26A...434..507S} {434, 507}

\bibitem[\protect\citeauthoryear{{Stasi{\'n}ska}}{{Stasi{\'n}ska}}{2006}]{Stasinska2006}
{Stasi{\'n}ska} G.,  2006, \mn@doi [\aap] {10.1051/0004-6361:20065516}, \href
  {http://adsabs.harvard.edu/abs/2006A%26A...454L.127S} {454, L127}

\bibitem[\protect\citeauthoryear{{Stasi{\'n}ska} et~al.,}{{Stasi{\'n}ska}
  et~al.}{2008}]{Stasinska2008}
{Stasi{\'n}ska} G.,  et~al., 2008, \mn@doi [\mnras]
  {10.1111/j.1745-3933.2008.00550.x}, \href
  {http://adsabs.harvard.edu/abs/2008MNRAS.391L..29S} {391, L29}

\bibitem[\protect\citeauthoryear{{Storchi-Bergmann}, {Calzetti}  \&
  {Kinney}}{{Storchi-Bergmann} et~al.}{1994}]{Storchi-Bergmann1994}
{Storchi-Bergmann} T.,  {Calzetti} D.,   {Kinney} A.~L.,  1994, \mn@doi [\apj]
  {10.1086/174345}, \href {http://adsabs.harvard.edu/abs/1994ApJ...429..572S}
  {429, 572}

\bibitem[\protect\citeauthoryear{{Sutherland} \& {Dopita}}{{Sutherland} \&
  {Dopita}}{1993}]{Sutherland1993}
{Sutherland} R.~S.,  {Dopita} M.~A.,  1993, \mn@doi [\apjs] {10.1086/191823},
  \href {http://adsabs.harvard.edu/abs/1993ApJS...88..253S} {88, 253}

\bibitem[\protect\citeauthoryear{{Tremonti} et~al.,}{{Tremonti}
  et~al.}{2004}]{Tremonti2004}
{Tremonti} C.~A.,  et~al., 2004, \mn@doi [\apj] {10.1086/423264}, \href
  {http://adsabs.harvard.edu/abs/2004ApJ...613..898T} {613, 898}

\bibitem[\protect\citeauthoryear{{Vale Asari}, {Stasi{\'n}ska}, {Morisset}  \&
  {Cid Fernandes}}{{Vale Asari} et~al.}{2016}]{ValeAsari2016}
{Vale Asari} N.,  {Stasi{\'n}ska} G.,  {Morisset} C.,   {Cid Fernandes} R.,
  2016, \mn@doi [\mnras] {10.1093/mnras/stw971}, \href
  {http://adsabs.harvard.edu/abs/2016MNRAS.460.1739V} {460, 1739}

\bibitem[\protect\citeauthoryear{{Vazdekis}, {Ricciardelli}, {Cenarro},
  {Rivero-Gonz{\'a}lez}, {D{\'{\i}}az-Garc{\'{\i}}a}  \&
  {Falc{\'o}n-Barroso}}{{Vazdekis} et~al.}{2012}]{Vazdekis2012}
{Vazdekis} A.,  {Ricciardelli} E.,  {Cenarro} A.~J.,  {Rivero-Gonz{\'a}lez}
  J.~G.,  {D{\'{\i}}az-Garc{\'{\i}}a} L.~A.,   {Falc{\'o}n-Barroso} J.,  2012,
  \mn@doi [\mnras] {10.1111/j.1365-2966.2012.21179.x}, \href
  {http://adsabs.harvard.edu/abs/2012MNRAS.424..157V} {424, 157}

\bibitem[\protect\citeauthoryear{{Voges} \& {Walterbos}}{{Voges} \&
  {Walterbos}}{2006}]{Voges2006}
{Voges} E.~S.,  {Walterbos} R.~A.~M.,  2006, \mn@doi [\apjl] {10.1086/505575},
  \href {http://adsabs.harvard.edu/abs/2006ApJ...644L..29V} {644, L29}

\bibitem[\protect\citeauthoryear{{Yan}}{{Yan}}{2018}]{Yan2018}
{Yan} R.,  2018, \mn@doi [\mnras] {10.1093/mnras/sty2143}, \href
  {http://ukads.nottingham.ac.uk/abs/2018MNRAS.481..476Y} {481, 476}

\bibitem[\protect\citeauthoryear{{Yan} \& {Blanton}}{{Yan} \&
  {Blanton}}{2012}]{Yan2012}
{Yan} R.,  {Blanton} M.~R.,  2012, \mn@doi [\apj] {10.1088/0004-637X/747/1/61},
  \href {http://adsabs.harvard.edu/abs/2012ApJ...747...61Y} {747, 61}

\bibitem[\protect\citeauthoryear{{Zhang} et~al.,}{{Zhang}
  et~al.}{2017}]{Zhang2017}
{Zhang} K.,  et~al., 2017, \mn@doi [\mnras] {10.1093/mnras/stw3308}, \href
  {http://adsabs.harvard.edu/abs/2017MNRAS.466.3217Z} {466, 3217}

\bibitem[\protect\citeauthoryear{{Zinchenko}, {Pilyugin}, {Grebel},
  {S{\'a}nchez}  \& {V{\'{\i}}lchez}}{{Zinchenko} et~al.}{2016}]{Zinchenko2016}
{Zinchenko} I.~A.,  {Pilyugin} L.~S.,  {Grebel} E.~K.,  {S{\'a}nchez} S.~F.,
  {V{\'{\i}}lchez} J.~M.,  2016, \mn@doi [\mnras] {10.1093/mnras/stw1857},
  \href {http://adsabs.harvard.edu/abs/2016MNRAS.462.2715Z} {462, 2715}

\makeatother
\end{thebibliography}

%

%%%%%%%%%%%%%%%%%%%%%%%%%%%%%%%%%%%%%%%%%%%%%%%%%%

%%%%%%%%%%%%%%%%% APPENDICES %%%%%%%%%%%%%%%%%%%%%
\section*{Supporting Information}
\indent Figures presented in appendix are available as supplementary online material.

\appendix

\section{Metallicity maps of galaxies before and after correction}
\indent Figures \textcolor{blue}{A1}-\textcolor{blue}{A23} show the metallicity maps of 23 galaxies in the
sample, before and after applying corrections. Upper panel shows the
uncorrected metallicity map obtained using O3N2 (left panel), O3S2
(middle panel) and N2S2H$\alpha$ (right panel). The middle panel shows
the metallicity maps obtained from O3N2 (left panel) and O3S2 (right
panel) diagnostics after applying corrections to the DIG/LIER/Seyfert
regions based on the spatially-resolved [SII]-BPT map (right
panel). The lower panel shows the metallicity maps obtained from O3N2
(left panel) and O3S2 (right panel) diagnostics after applying
corrections to the DIG/LIER regions based on the spatially-resolved
[NII]-BPT map (right panel).

\begin{figure*}
	\centering
	\includegraphics[width=0.28\textwidth, trim={0 1.2cm 0 5.5cm}, clip]{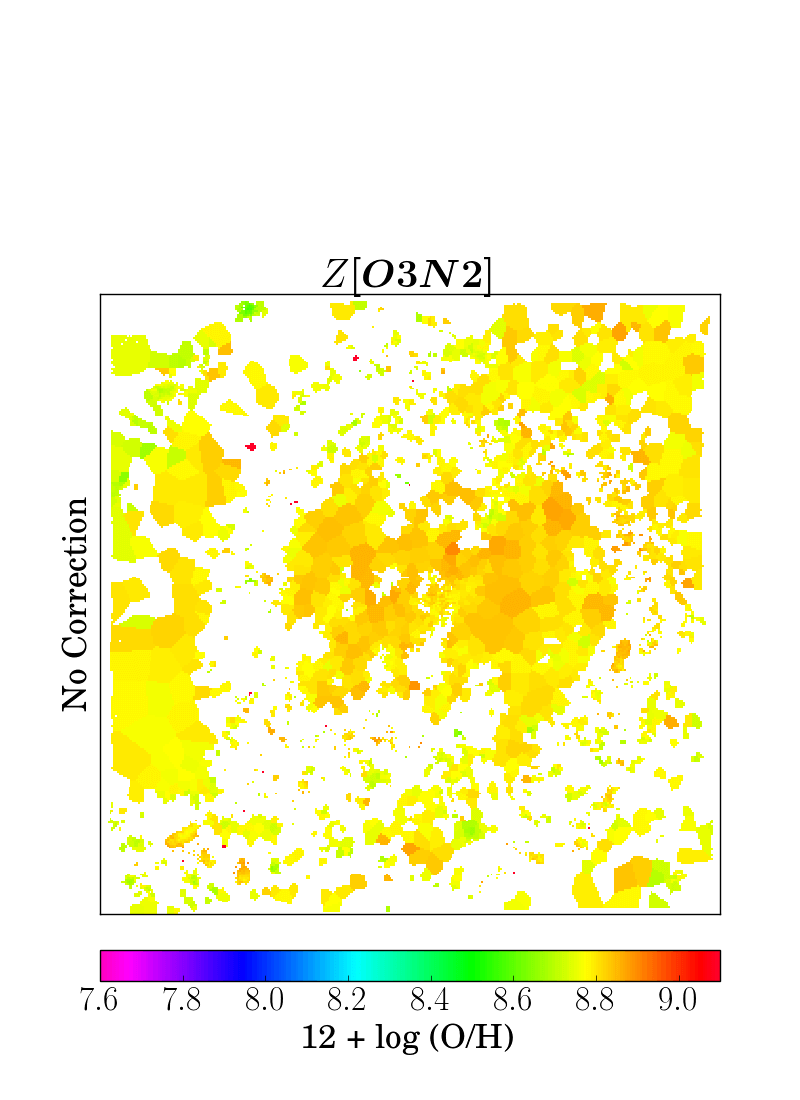}
	\includegraphics[width=0.28\textwidth, trim={0 1.2cm 0 5.5cm}, clip]{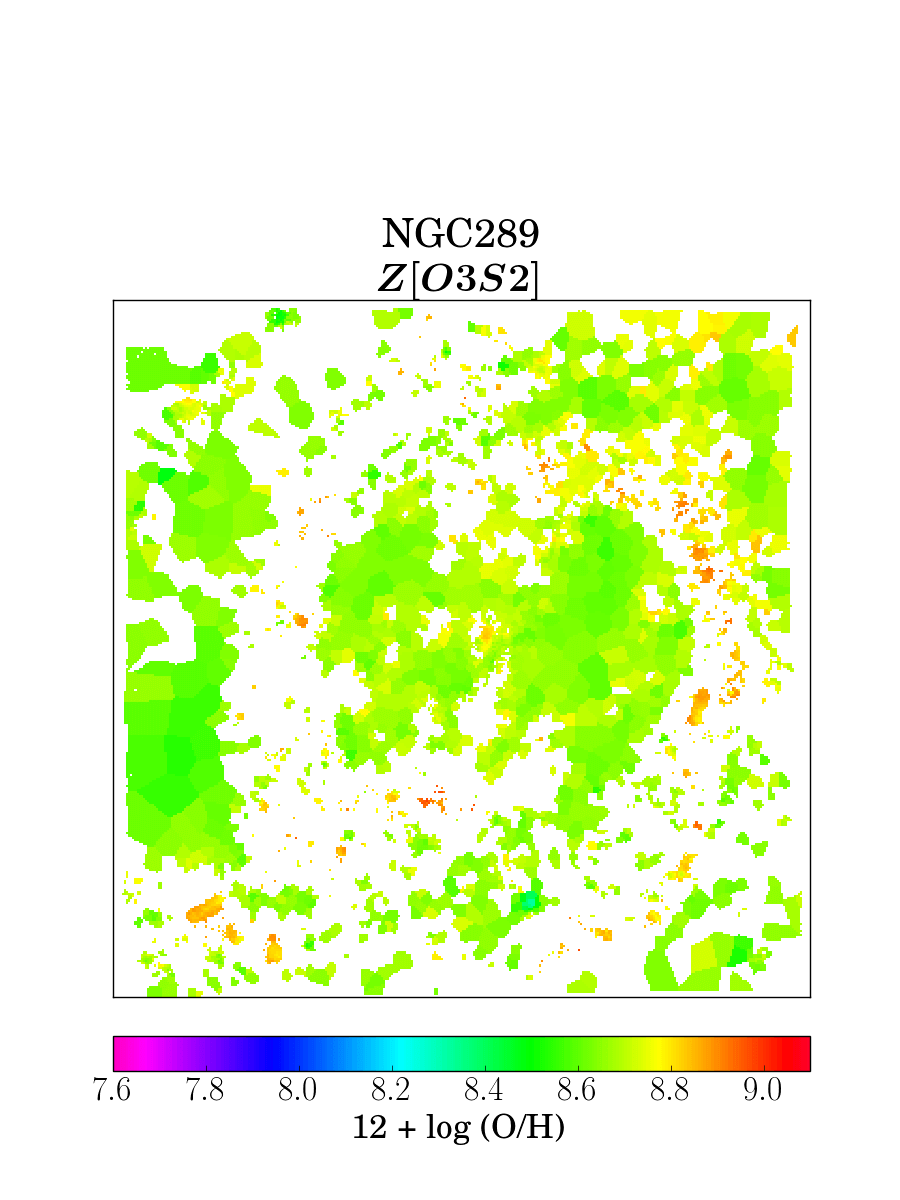}
	\includegraphics[width=0.28\textwidth, trim={0 1.2cm 0 5.5cm}, clip]{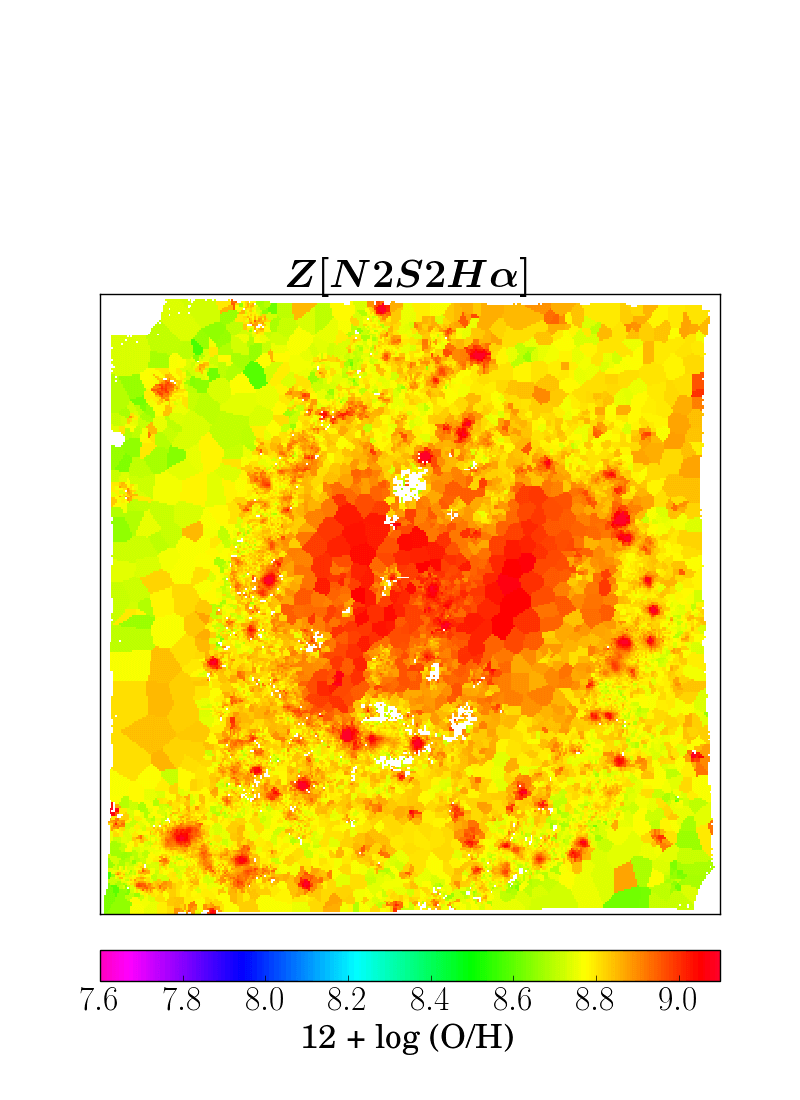}
	\includegraphics[width=0.28\textwidth, trim={0 1.2cm 0 5.5cm}, clip]{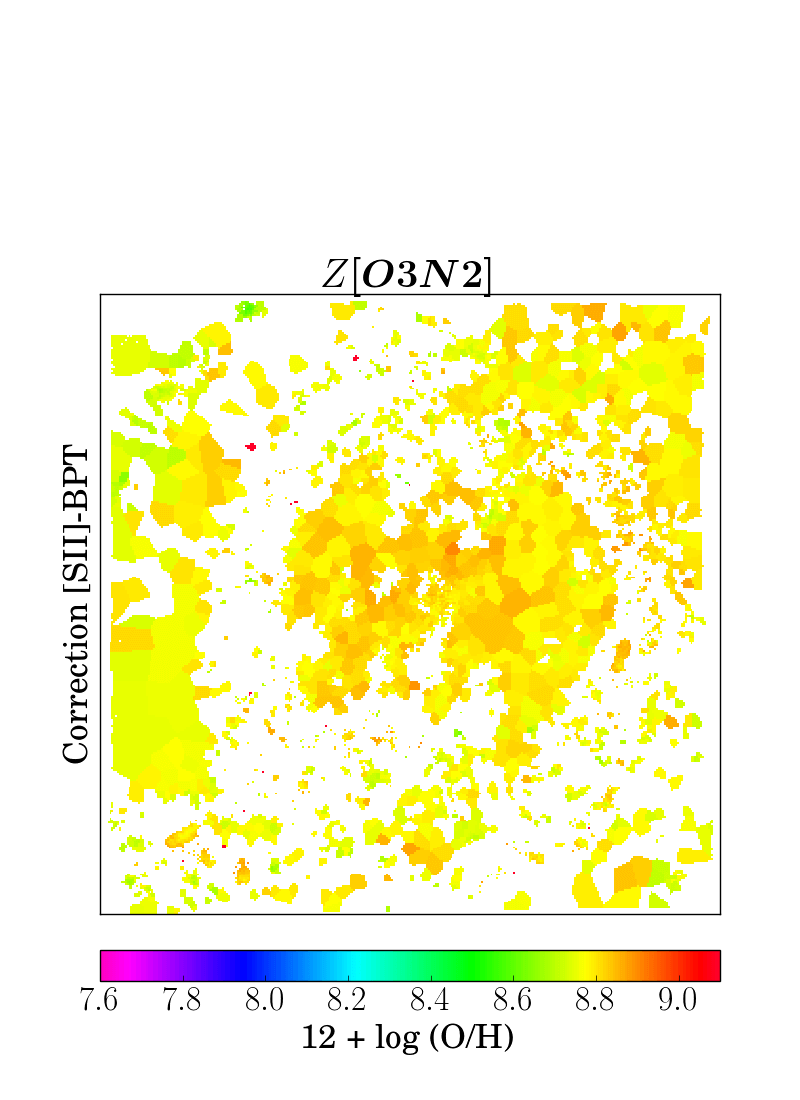}
	\includegraphics[width=0.28\textwidth, trim={0 1.2cm 0 5.5cm}, clip]{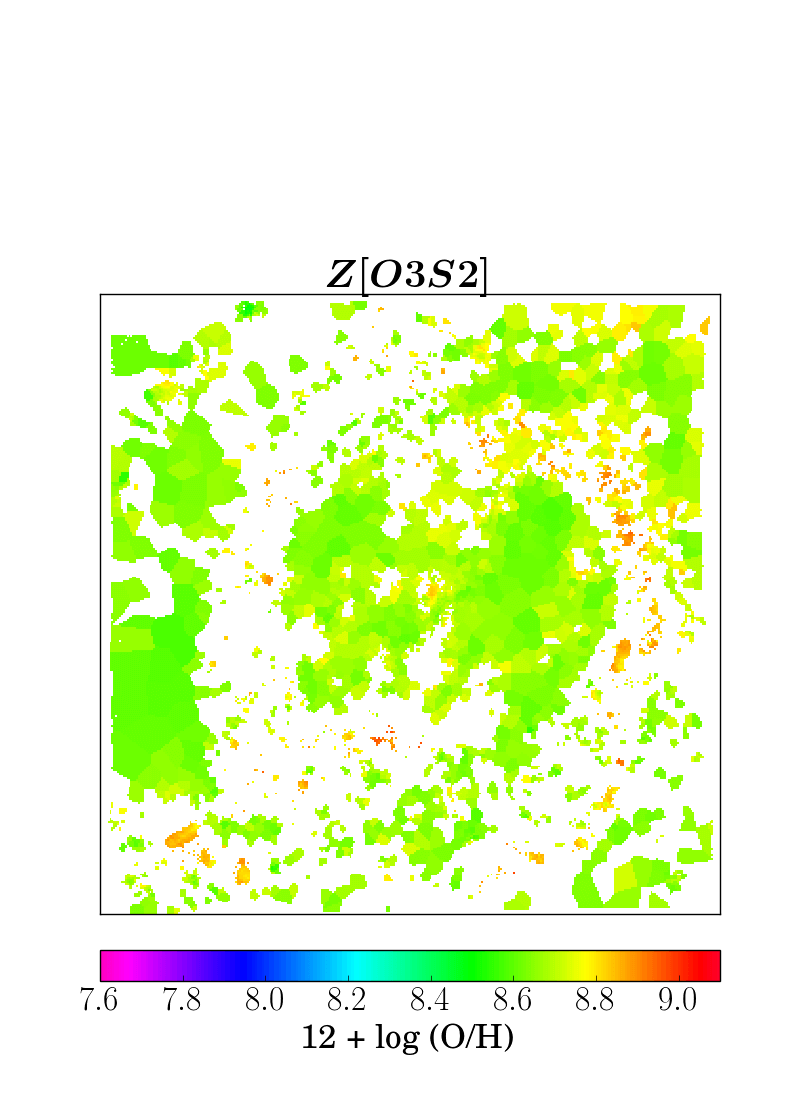}
	\includegraphics[width=0.28\textwidth, trim={2.8cm 0 2.8cm 0}, clip ]{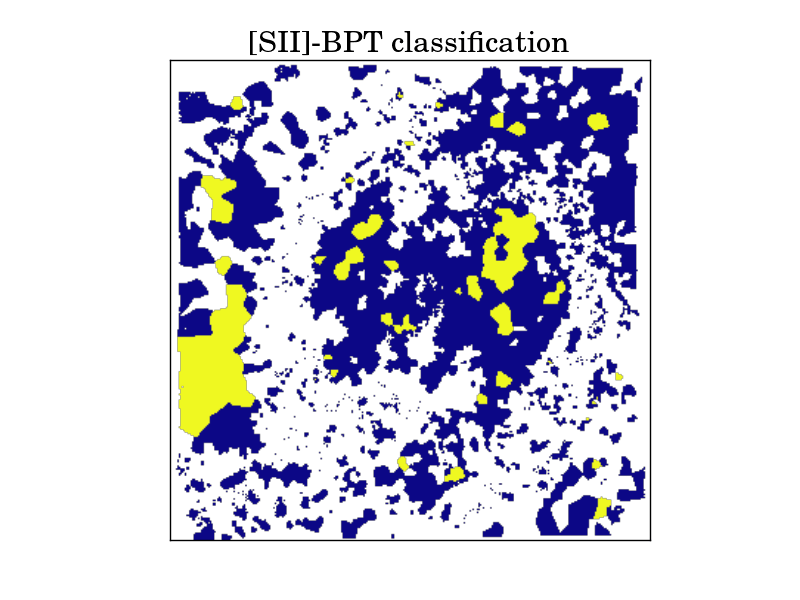}
	\includegraphics[width=0.28\textwidth, trim={0 1.2cm 0 5.5cm}, clip]{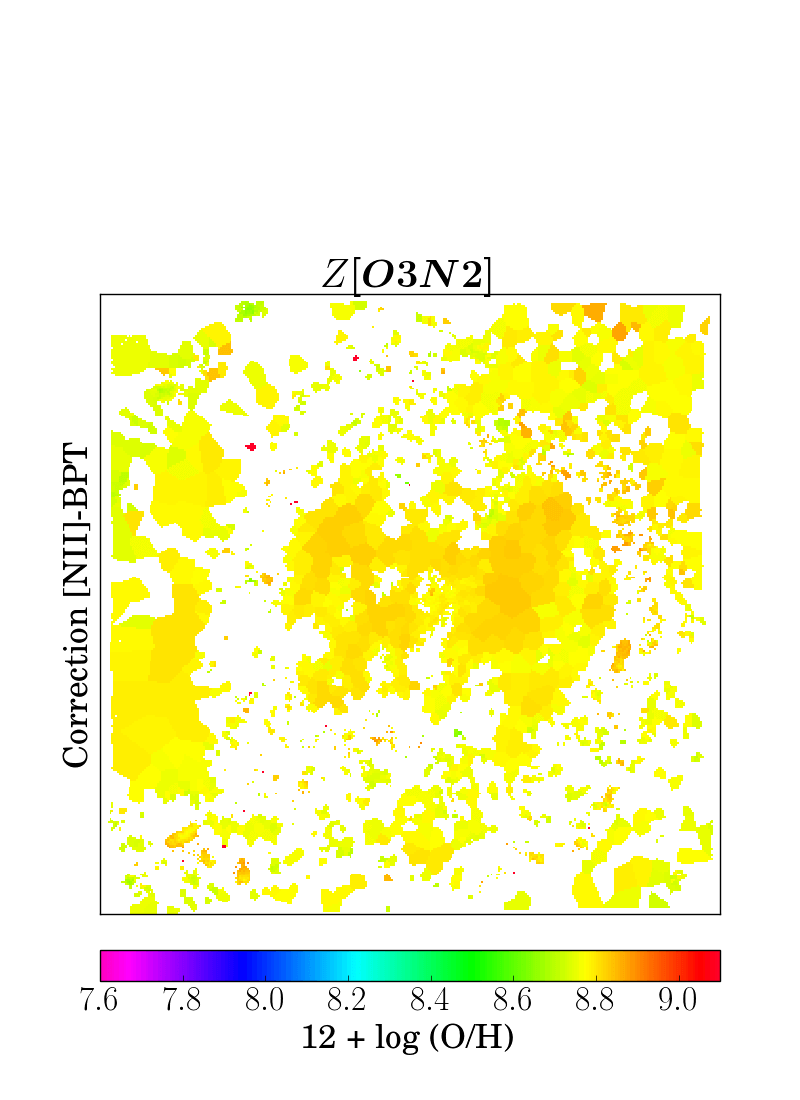}
	\includegraphics[width=0.28\textwidth, trim={0 1.2cm 0 5.5cm}, clip]{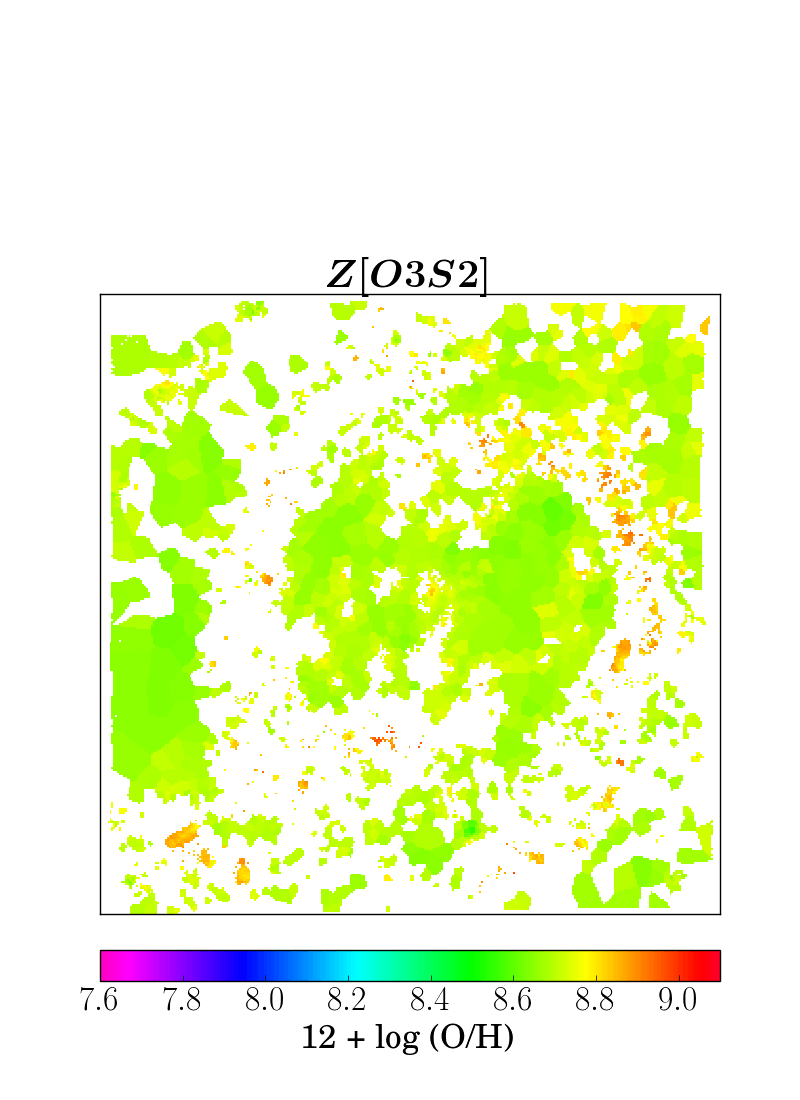}
	\includegraphics[width=0.28\textwidth, trim={2.8cm 0 2.8cm 0}, clip ]{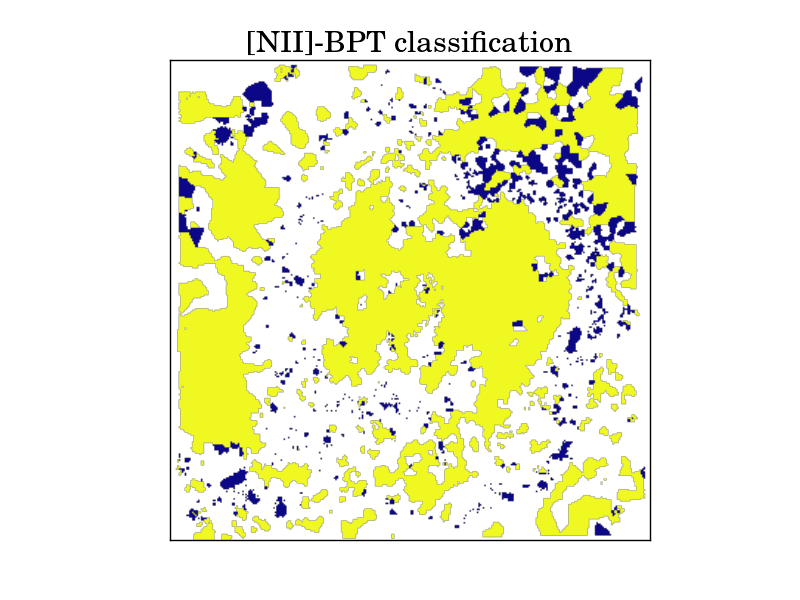}
	\caption{ Maps correspond to galaxy NGC289, see caption of Figure \ref{fig:NGC1042} for details.}
	\label{fig:NGC289}
\end{figure*}
\begin{figure*}
	\centering
	\includegraphics[width=0.28\textwidth, trim={0 1.2cm 0 5.5cm}, clip]{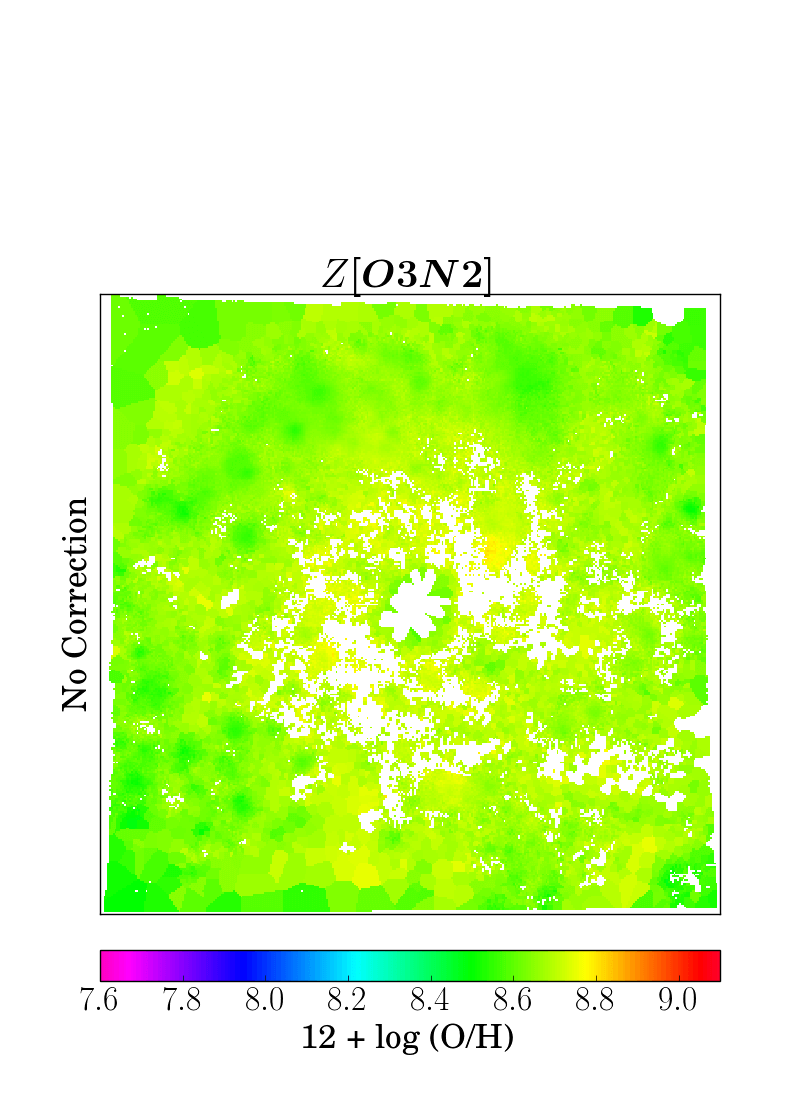}
	\includegraphics[width=0.28\textwidth, trim={0 1.2cm 0 5.5cm}, clip]{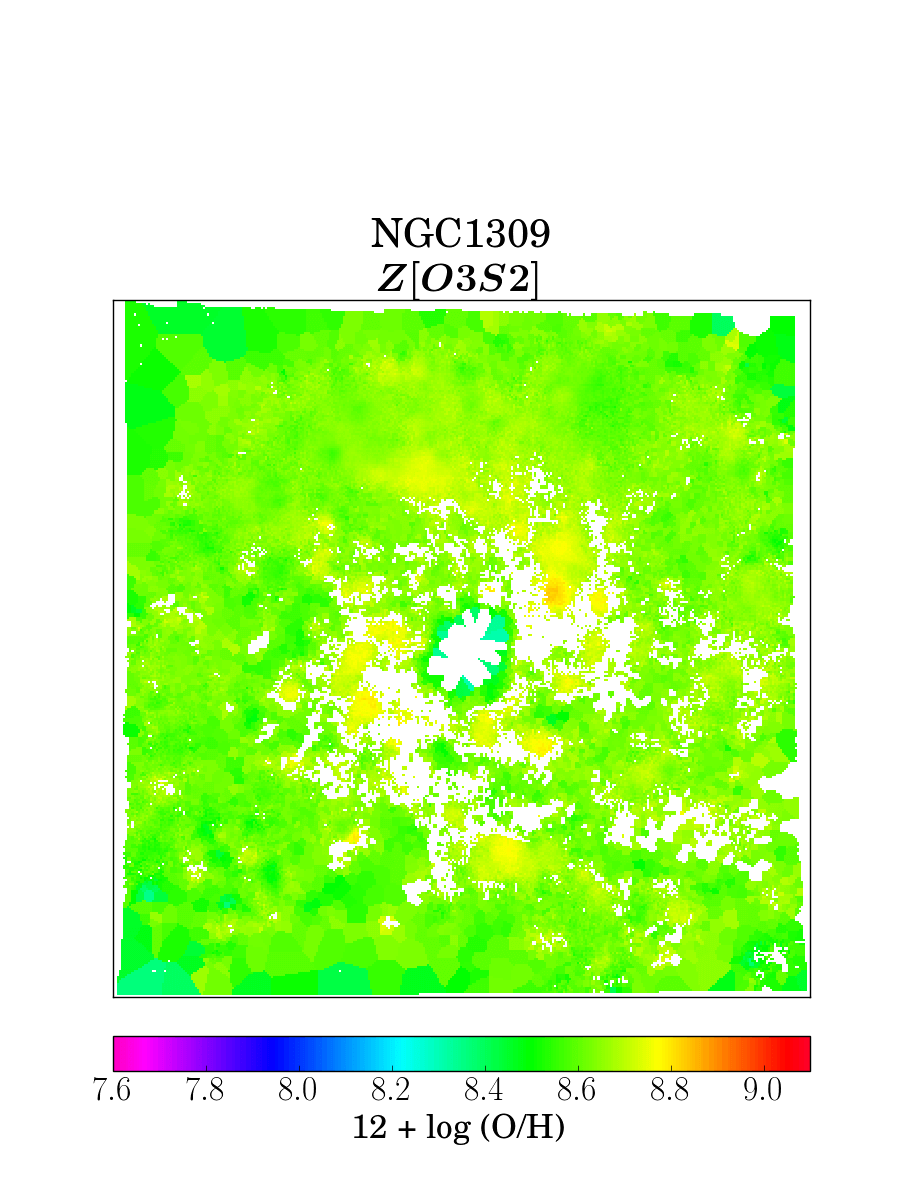}
	\includegraphics[width=0.28\textwidth, trim={0 1.2cm 0 5.5cm}, clip]{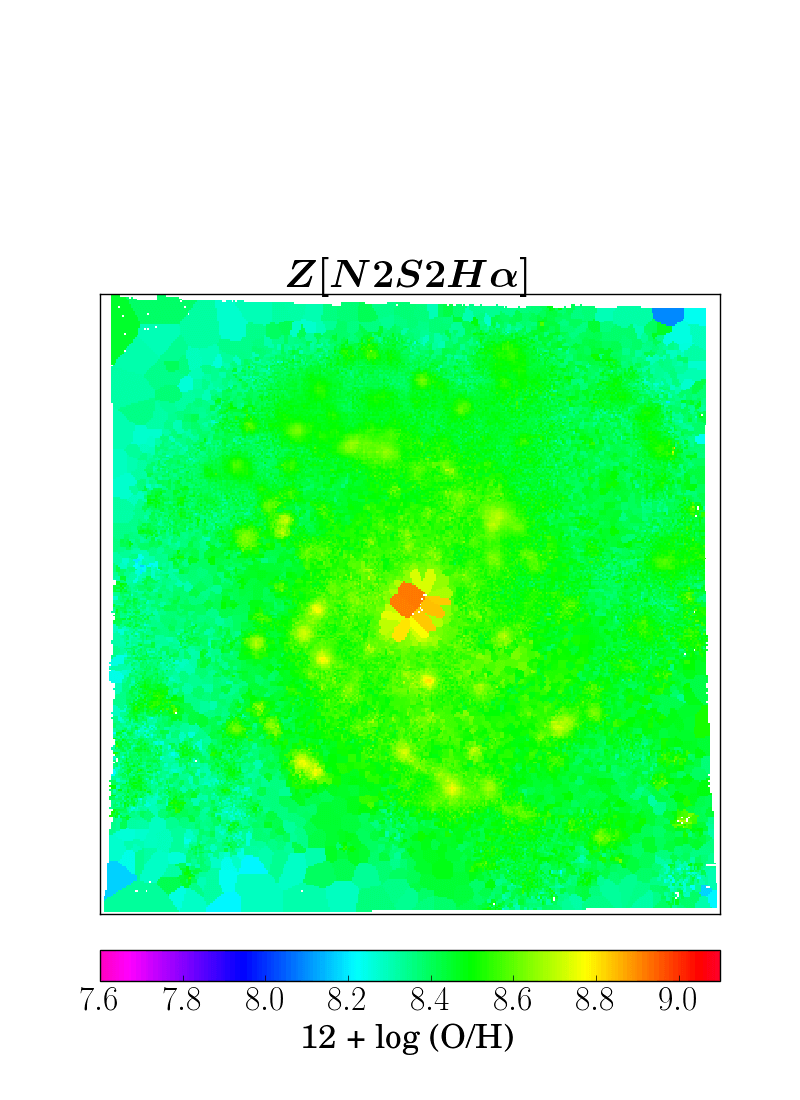}
	\includegraphics[width=0.28\textwidth, trim={0 1.2cm 0 5.5cm}, clip]{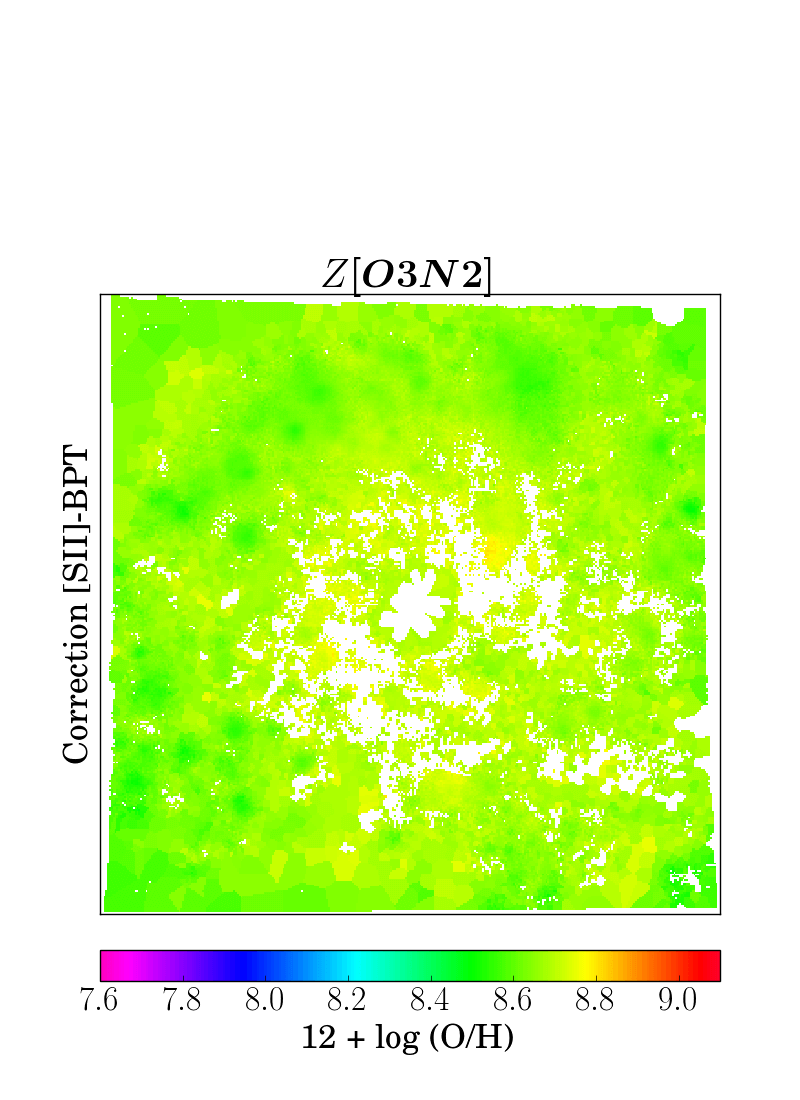}
	\includegraphics[width=0.28\textwidth, trim={0 1.2cm 0 5.5cm}, clip]{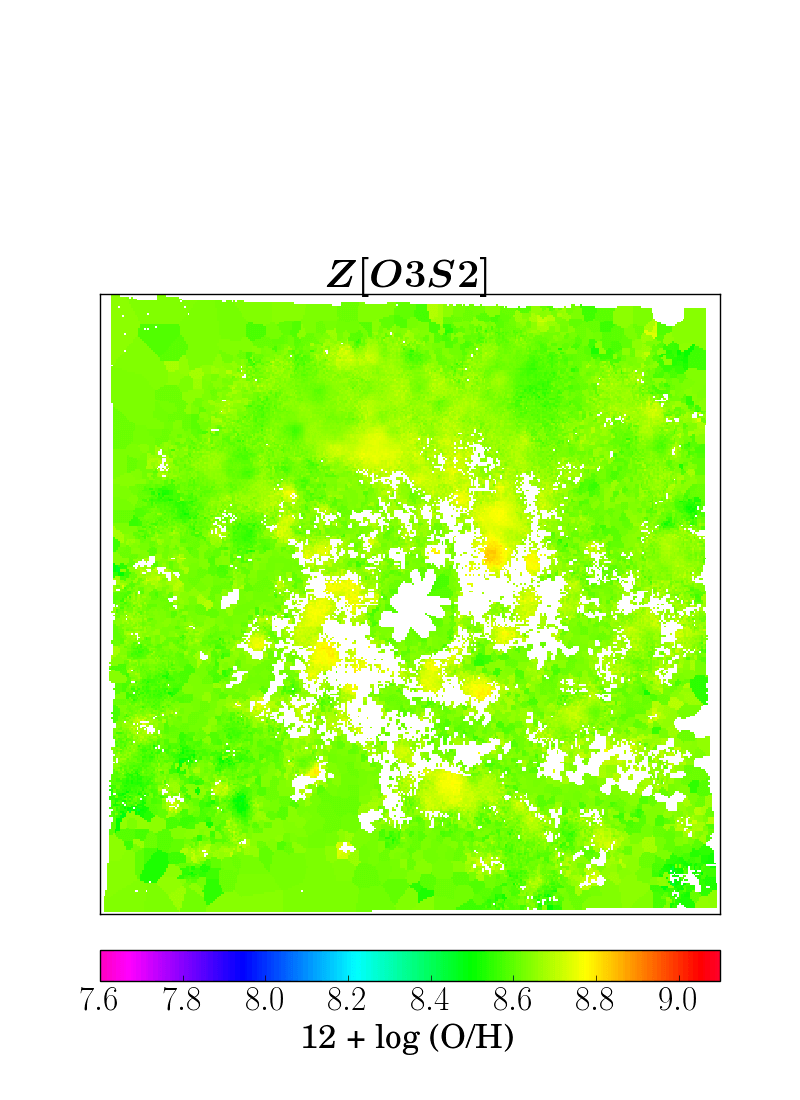}
	\includegraphics[width=0.28\textwidth, trim={2.8cm 0 2.8cm 0}, clip ]{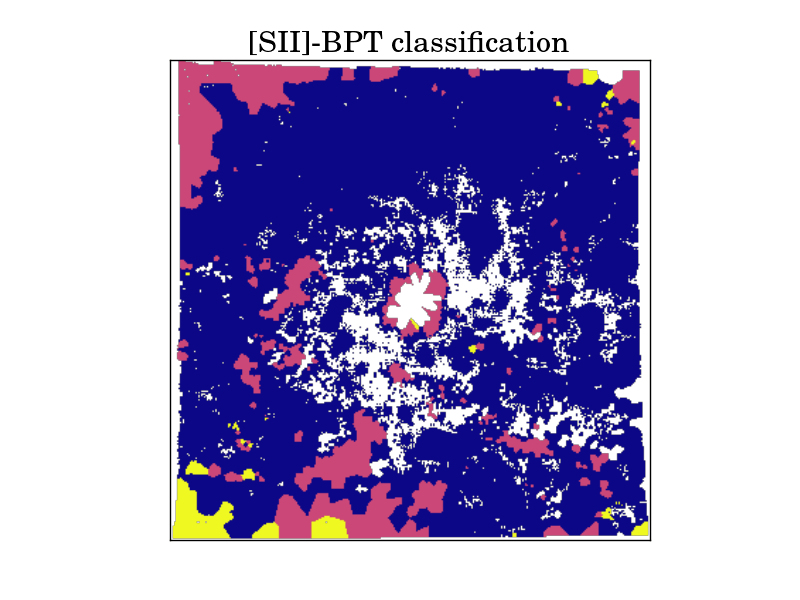}
	\includegraphics[width=0.28\textwidth, trim={0 1.2cm 0 5.5cm}, clip]{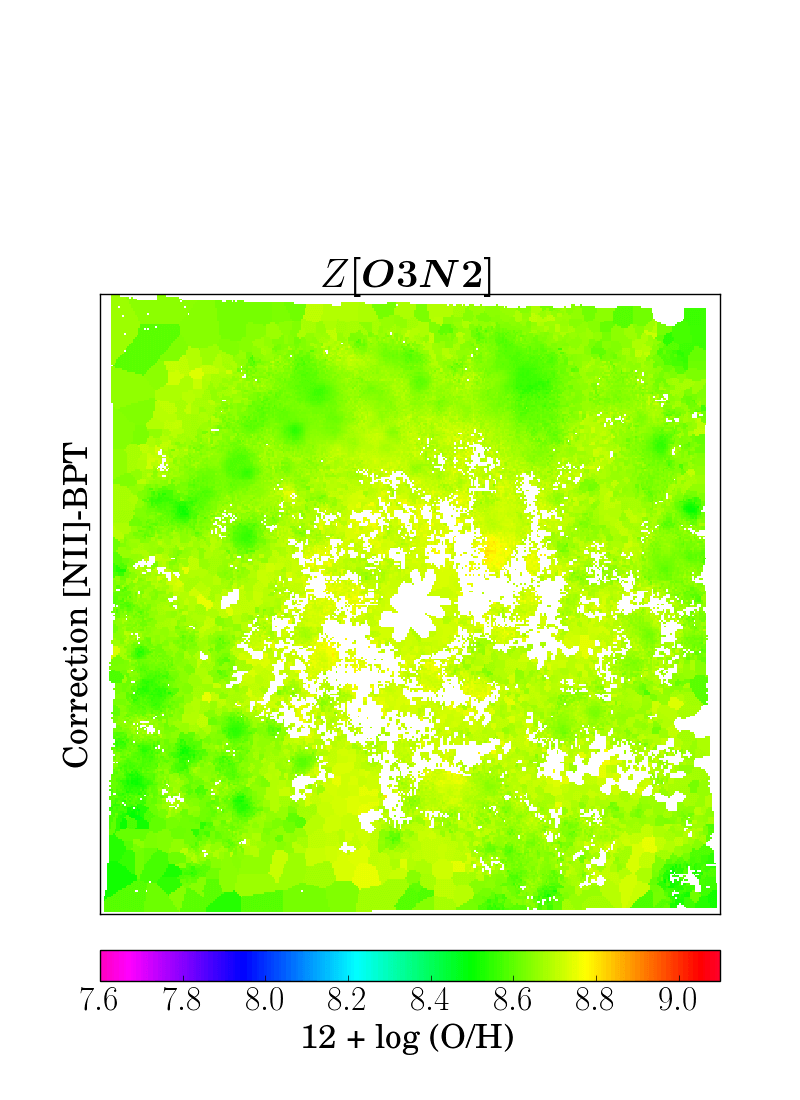}
	\includegraphics[width=0.28\textwidth, trim={0 1.2cm 0 5.5cm}, clip]{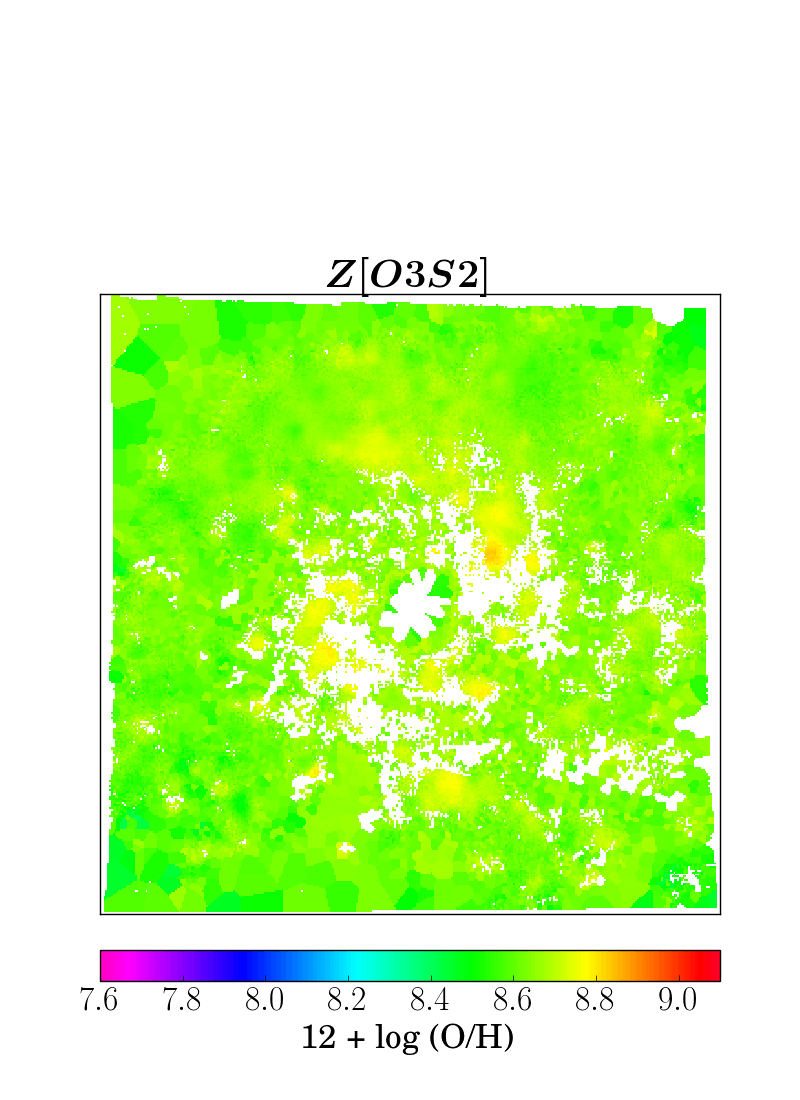}
	\includegraphics[width=0.28\textwidth, trim={2.8cm 0 2.8cm 0}, clip ]{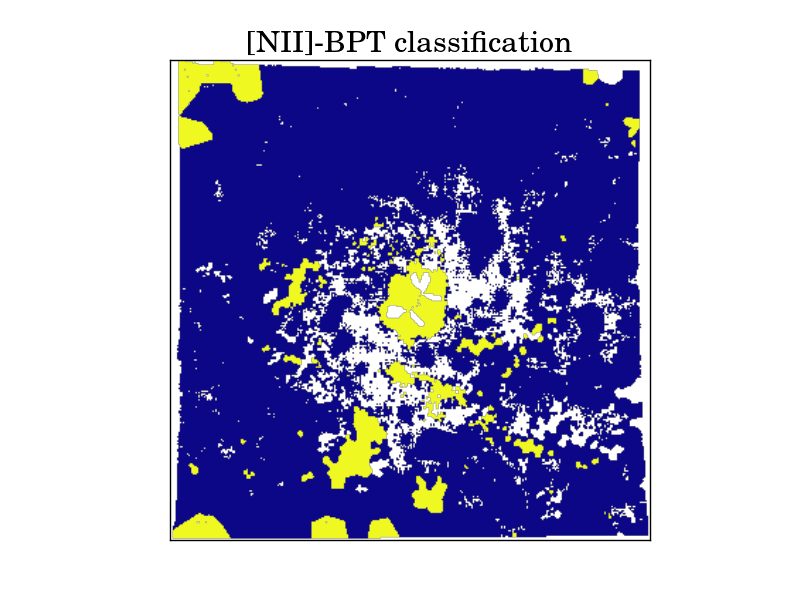}
	\caption{ Maps correspond to galaxy NGC1309, see caption of Figure \ref{fig:NGC1042} for details.}
	\label{fig:NGC1309}
\end{figure*}
\begin{figure*}
	\centering
	\includegraphics[width=0.28\textwidth, trim={0 1.2cm 0 5.5cm}, clip]{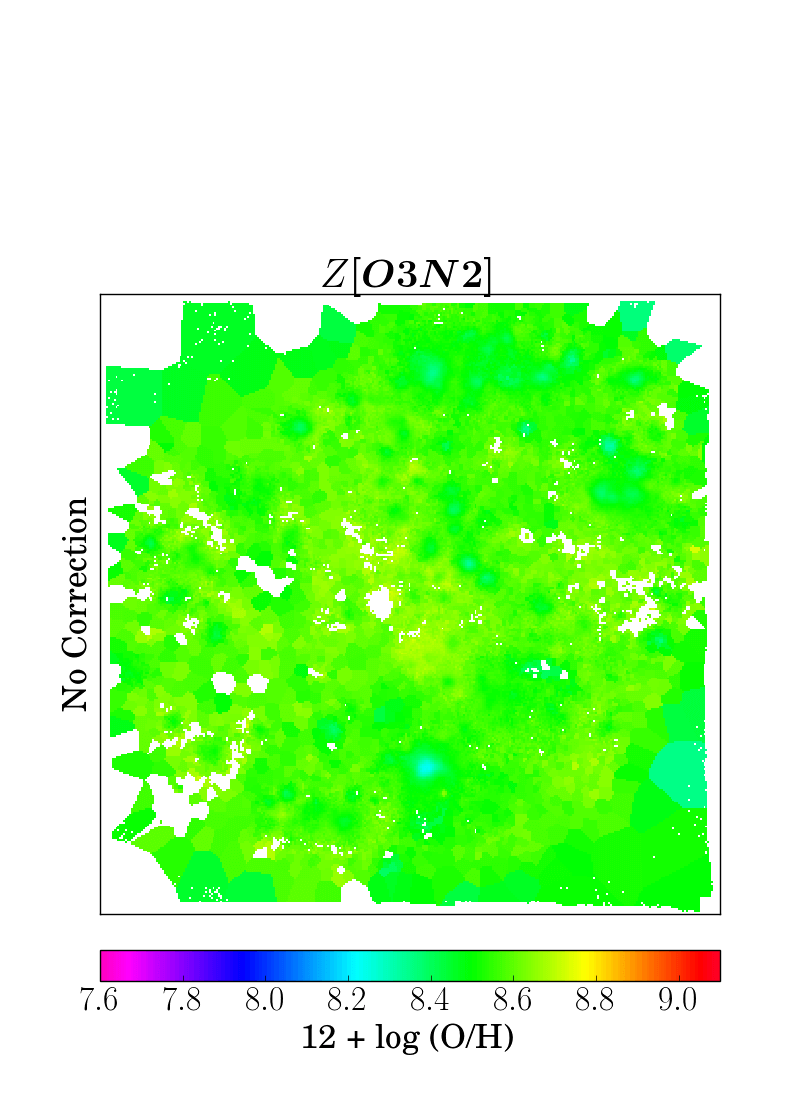}
	\includegraphics[width=0.28\textwidth, trim={0 1.2cm 0 5.5cm}, clip]{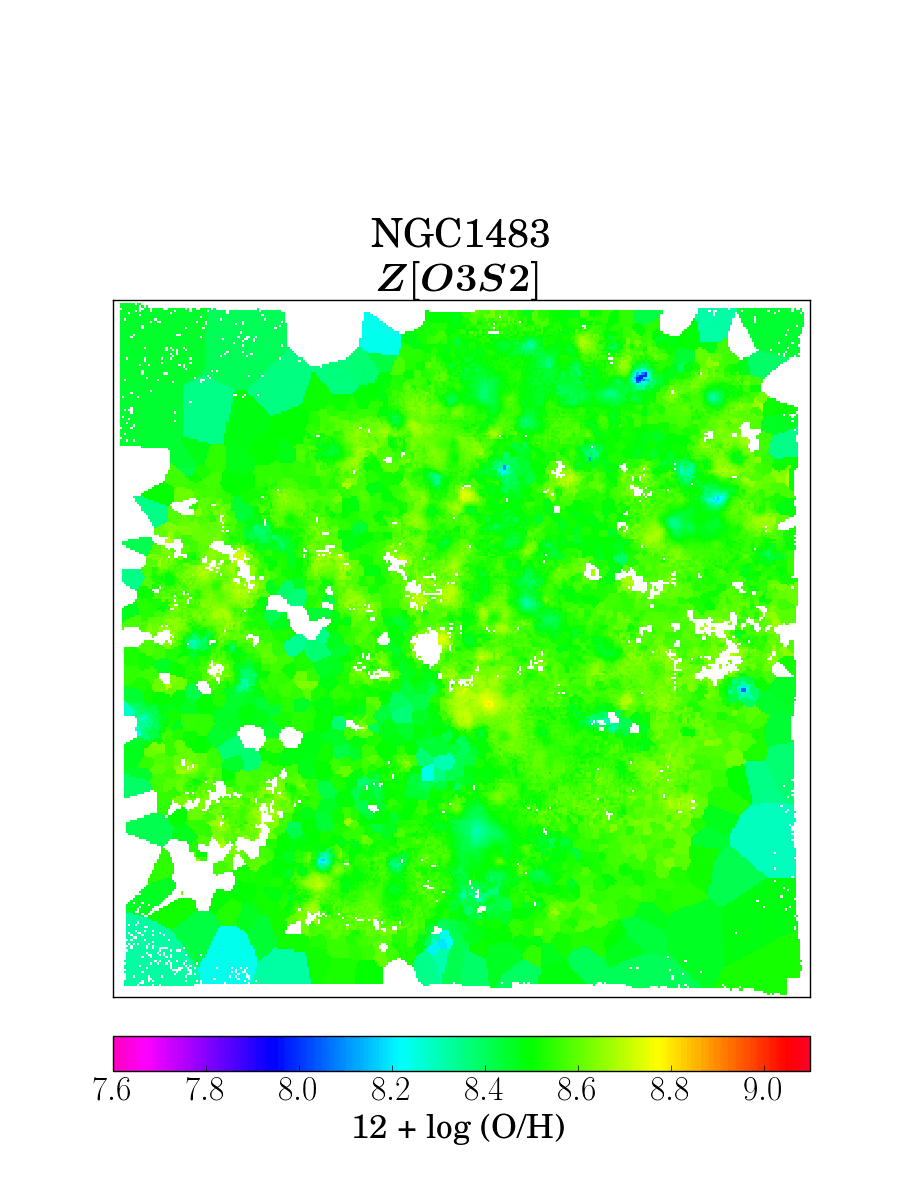}
	\includegraphics[width=0.28\textwidth, trim={0 1.2cm 0 5.5cm}, clip]{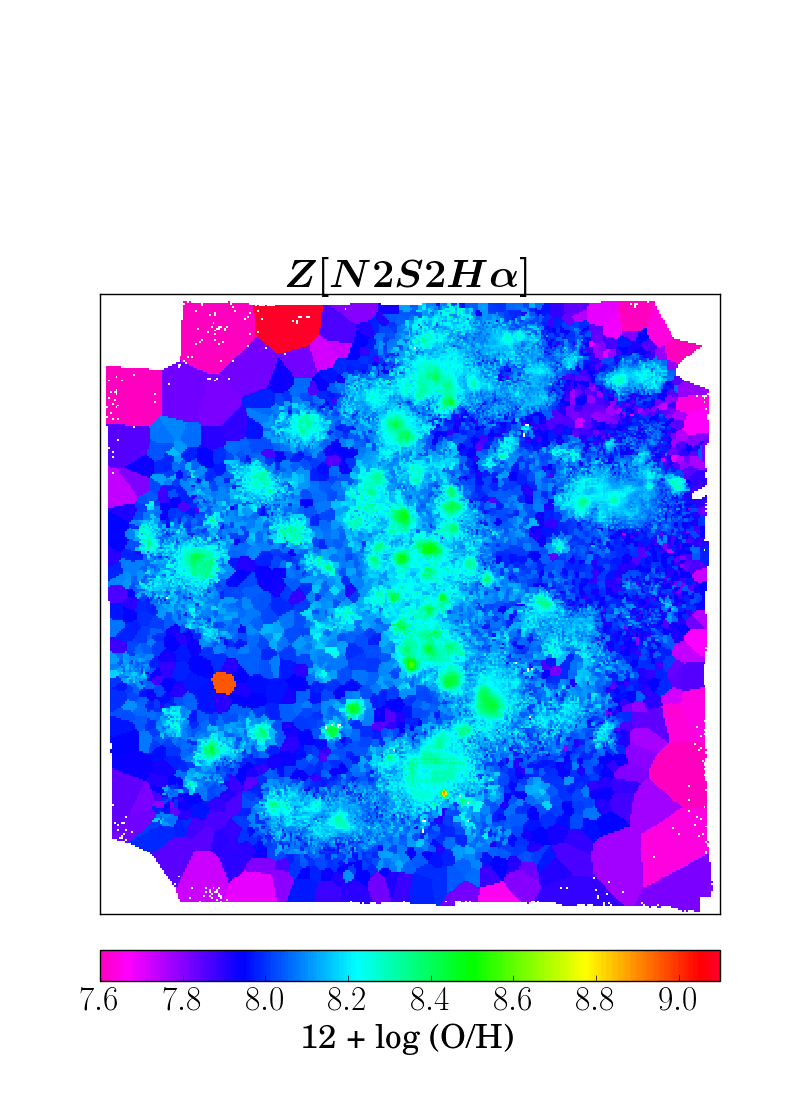}
	\includegraphics[width=0.28\textwidth, trim={0 1.2cm 0 5.5cm}, clip]{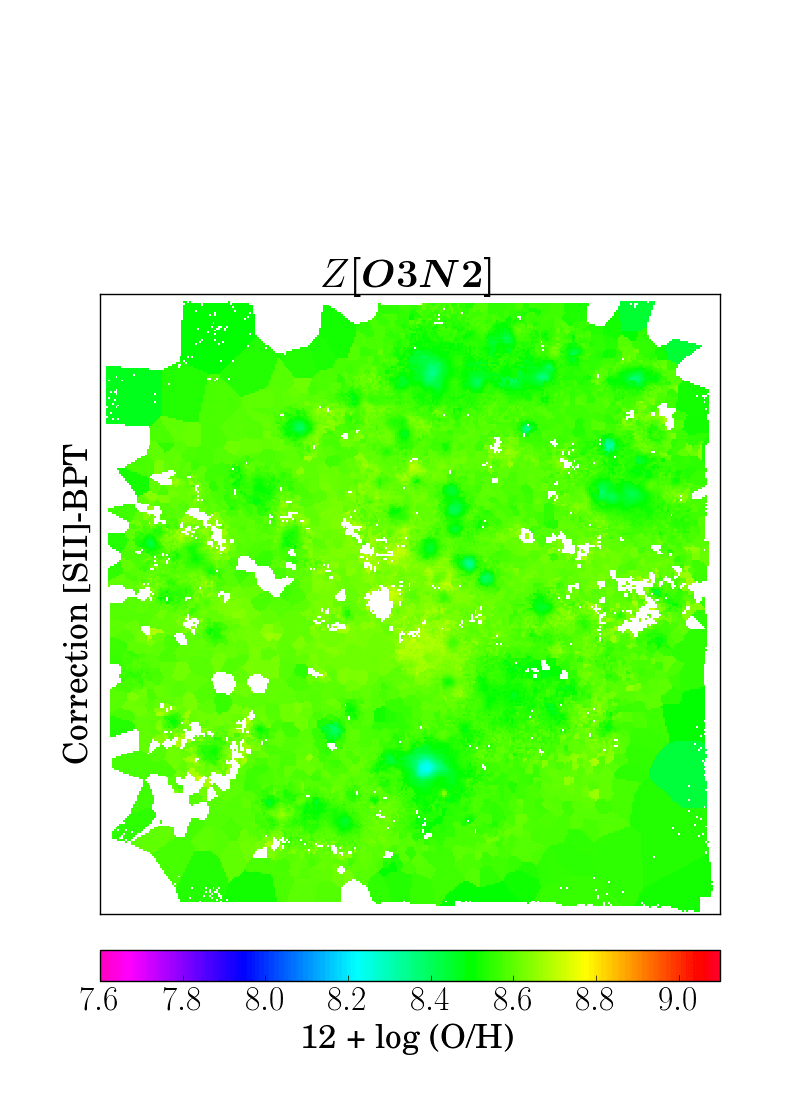}
	\includegraphics[width=0.28\textwidth, trim={0 1.2cm 0 5.5cm}, clip]{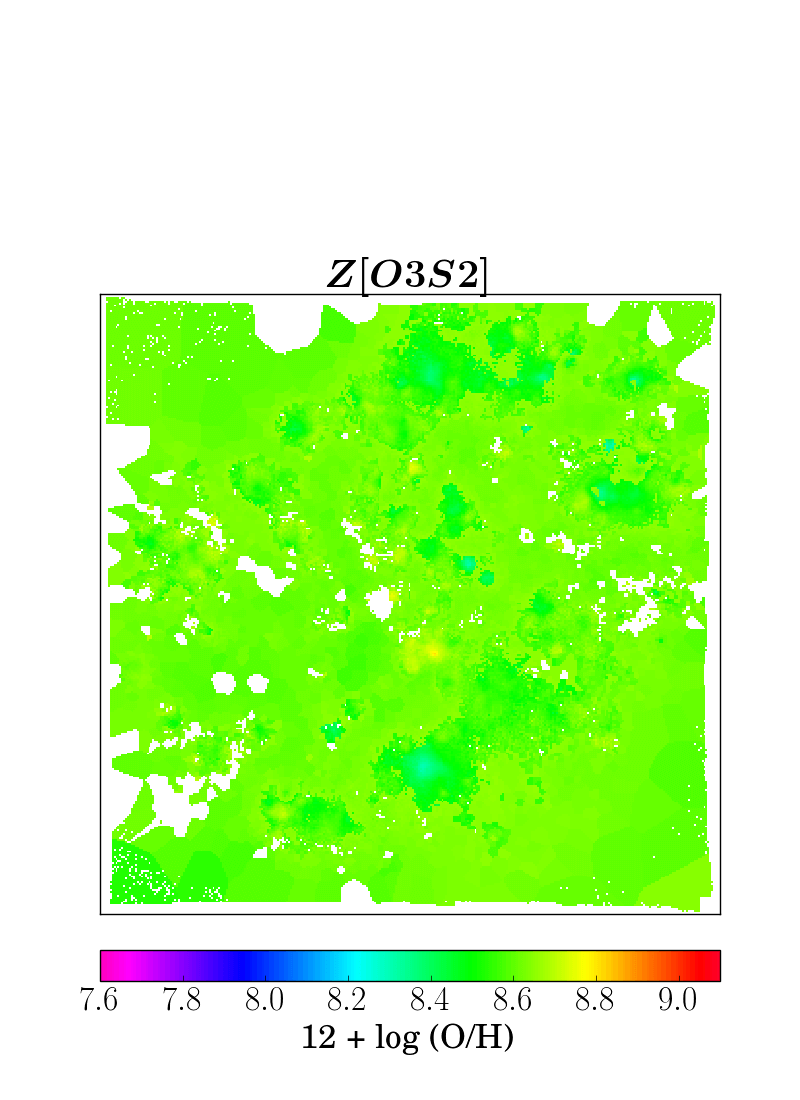}
	\includegraphics[width=0.28\textwidth, trim={2.8cm 0 2.8cm 0}, clip ]{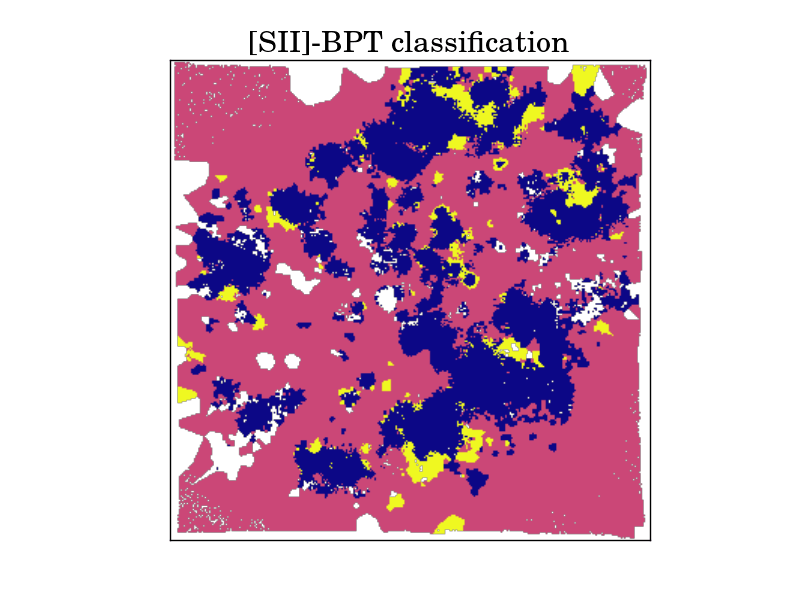}
	\includegraphics[width=0.28\textwidth, trim={0 1.2cm 0 5.5cm}, clip]{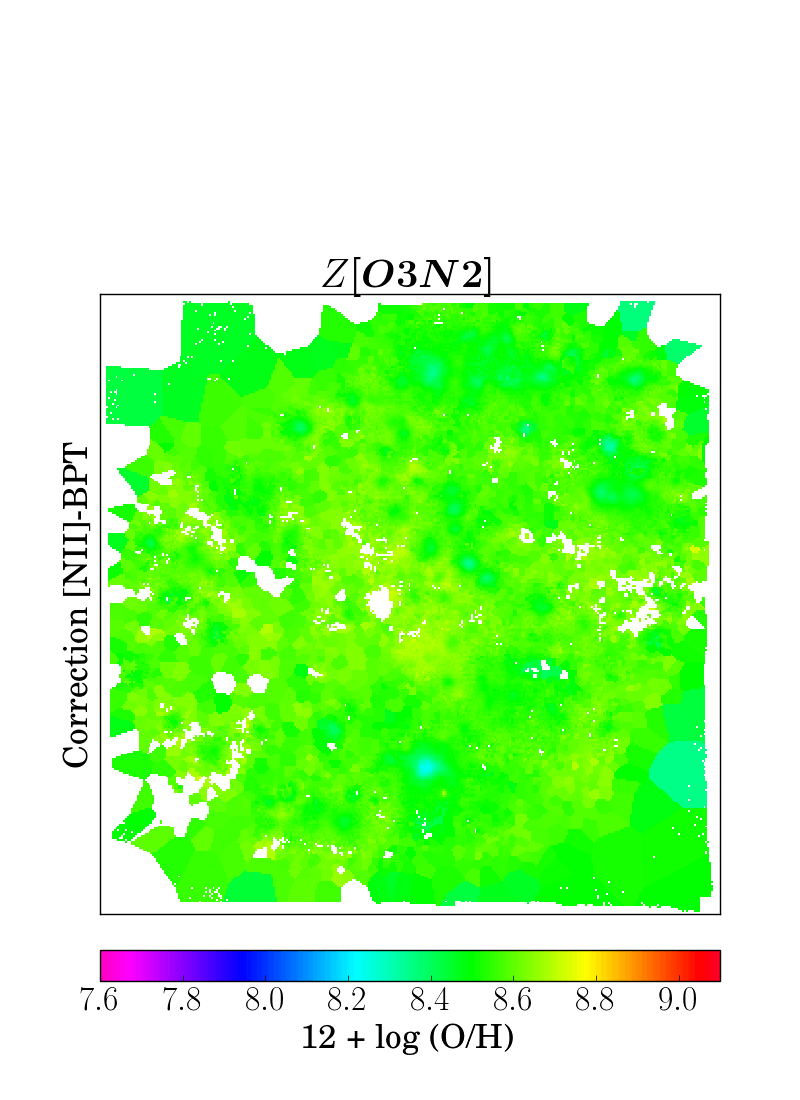}
	\includegraphics[width=0.28\textwidth, trim={0 1.2cm 0 5.5cm}, clip]{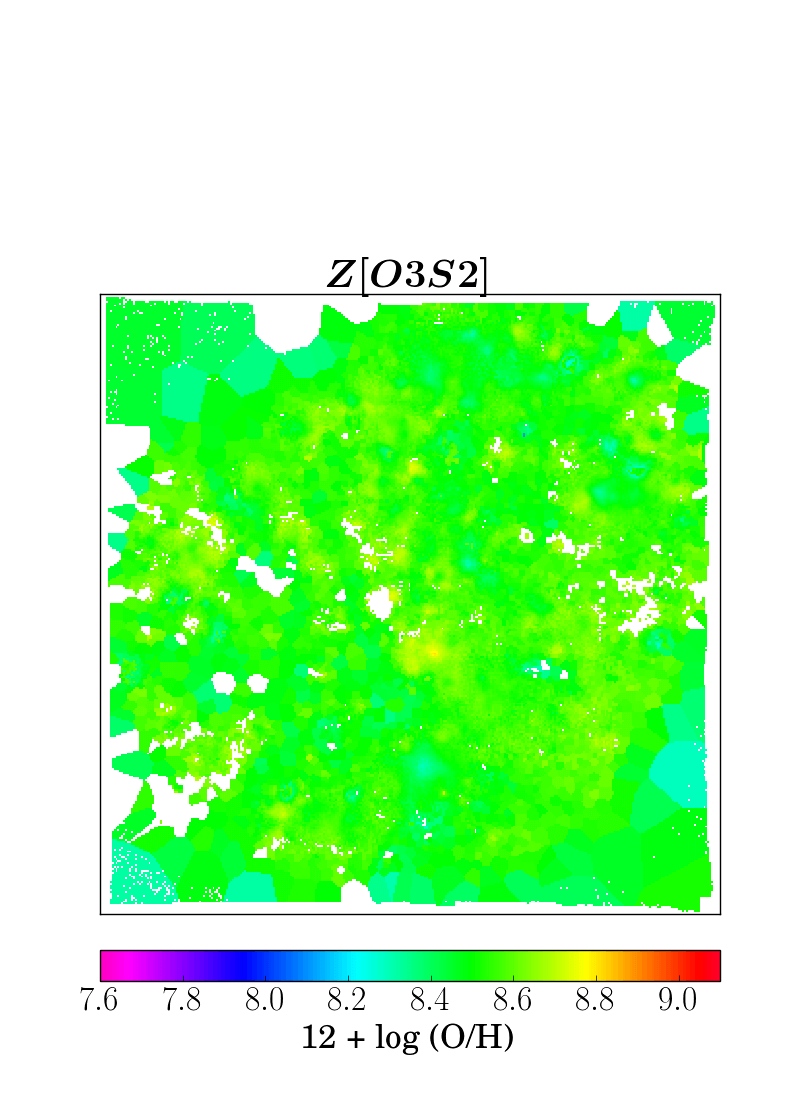}
	\includegraphics[width=0.28\textwidth, trim={2.8cm 0 2.8cm 0}, clip ]{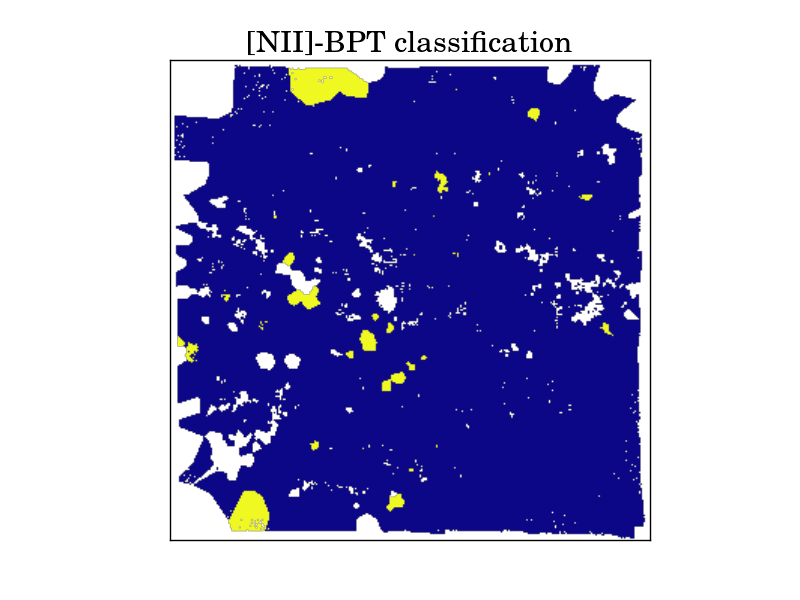}
	\caption{ Maps correspond to galaxy NGC1483, see caption of Figure \ref{fig:NGC1042} for details.}
	\label{fig:NGC1483}
\end{figure*}
\begin{figure*}
	\centering
	\includegraphics[width=0.28\textwidth, trim={0 1.2cm 0 5.5cm}, clip]{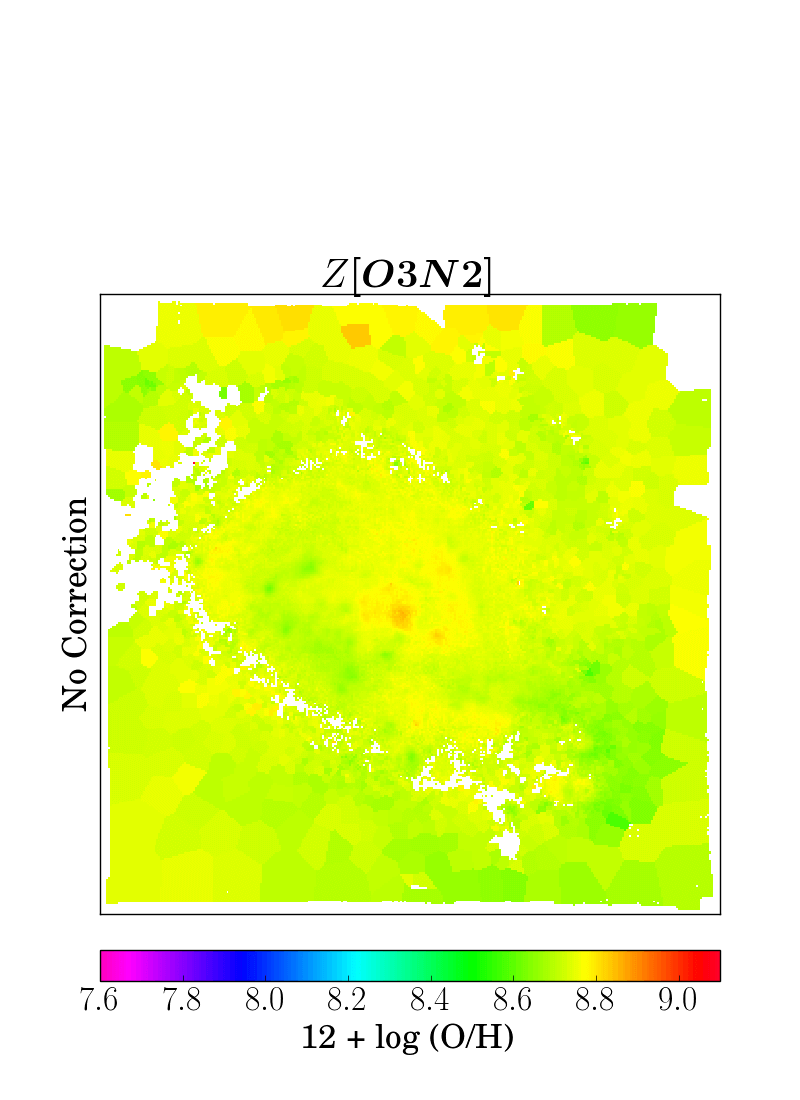}
	\includegraphics[width=0.28\textwidth, trim={0 1.2cm 0 5.5cm}, clip]{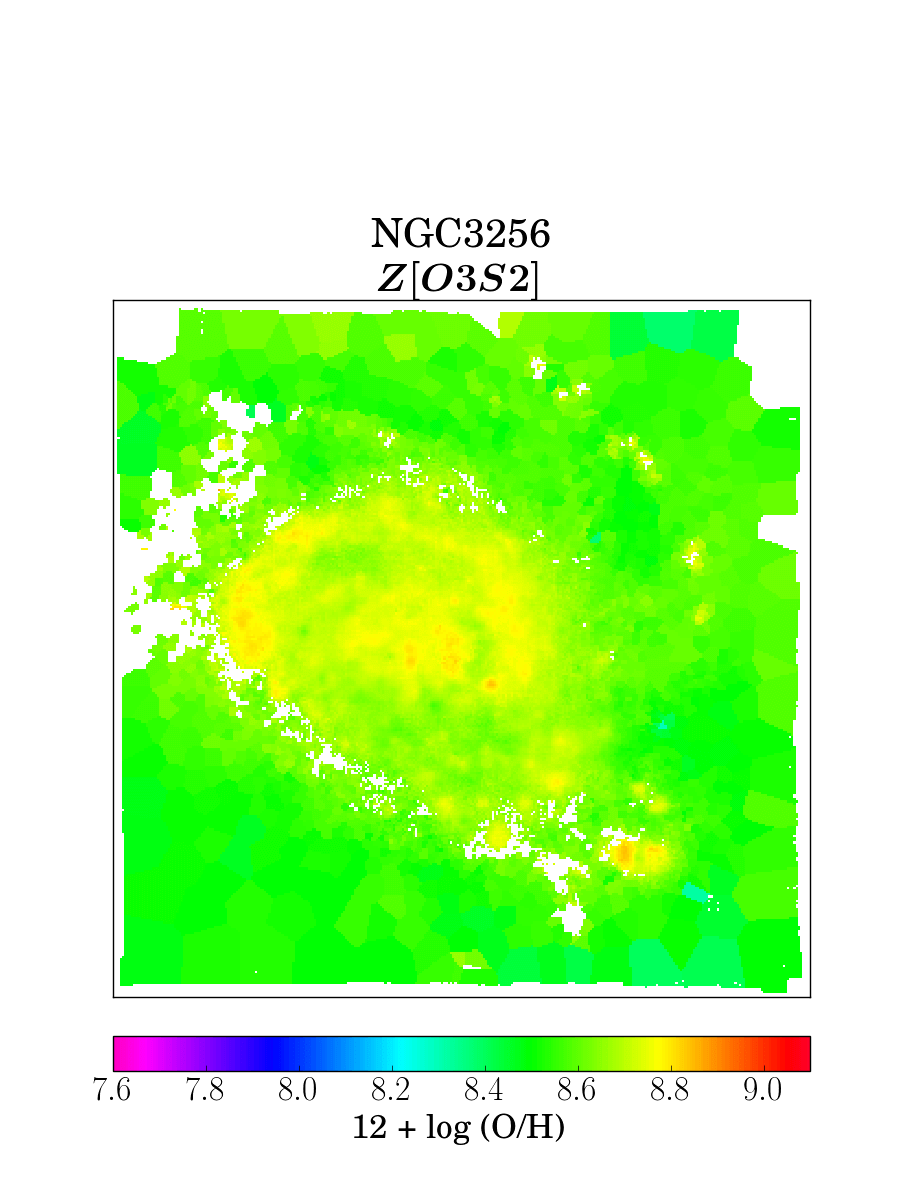}
	\includegraphics[width=0.28\textwidth, trim={0 1.2cm 0 5.5cm}, clip]{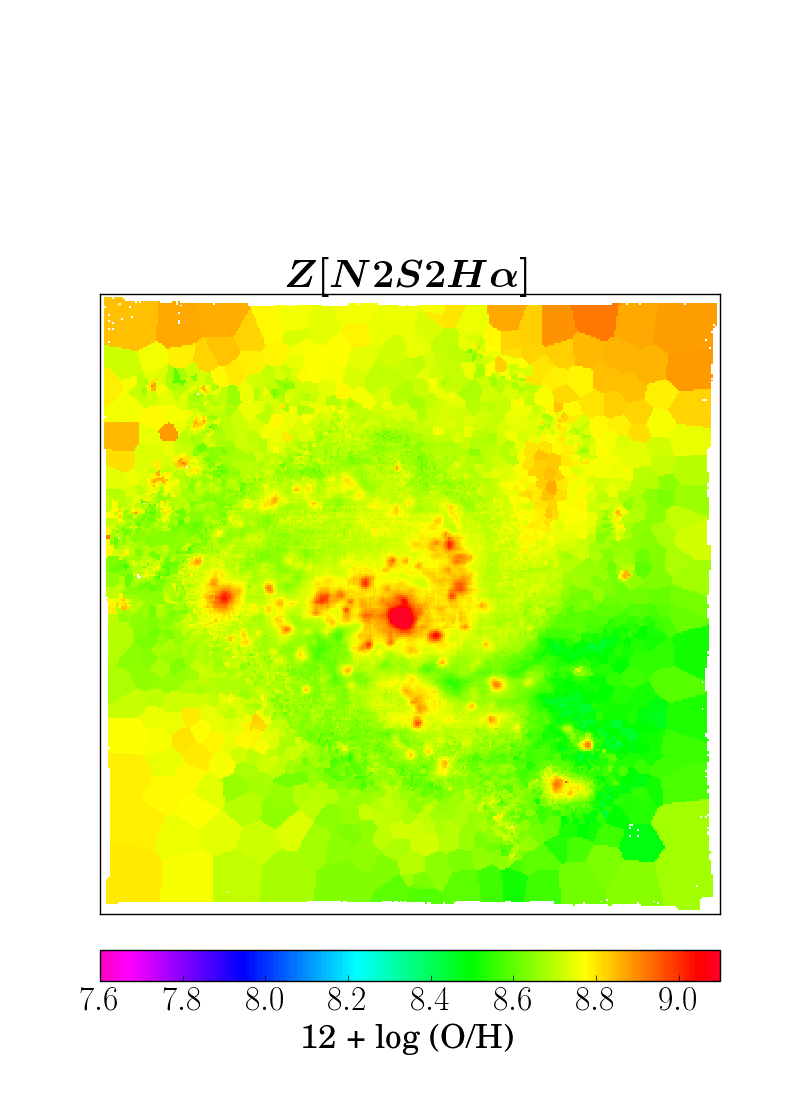}
	\includegraphics[width=0.28\textwidth, trim={0 1.2cm 0 5.5cm}, clip]{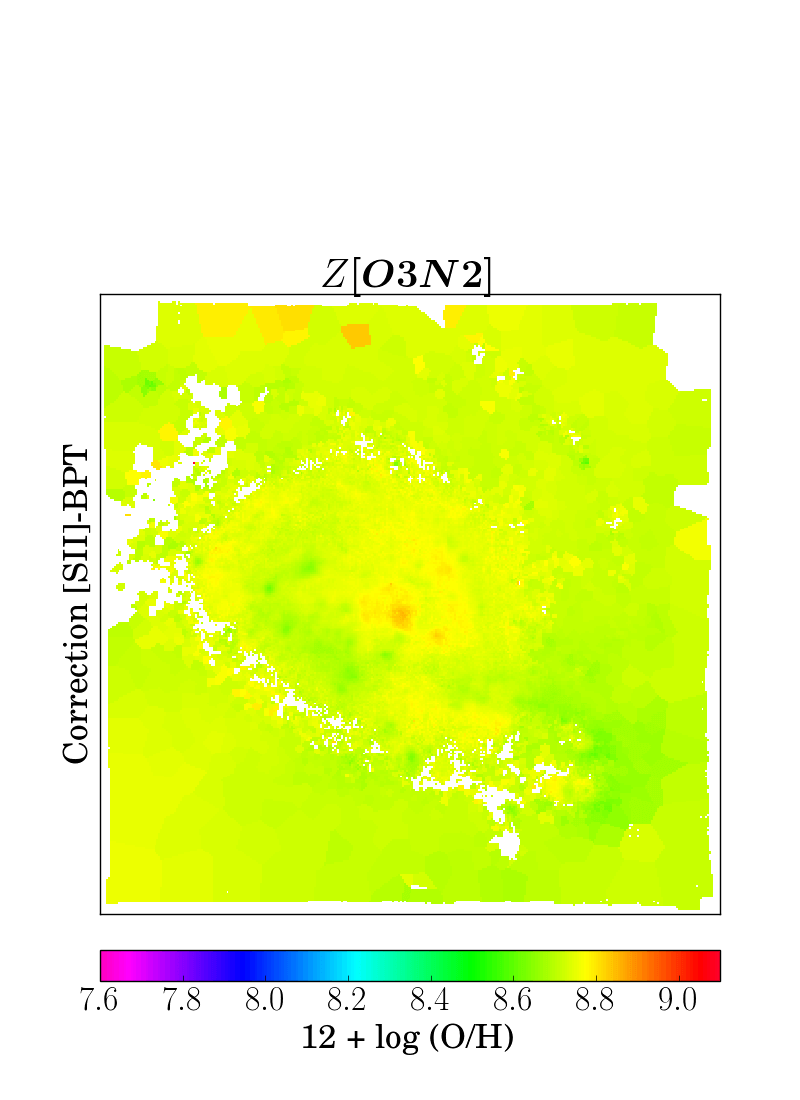}
	\includegraphics[width=0.28\textwidth, trim={0 1.2cm 0 5.5cm}, clip]{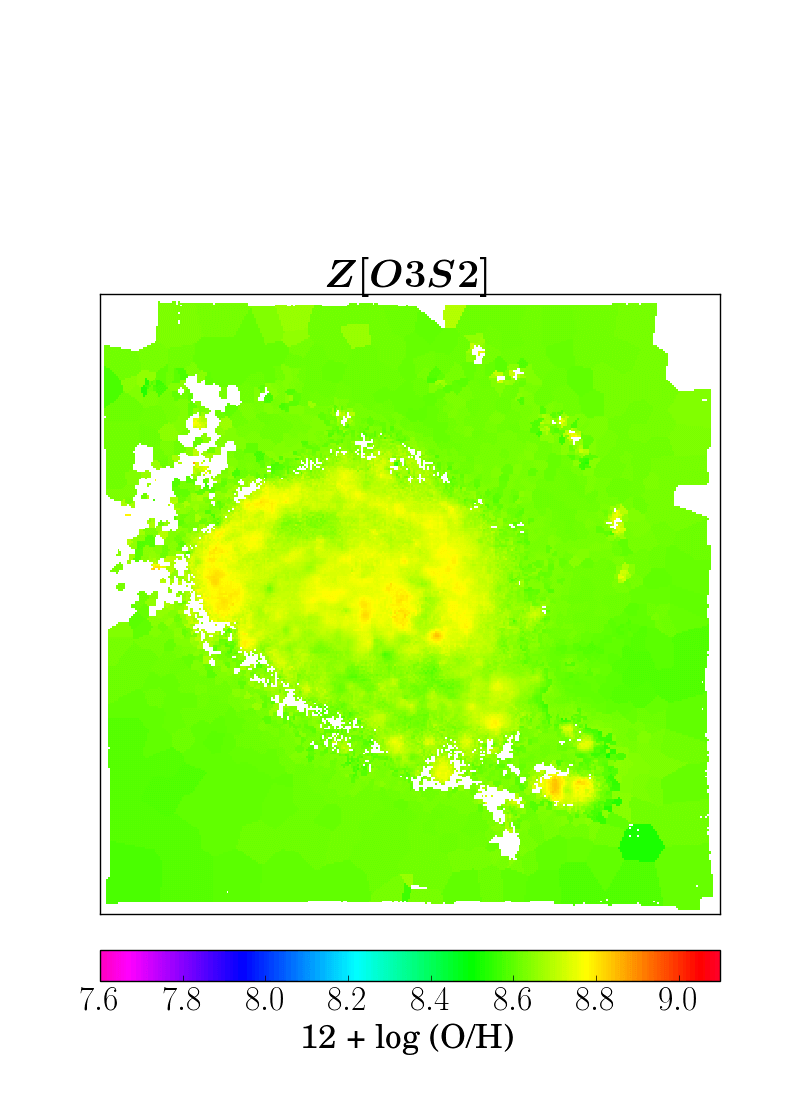}
	\includegraphics[width=0.28\textwidth, trim={2.8cm 0 2.8cm 0}, clip ]{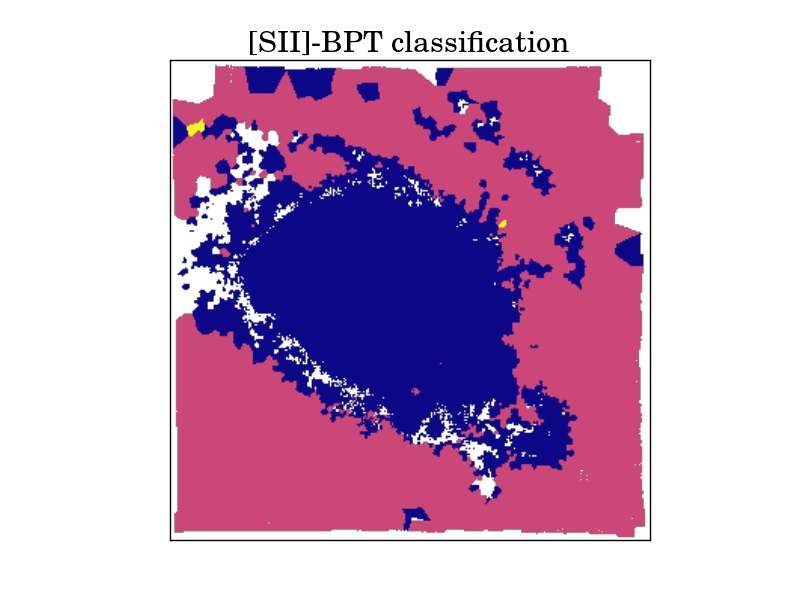}
	\includegraphics[width=0.28\textwidth, trim={0 1.2cm 0 5.5cm}, clip]{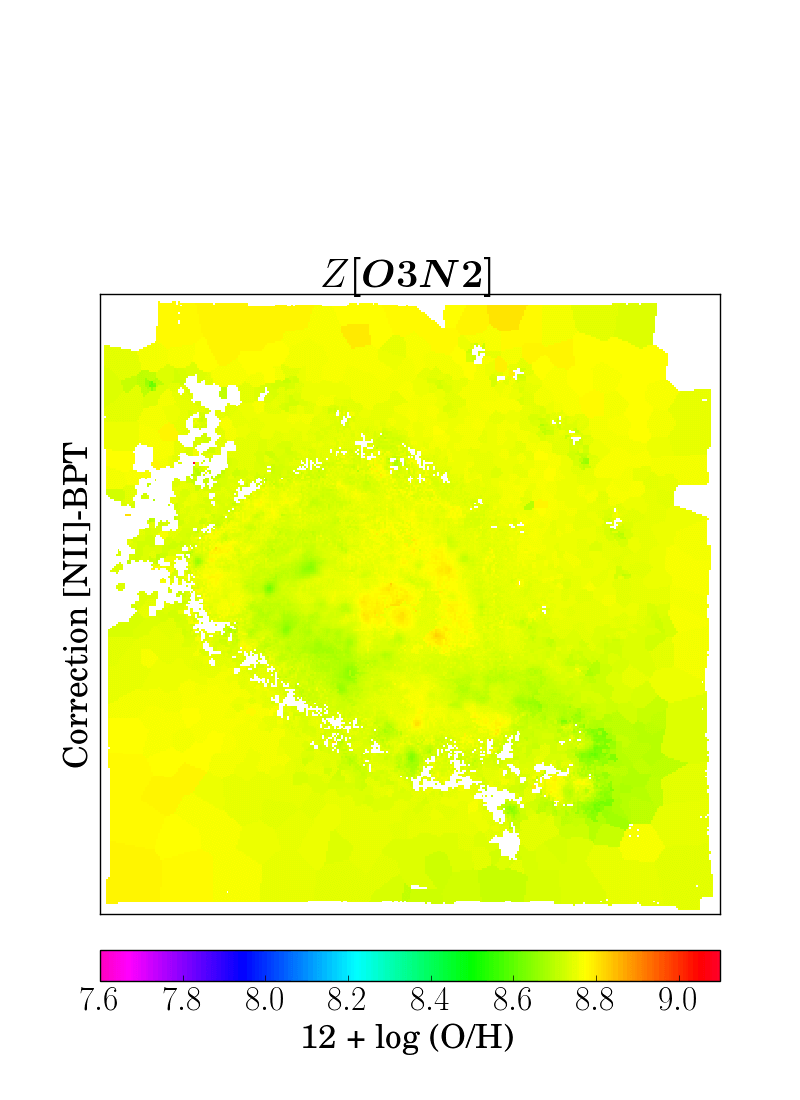}
	\includegraphics[width=0.28\textwidth, trim={0 1.2cm 0 5.5cm}, clip]{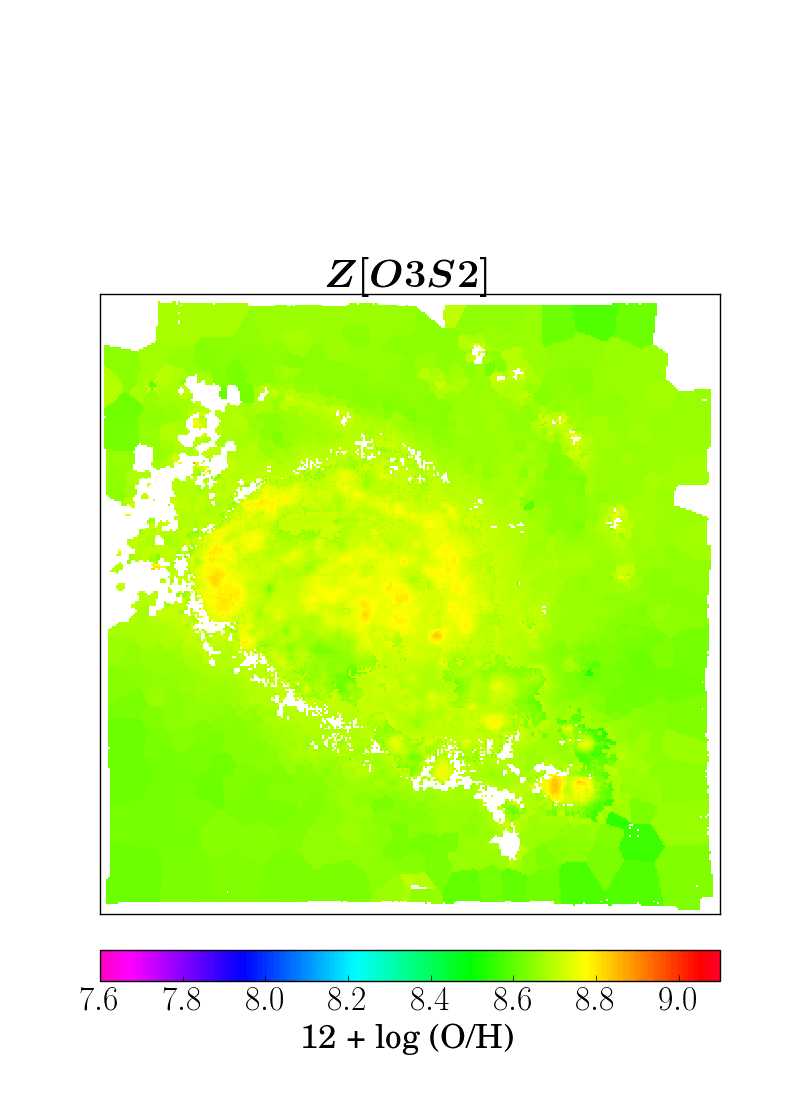}
	\includegraphics[width=0.28\textwidth, trim={2.8cm 0 2.8cm 0}, clip ]{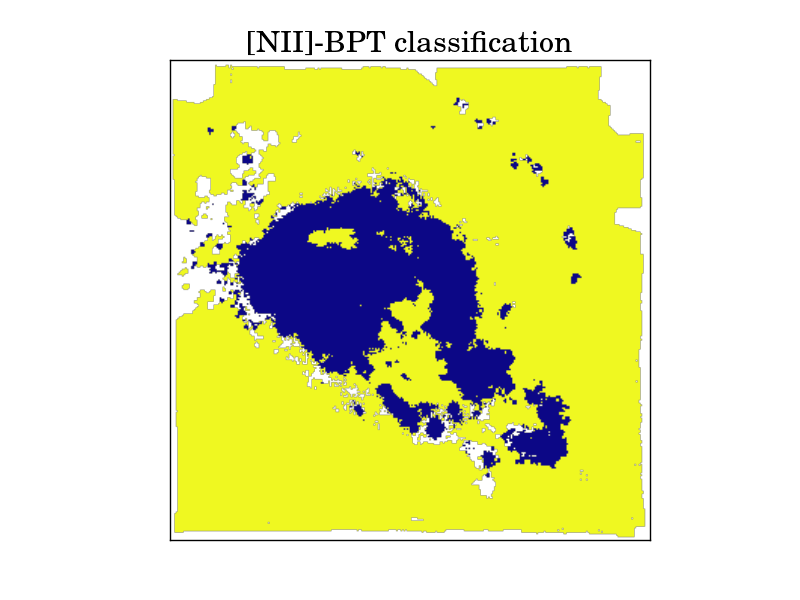}
	\caption{ Maps correspond to galaxy NGC3256, see caption of Figure \ref{fig:NGC1042} for details.}
	\label{fig:NGC3256}
\end{figure*}
\begin{figure*}
	\centering
	\includegraphics[width=0.28\textwidth, trim={0 1.2cm 0 5.5cm}, clip]{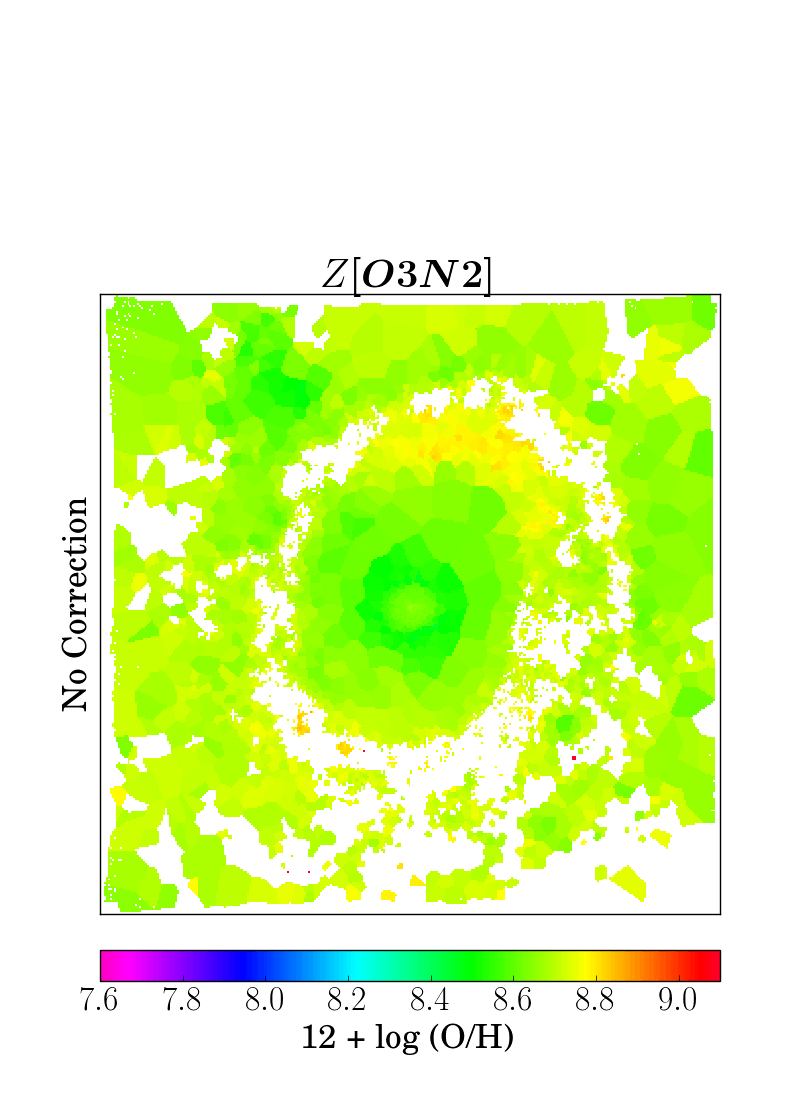}
	\includegraphics[width=0.28\textwidth, trim={0 1.2cm 0 5.5cm}, clip]{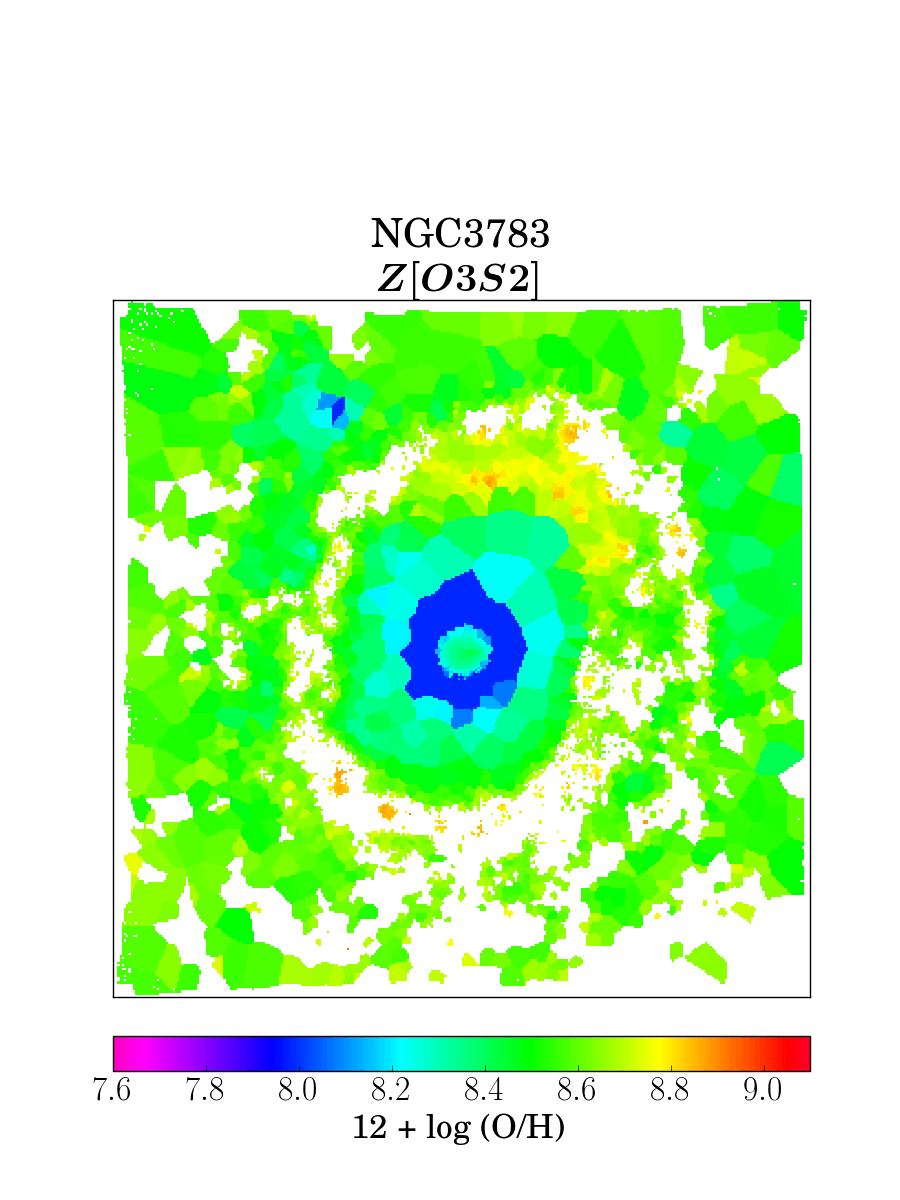}
	\includegraphics[width=0.28\textwidth, trim={0 1.2cm 0 5.5cm}, clip]{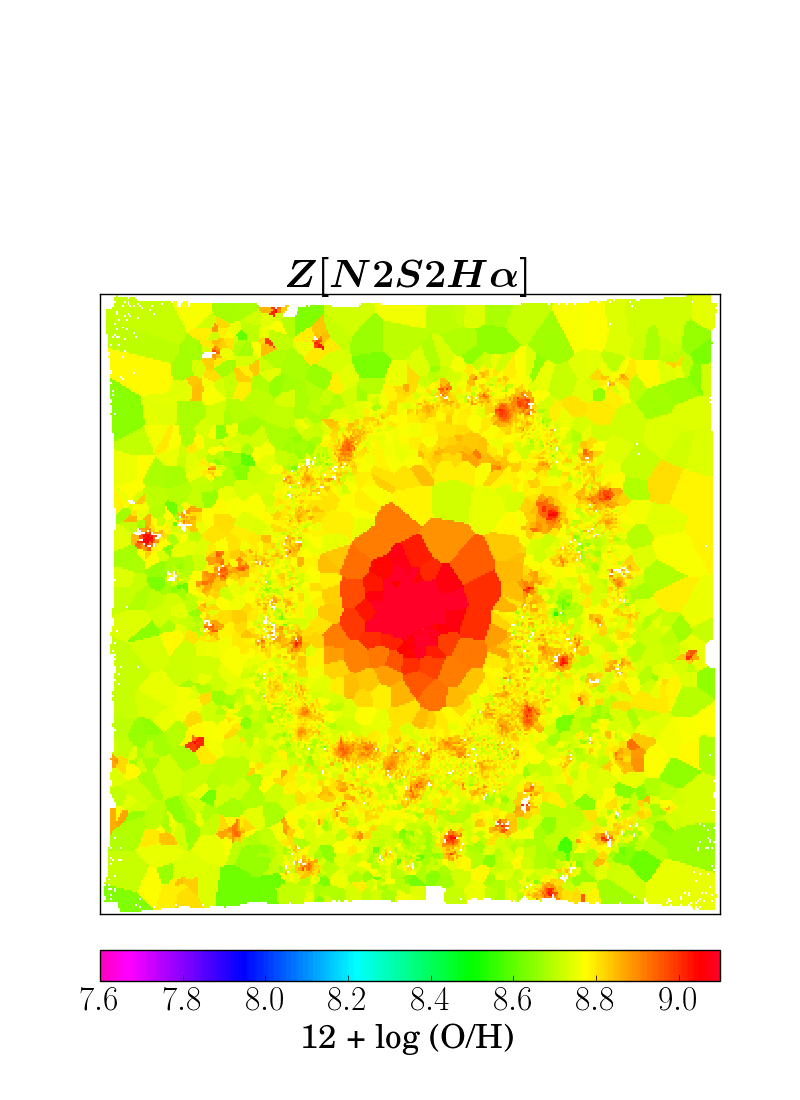}
	\includegraphics[width=0.28\textwidth, trim={0 1.2cm 0 5.5cm}, clip]{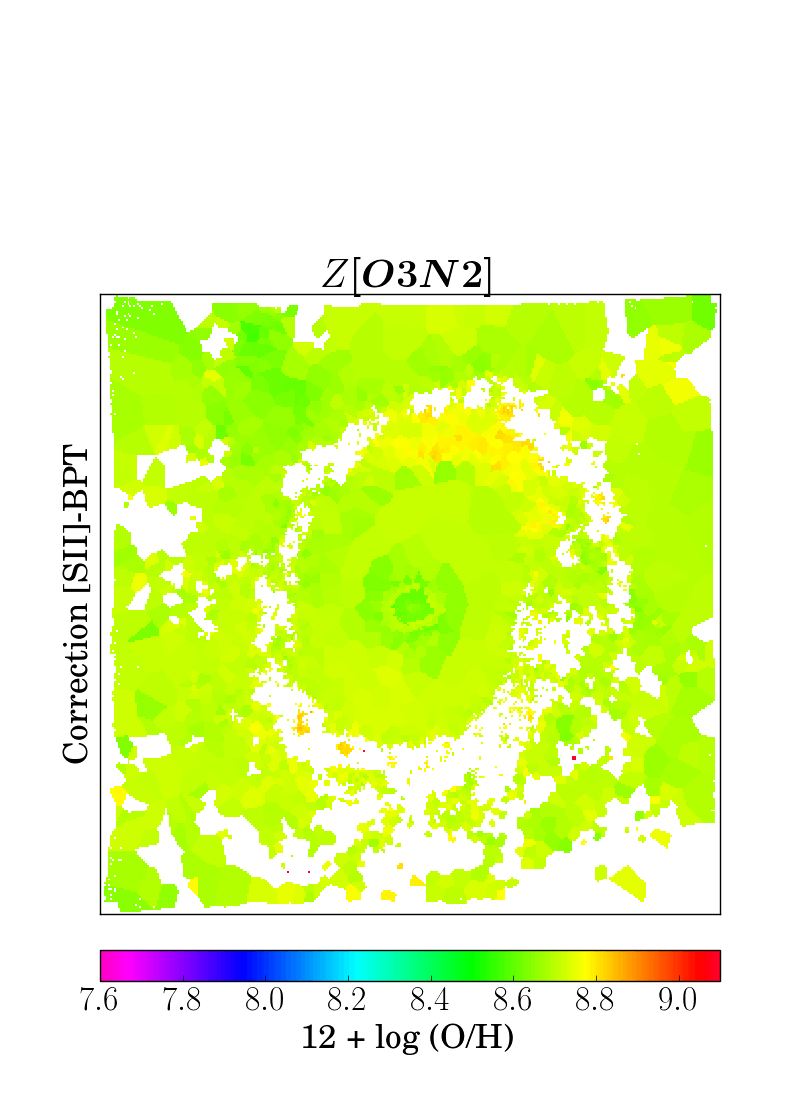}
	\includegraphics[width=0.28\textwidth, trim={0 1.2cm 0 5.5cm}, clip]{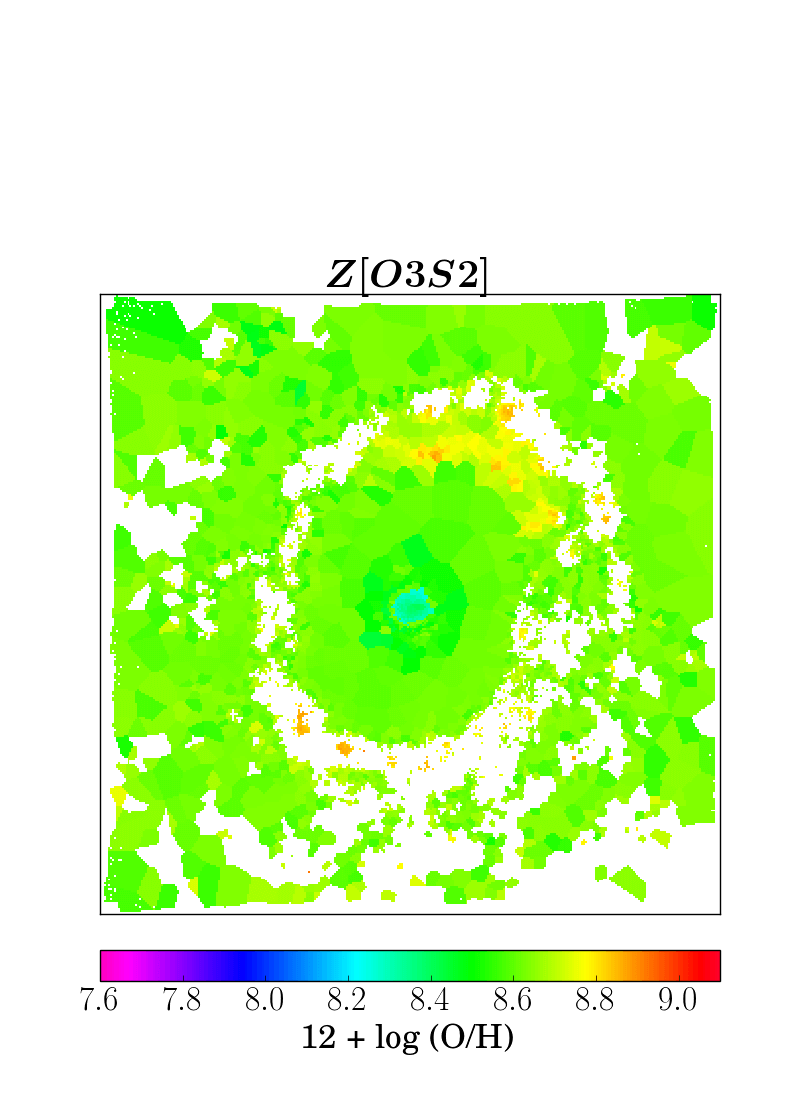}
	\includegraphics[width=0.28\textwidth, trim={2.8cm 0 2.8cm 0}, clip ]{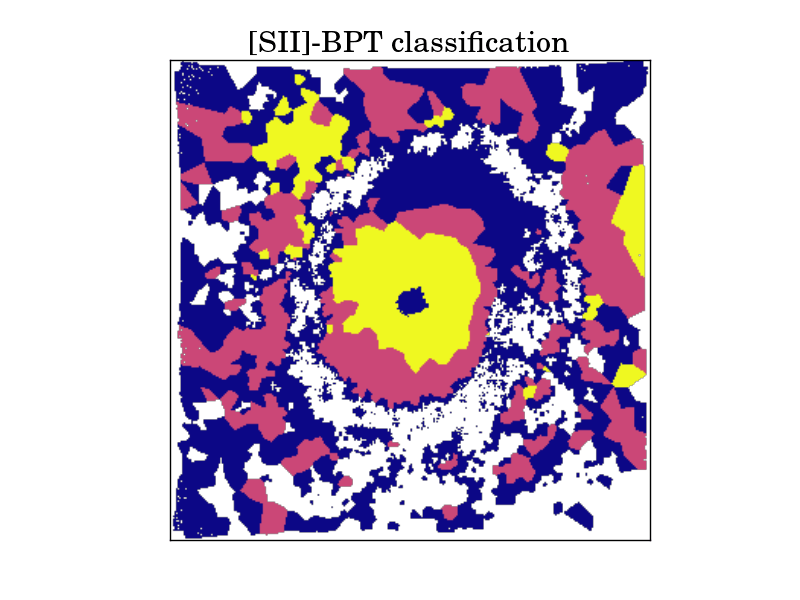}
	\includegraphics[width=0.28\textwidth, trim={0 1.2cm 0 5.5cm}, clip]{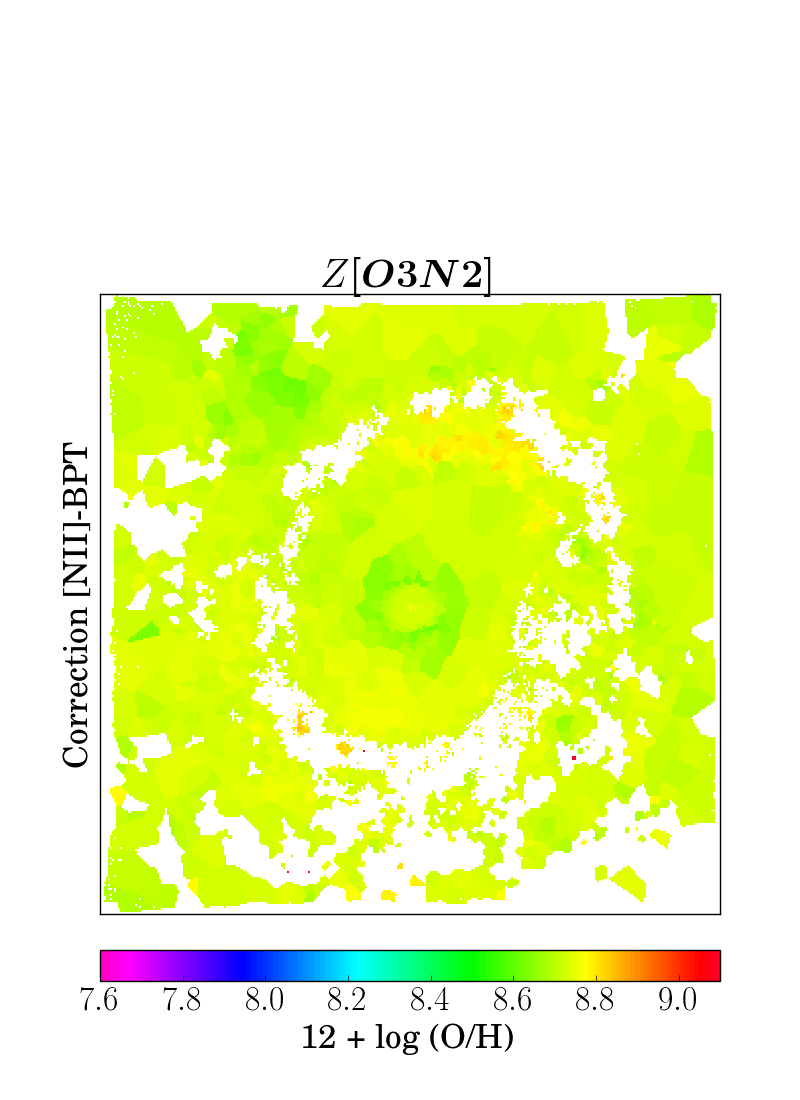}
	\includegraphics[width=0.28\textwidth, trim={0 1.2cm 0 5.5cm}, clip]{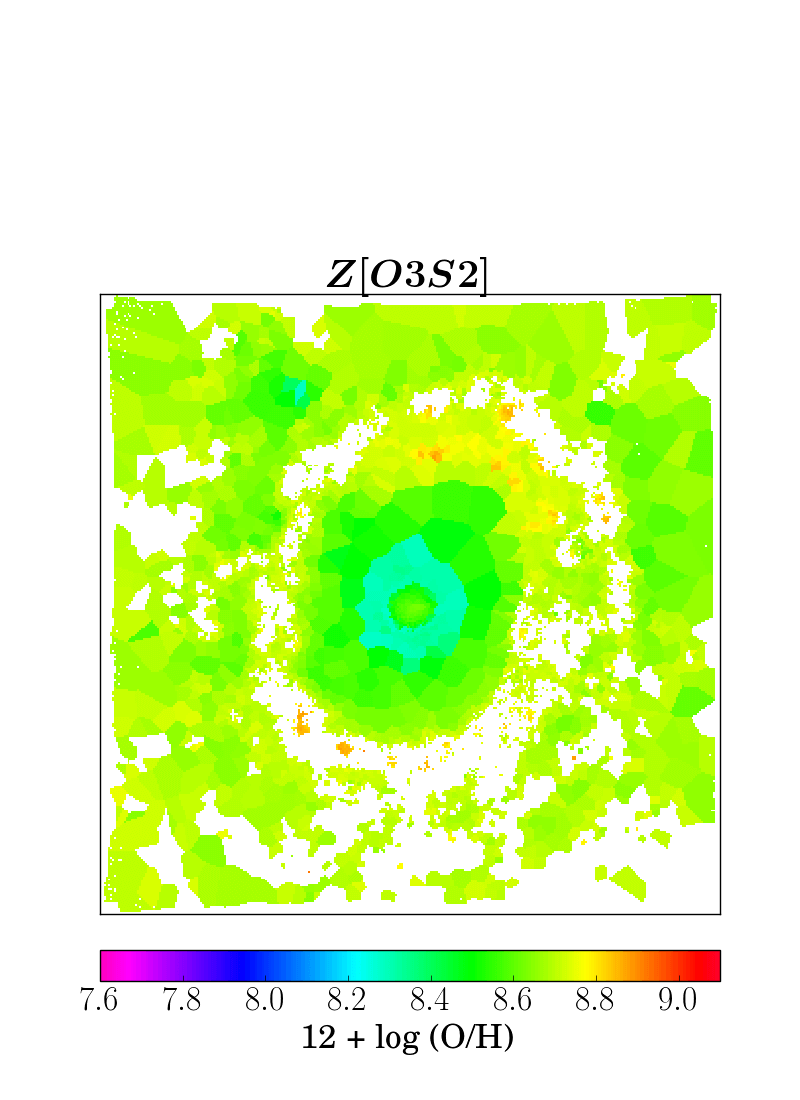}
	\includegraphics[width=0.28\textwidth, trim={2.8cm 0 2.8cm 0}, clip ]{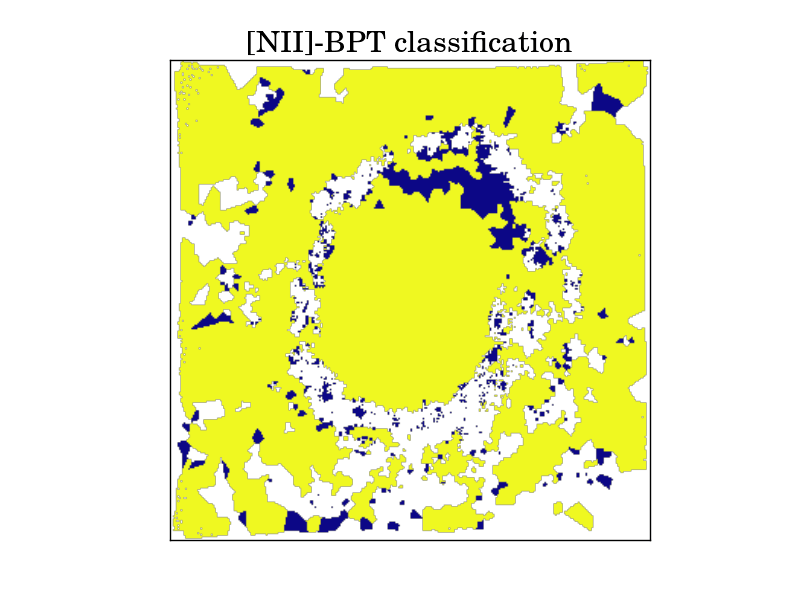}
	\caption{ Maps correspond to galaxy NGC3783, see caption of Figure \ref{fig:NGC1042} for details.}
	\label{fig:NGC3783}
\end{figure*}
\begin{figure*}
	\centering
	\includegraphics[width=0.28\textwidth, trim={0 1.2cm 0 5.5cm}, clip]{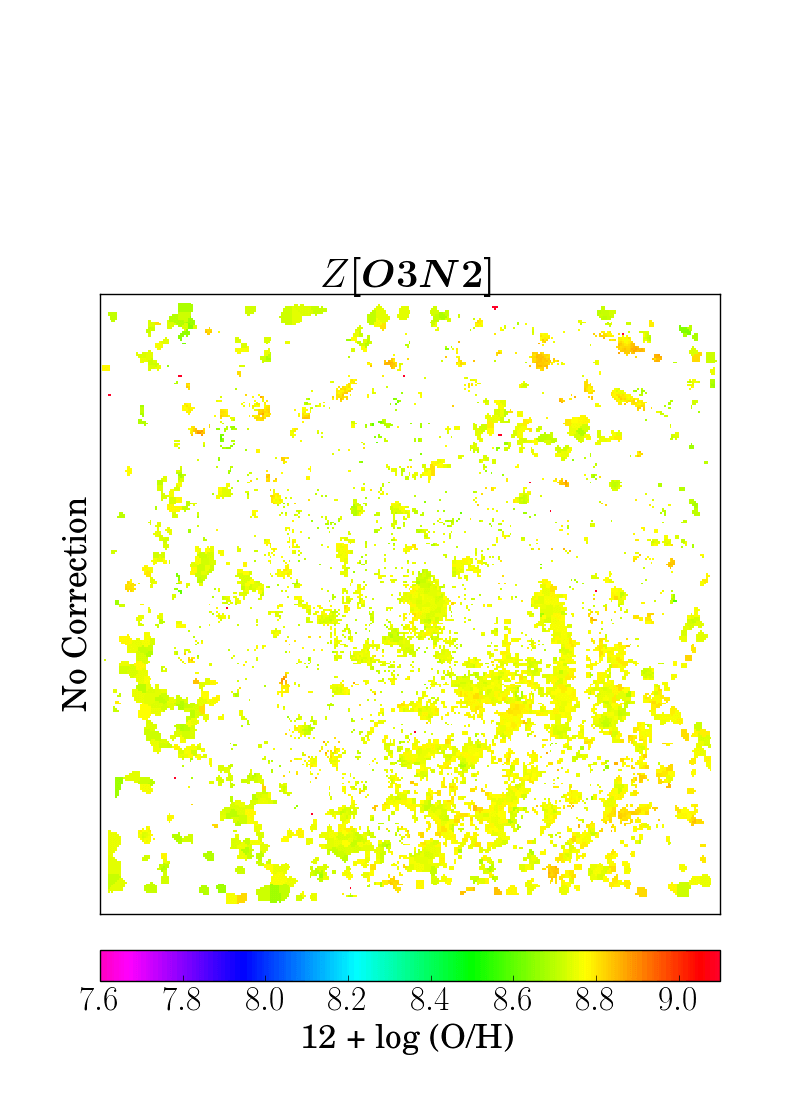}
	\includegraphics[width=0.28\textwidth, trim={0 1.2cm 0 5.5cm}, clip]{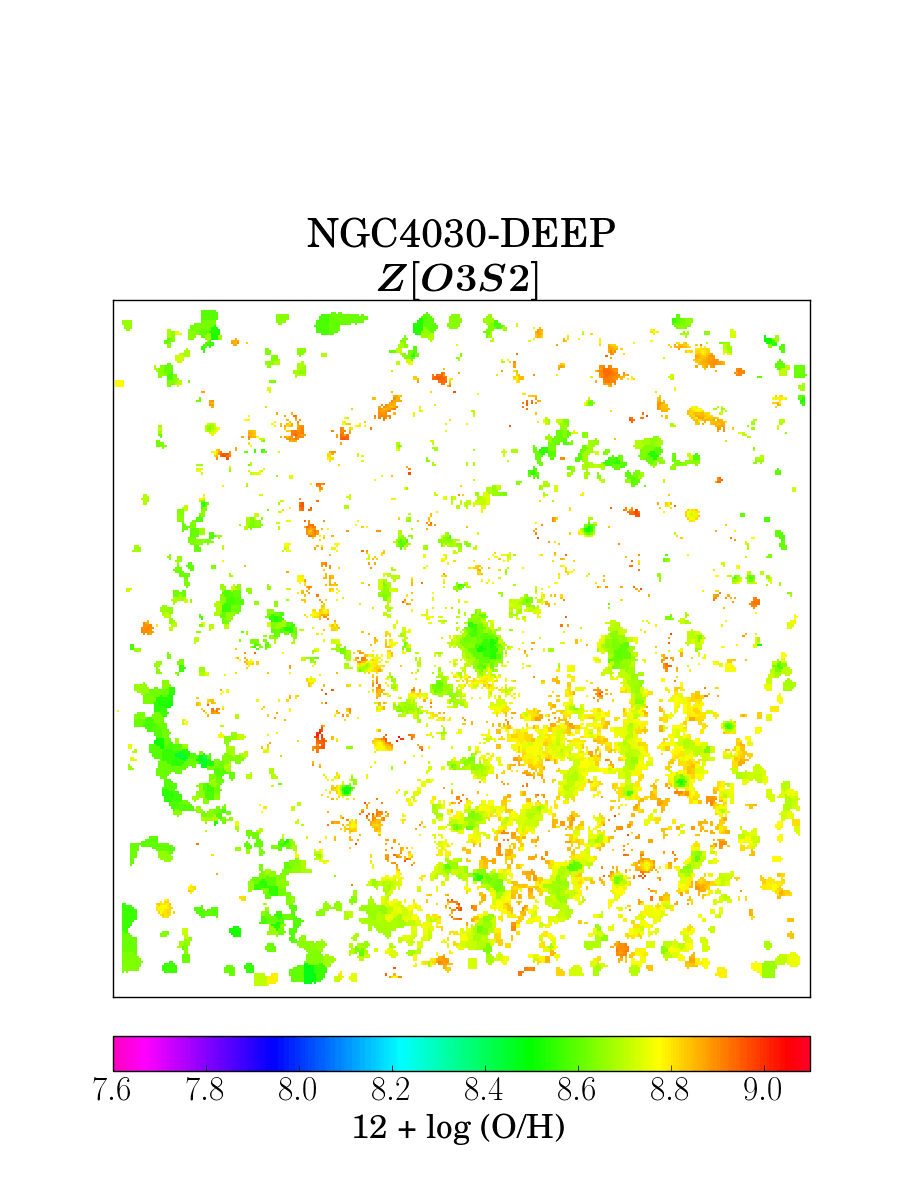}
	\includegraphics[width=0.28\textwidth, trim={0 1.2cm 0 5.5cm}, clip]{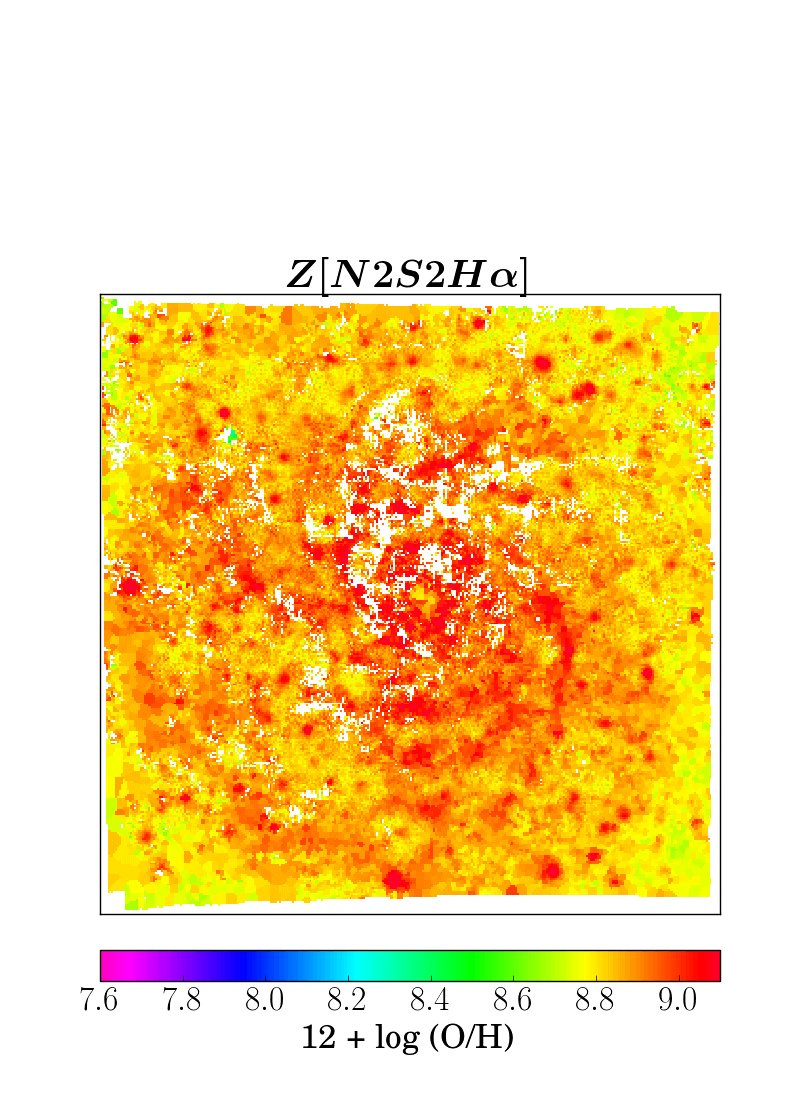}
	\includegraphics[width=0.28\textwidth, trim={0 1.2cm 0 5.5cm}, clip]{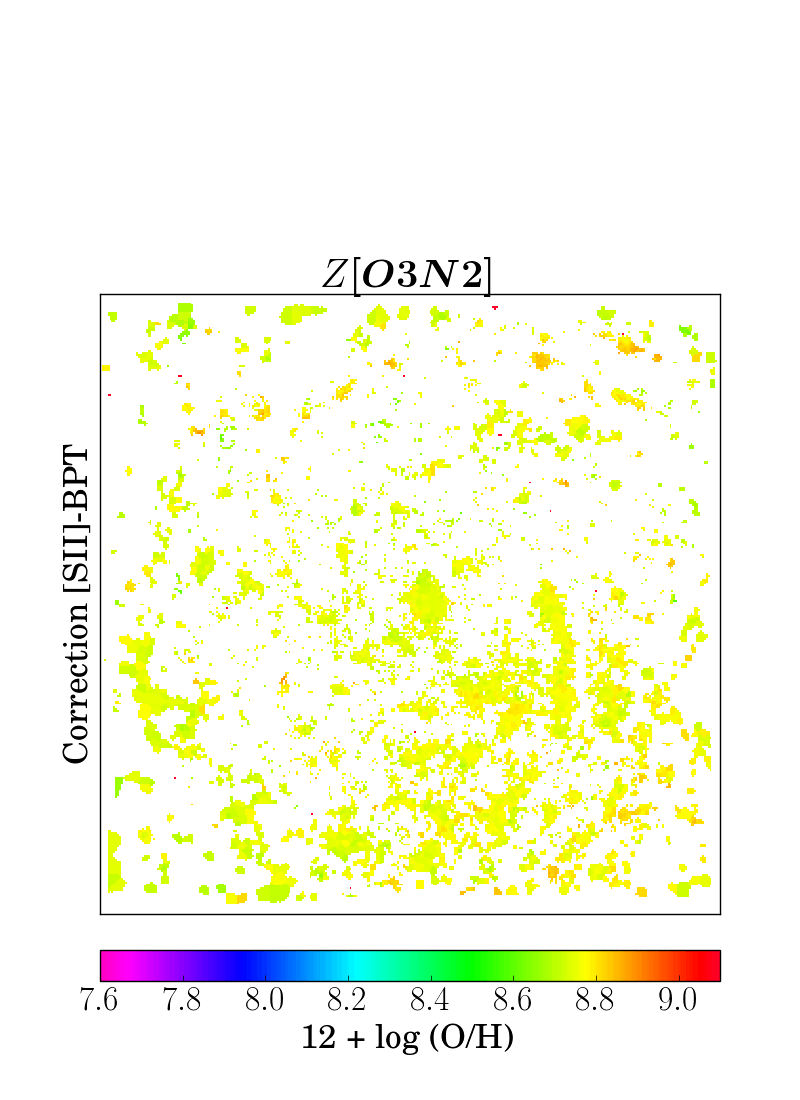}
	\includegraphics[width=0.28\textwidth, trim={0 1.2cm 0 5.5cm}, clip]{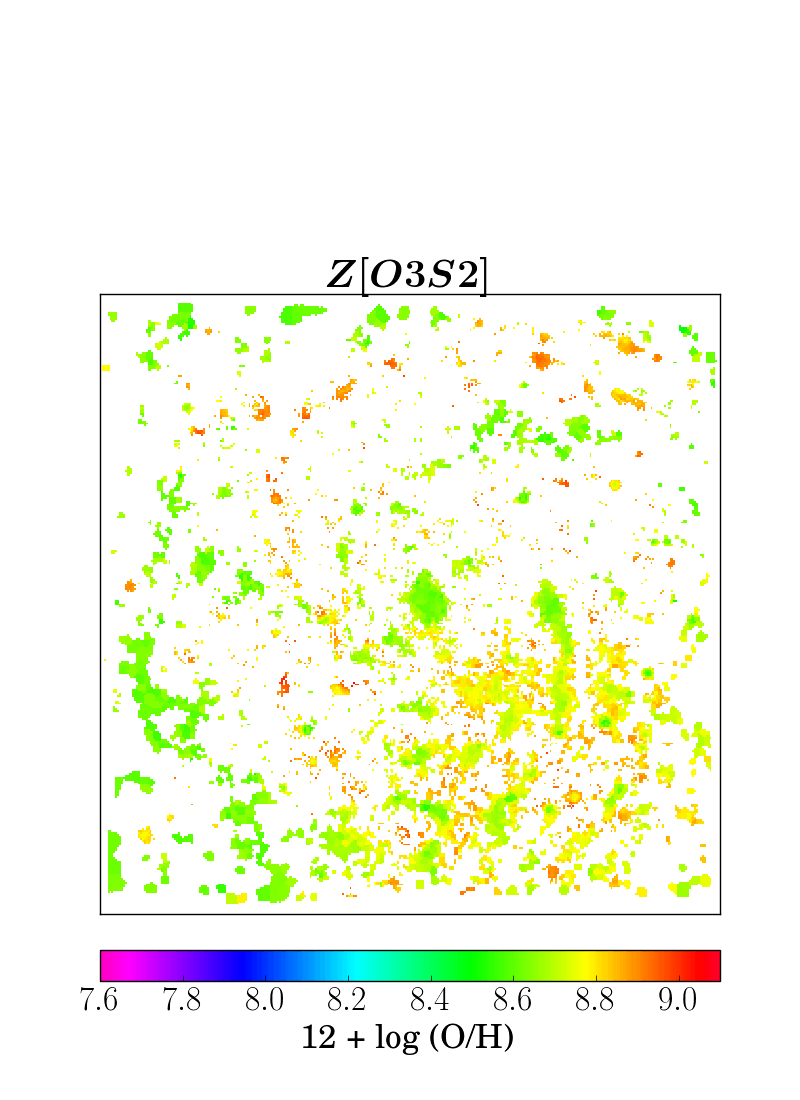}
	\includegraphics[width=0.28\textwidth, trim={2.8cm 0 2.8cm 0}, clip ]{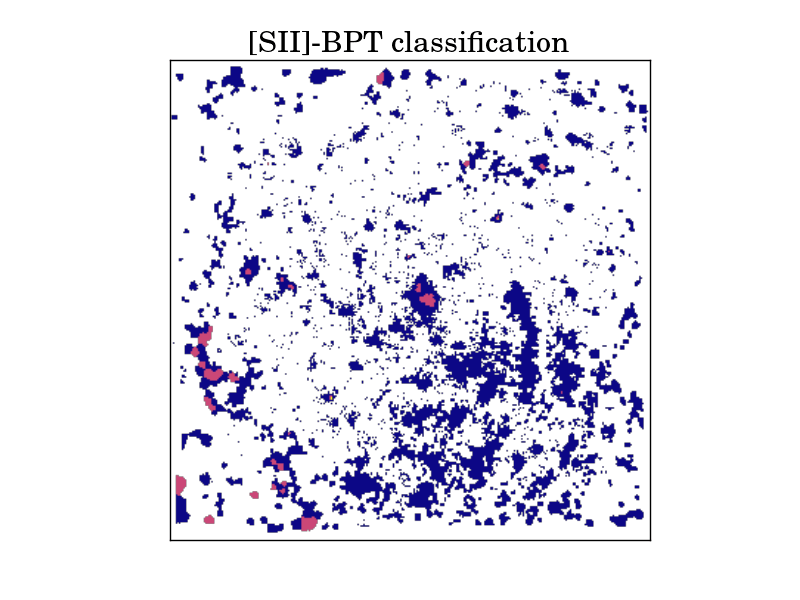}
	\includegraphics[width=0.28\textwidth, trim={0 1.2cm 0 5.5cm}, clip]{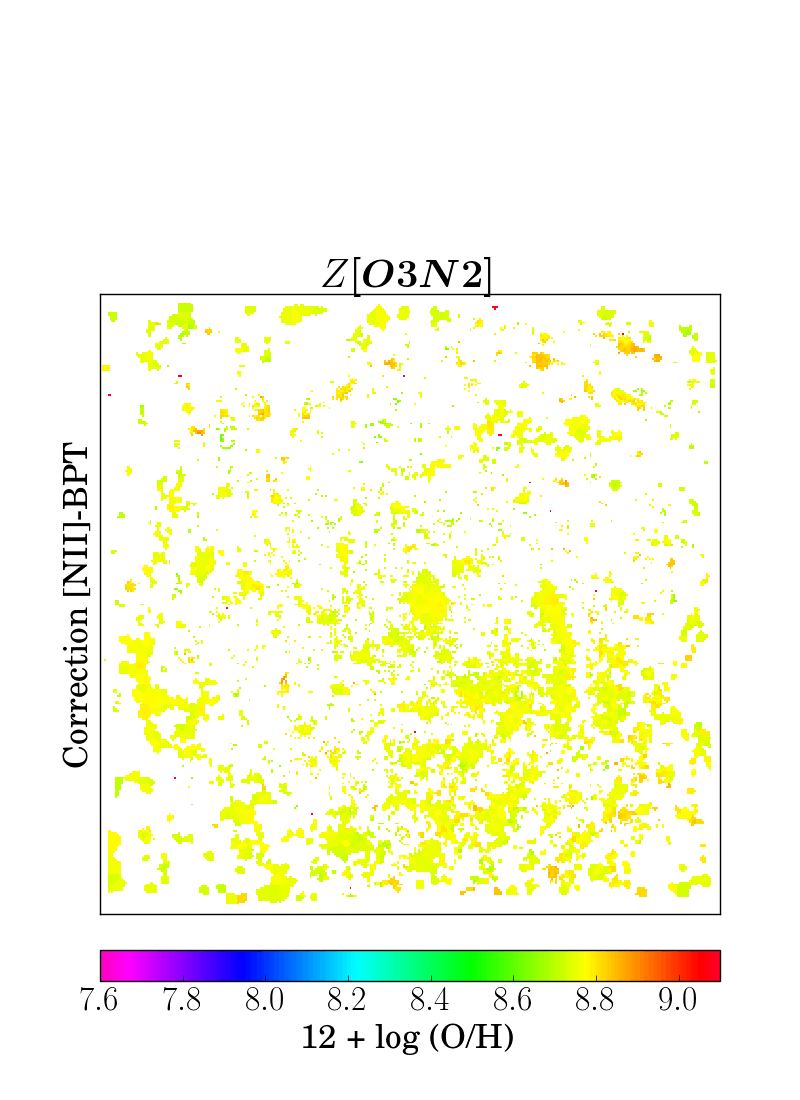}
	\includegraphics[width=0.28\textwidth, trim={0 1.2cm 0 5.5cm}, clip]{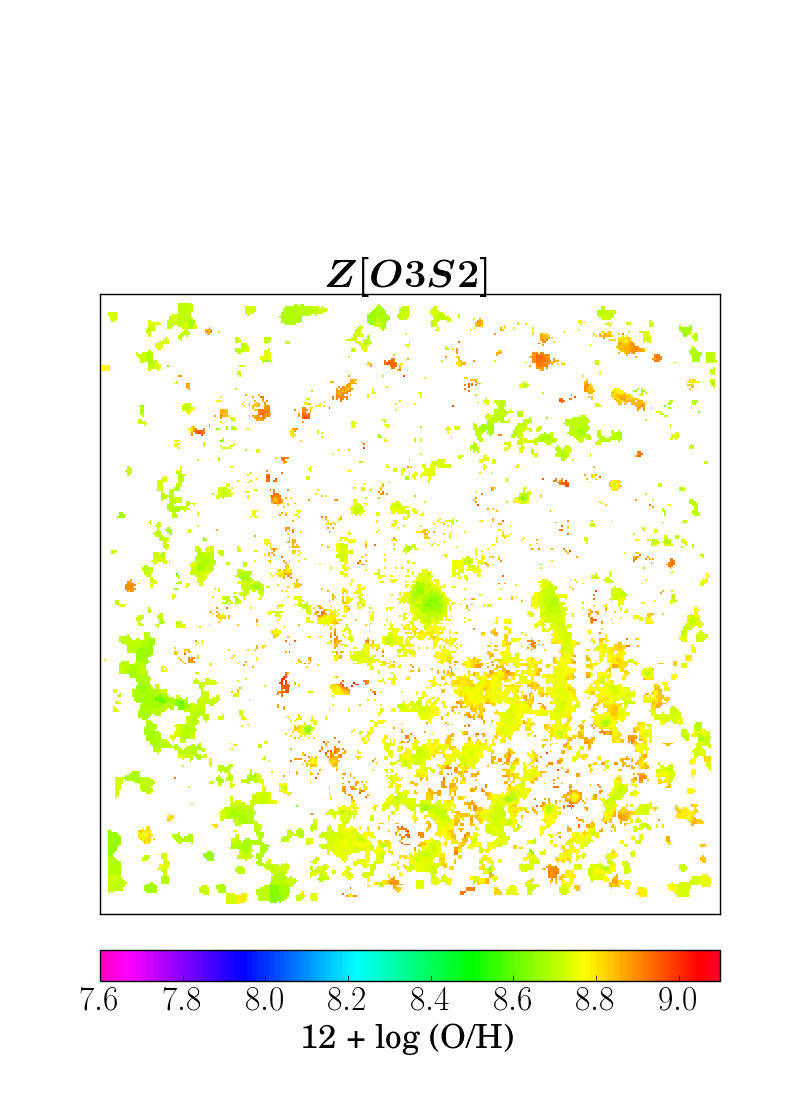}
	\includegraphics[width=0.28\textwidth, trim={2.8cm 0 2.8cm 0}, clip ]{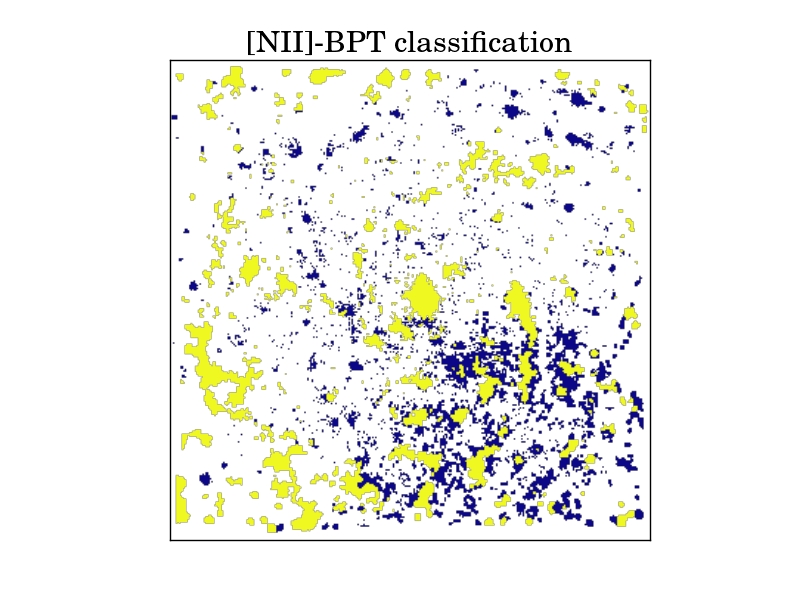}
	\caption{ Maps correspond to galaxy NGC4030-DEEP, see caption of Figure \ref{fig:NGC1042} for details.}
	\label{fig:NGC4030-DEEP}
\end{figure*}
\begin{figure*}
	\centering
	\includegraphics[width=0.28\textwidth, trim={0 1.2cm 0 5.5cm}, clip]{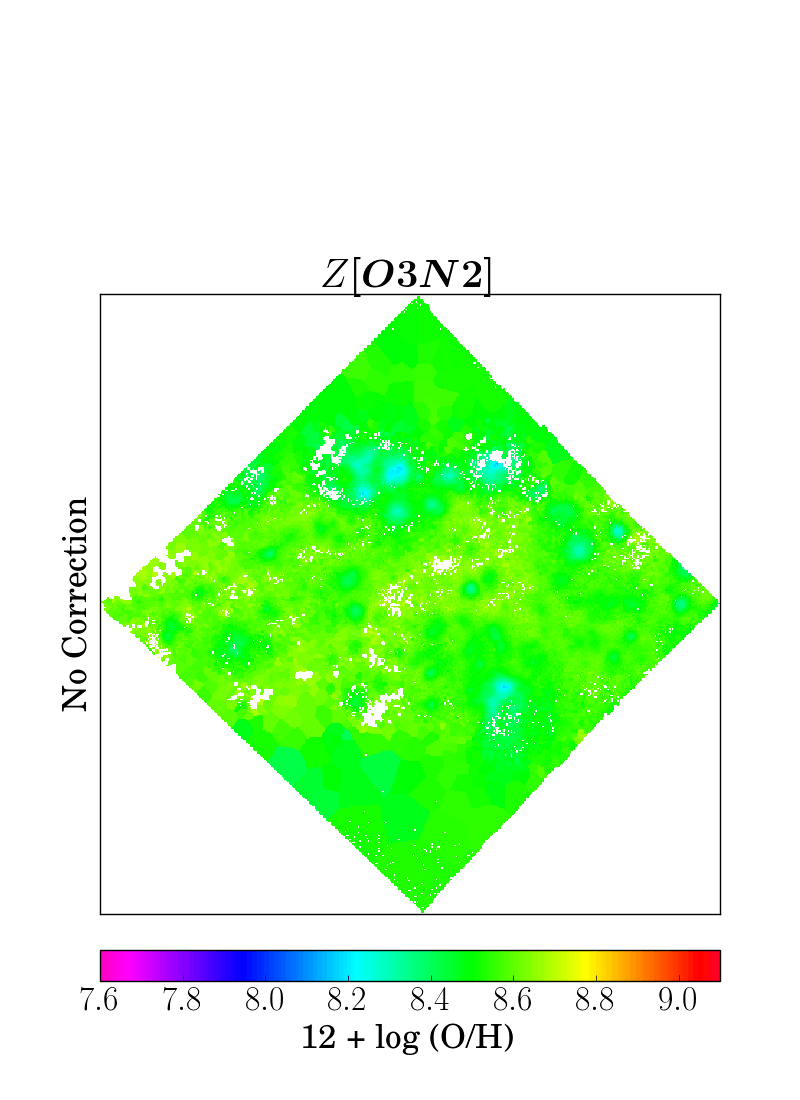}
	\includegraphics[width=0.28\textwidth, trim={0 1.2cm 0 5.5cm}, clip]{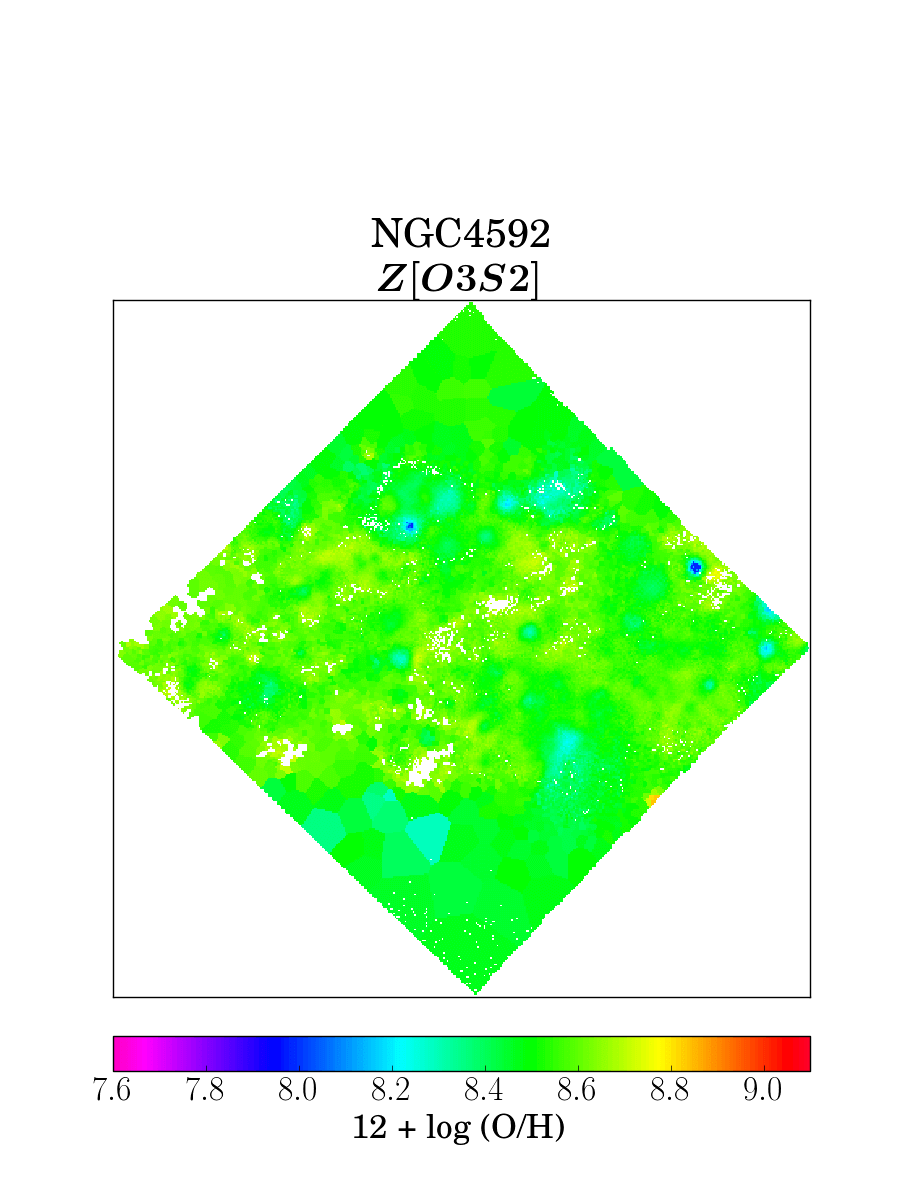}
	\includegraphics[width=0.28\textwidth, trim={0 1.2cm 0 5.5cm}, clip]{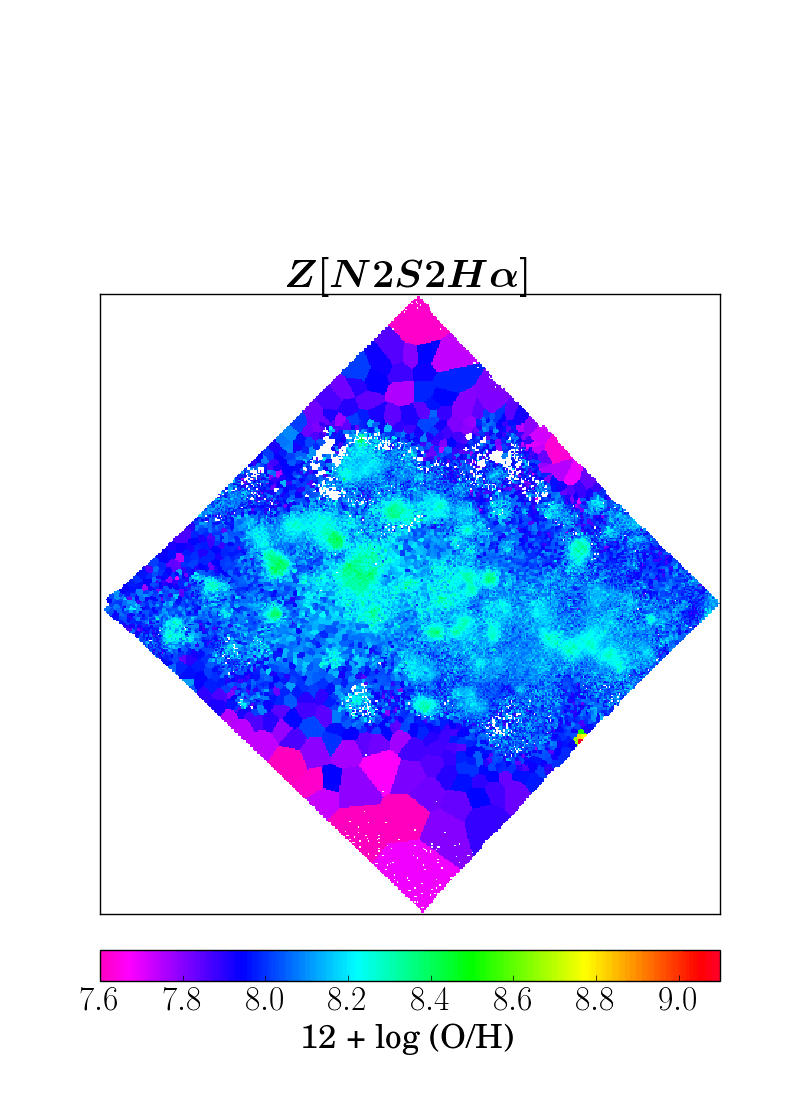}
	\includegraphics[width=0.28\textwidth, trim={0 1.2cm 0 5.5cm}, clip]{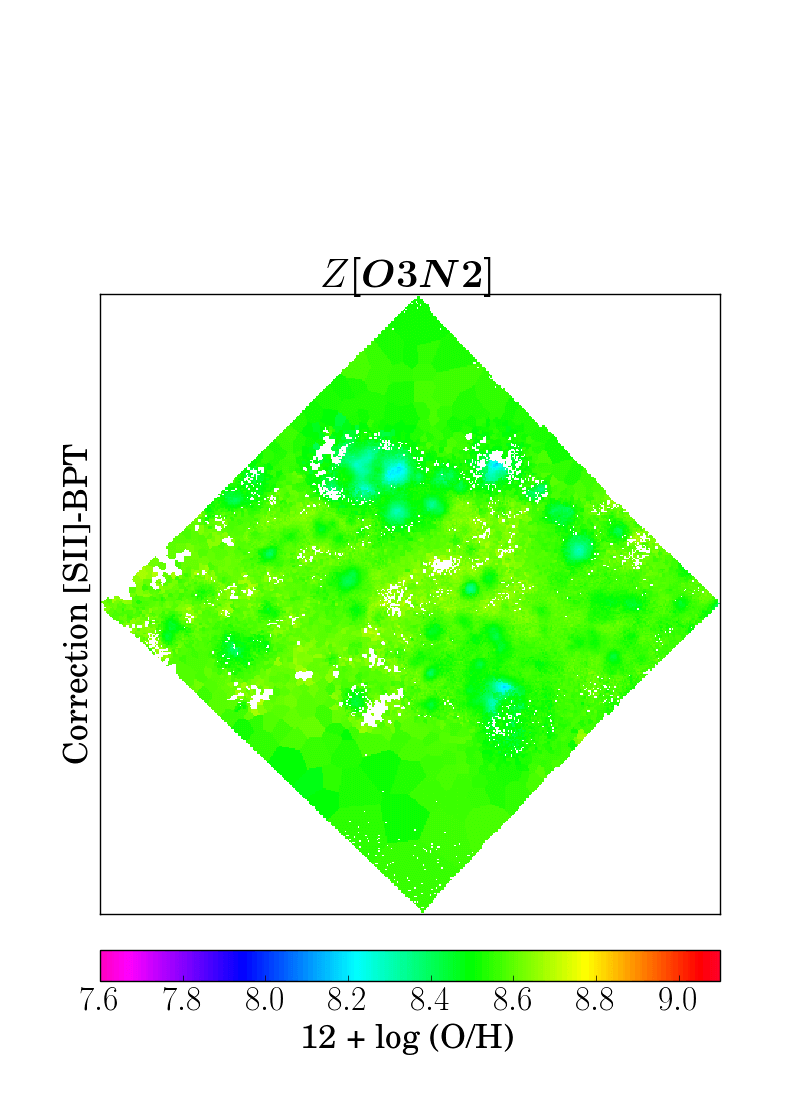}
	\includegraphics[width=0.28\textwidth, trim={0 1.2cm 0 5.5cm}, clip]{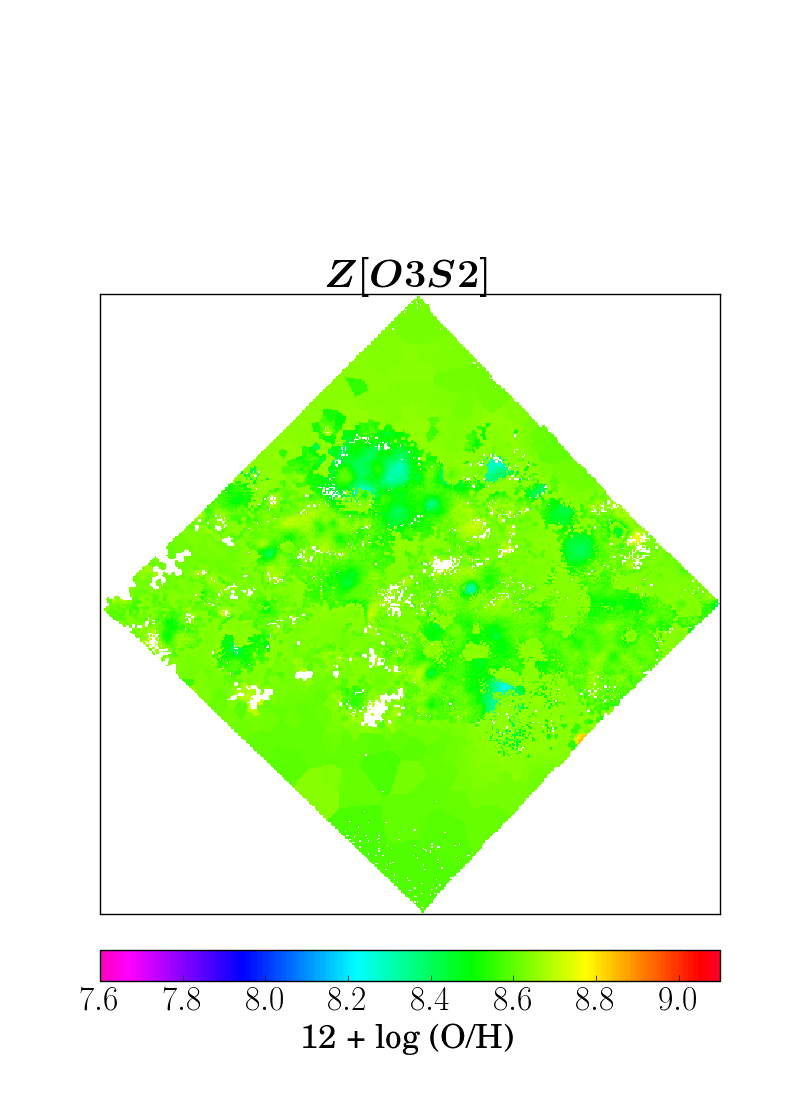}
	\includegraphics[width=0.28\textwidth, trim={2.8cm 0 2.8cm 0}, clip ]{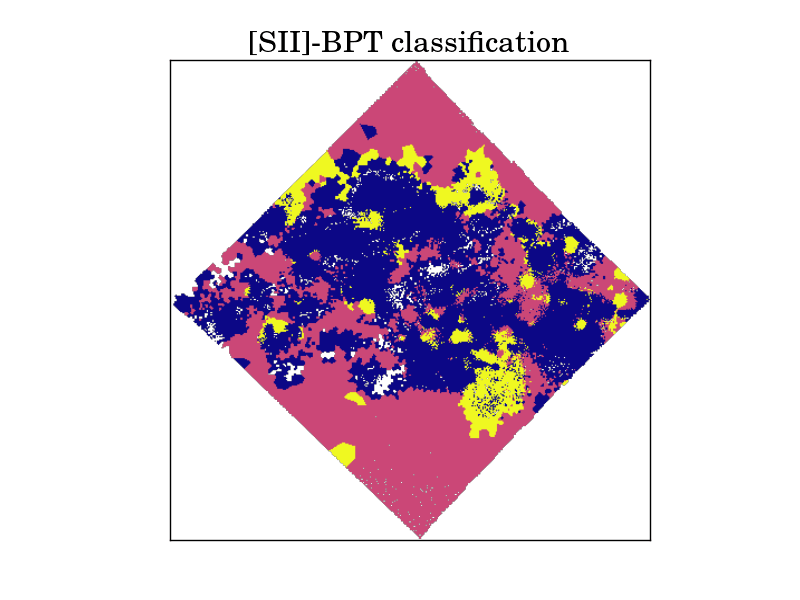}
	\includegraphics[width=0.28\textwidth, trim={0 1.2cm 0 5.5cm}, clip]{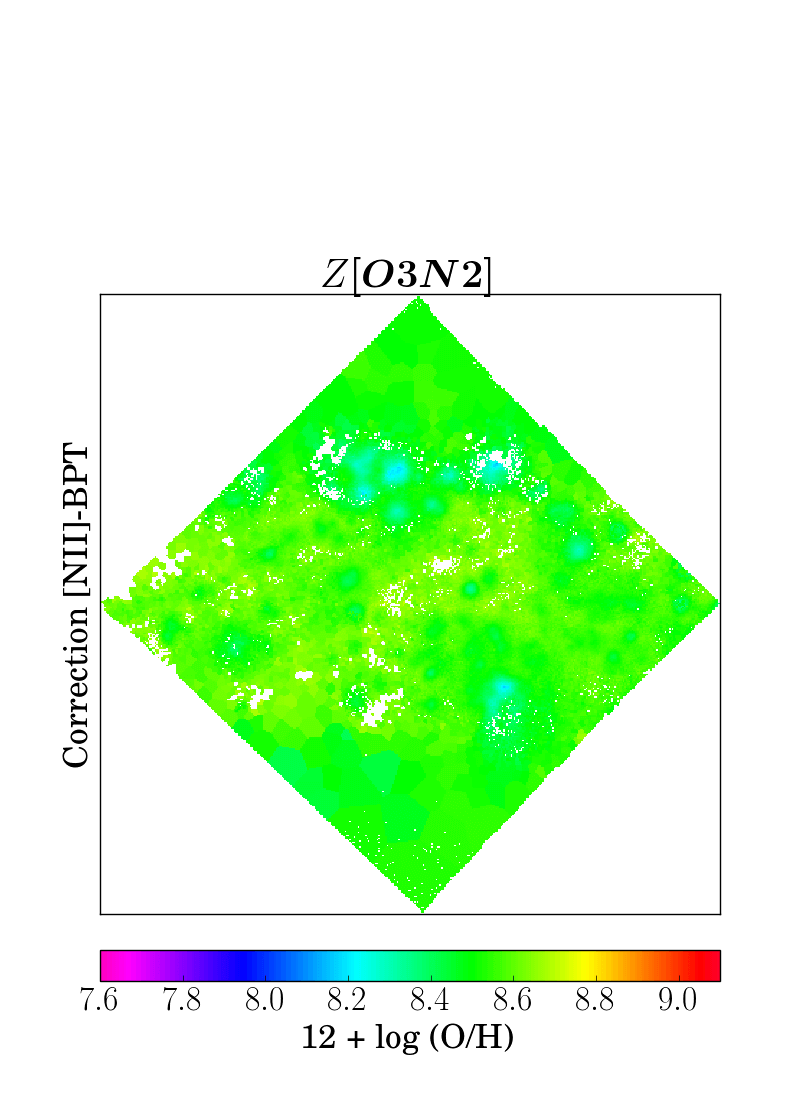}
	\includegraphics[width=0.28\textwidth, trim={0 1.2cm 0 5.5cm}, clip]{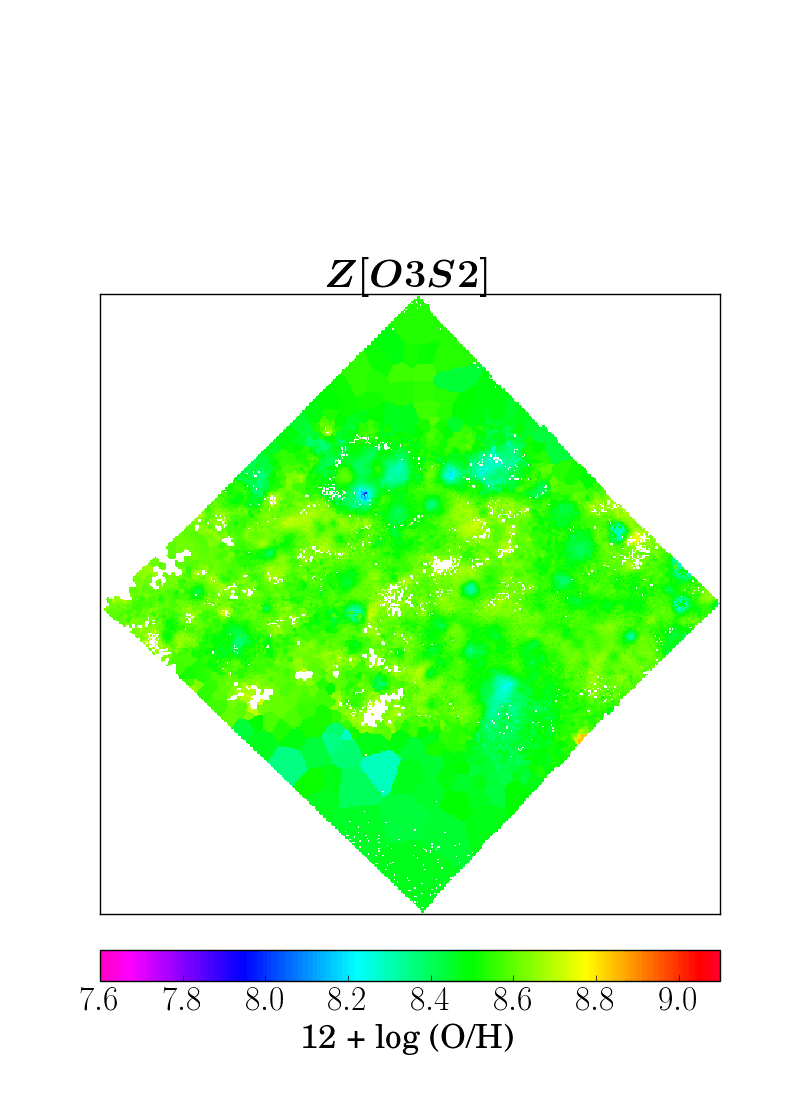}
	\includegraphics[width=0.28\textwidth, trim={2.8cm 0 2.8cm 0}, clip ]{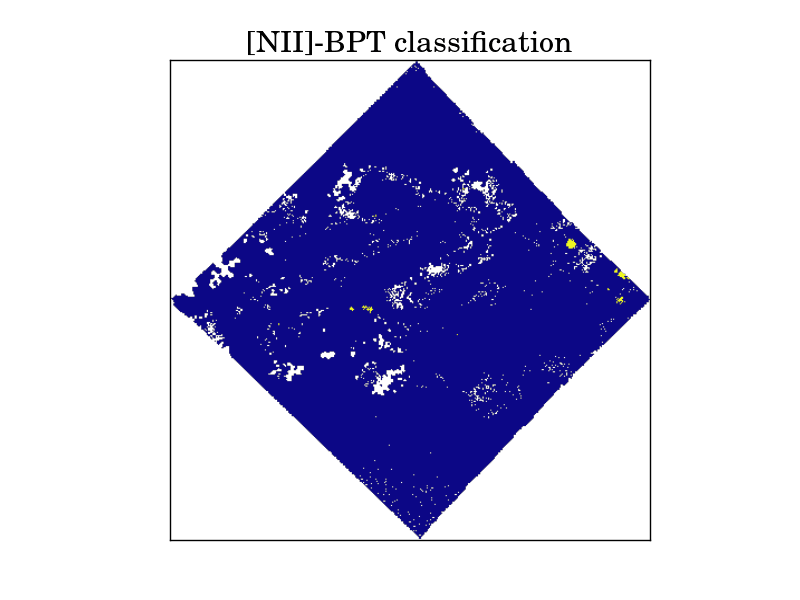}
	\caption{ Maps correspond to galaxy NGC4592, see caption of Figure \ref{fig:NGC1042} for details.}
	\label{fig:NGC4592}
\end{figure*}
\begin{figure*}
	\centering
	\includegraphics[width=0.28\textwidth, trim={0 1.2cm 0 5.5cm}, clip]{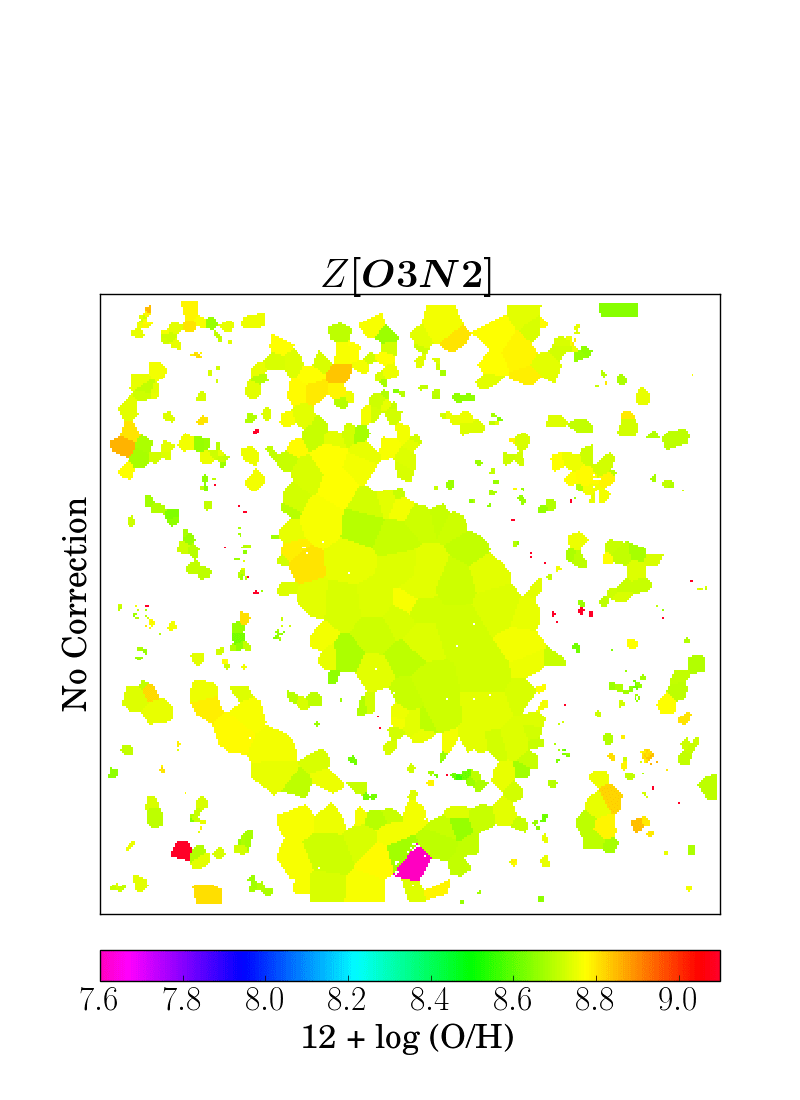}
	\includegraphics[width=0.28\textwidth, trim={0 1.2cm 0 5.5cm}, clip]{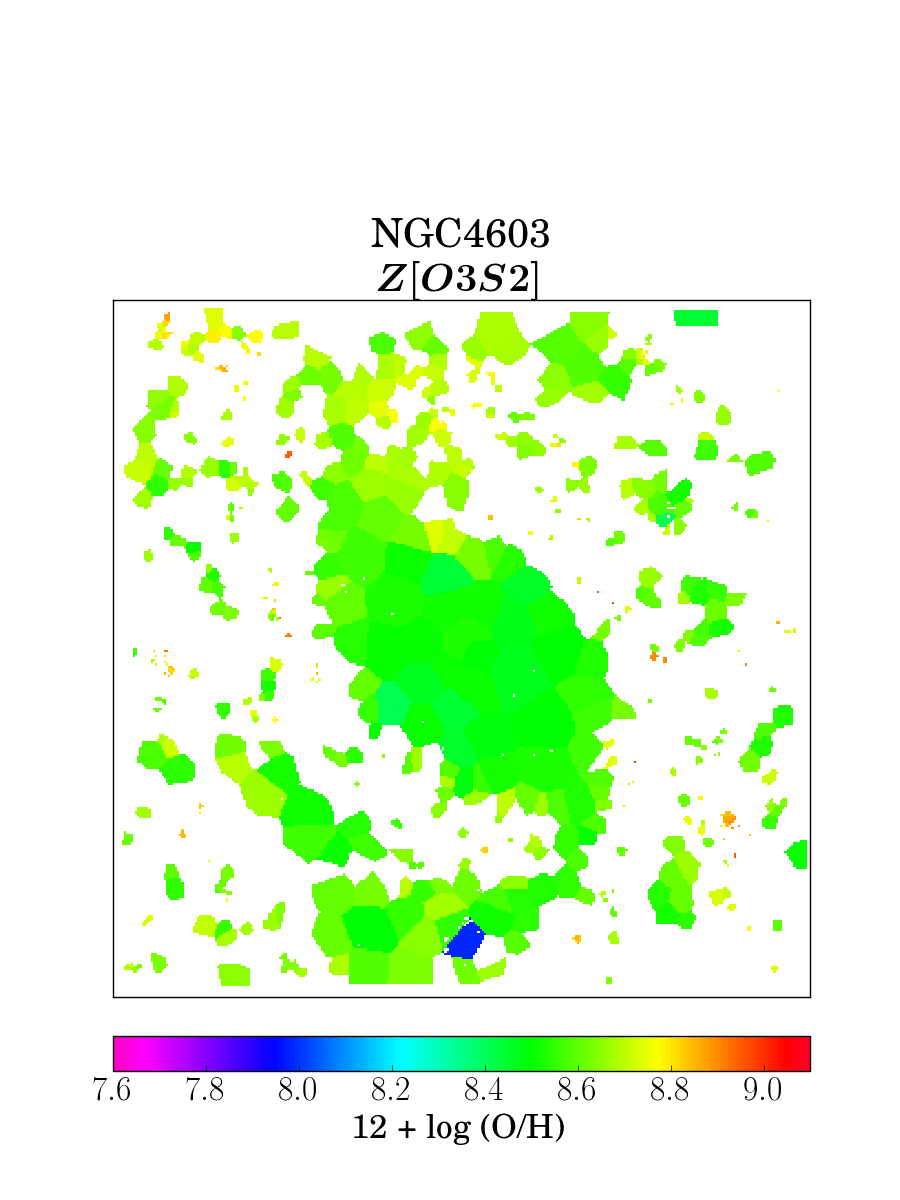}
	\includegraphics[width=0.28\textwidth, trim={0 1.2cm 0 5.5cm}, clip]{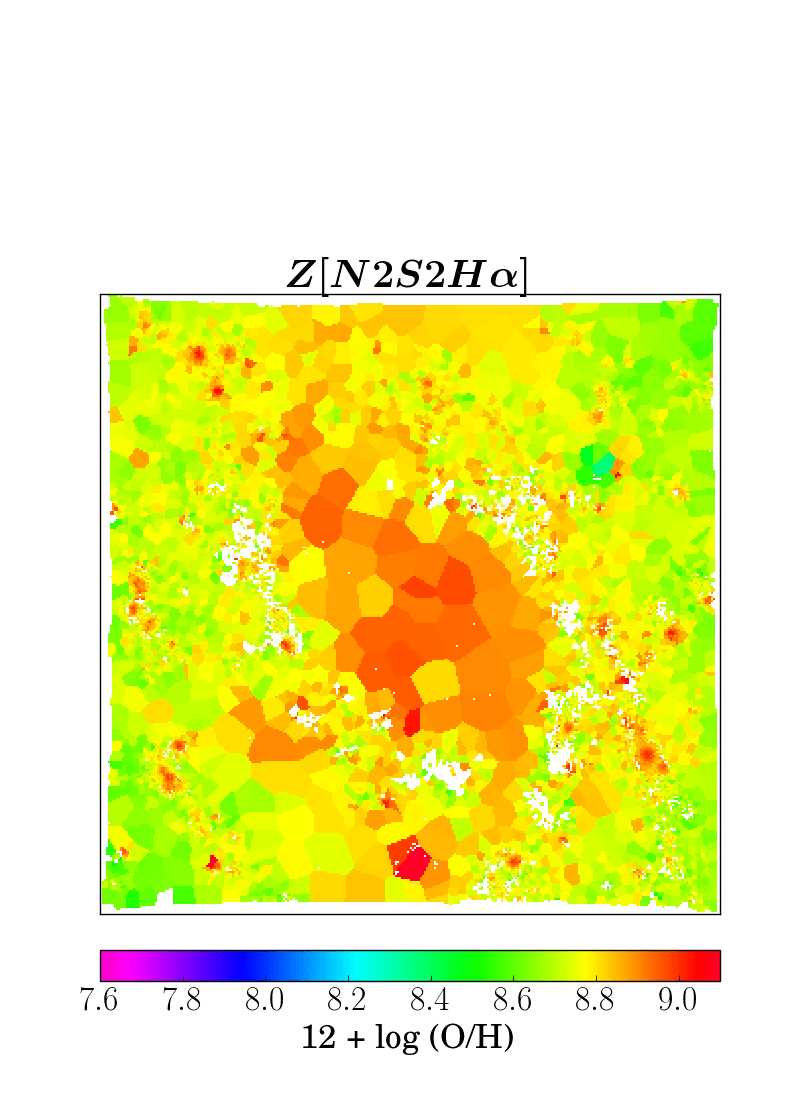}
	\includegraphics[width=0.28\textwidth, trim={0 1.2cm 0 5.5cm}, clip]{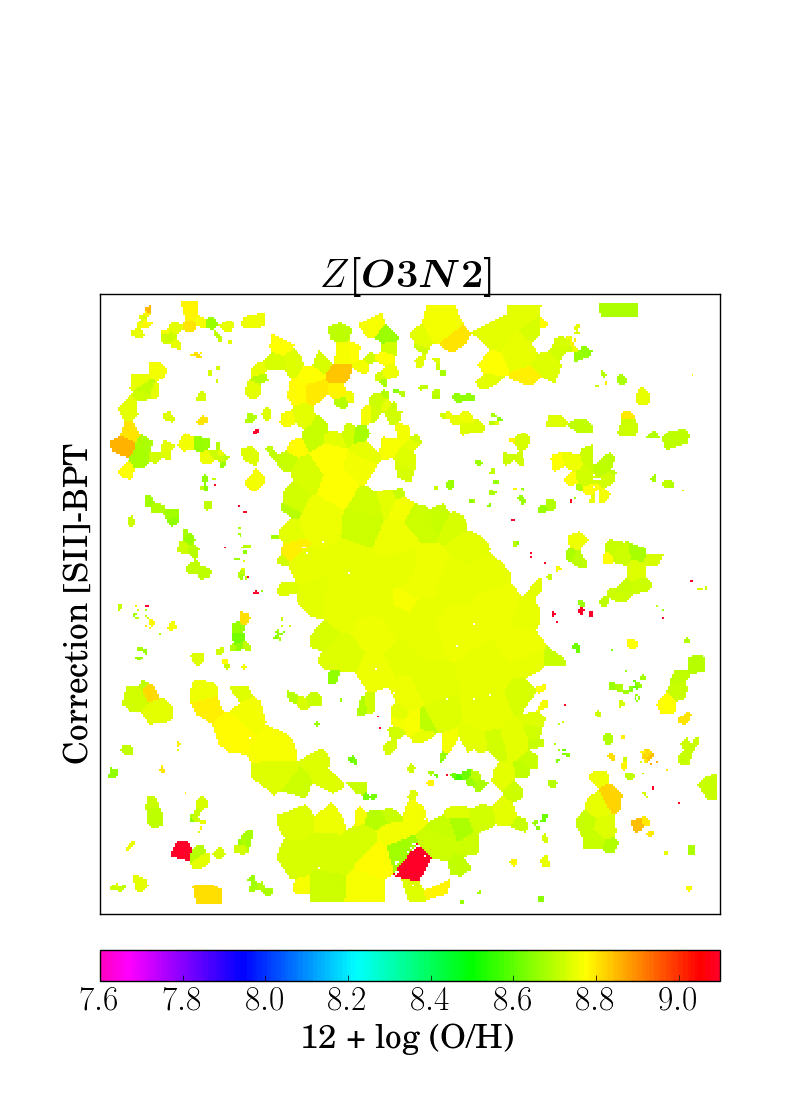}
	\includegraphics[width=0.28\textwidth, trim={0 1.2cm 0 5.5cm}, clip]{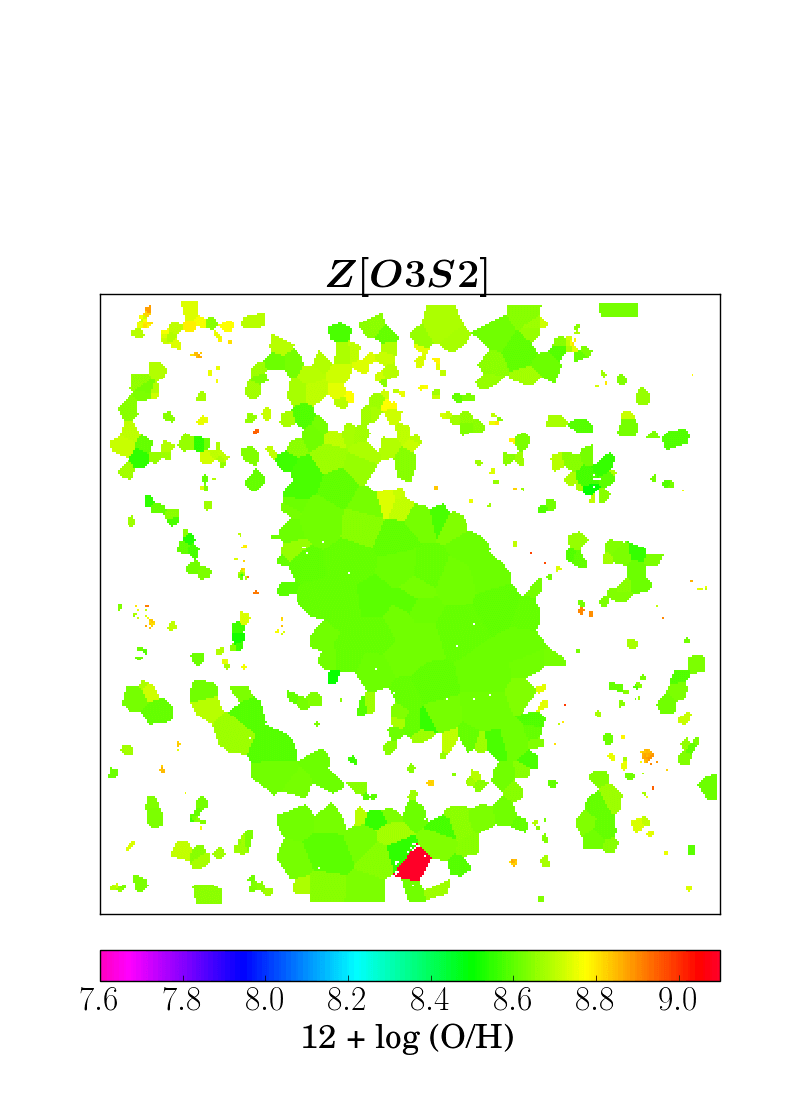}
	\includegraphics[width=0.28\textwidth, trim={2.8cm 0 2.8cm 0}, clip ]{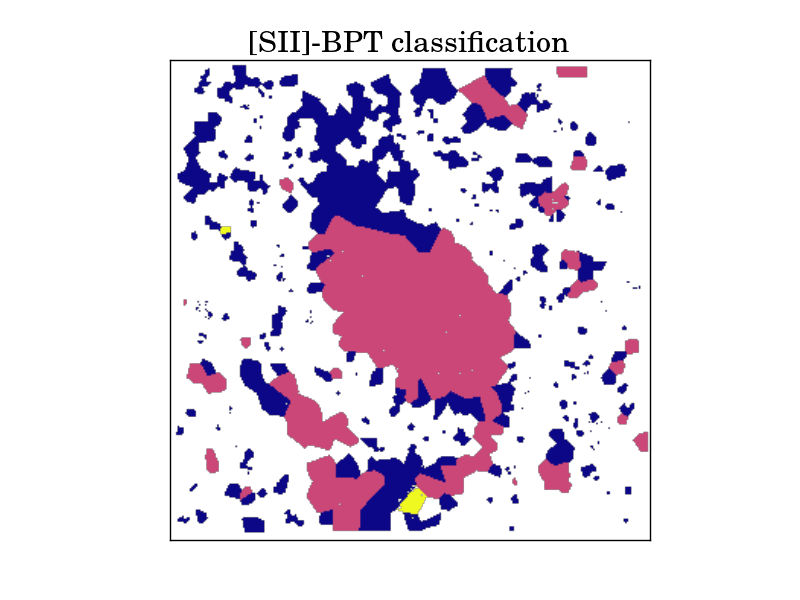}
	\includegraphics[width=0.28\textwidth, trim={0 1.2cm 0 5.5cm}, clip]{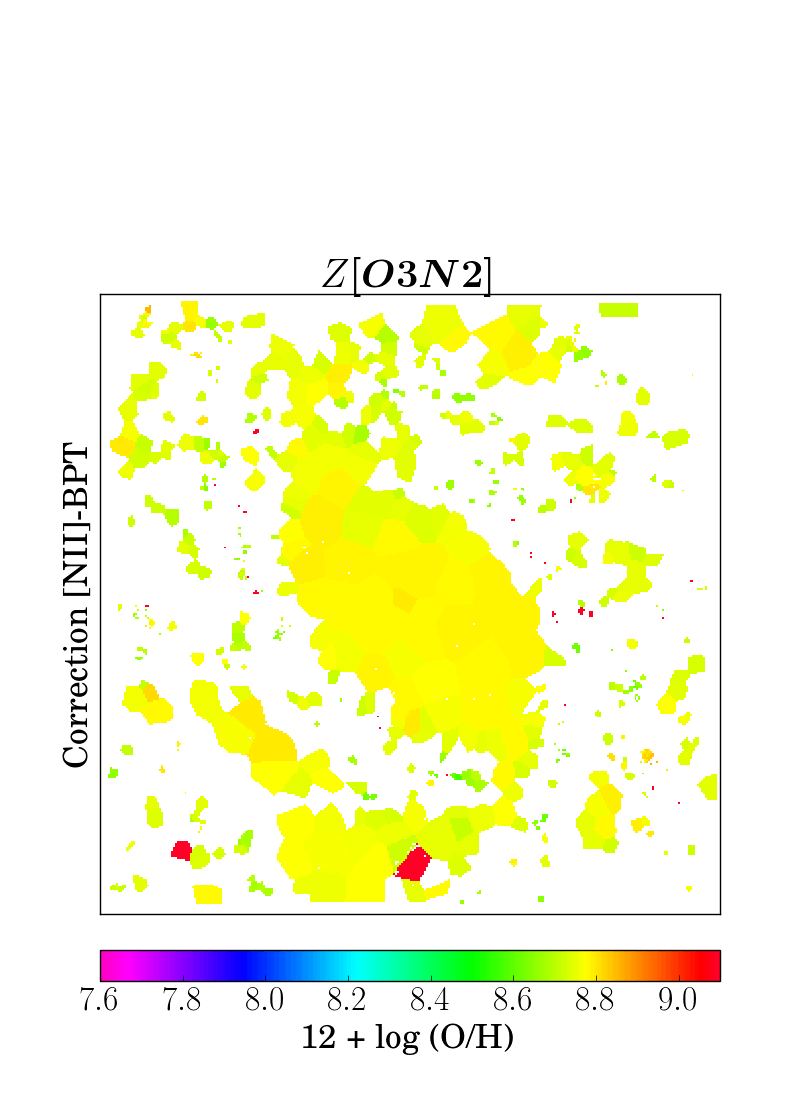}
	\includegraphics[width=0.28\textwidth, trim={0 1.2cm 0 5.5cm}, clip]{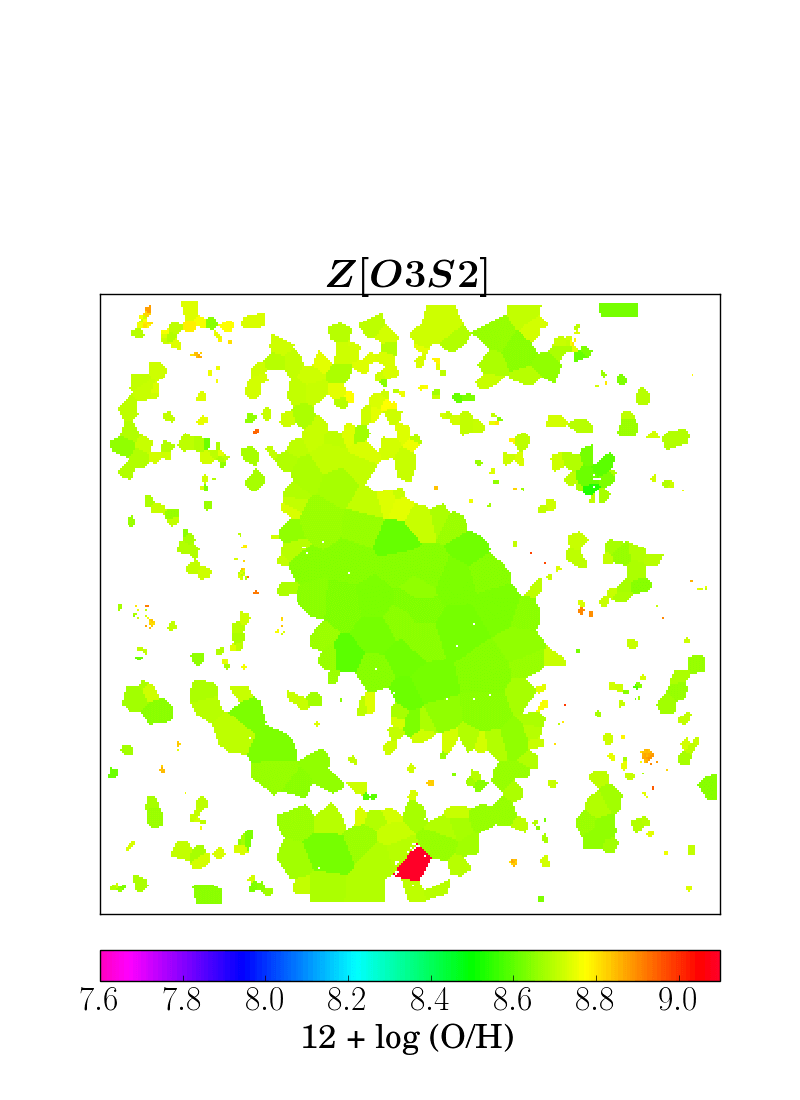}
	\includegraphics[width=0.28\textwidth, trim={2.8cm 0 2.8cm 0}, clip ]{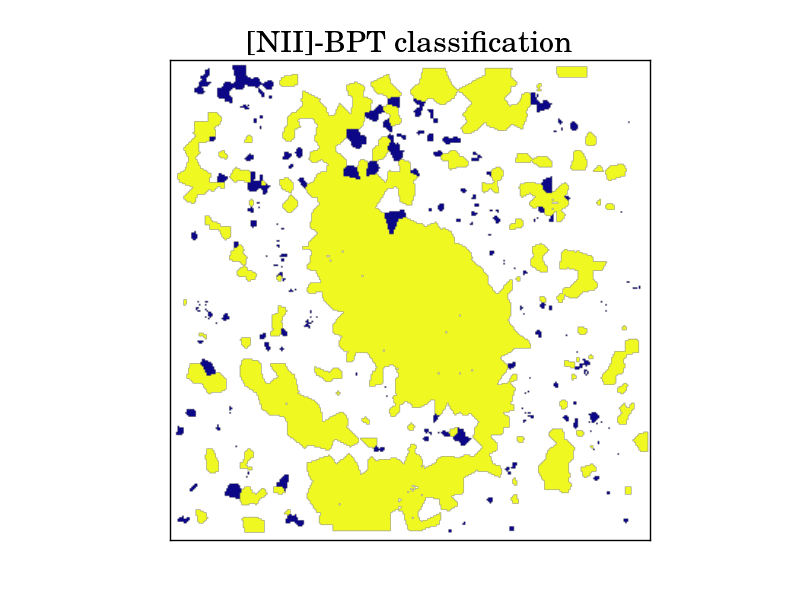}
	\caption{ Maps correspond to galaxy NGC4603, see caption of Figure \ref{fig:NGC1042} for details.}
	\label{fig:NGC4603}
\end{figure*}
\begin{figure*}
	\centering
	\includegraphics[width=0.28\textwidth, trim={0 1.2cm 0 5.5cm}, clip]{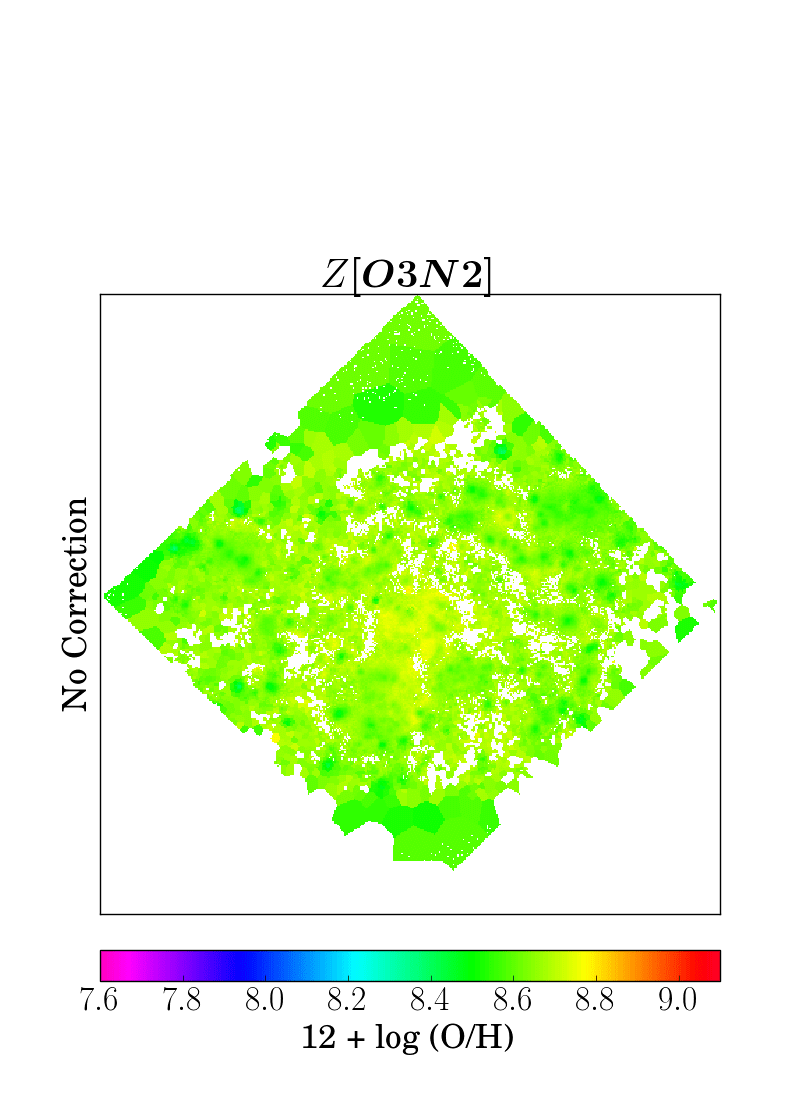}
	\includegraphics[width=0.28\textwidth, trim={0 1.2cm 0 5.5cm}, clip]{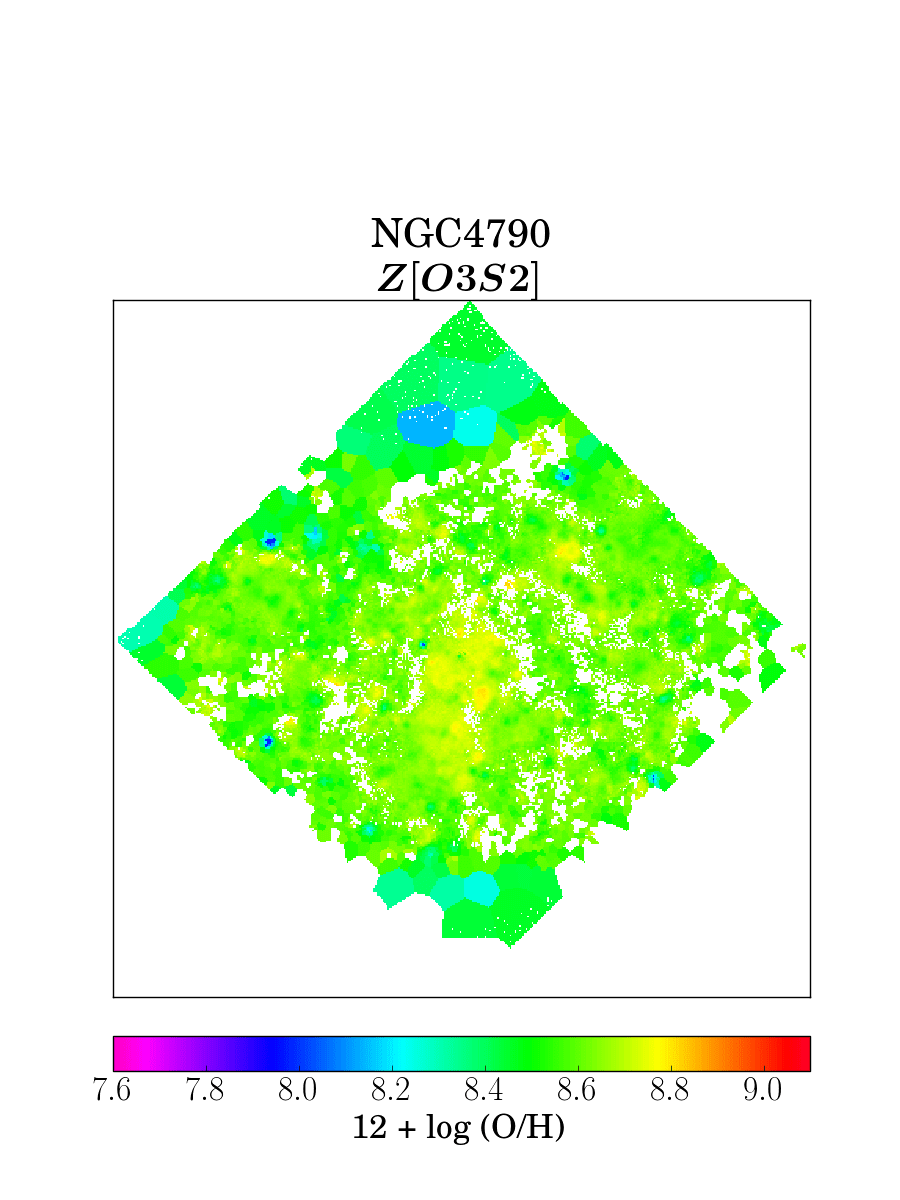}
	\includegraphics[width=0.28\textwidth, trim={0 1.2cm 0 5.5cm}, clip]{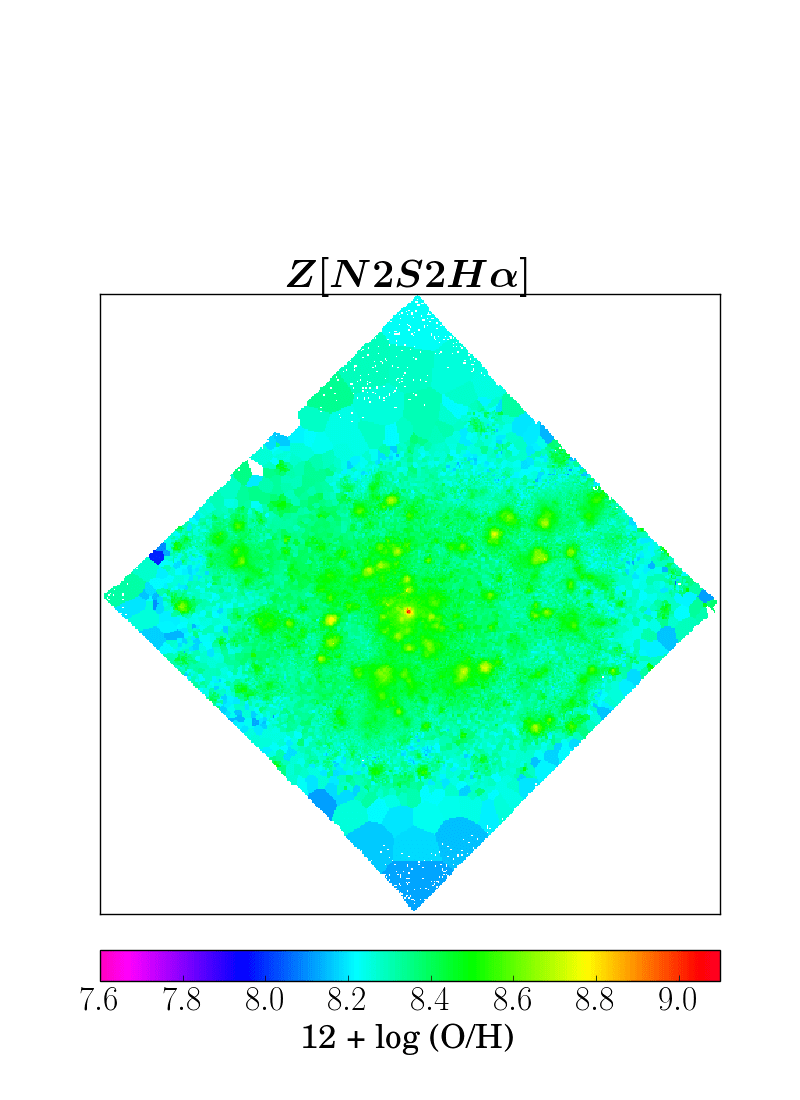}
	\includegraphics[width=0.28\textwidth, trim={0 1.2cm 0 5.5cm}, clip]{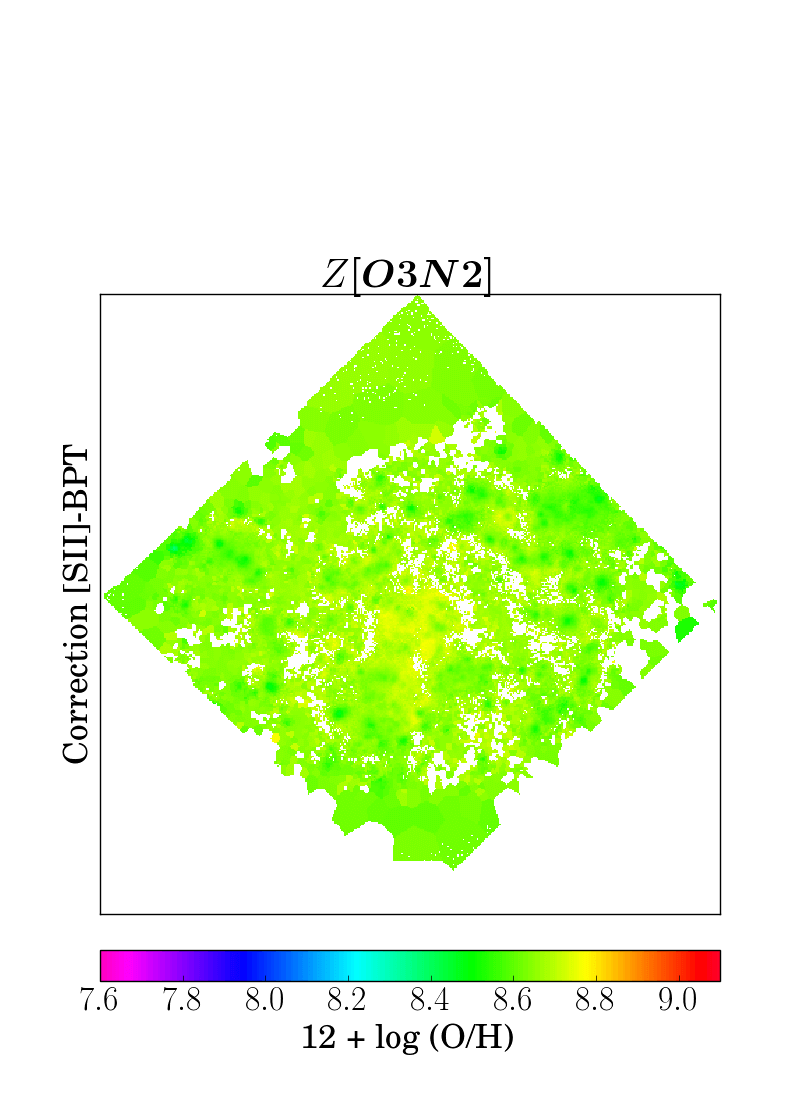}
	\includegraphics[width=0.28\textwidth, trim={0 1.2cm 0 5.5cm}, clip]{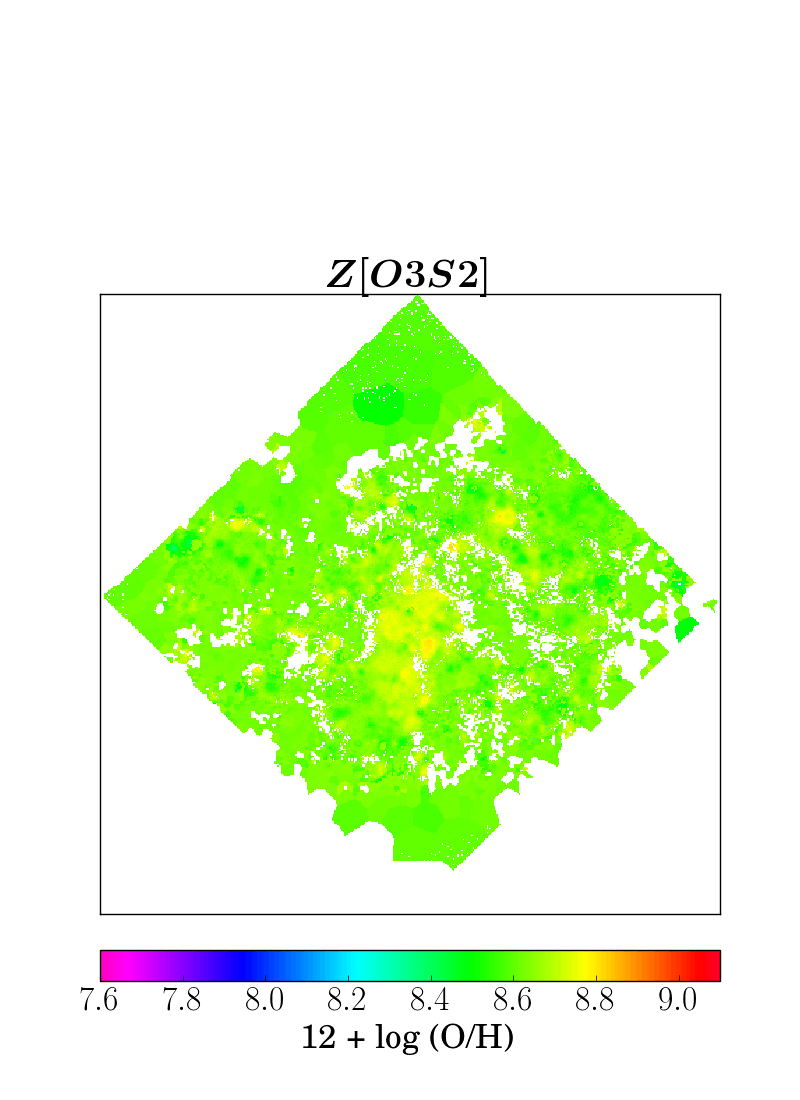}
	\includegraphics[width=0.28\textwidth, trim={2.8cm 0 2.8cm 0}, clip ]{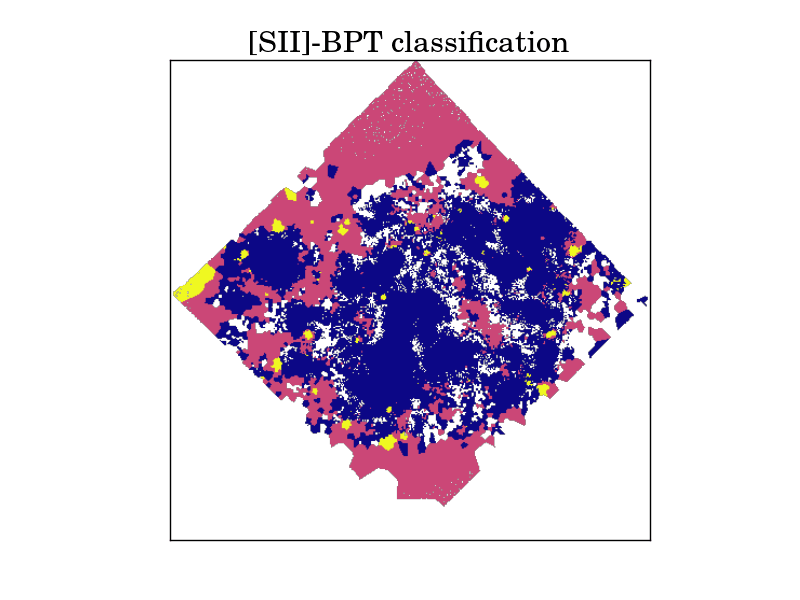}
	\includegraphics[width=0.28\textwidth, trim={0 1.2cm 0 5.5cm}, clip]{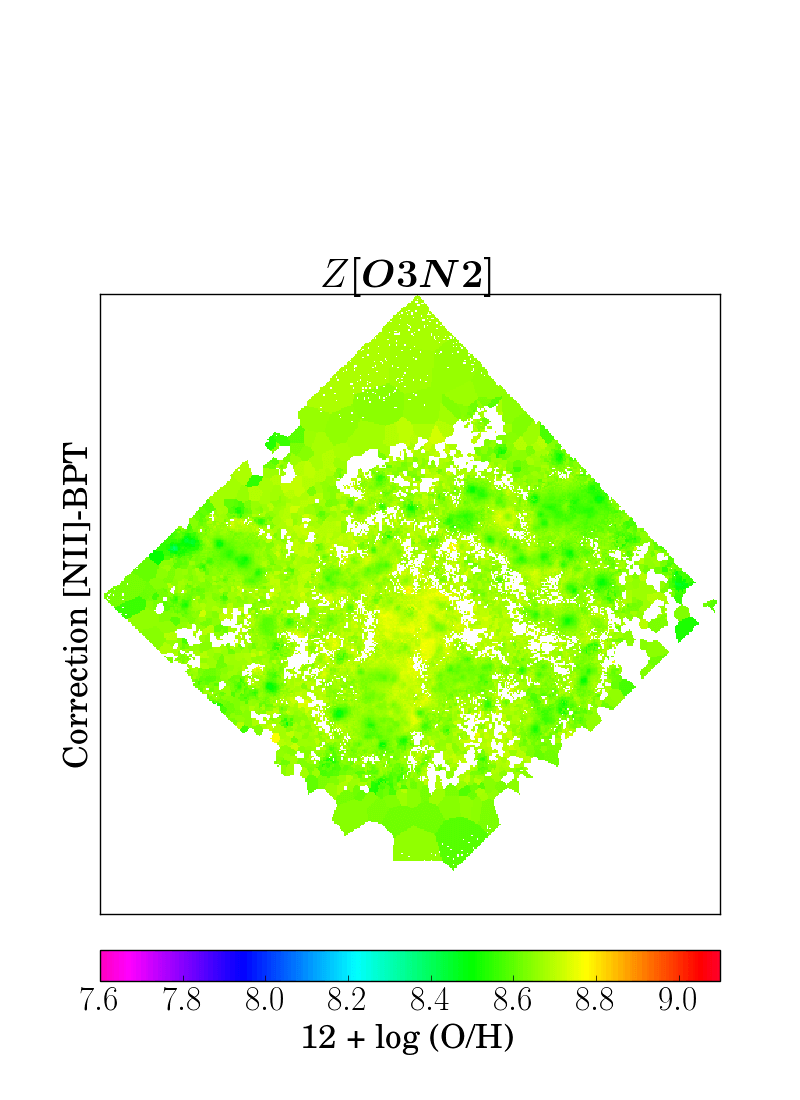}
	\includegraphics[width=0.28\textwidth, trim={0 1.2cm 0 5.5cm}, clip]{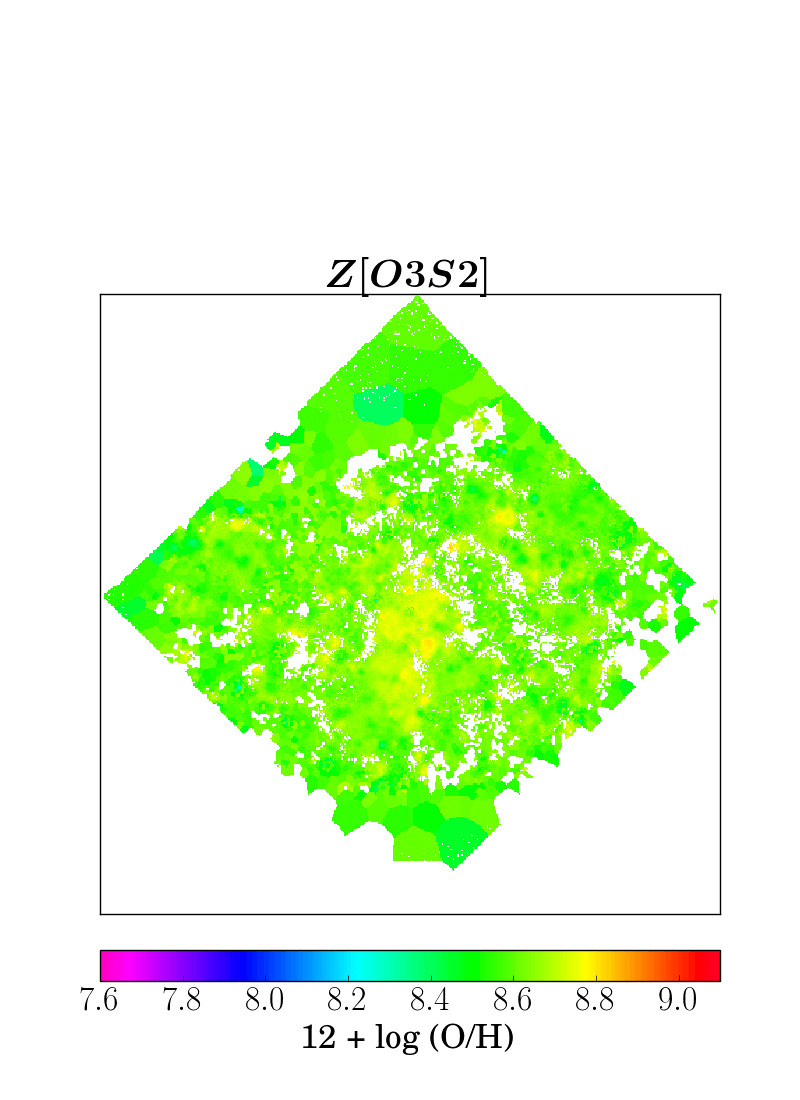}
	\includegraphics[width=0.28\textwidth, trim={2.8cm 0 2.8cm 0}, clip ]{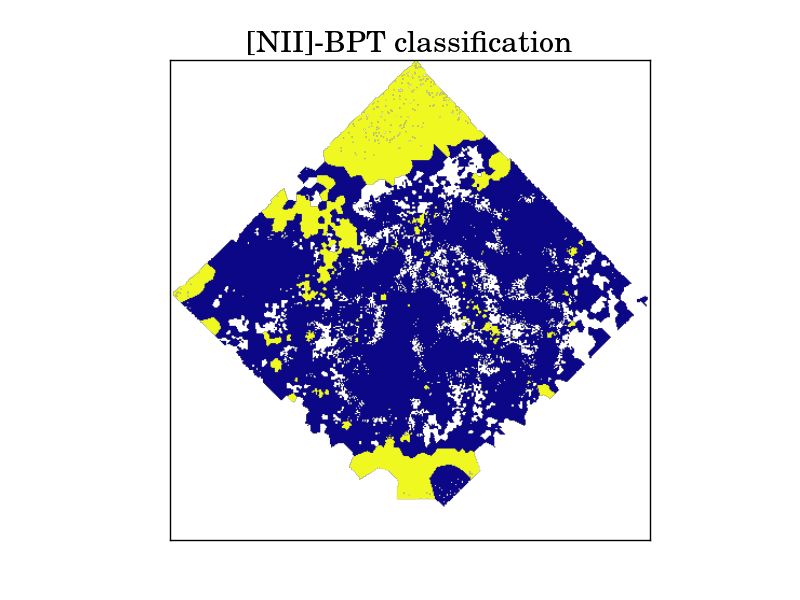}
	\caption{ Maps correspond to galaxy NGC4790, see caption of Figure \ref{fig:NGC1042} for details.}
	\label{fig:NGC4790}
\end{figure*}
\begin{figure*}
	\centering
	\includegraphics[width=0.28\textwidth, trim={0 1.2cm 0 5.5cm}, clip]{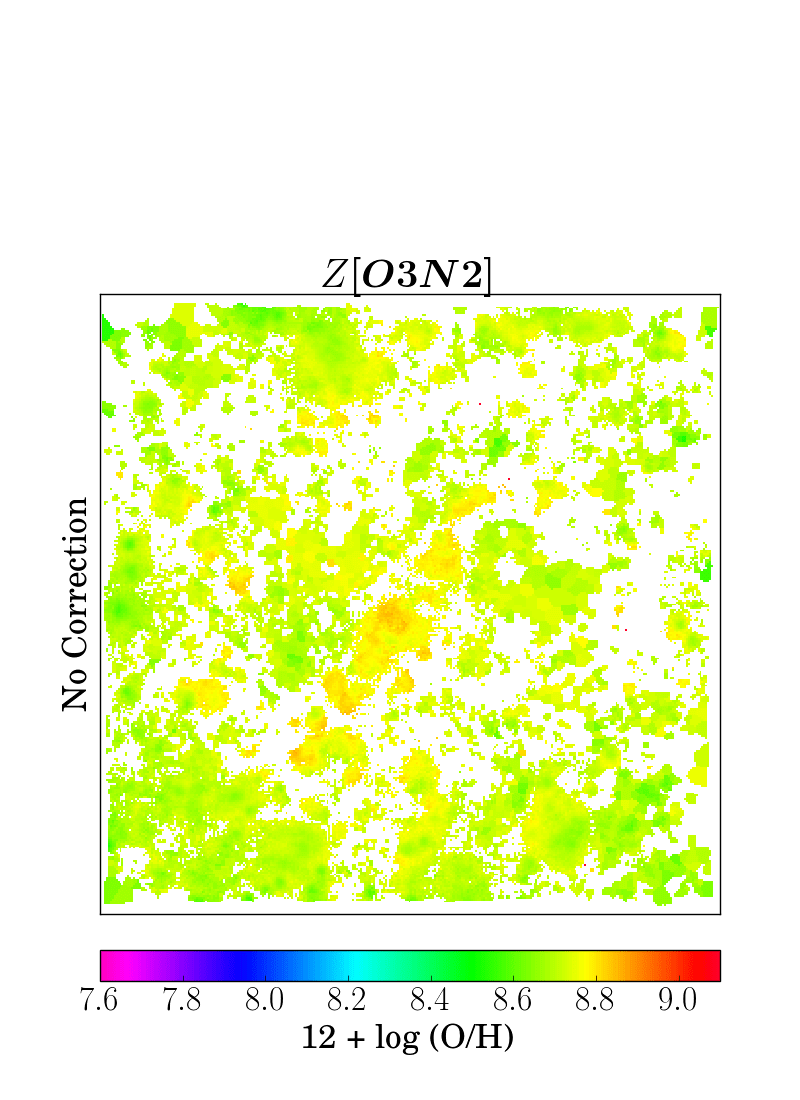}
	\includegraphics[width=0.28\textwidth, trim={0 1.2cm 0 5.5cm}, clip]{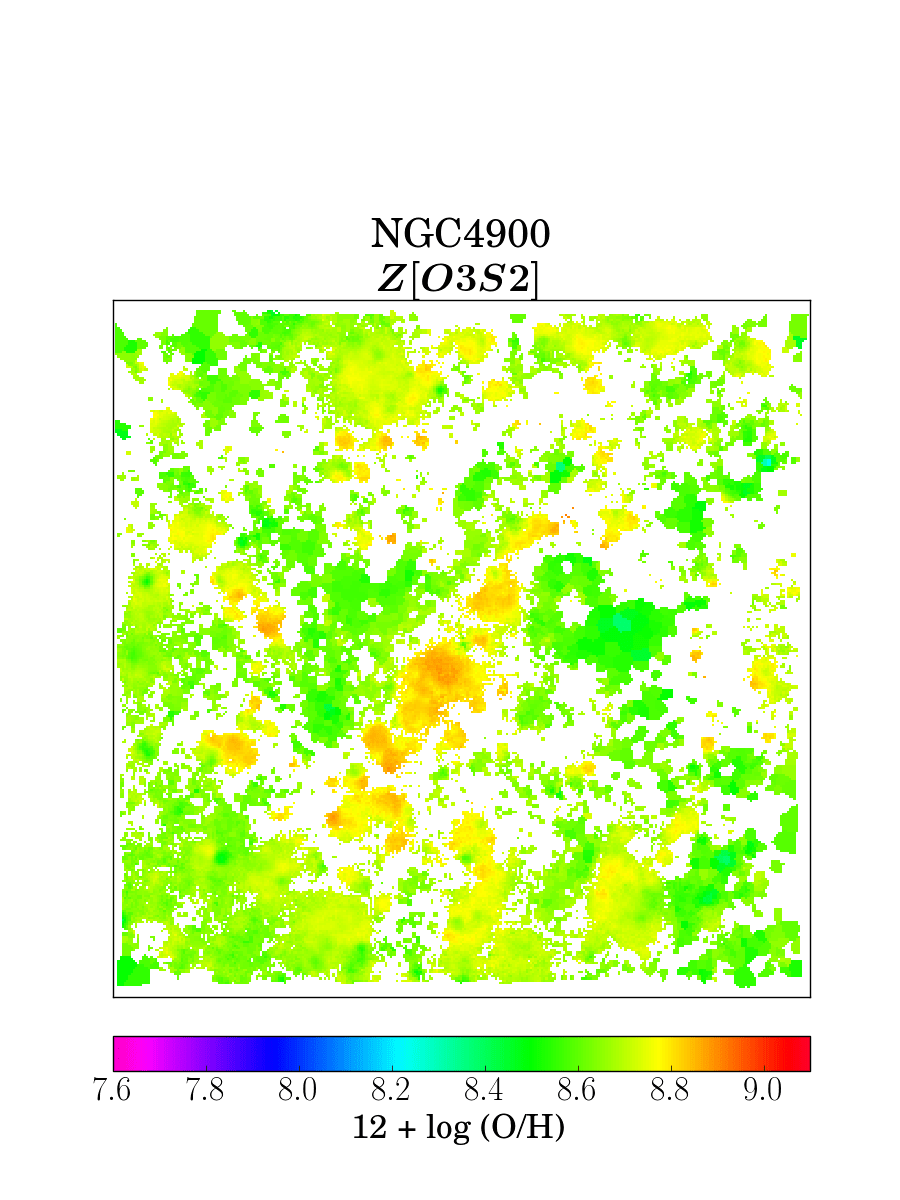}
	\includegraphics[width=0.28\textwidth, trim={0 1.2cm 0 5.5cm}, clip]{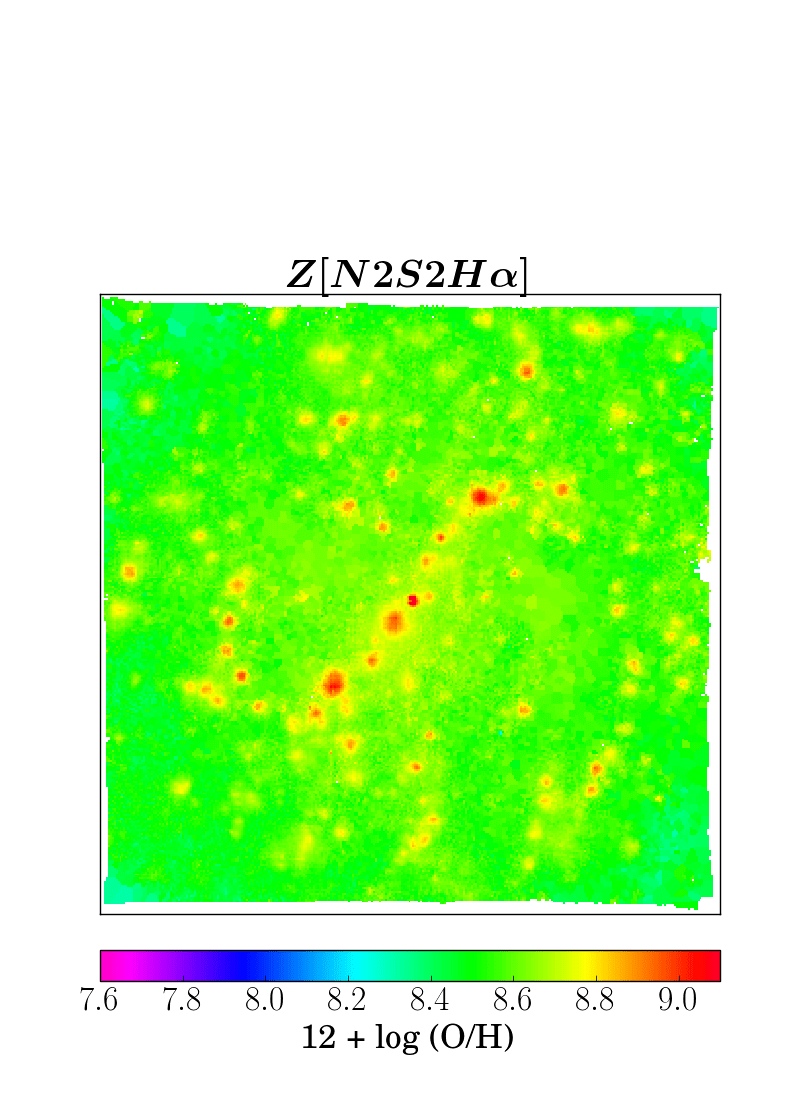}
	\includegraphics[width=0.28\textwidth, trim={0 1.2cm 0 5.5cm}, clip]{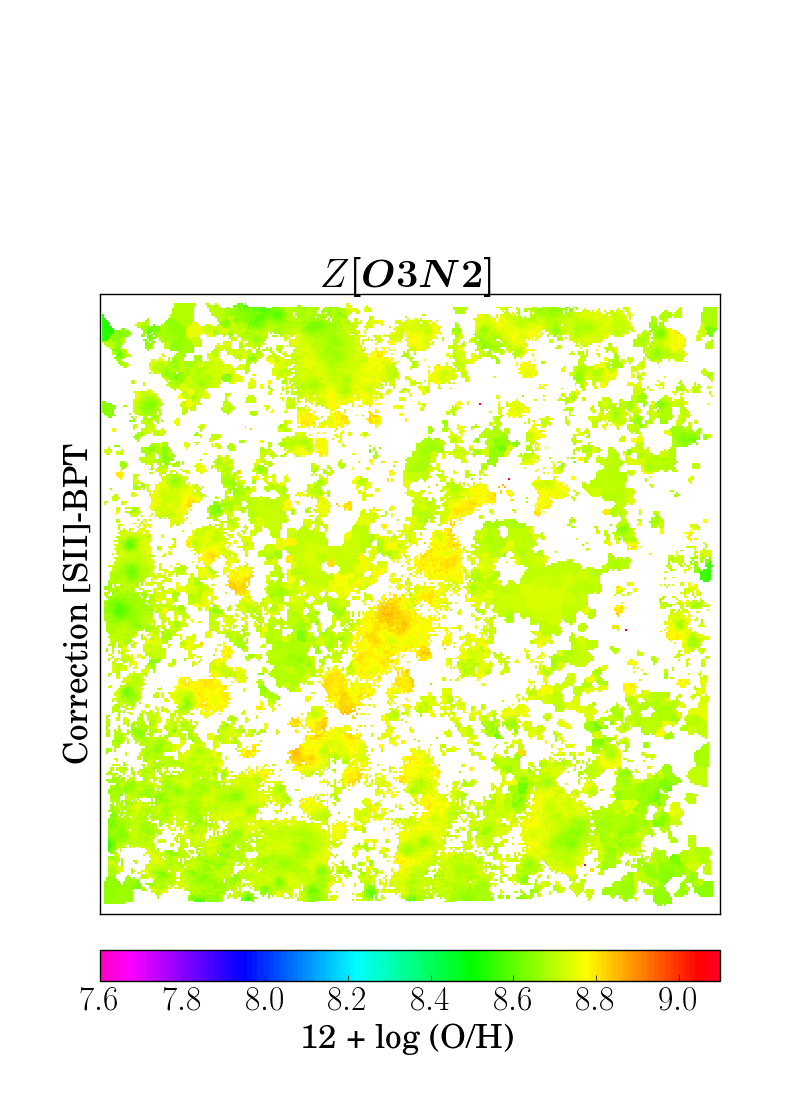}
	\includegraphics[width=0.28\textwidth, trim={0 1.2cm 0 5.5cm}, clip]{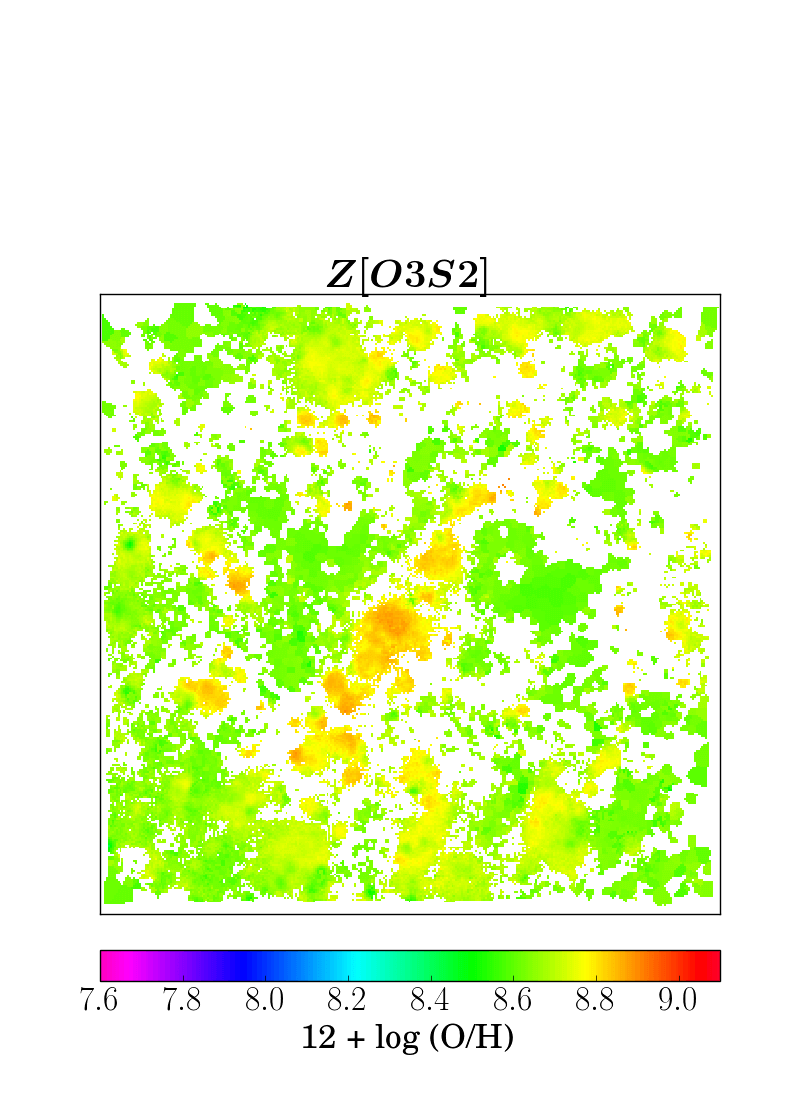}
	\includegraphics[width=0.28\textwidth, trim={2.8cm 0 2.8cm 0}, clip ]{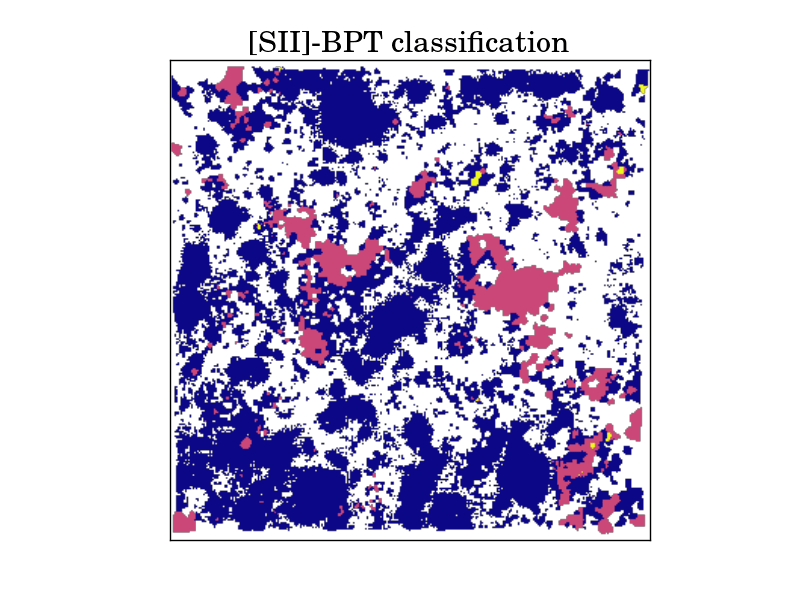}
	\includegraphics[width=0.28\textwidth, trim={0 1.2cm 0 5.5cm}, clip]{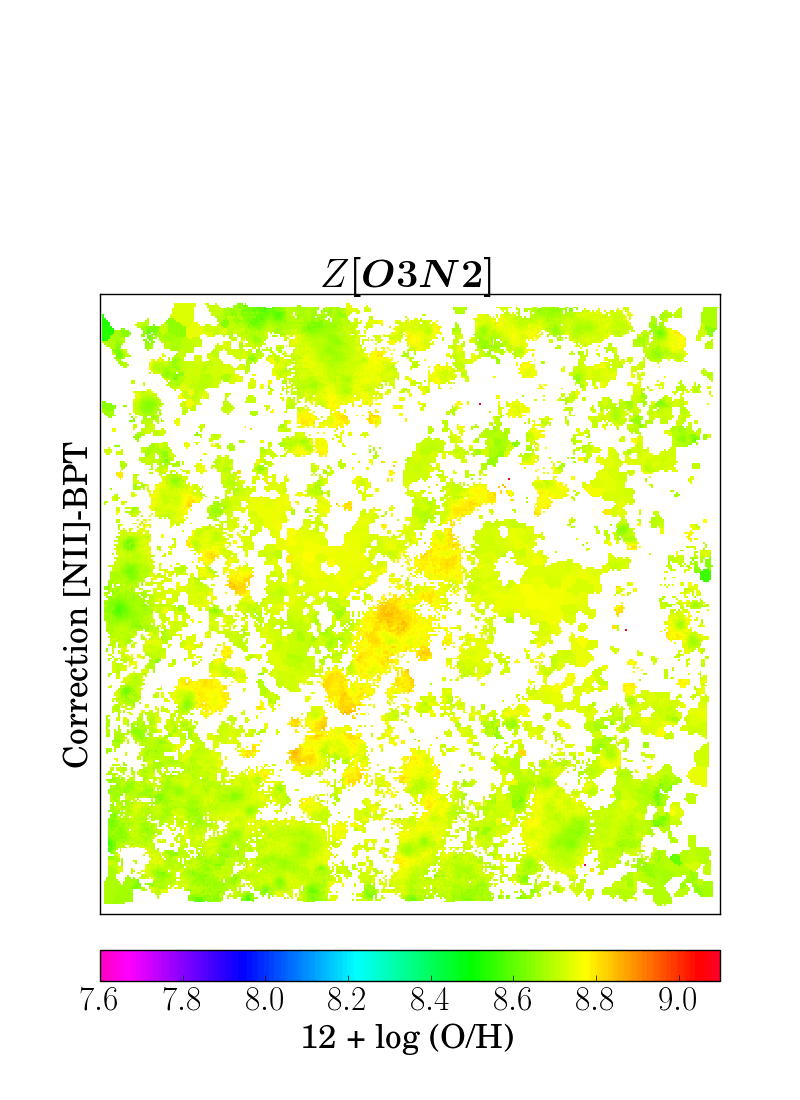}
	\includegraphics[width=0.28\textwidth, trim={0 1.2cm 0 5.5cm}, clip]{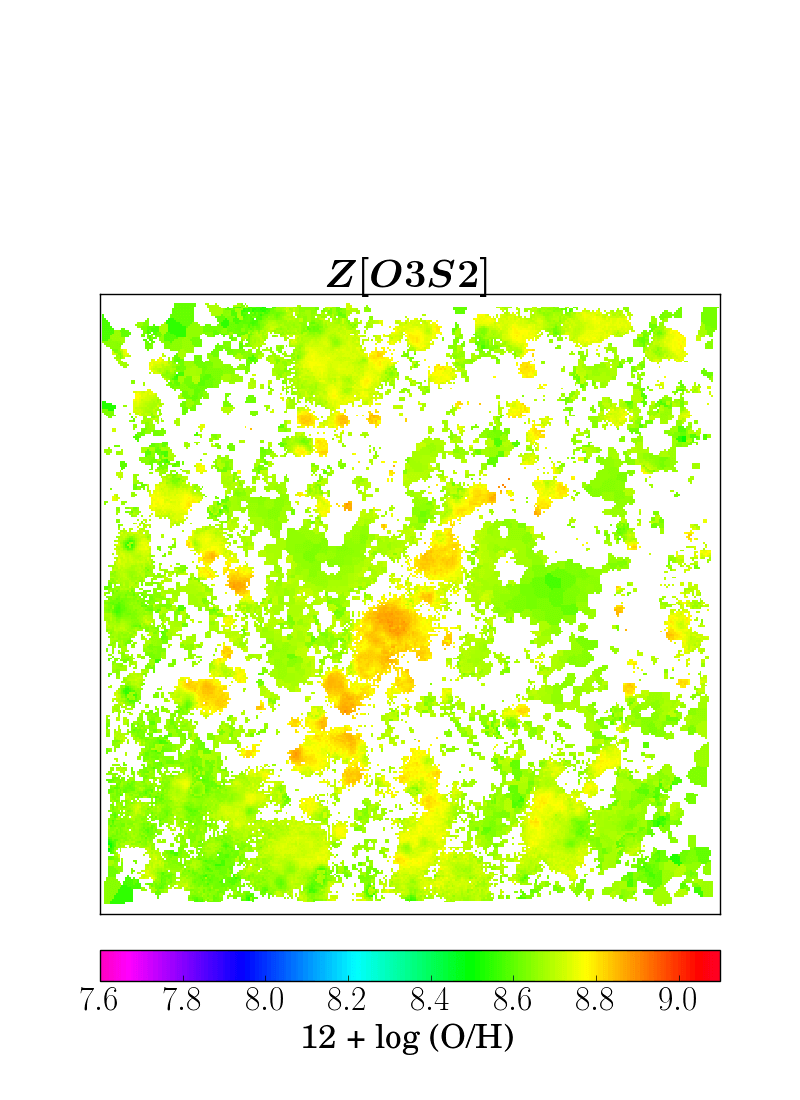}
	\includegraphics[width=0.28\textwidth, trim={2.8cm 0 2.8cm 0}, clip ]{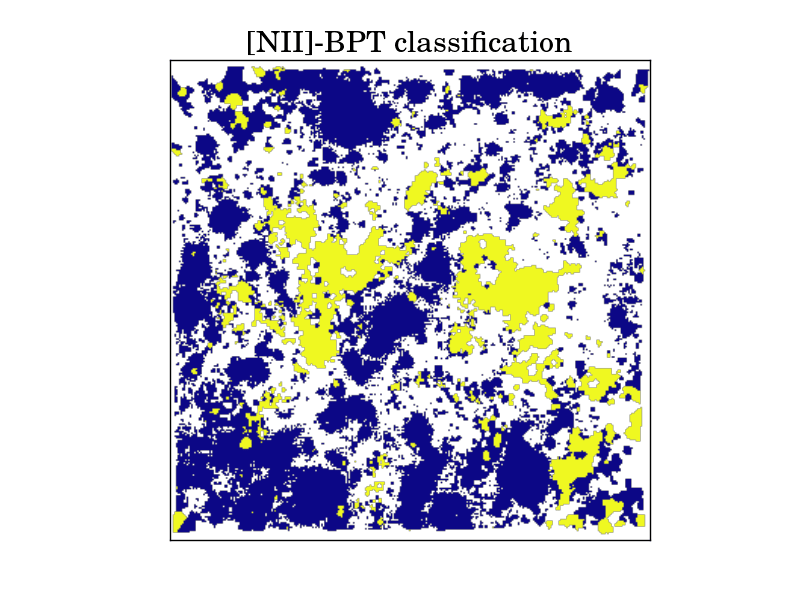}
	\caption{ Maps correspond to galaxy NGC4900, see caption of Figure \ref{fig:NGC1042} for details.}
	\label{fig:NGC4900}
\end{figure*}
\begin{figure*}
	\centering
	\includegraphics[width=0.28\textwidth, trim={0 1.2cm 0 5.5cm}, clip]{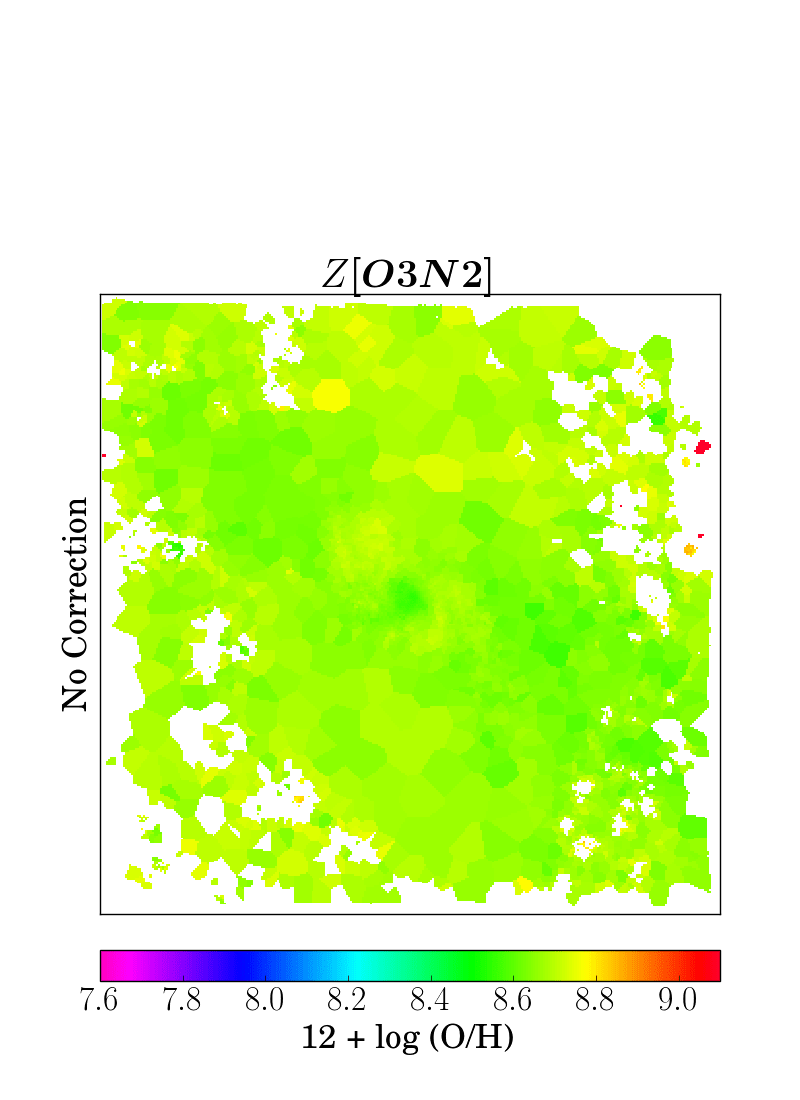}
	\includegraphics[width=0.28\textwidth, trim={0 1.2cm 0 5.5cm}, clip]{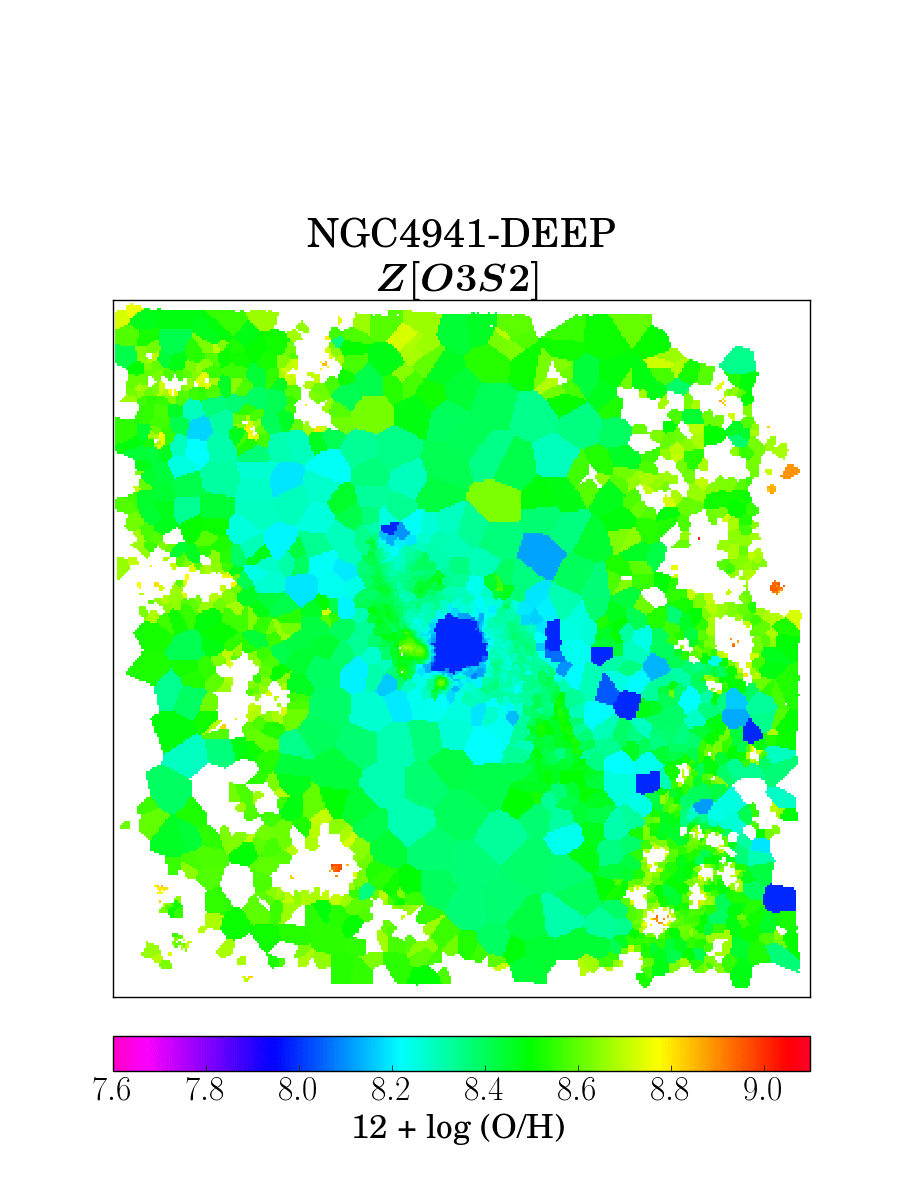}
	\includegraphics[width=0.28\textwidth, trim={0 1.2cm 0 5.5cm}, clip]{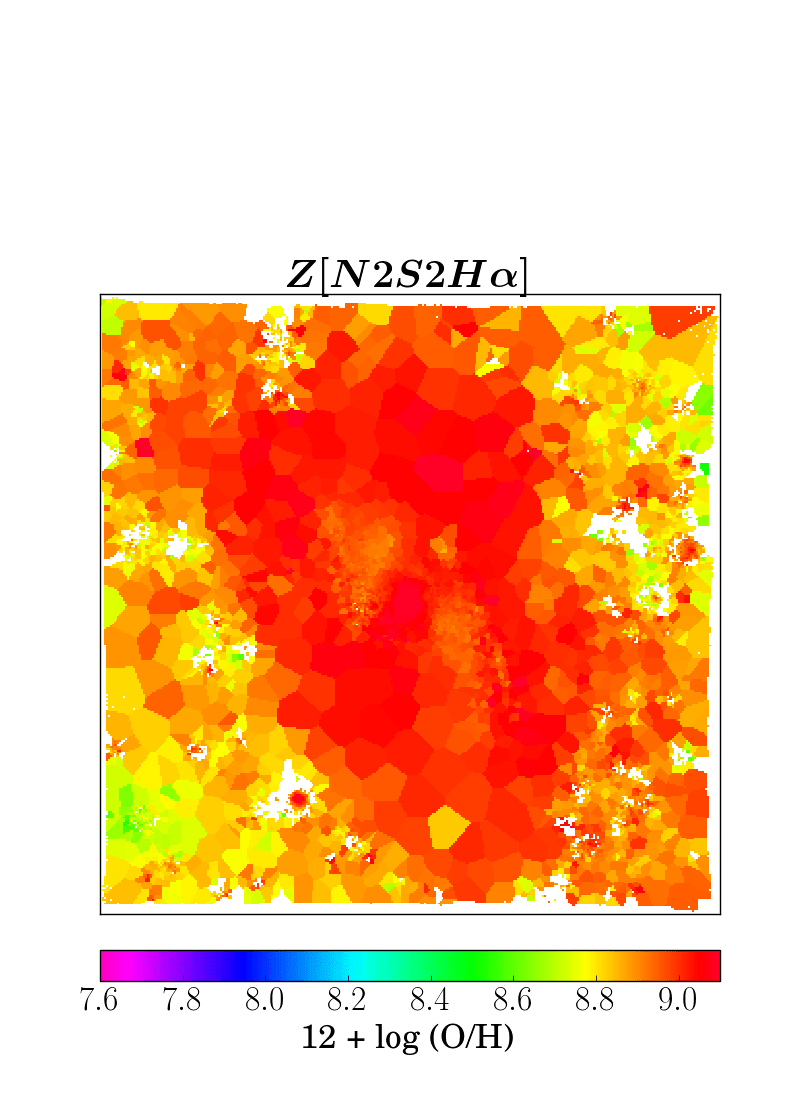}
	\includegraphics[width=0.28\textwidth, trim={0 1.2cm 0 5.5cm}, clip]{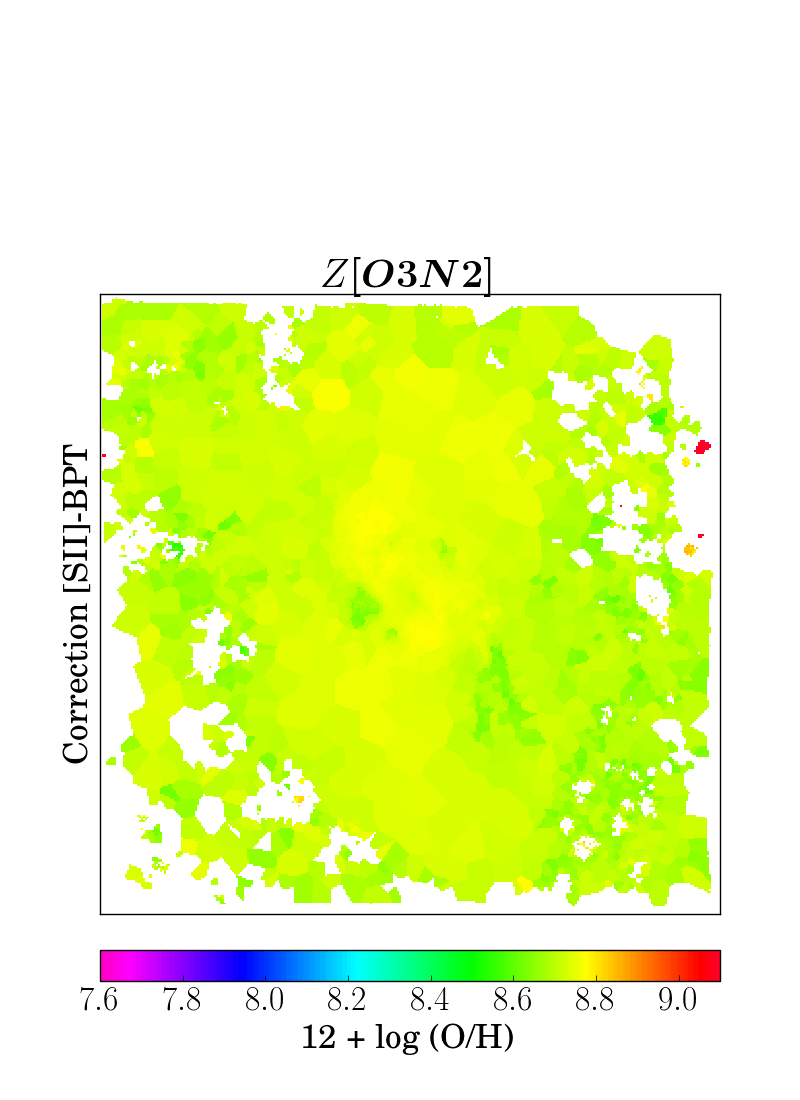}
	\includegraphics[width=0.28\textwidth, trim={0 1.2cm 0 5.5cm}, clip]{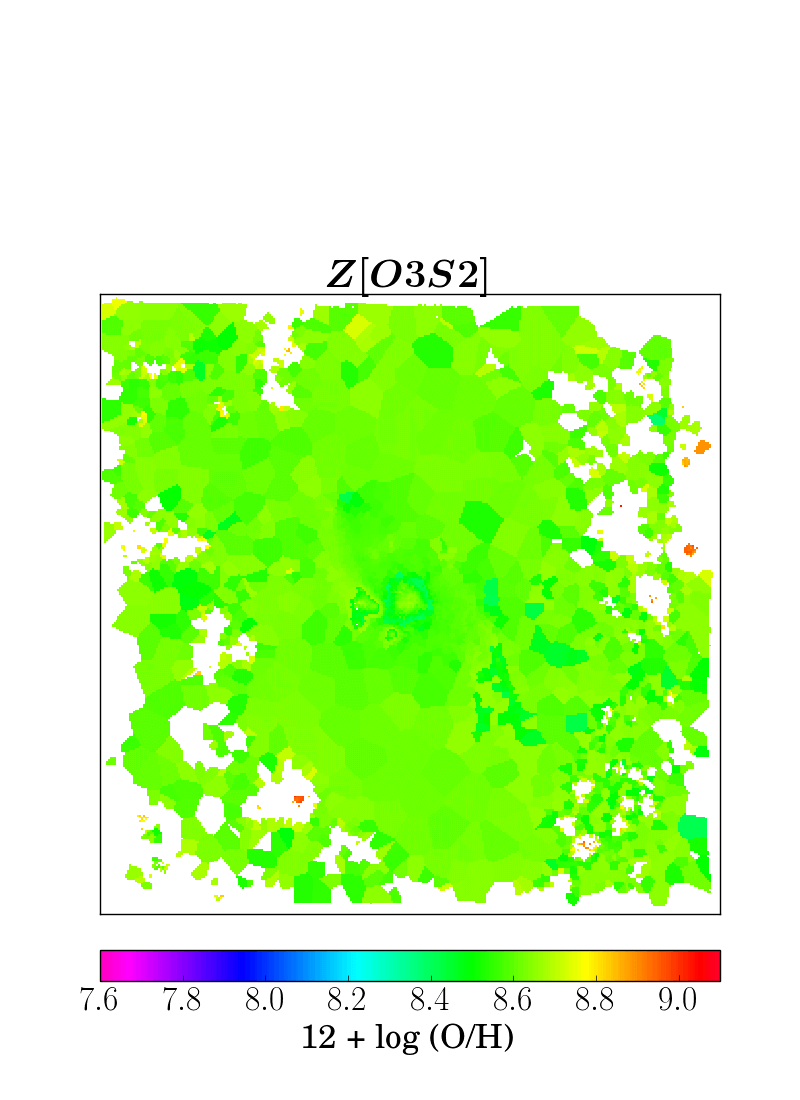}
	\includegraphics[width=0.28\textwidth, trim={2.8cm 0 2.8cm 0}, clip ]{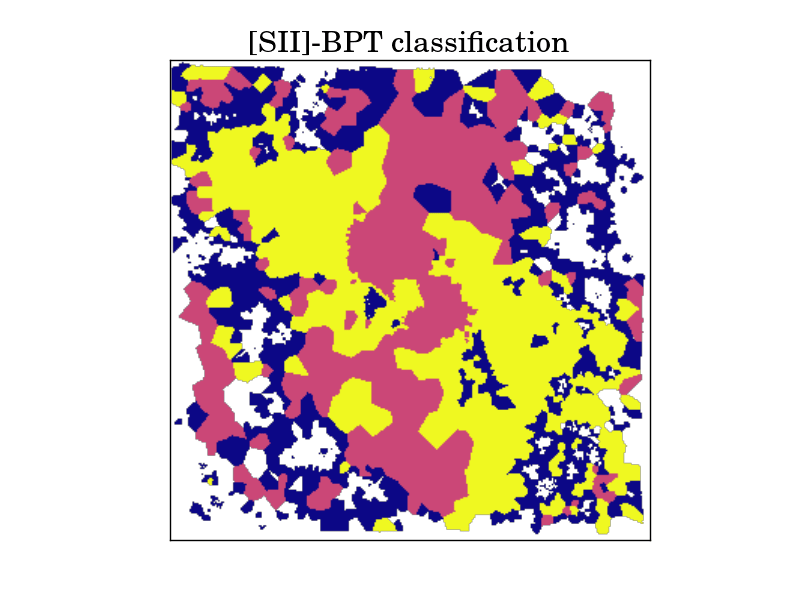}
	\includegraphics[width=0.28\textwidth, trim={0 1.2cm 0 5.5cm}, clip]{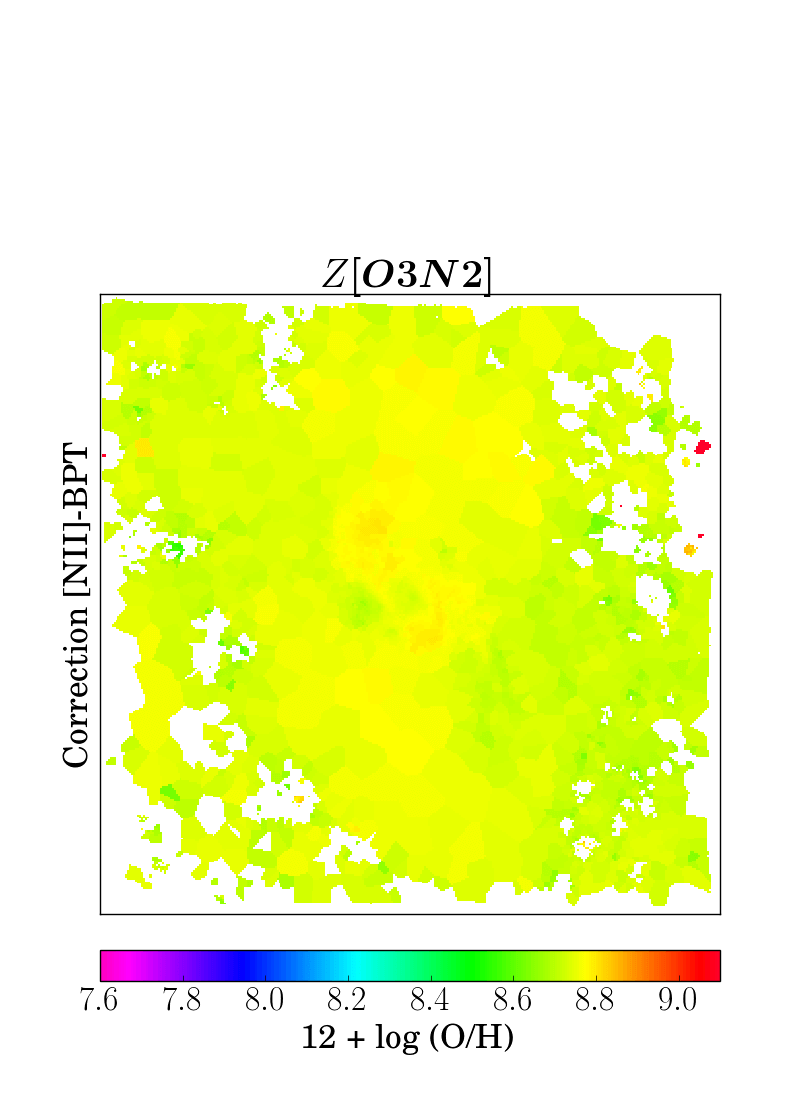}
	\includegraphics[width=0.28\textwidth, trim={0 1.2cm 0 5.5cm}, clip]{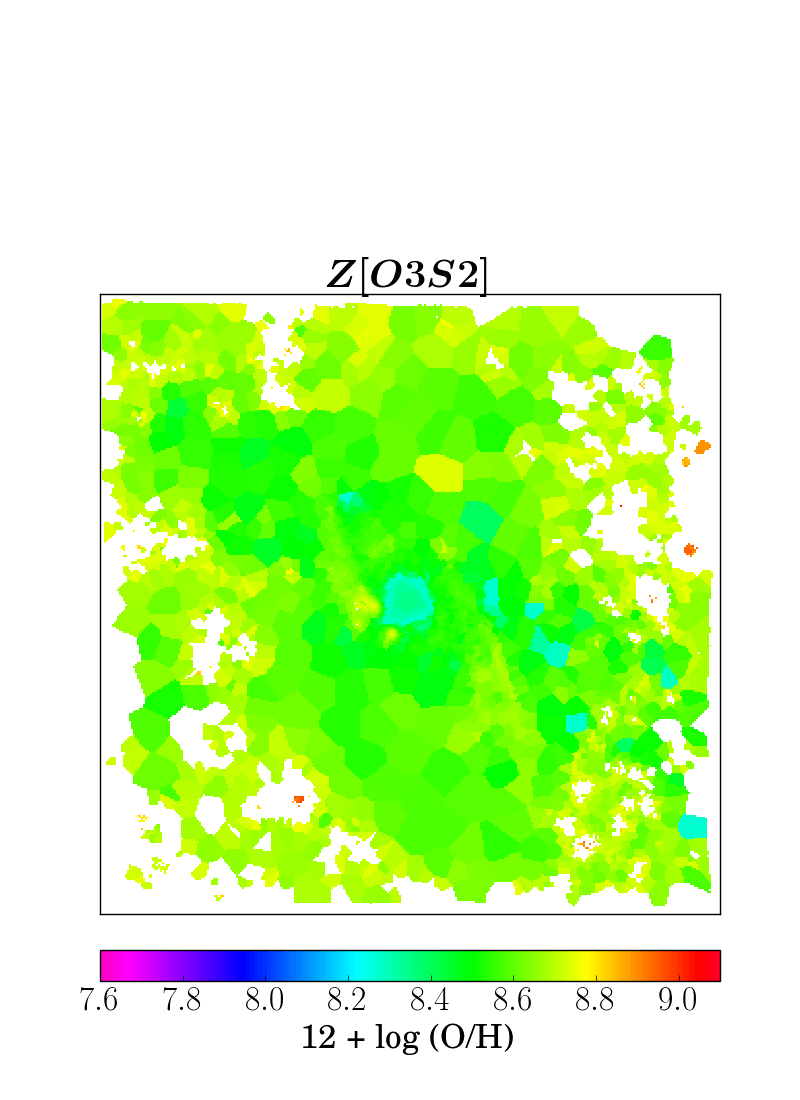}
	\includegraphics[width=0.28\textwidth, trim={2.8cm 0 2.8cm 0}, clip ]{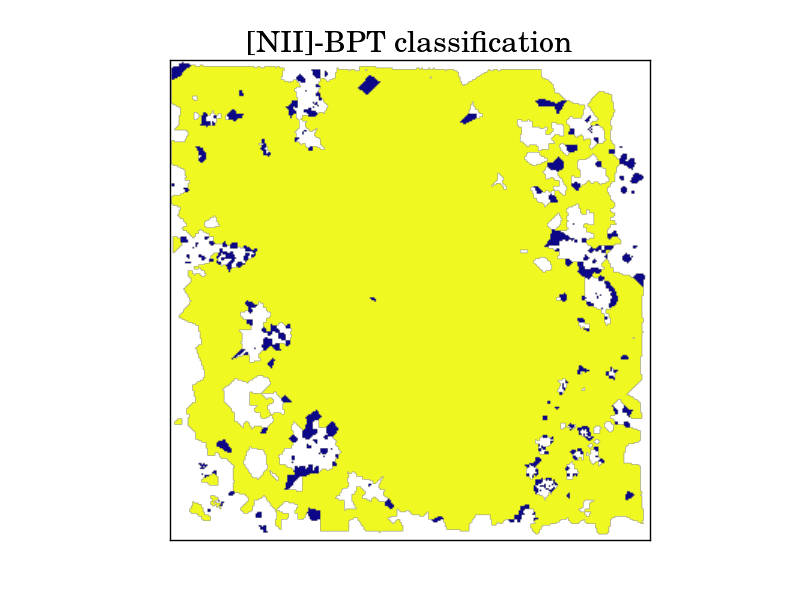}
	\caption{ Maps correspond to galaxy NGC4941-DEEP, see caption of Figure \ref{fig:NGC1042} for details.}
	\label{fig:NGC4941-DEEP}
\end{figure*}
\begin{figure*}
	\centering
	\includegraphics[width=0.28\textwidth, trim={0 1.2cm 0 5.5cm}, clip]{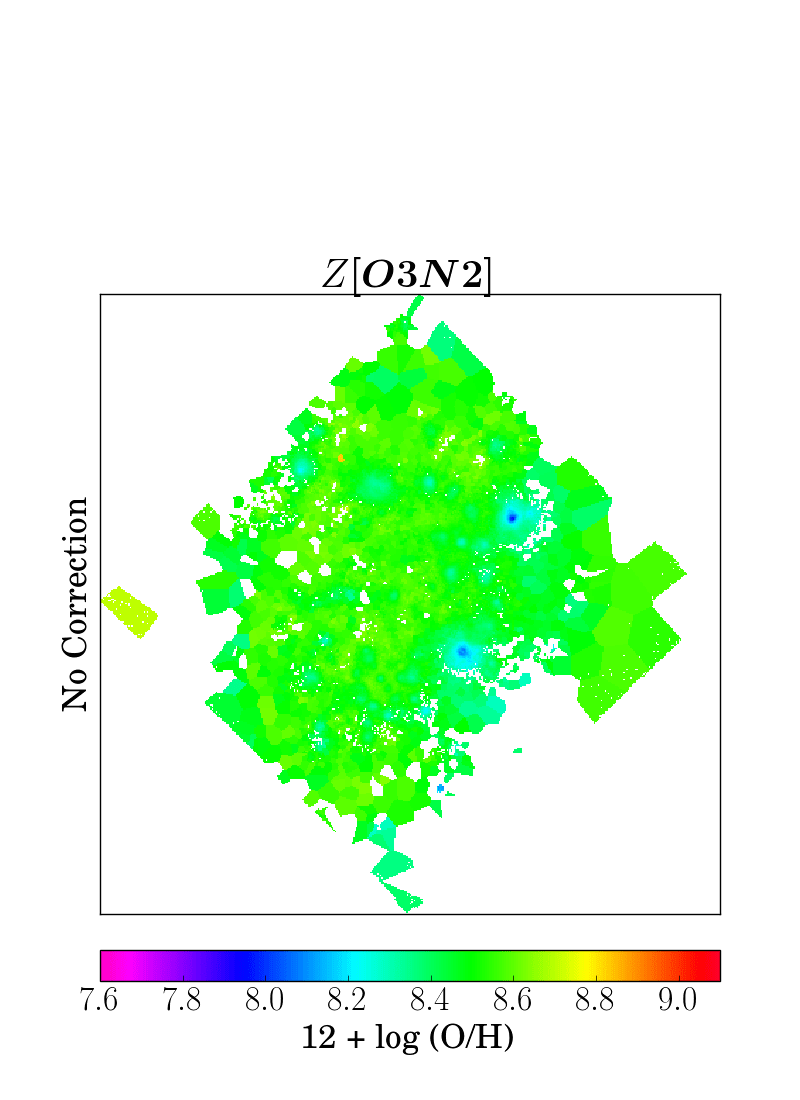}
	\includegraphics[width=0.28\textwidth, trim={0 1.2cm 0 5.5cm}, clip]{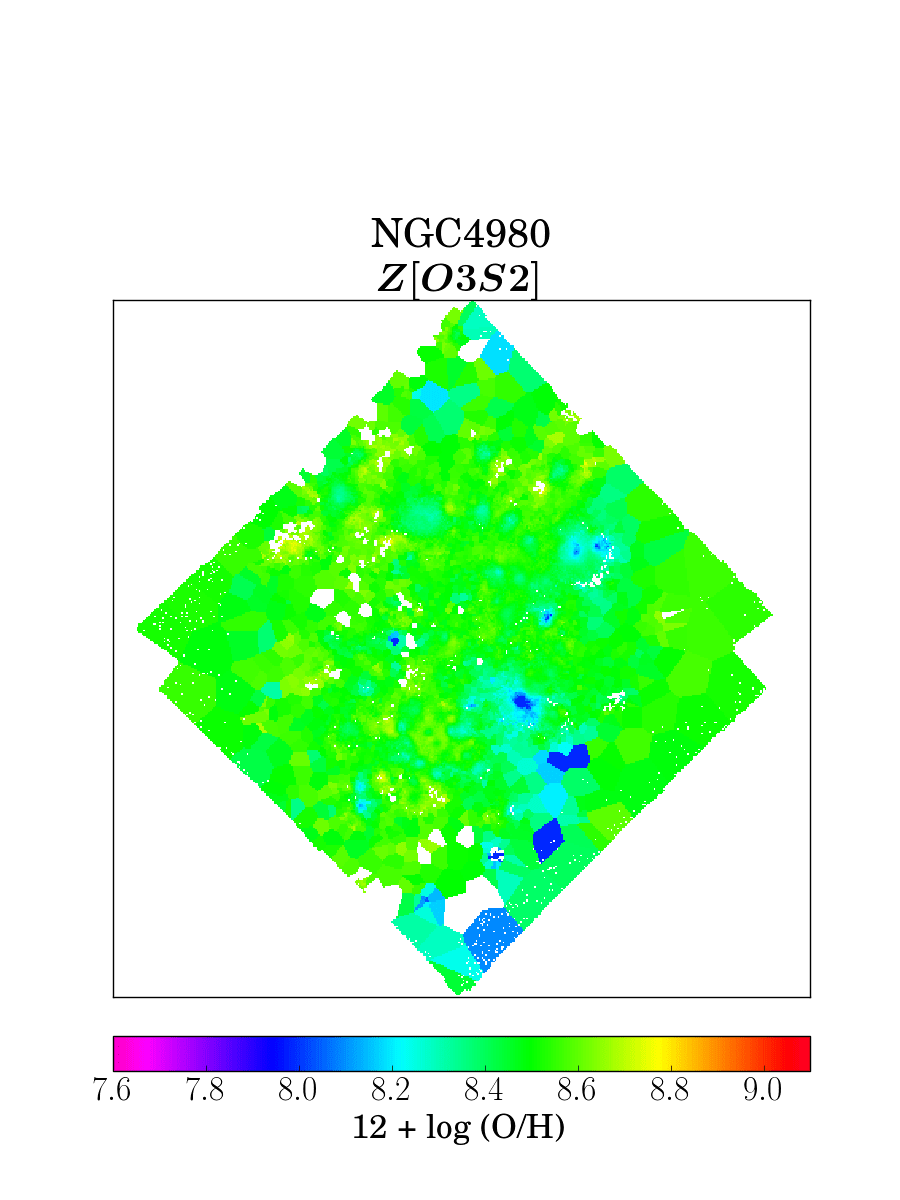}
	\includegraphics[width=0.28\textwidth, trim={0 1.2cm 0 5.5cm}, clip]{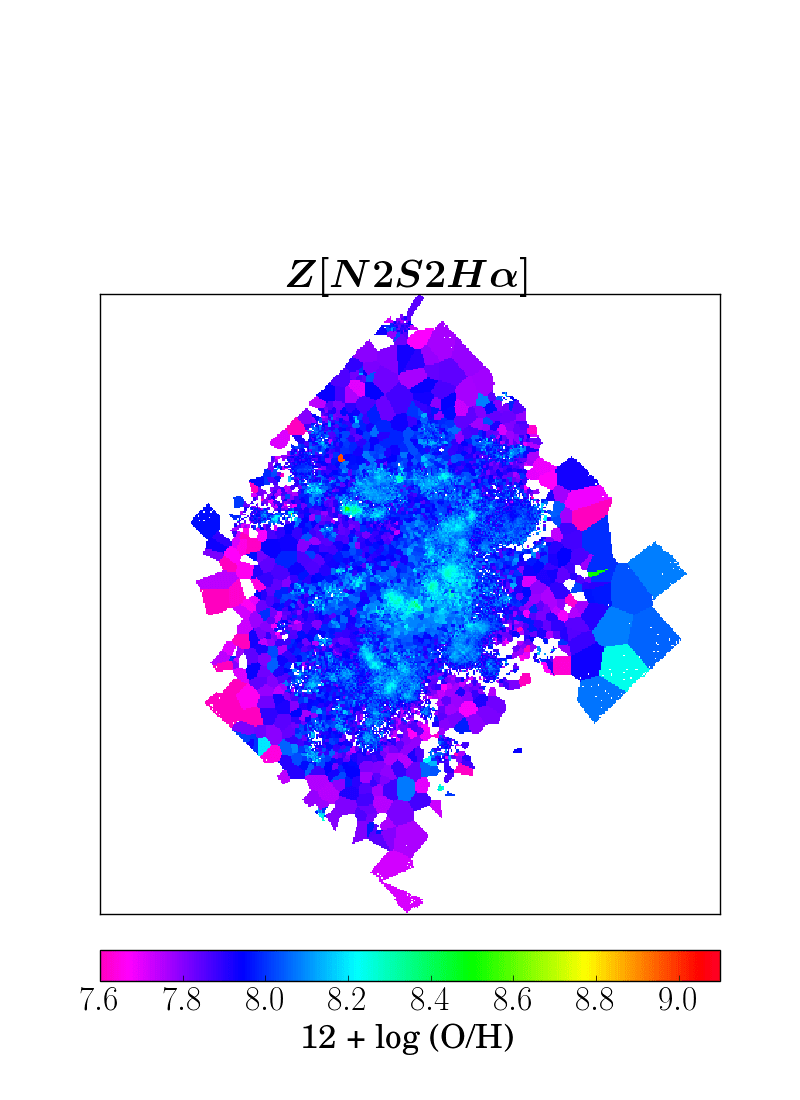}
	\includegraphics[width=0.28\textwidth, trim={0 1.2cm 0 5.5cm}, clip]{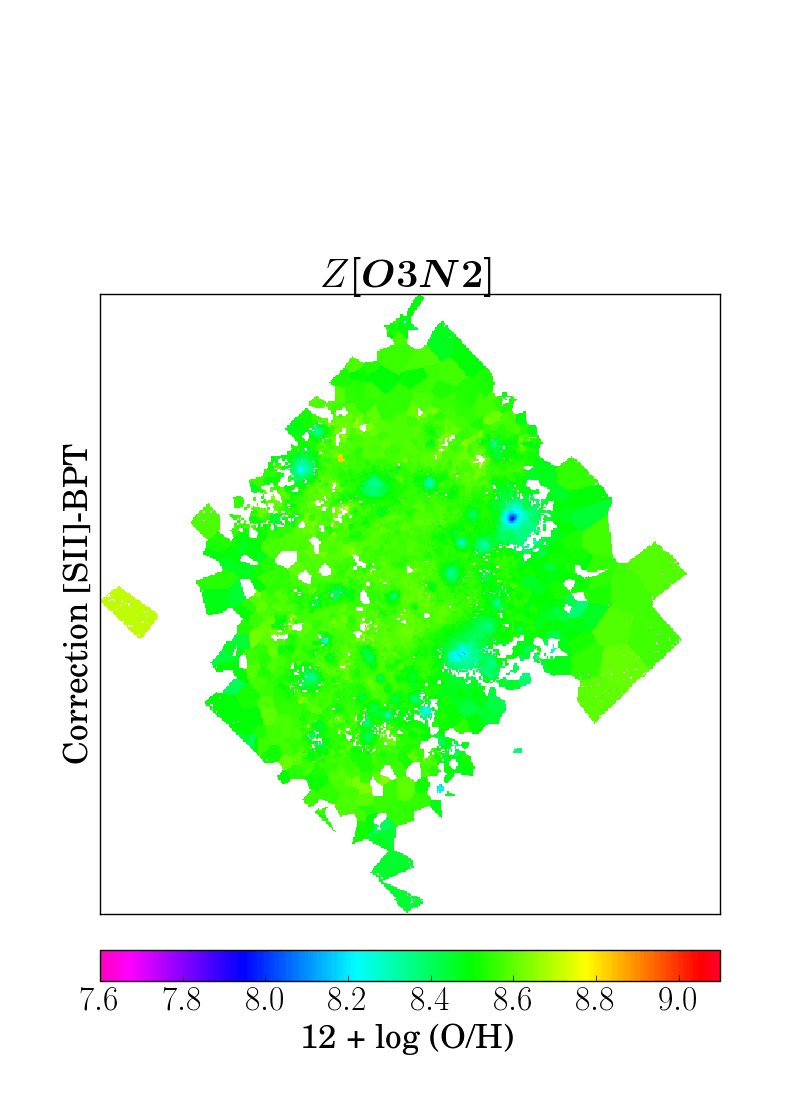}
	\includegraphics[width=0.28\textwidth, trim={0 1.2cm 0 5.5cm}, clip]{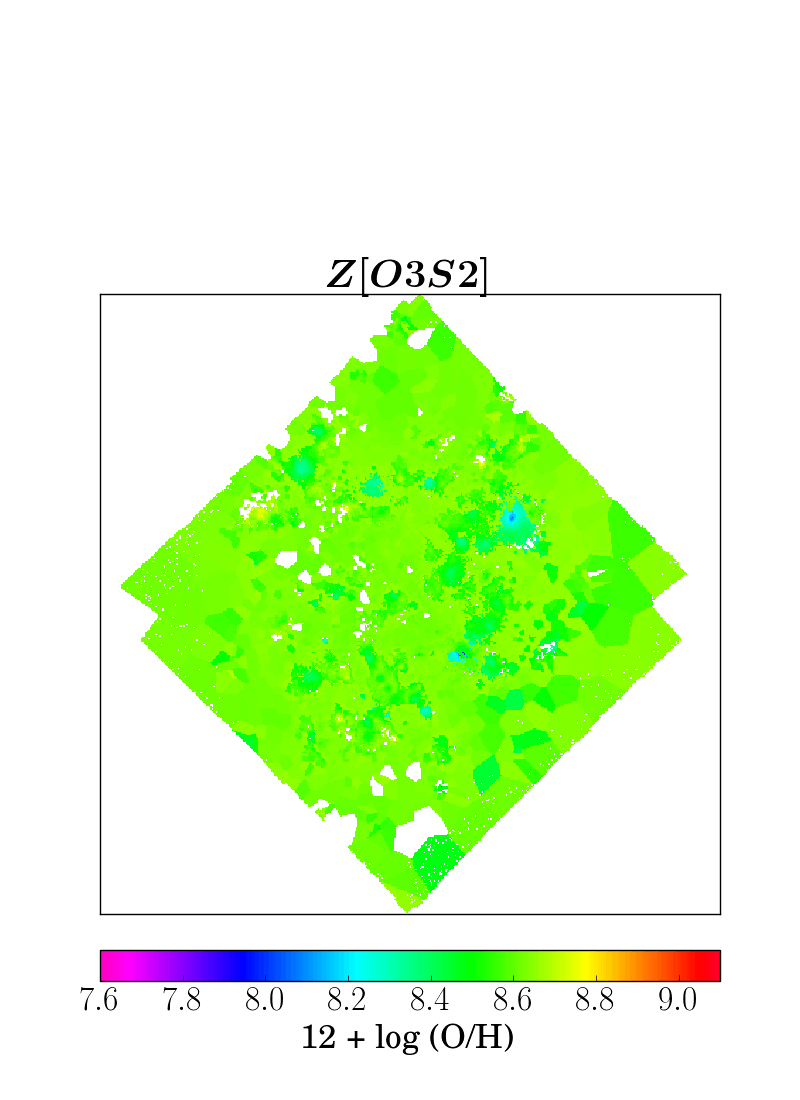}
	\includegraphics[width=0.28\textwidth, trim={2.8cm 0 2.8cm 0}, clip ]{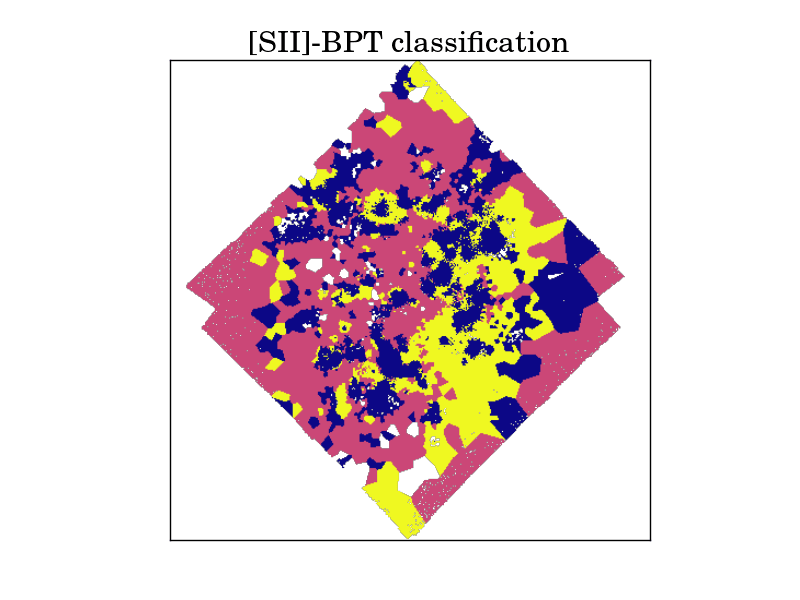}
	\includegraphics[width=0.28\textwidth, trim={0 1.2cm 0 5.5cm}, clip]{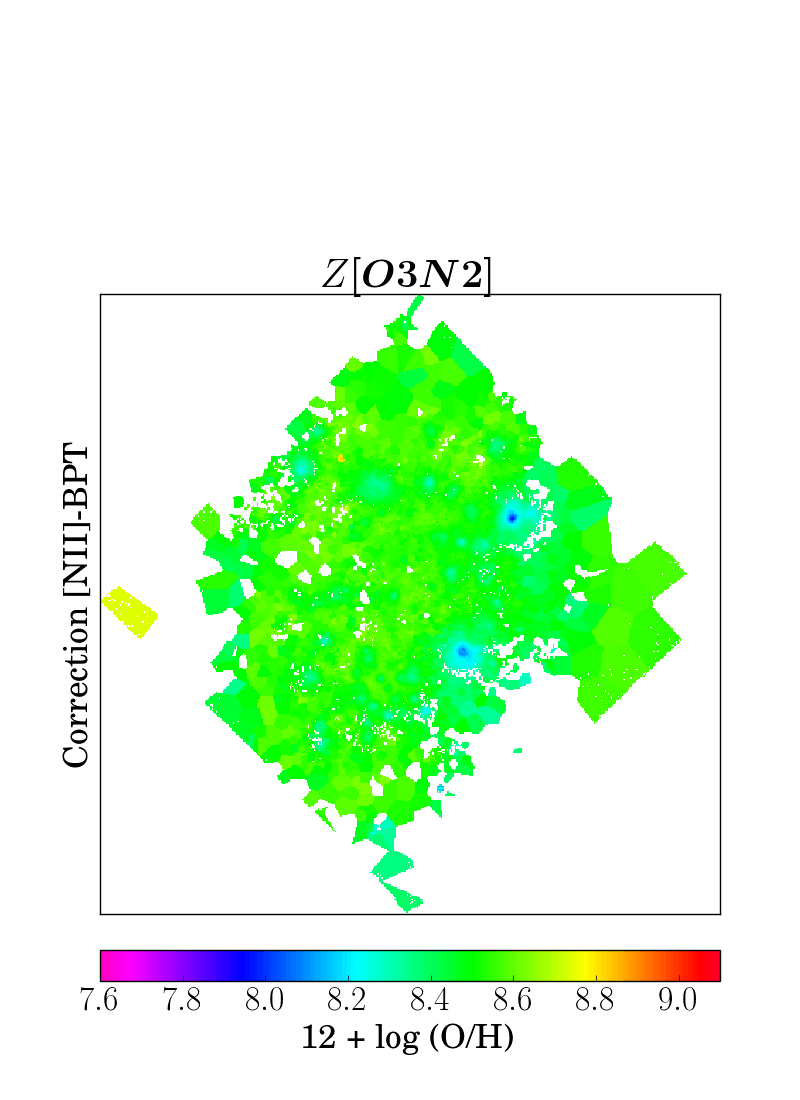}
	\includegraphics[width=0.28\textwidth, trim={0 1.2cm 0 5.5cm}, clip]{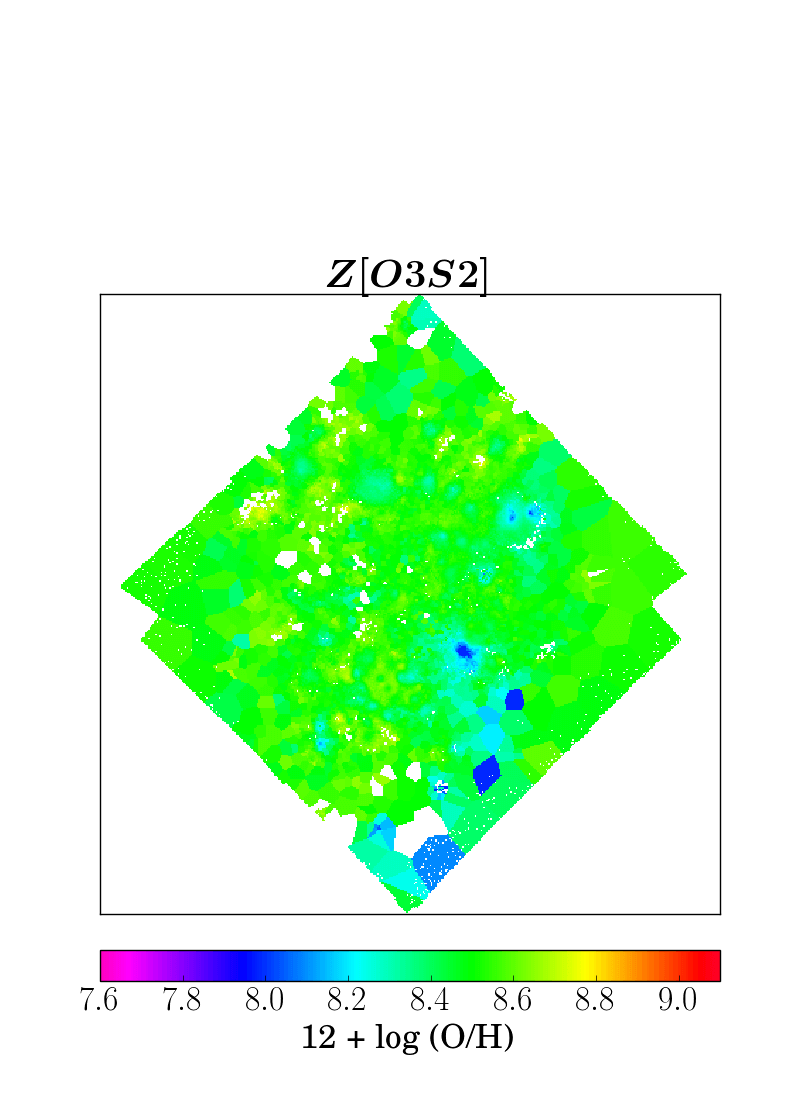}
	\includegraphics[width=0.28\textwidth, trim={2.8cm 0 2.8cm 0}, clip ]{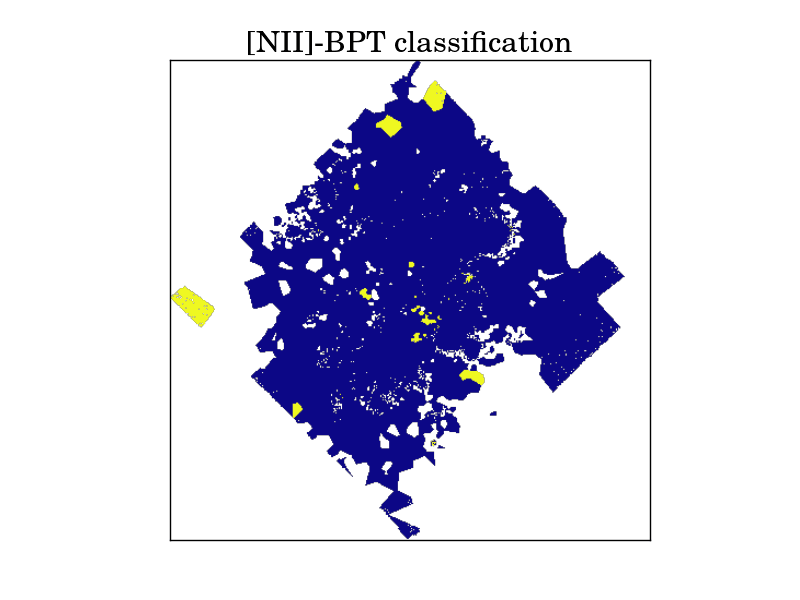}
	\caption{ Maps correspond to galaxy NGC4980, see caption of Figure \ref{fig:NGC1042} for details.}
	\label{fig:NGC4980}
\end{figure*}
\begin{figure*}
	\centering
	\includegraphics[width=0.28\textwidth, trim={0 1.2cm 0 5.5cm}, clip]{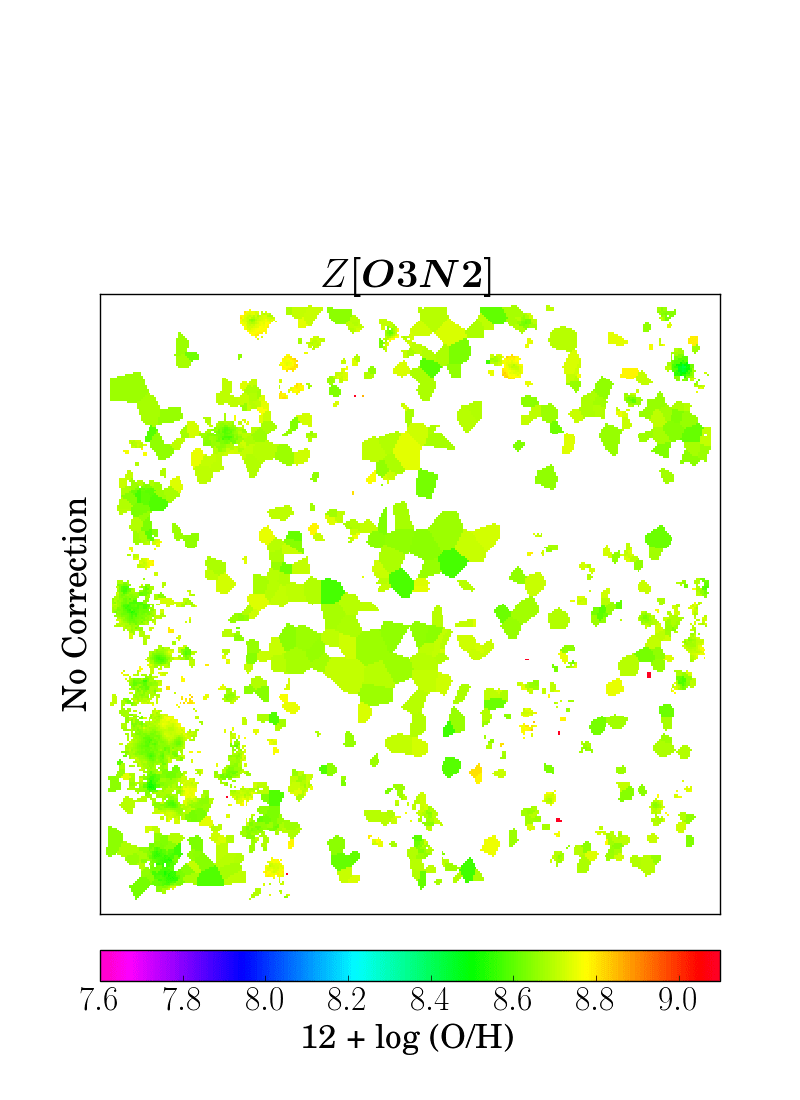}
	\includegraphics[width=0.28\textwidth, trim={0 1.2cm 0 5.5cm}, clip]{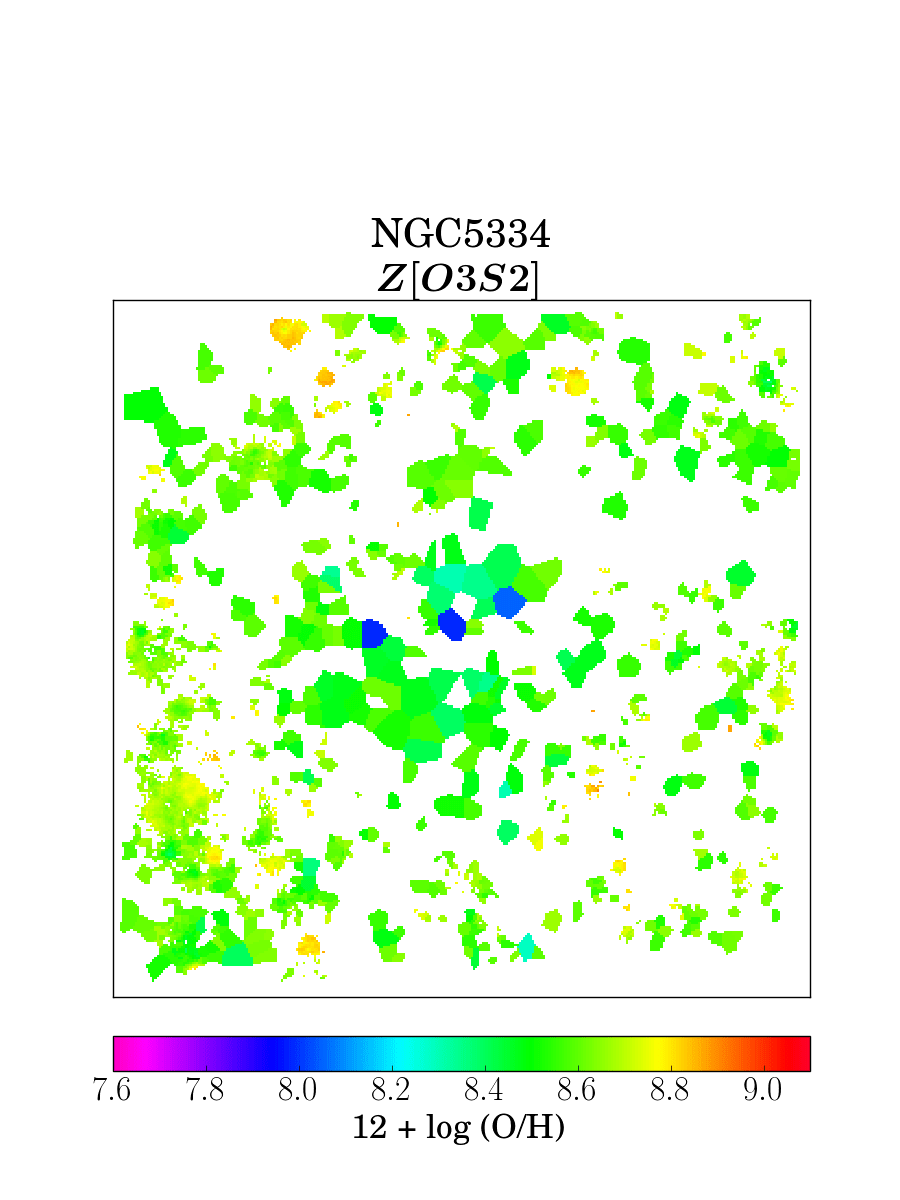}
	\includegraphics[width=0.28\textwidth, trim={0 1.2cm 0 5.5cm}, clip]{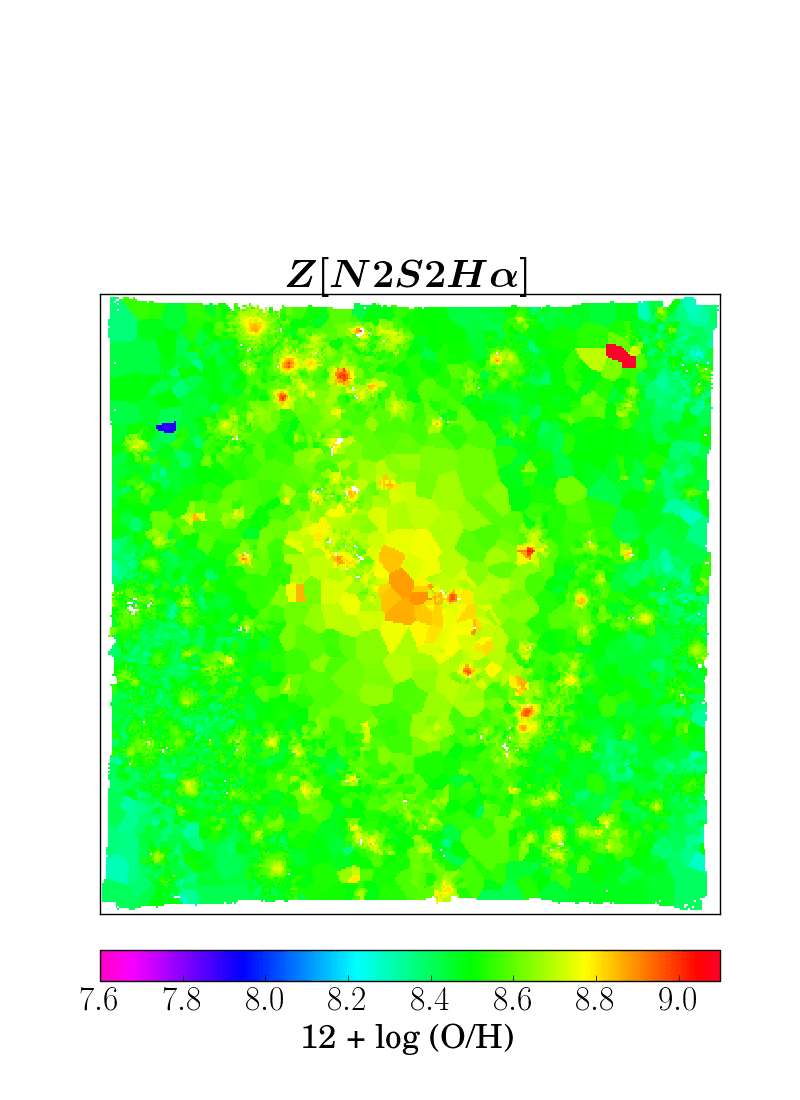}
	\includegraphics[width=0.28\textwidth, trim={0 1.2cm 0 5.5cm}, clip]{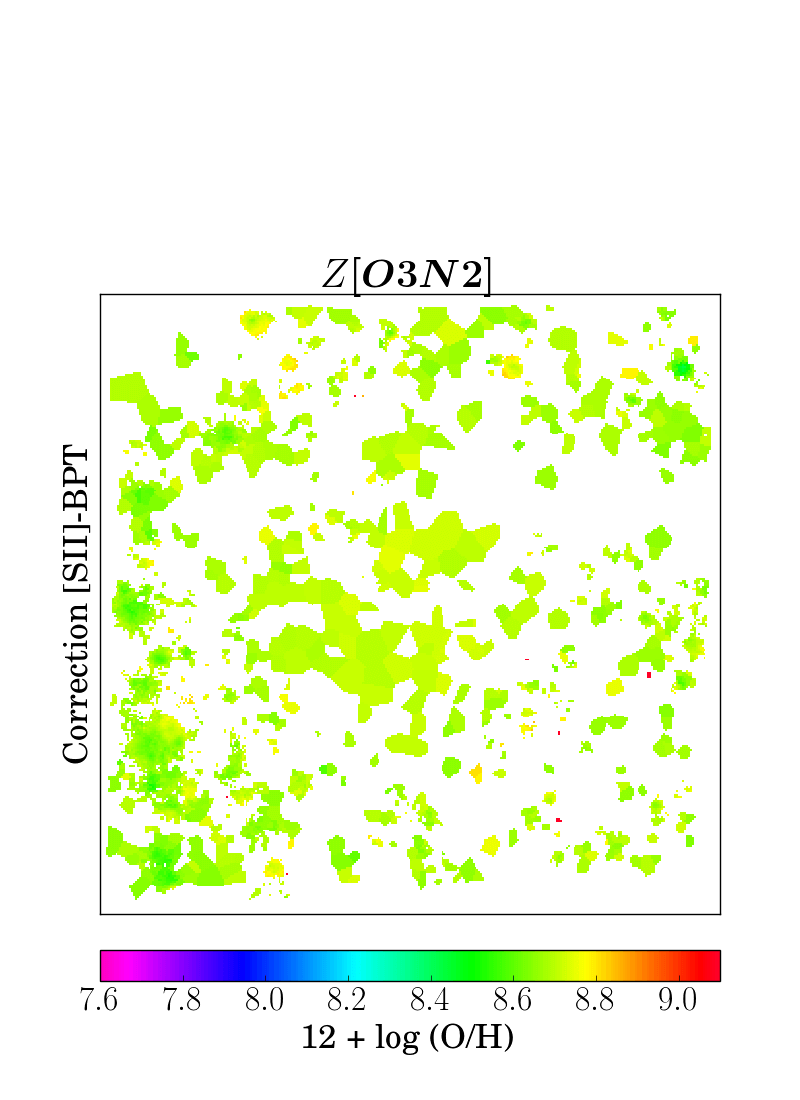}
	\includegraphics[width=0.28\textwidth, trim={0 1.2cm 0 5.5cm}, clip]{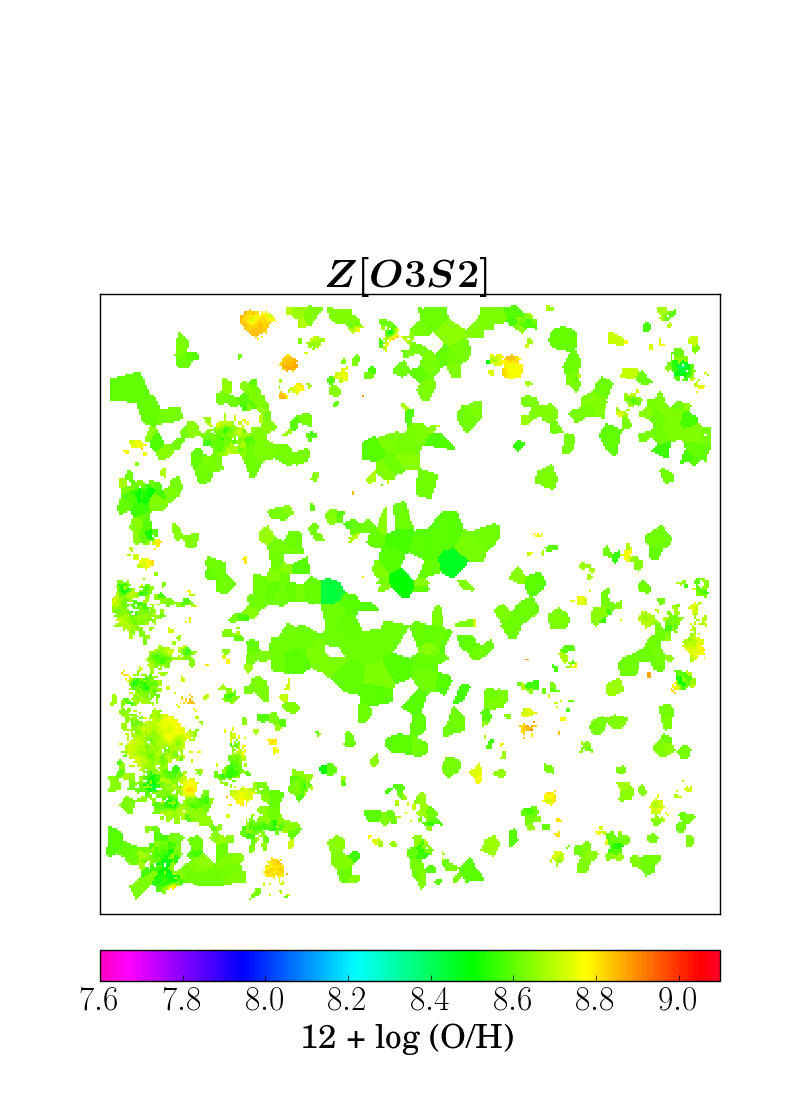}
	\includegraphics[width=0.28\textwidth, trim={2.8cm 0 2.8cm 0}, clip ]{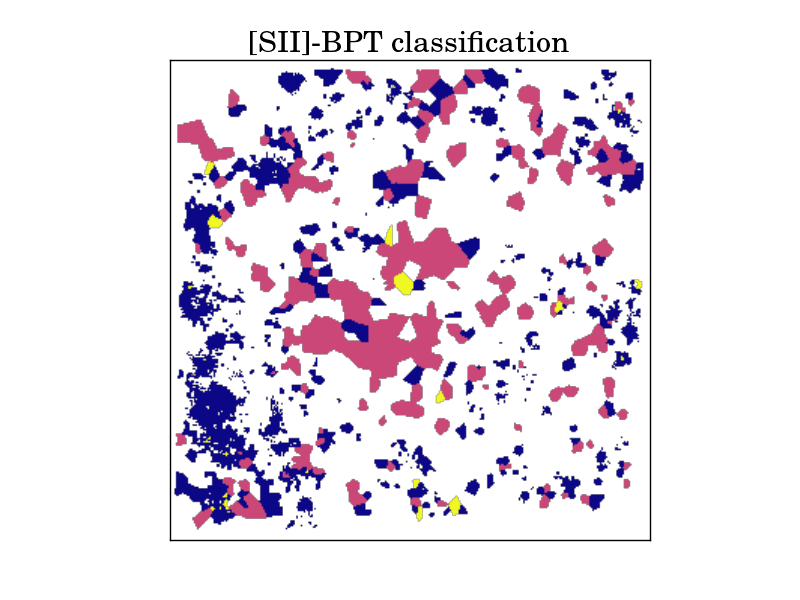}
	\includegraphics[width=0.28\textwidth, trim={0 1.2cm 0 5.5cm}, clip]{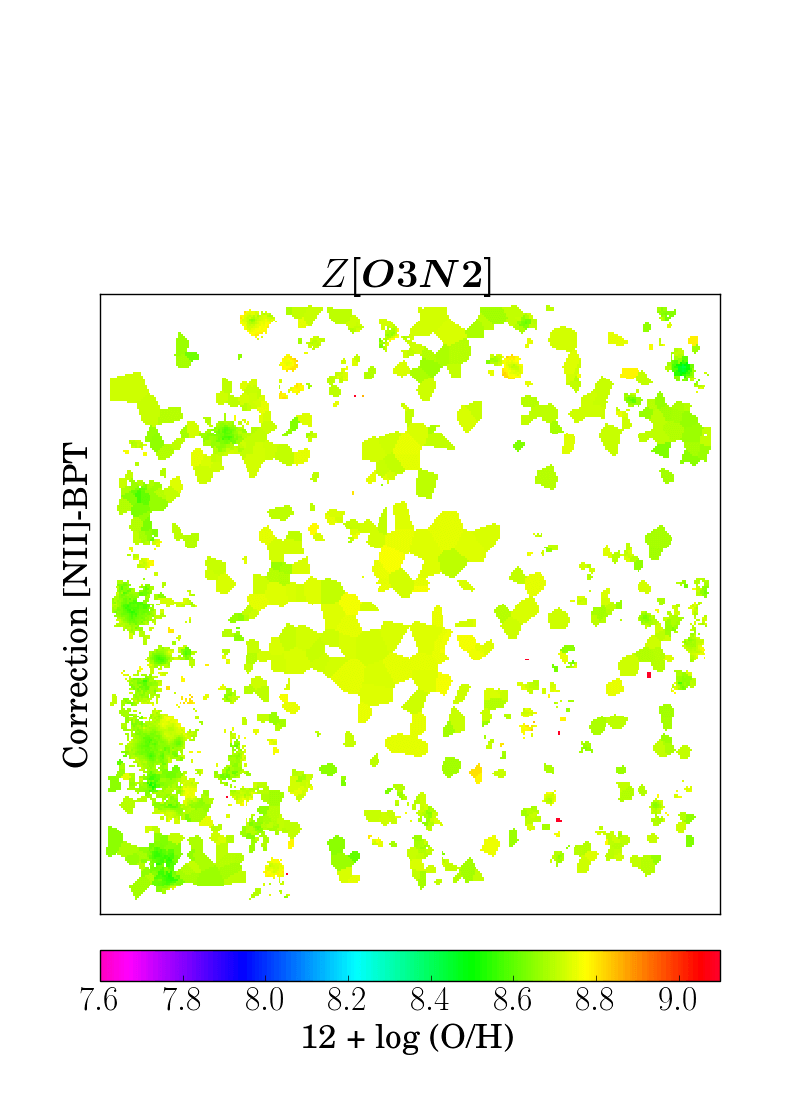}
	\includegraphics[width=0.28\textwidth, trim={0 1.2cm 0 5.5cm}, clip]{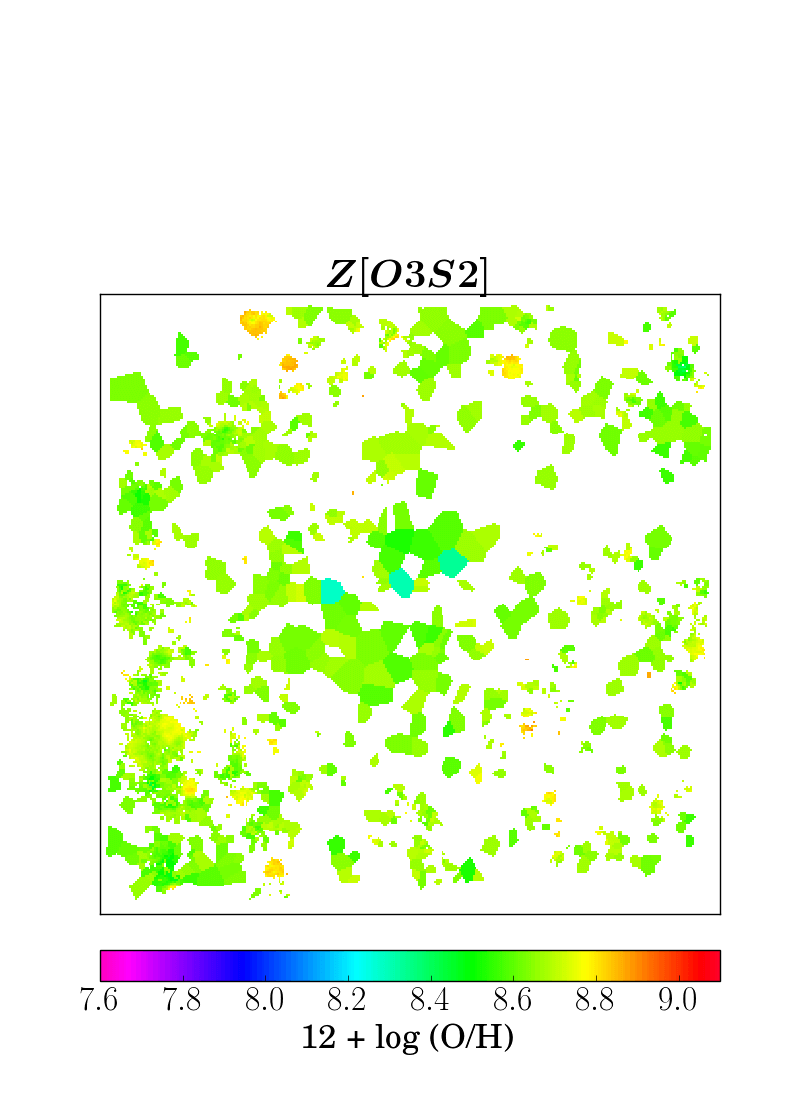}
	\includegraphics[width=0.28\textwidth, trim={2.8cm 0 2.8cm 0}, clip ]{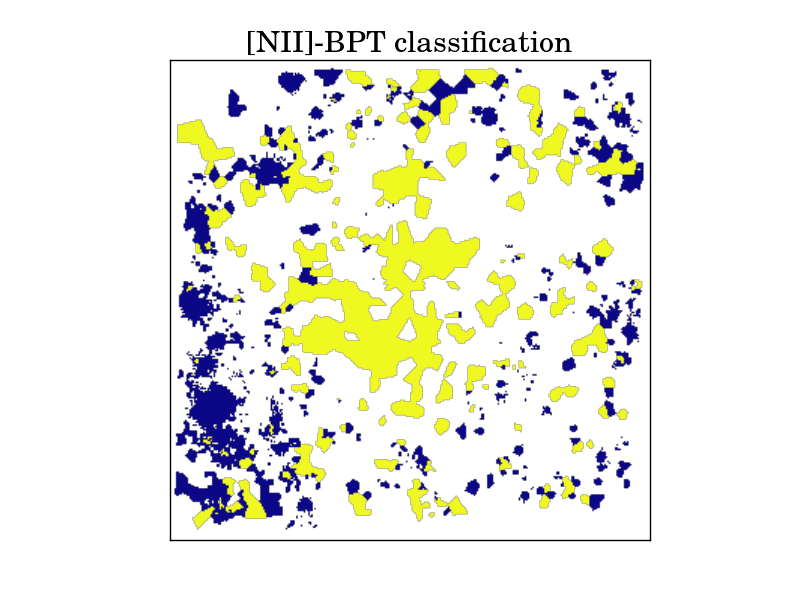}
	\caption{ Maps correspond to galaxy NGC5334, see caption of Figure \ref{fig:NGC1042} for details.}
	\label{fig:NGC5334}
\end{figure*}
\begin{figure*}
	\centering
	\includegraphics[width=0.28\textwidth, trim={0 1.2cm 0 5.5cm}, clip]{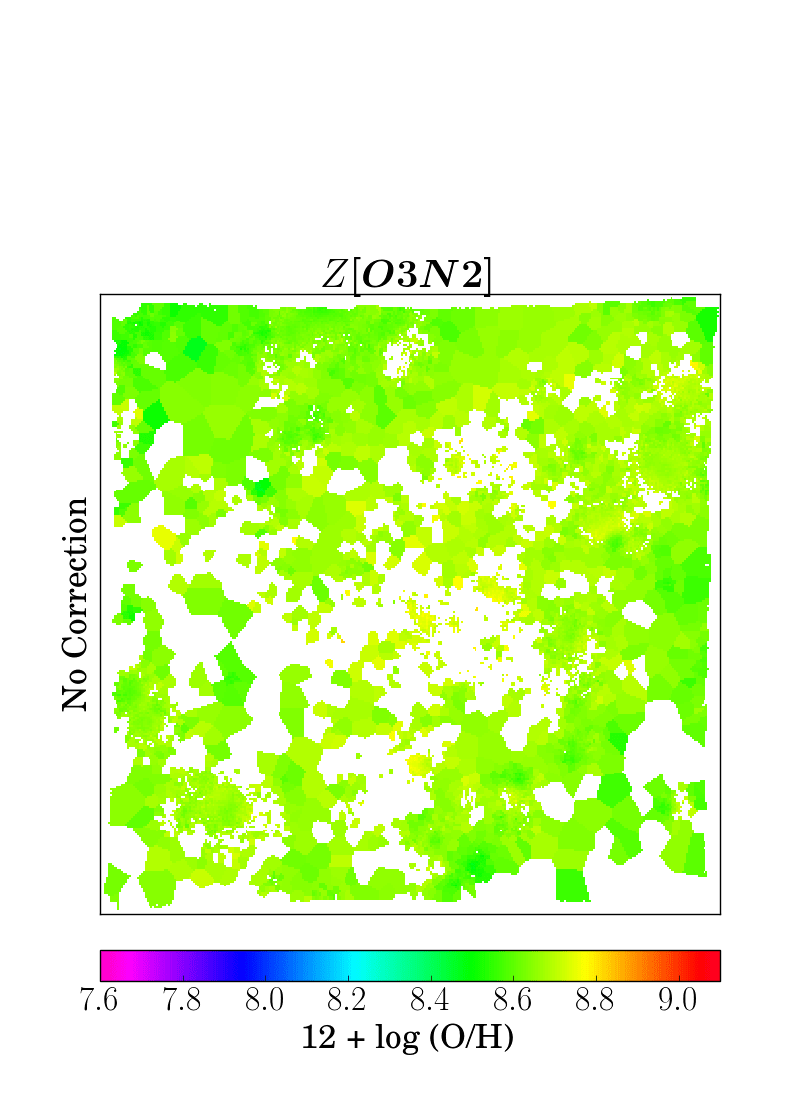}
	\includegraphics[width=0.28\textwidth, trim={0 1.2cm 0 5.5cm}, clip]{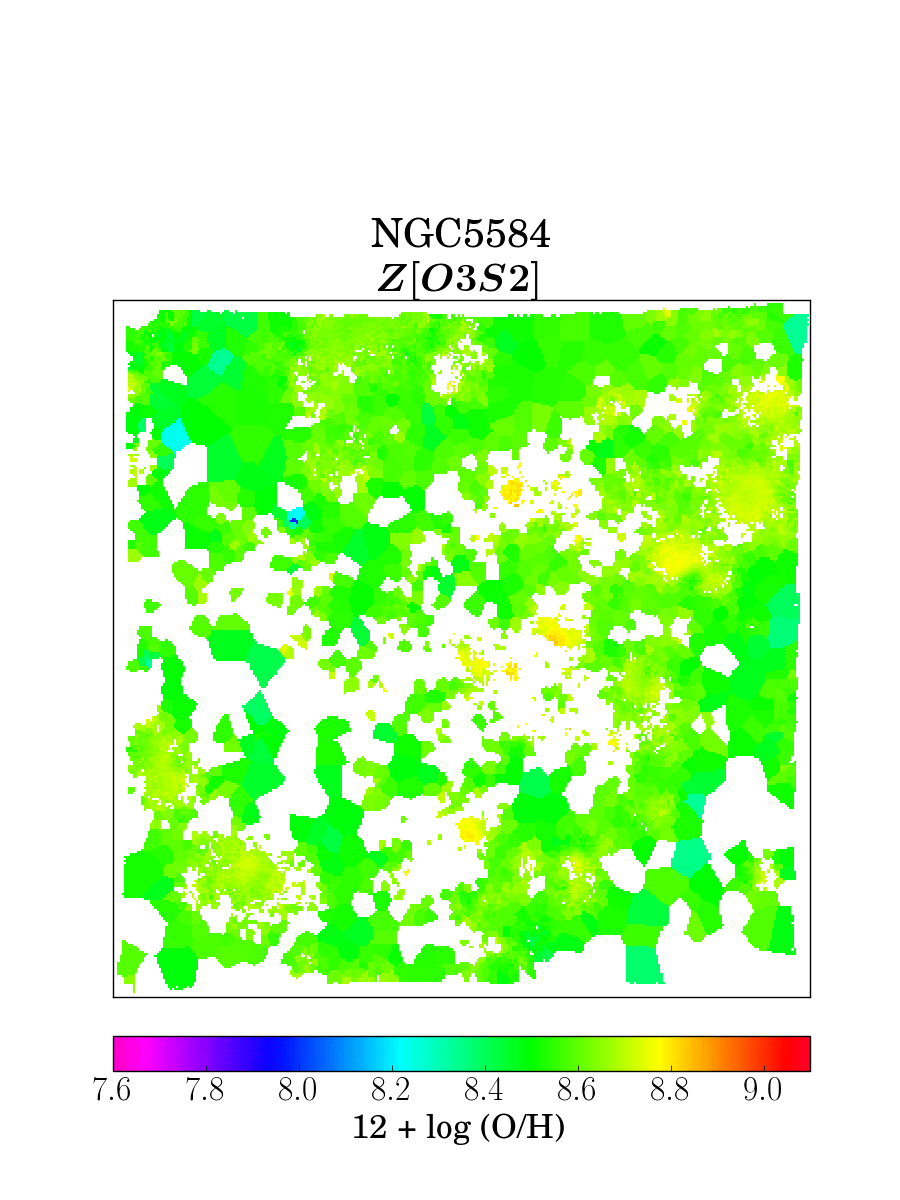}
	\includegraphics[width=0.28\textwidth, trim={0 1.2cm 0 5.5cm}, clip]{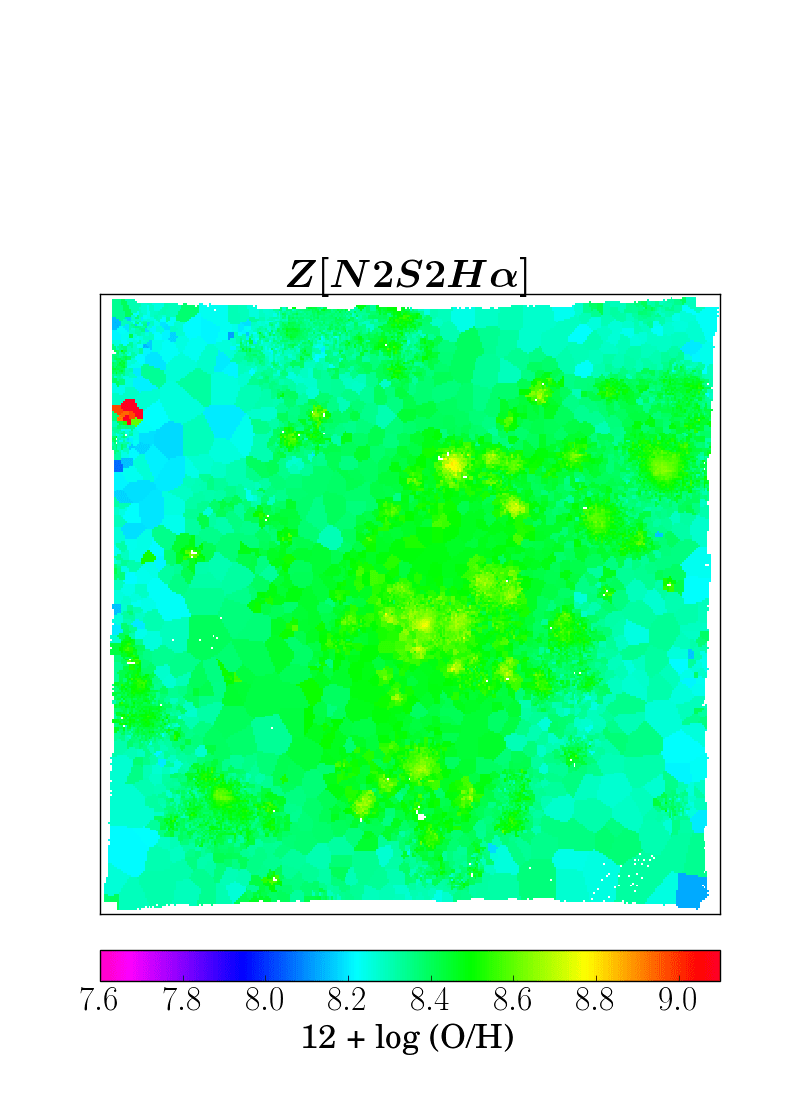}
	\includegraphics[width=0.28\textwidth, trim={0 1.2cm 0 5.5cm}, clip]{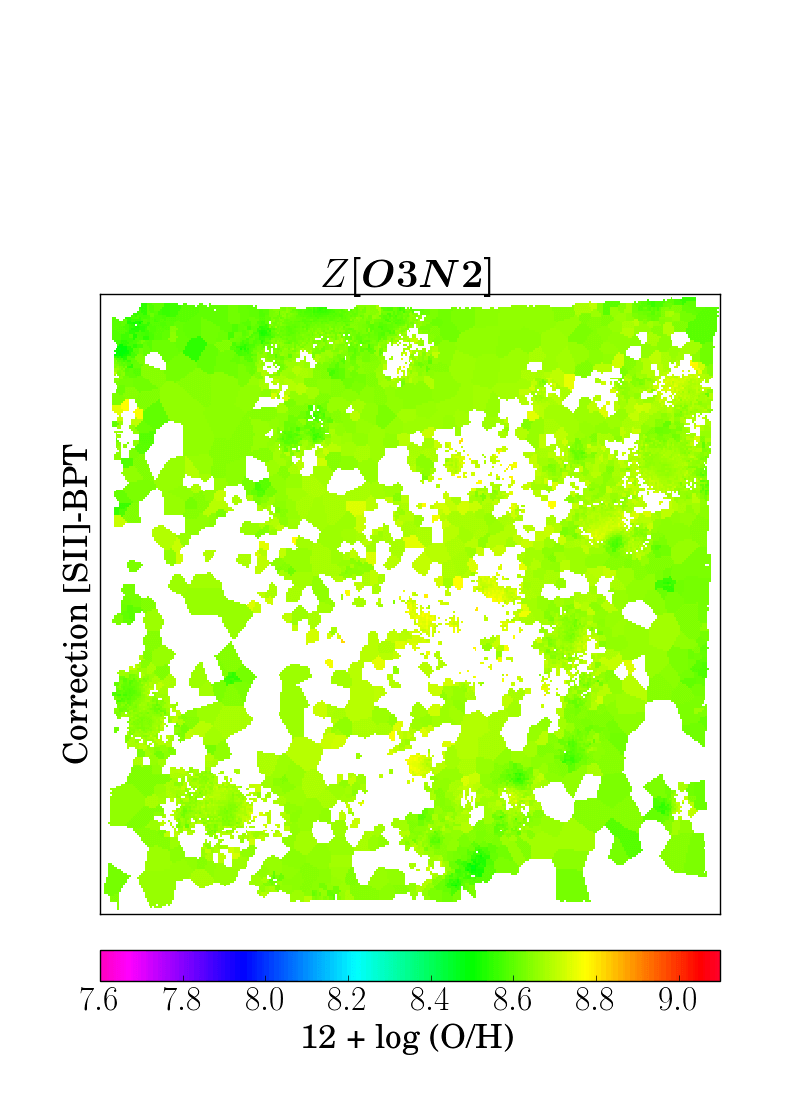}
	\includegraphics[width=0.28\textwidth, trim={0 1.2cm 0 5.5cm}, clip]{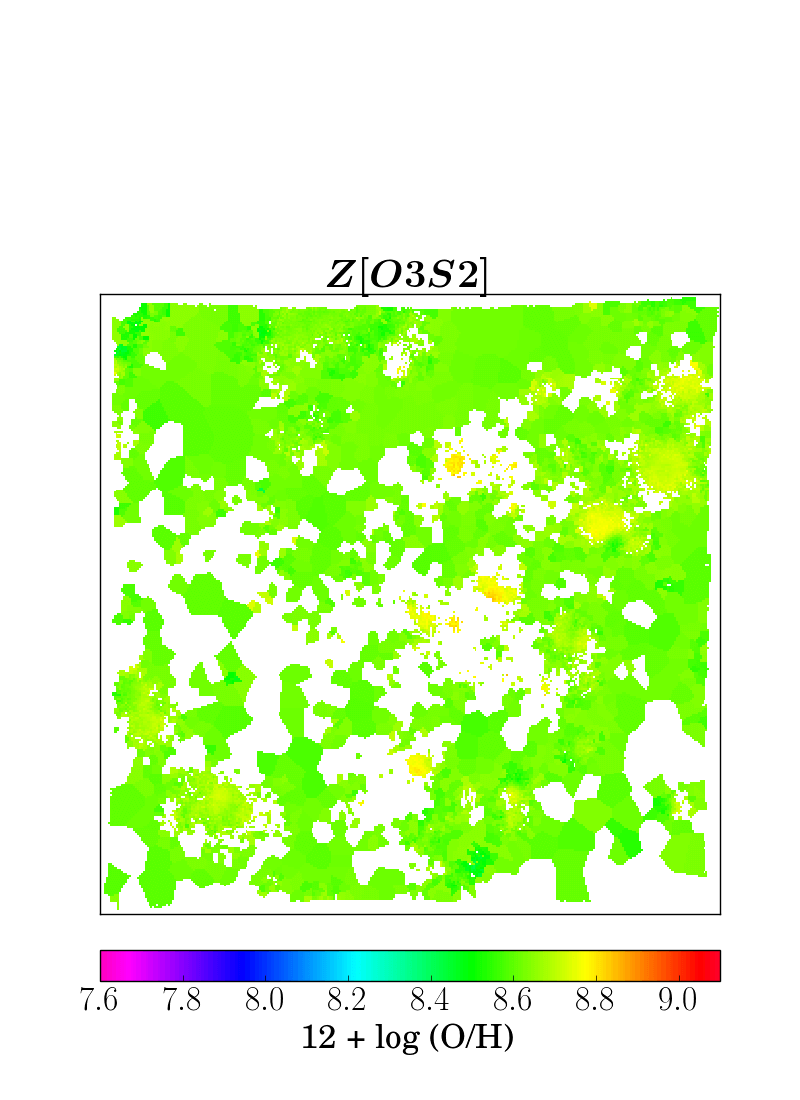}
	\includegraphics[width=0.28\textwidth, trim={2.8cm 0 2.8cm 0}, clip ]{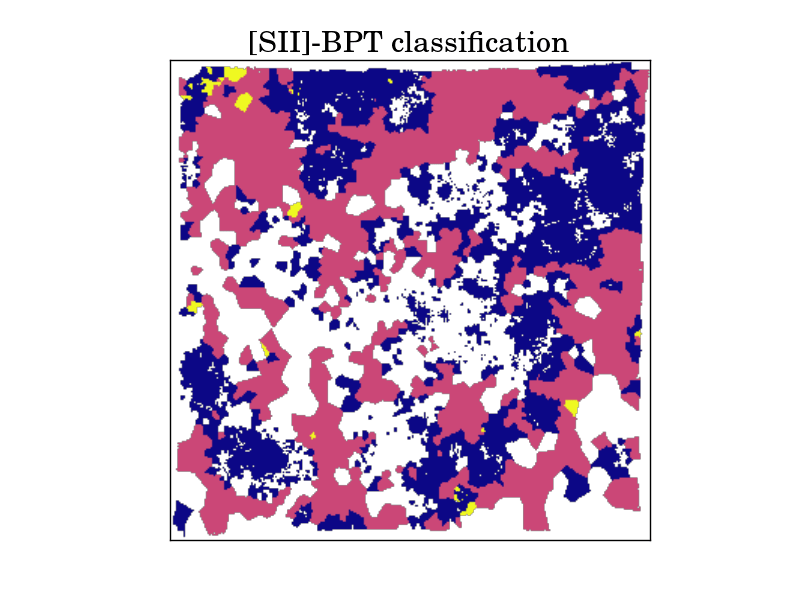}
	\includegraphics[width=0.28\textwidth, trim={0 1.2cm 0 5.5cm}, clip]{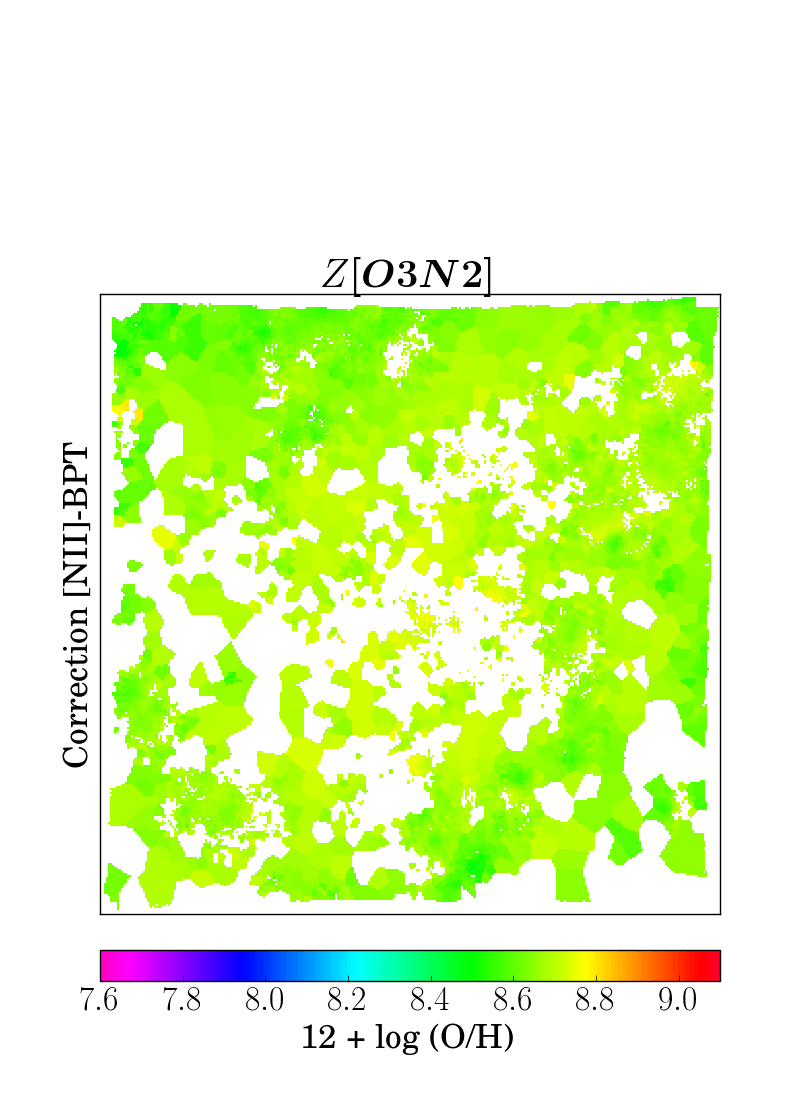}
	\includegraphics[width=0.28\textwidth, trim={0 1.2cm 0 5.5cm}, clip]{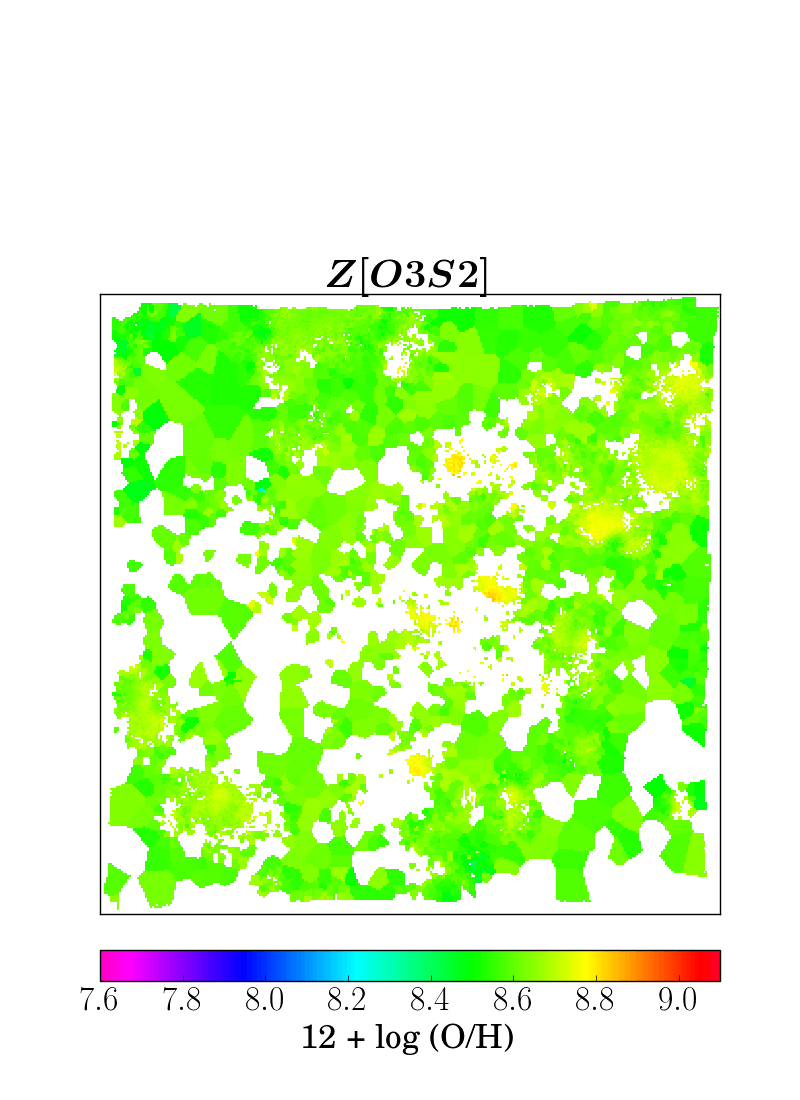}
	\includegraphics[width=0.28\textwidth, trim={2.8cm 0 2.8cm 0}, clip ]{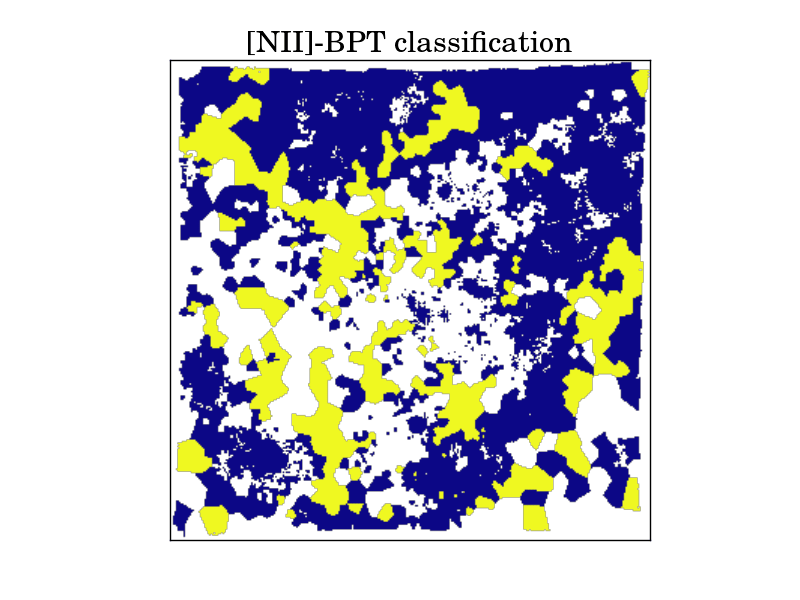}
	\caption{ Maps correspond to galaxy NGC5584, see caption of Figure \ref{fig:NGC1042} for details.}
	\label{fig:NGC5584}
\end{figure*}
\begin{figure*}
	\centering
	\includegraphics[width=0.28\textwidth, trim={0 1.2cm 0 5.5cm}, clip]{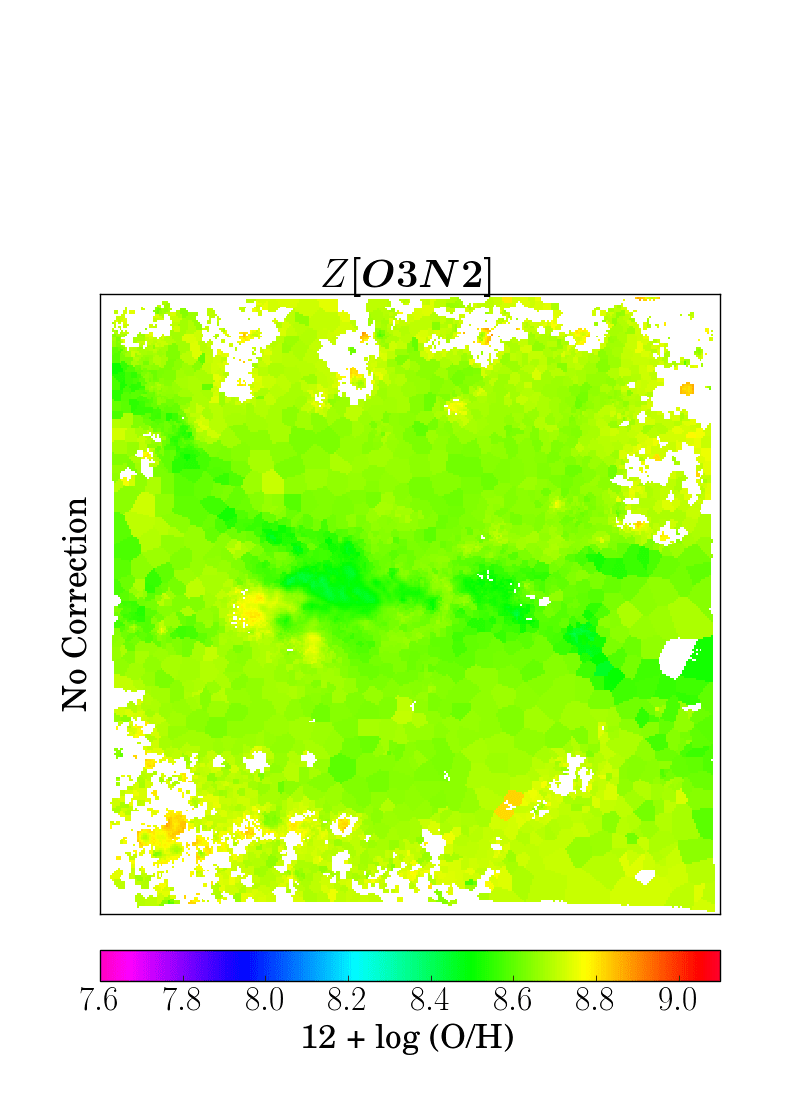}
	\includegraphics[width=0.28\textwidth, trim={0 1.2cm 0 5.5cm}, clip]{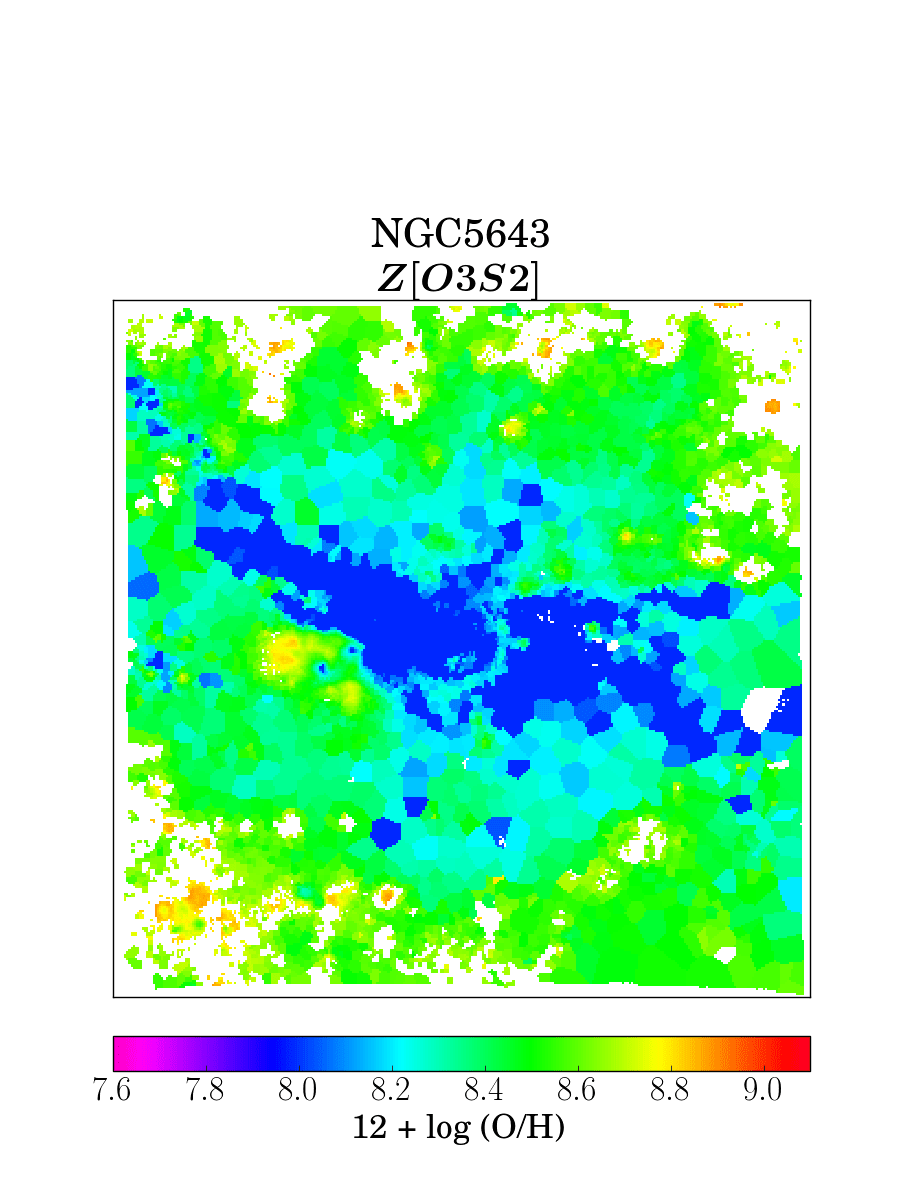}
	\includegraphics[width=0.28\textwidth, trim={0 1.2cm 0 5.5cm}, clip]{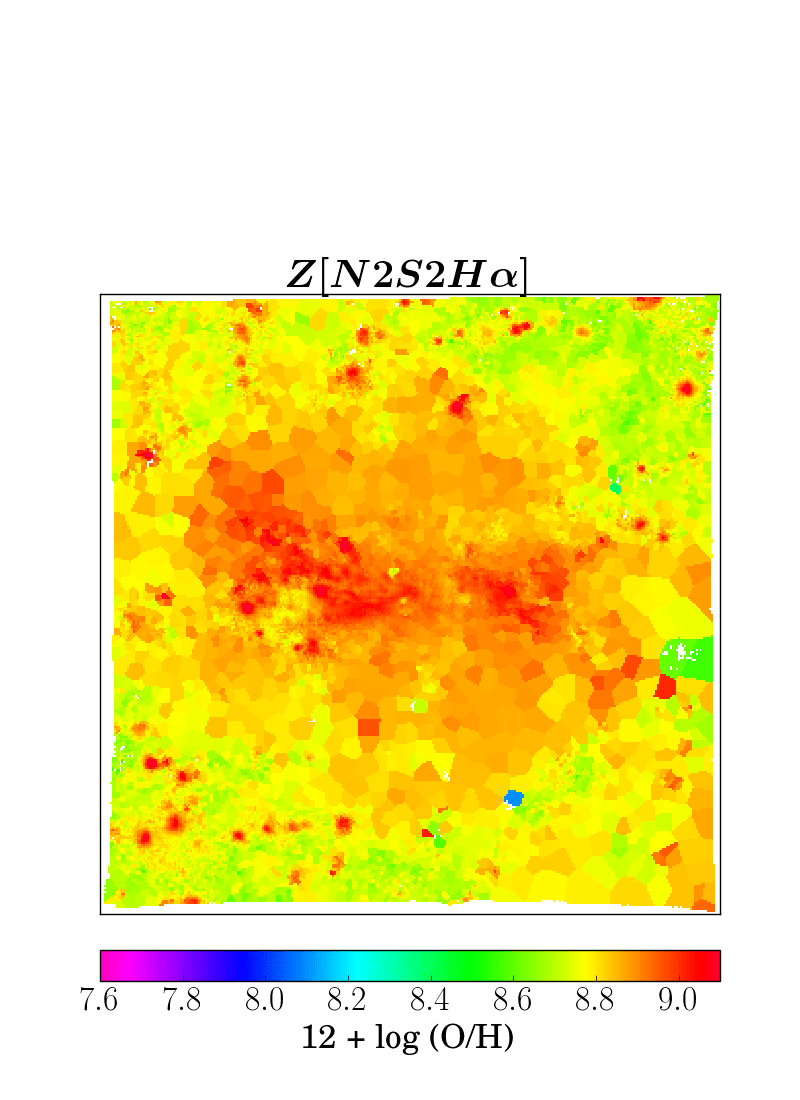}
	\includegraphics[width=0.28\textwidth, trim={0 1.2cm 0 5.5cm}, clip]{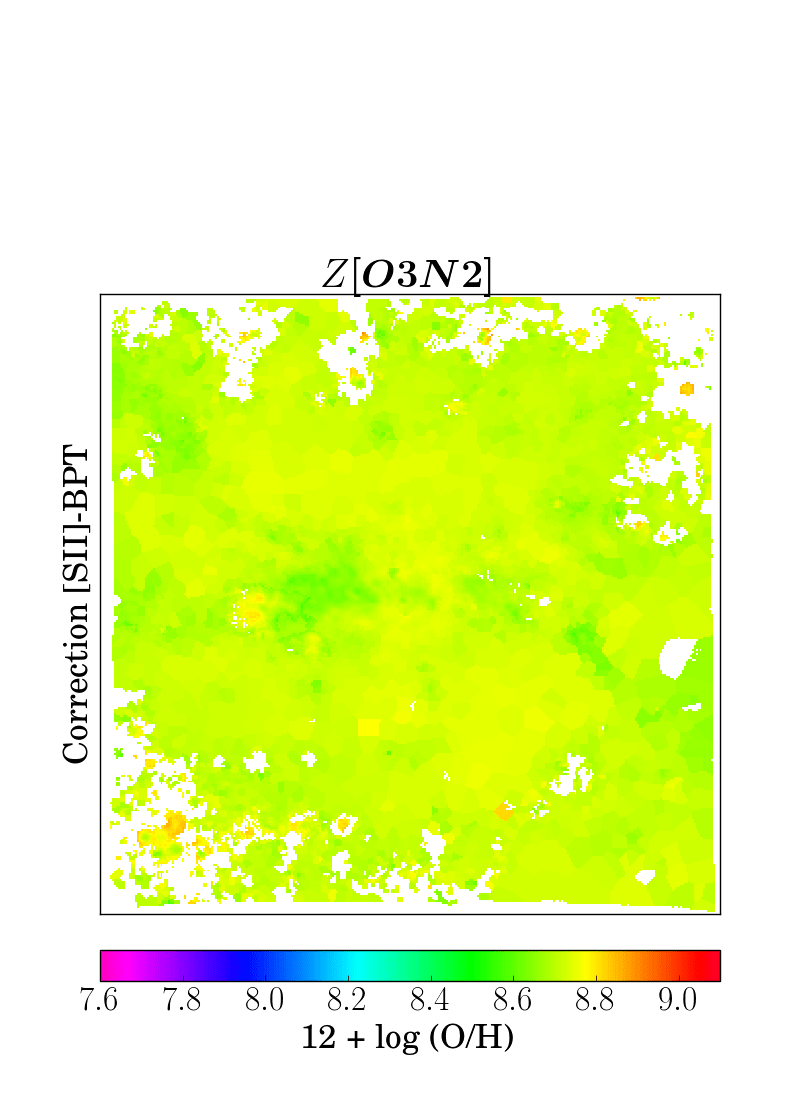}
	\includegraphics[width=0.28\textwidth, trim={0 1.2cm 0 5.5cm}, clip]{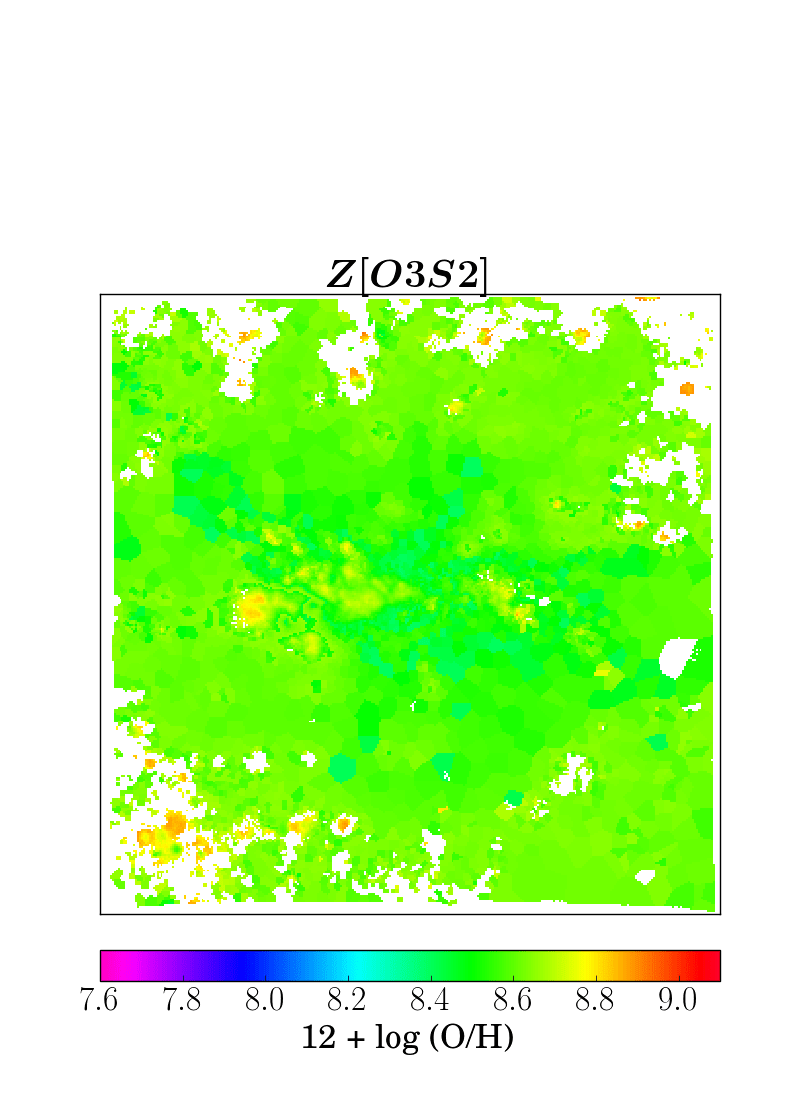}
	\includegraphics[width=0.28\textwidth, trim={2.8cm 0 2.8cm 0}, clip ]{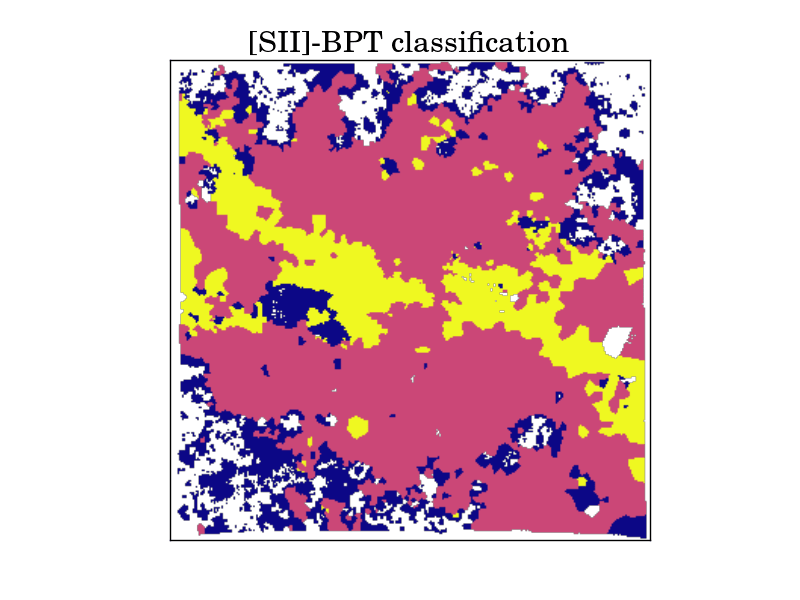}
	\includegraphics[width=0.28\textwidth, trim={0 1.2cm 0 5.5cm}, clip]{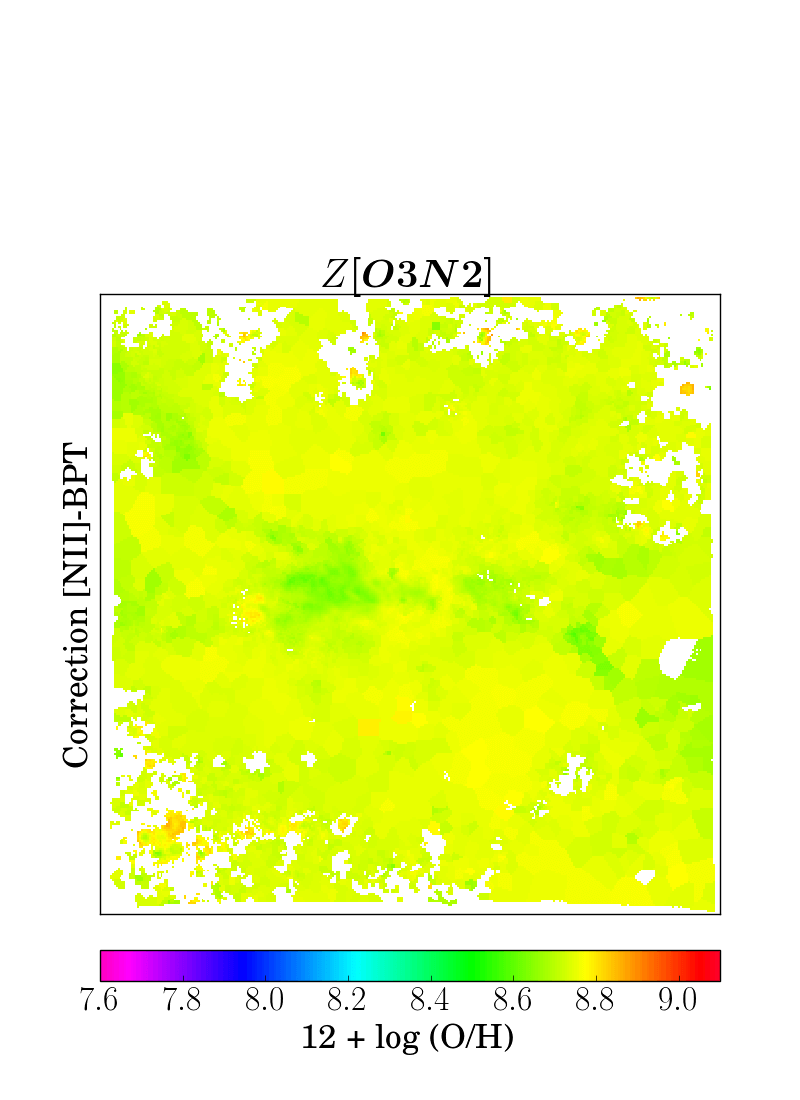}
	\includegraphics[width=0.28\textwidth, trim={0 1.2cm 0 5.5cm}, clip]{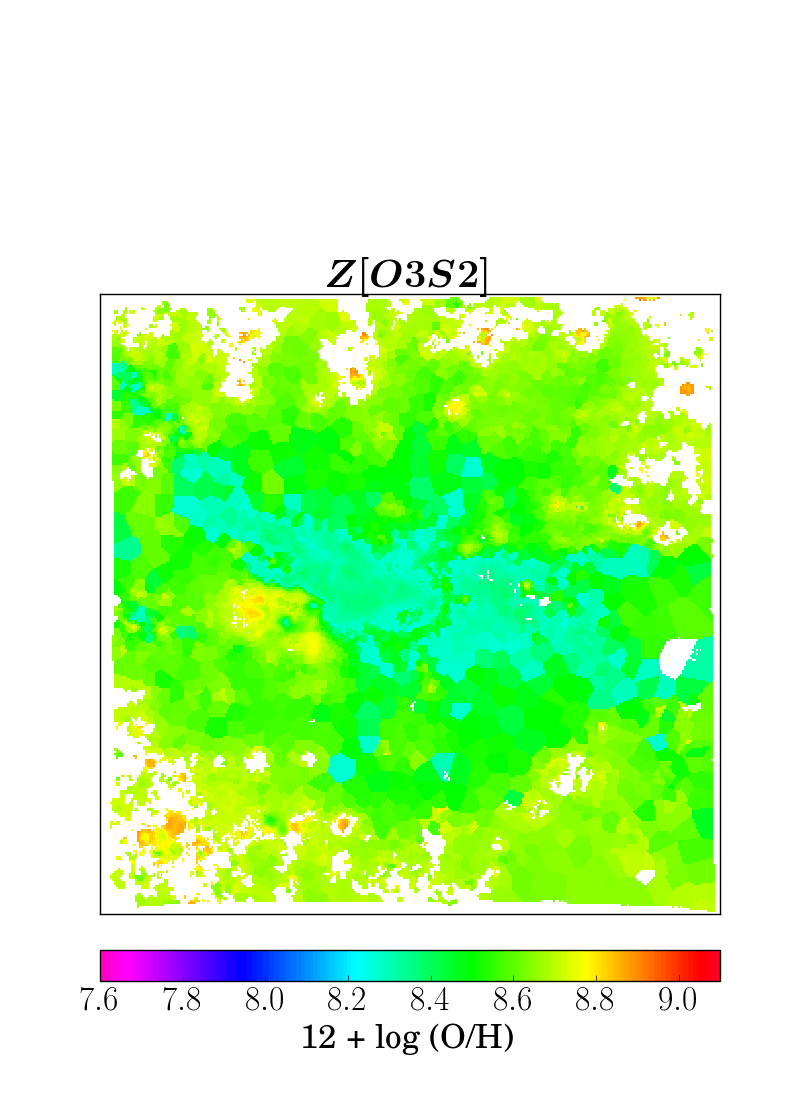}
	\includegraphics[width=0.28\textwidth, trim={2.8cm 0 2.8cm 0}, clip ]{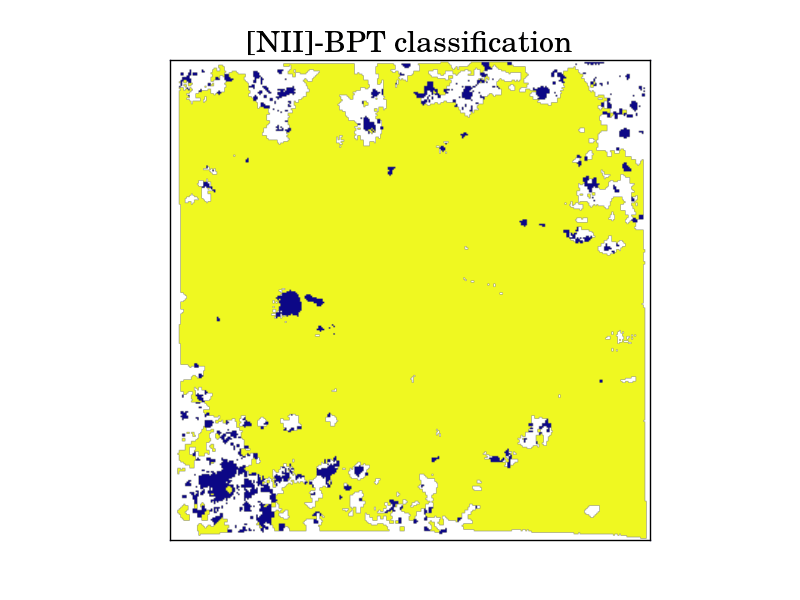}
	\caption{ Maps correspond to galaxy NGC5643, see caption of Figure \ref{fig:NGC1042} for details.}
	\label{fig:NGC5643}
\end{figure*}
\begin{figure*}
	\centering
	\includegraphics[width=0.28\textwidth, trim={0 1.2cm 0 5.5cm}, clip]{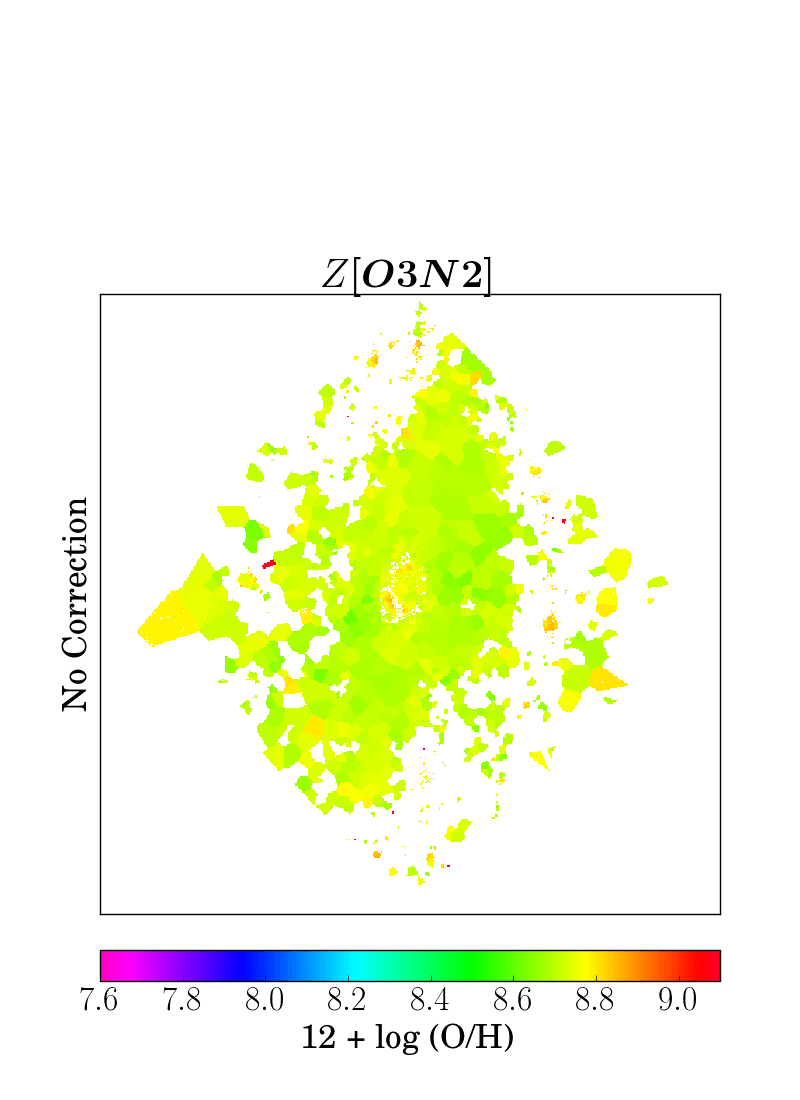}
	\includegraphics[width=0.28\textwidth, trim={0 1.2cm 0 5.5cm}, clip]{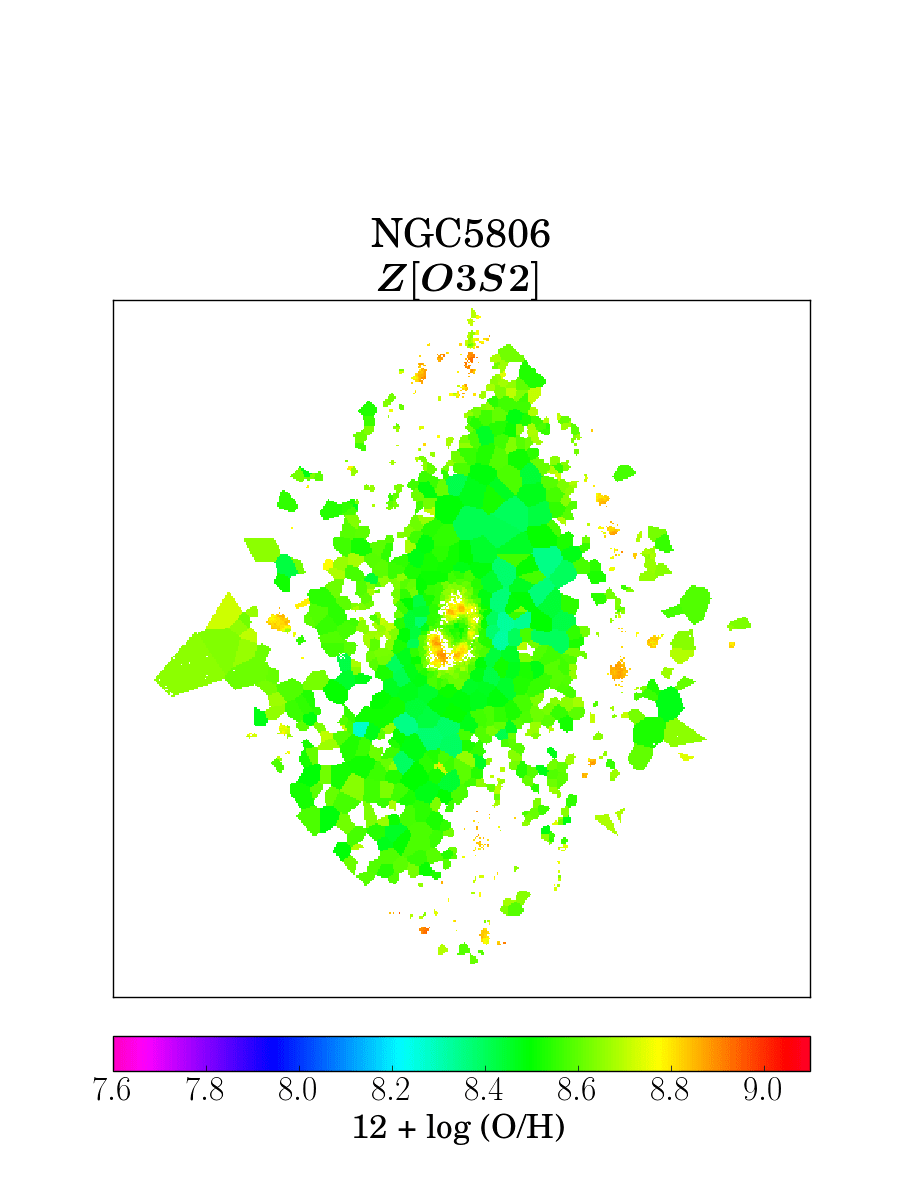}
	\includegraphics[width=0.28\textwidth, trim={0 1.2cm 0 5.5cm}, clip]{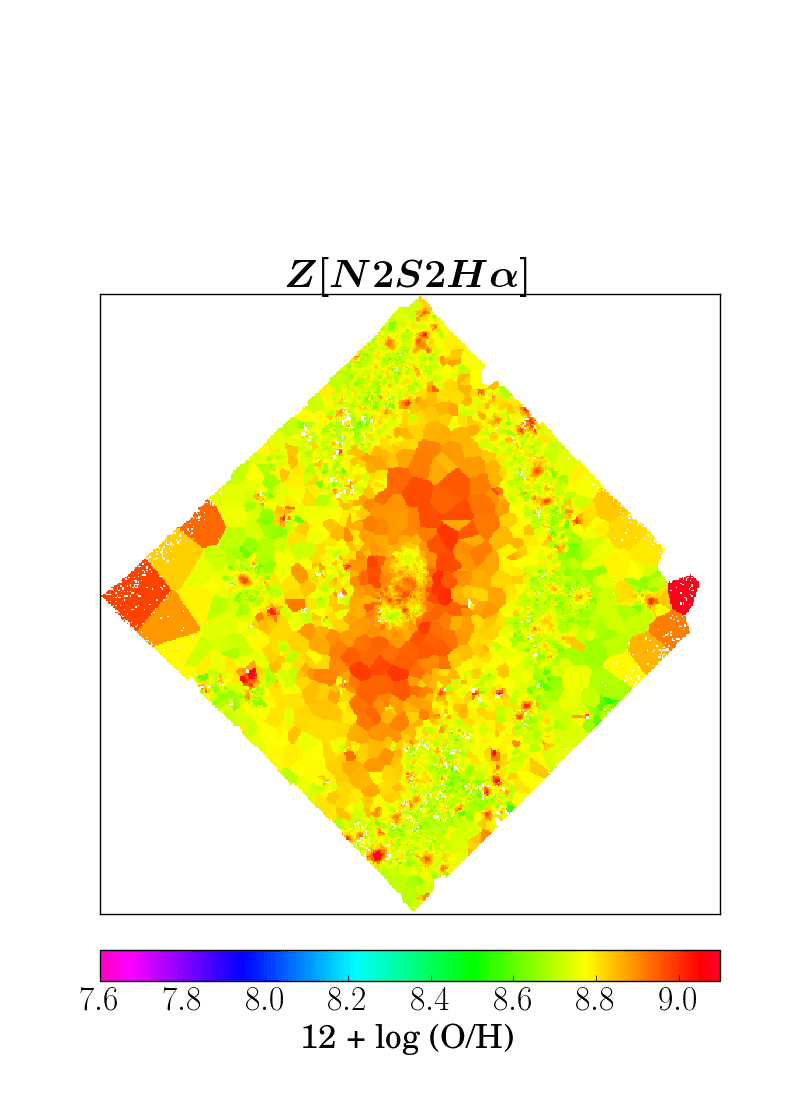}
	\includegraphics[width=0.28\textwidth, trim={0 1.2cm 0 5.5cm}, clip]{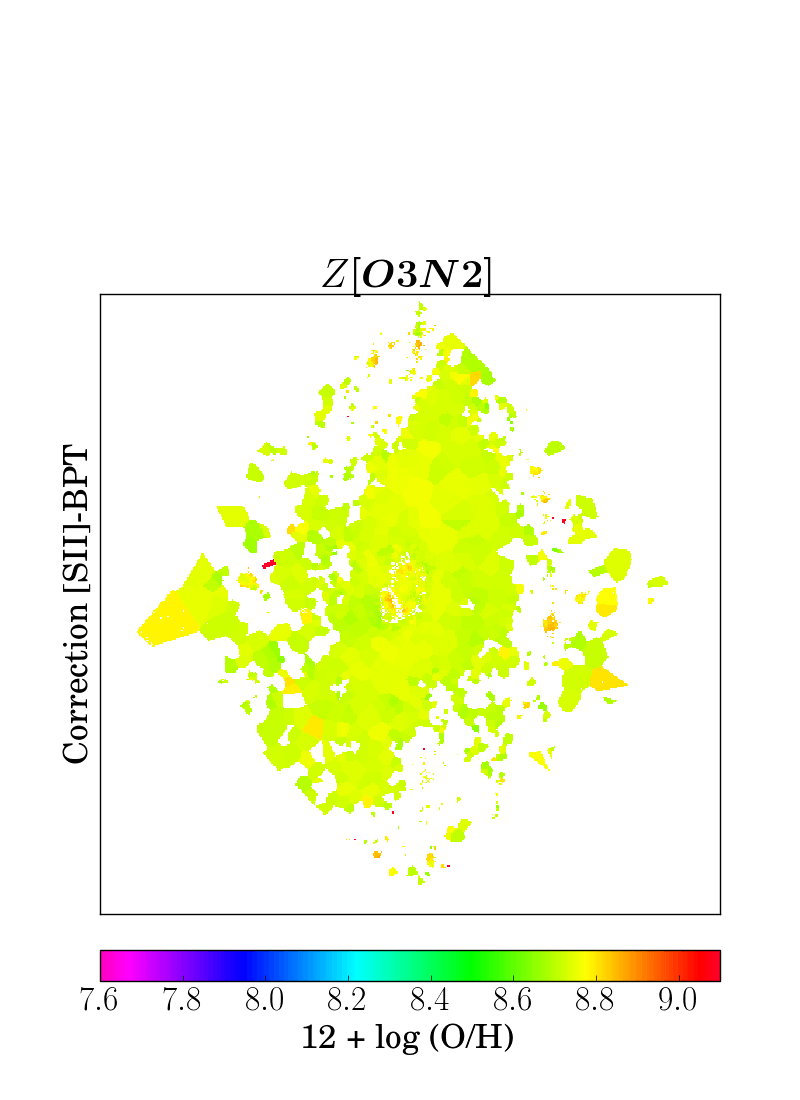}
	\includegraphics[width=0.28\textwidth, trim={0 1.2cm 0 5.5cm}, clip]{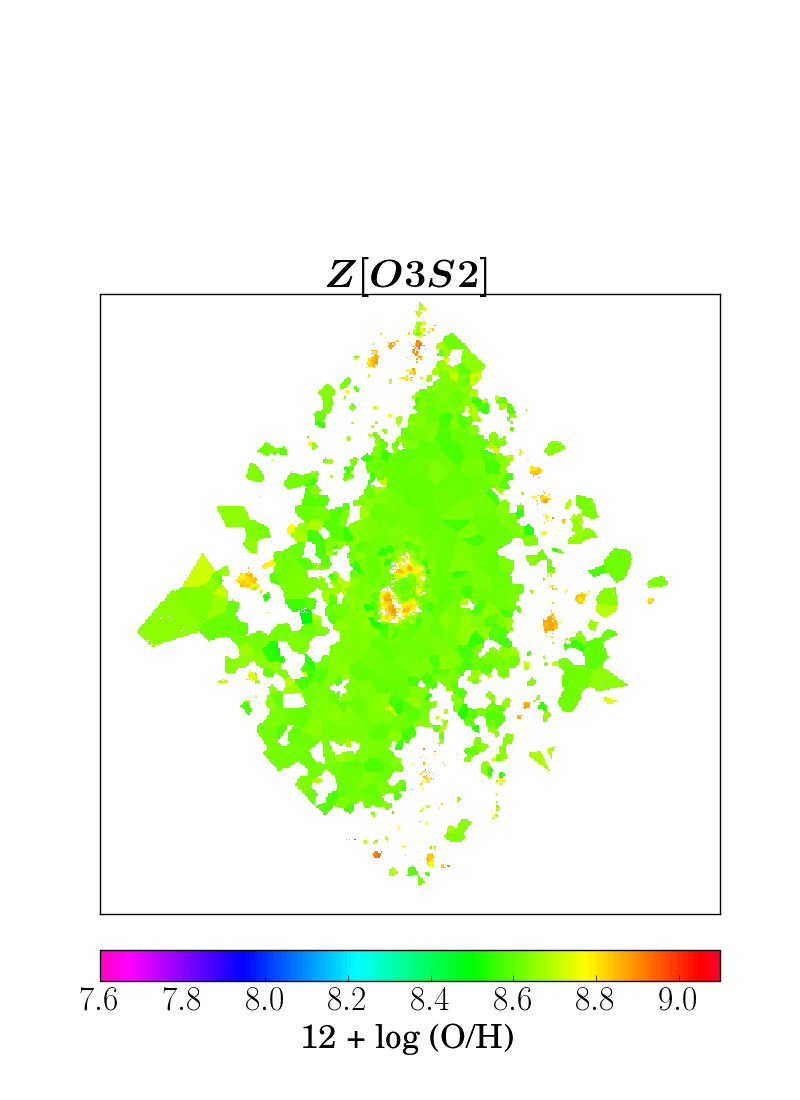}
	\includegraphics[width=0.28\textwidth, trim={2.8cm 0 2.8cm 0}, clip ]{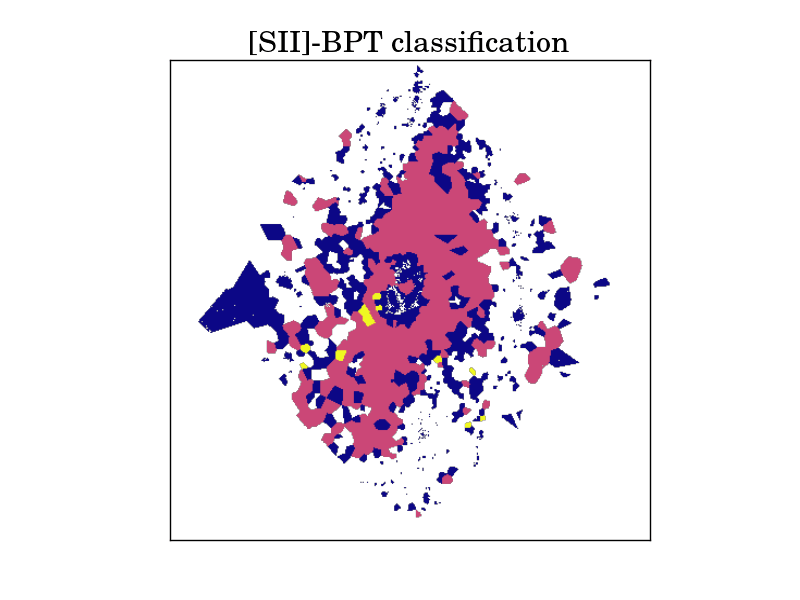}
	\includegraphics[width=0.28\textwidth, trim={0 1.2cm 0 5.5cm}, clip]{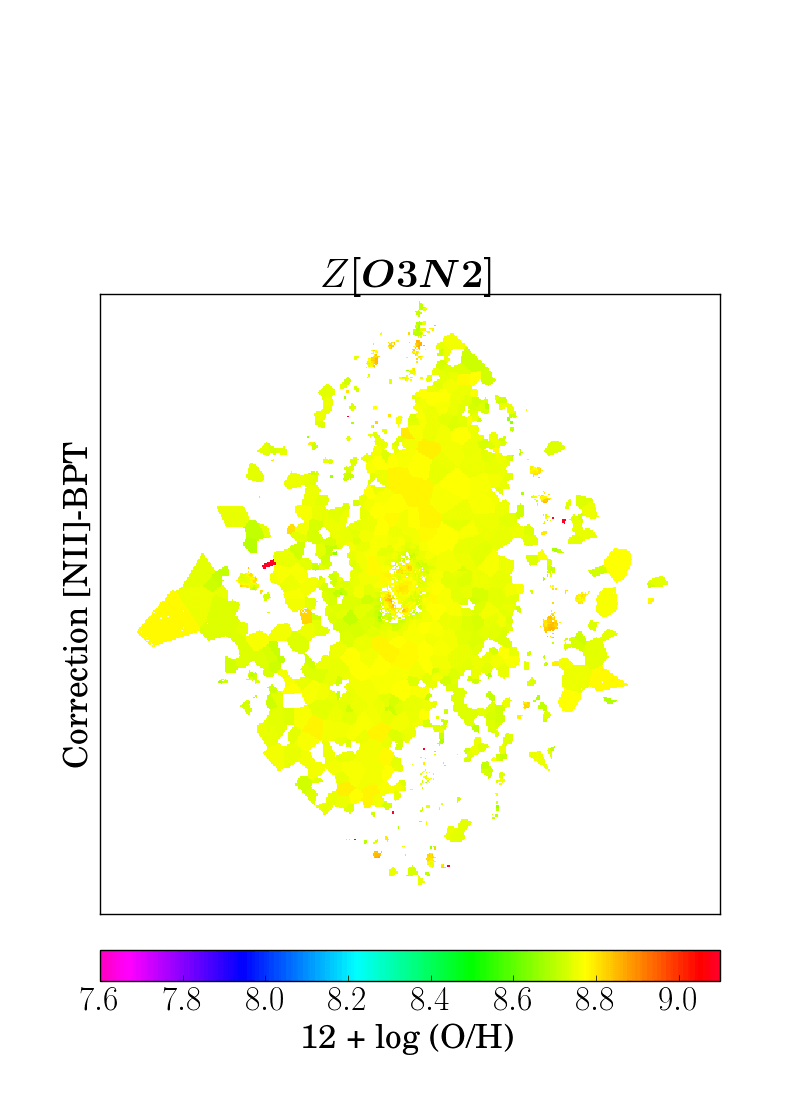}
	\includegraphics[width=0.28\textwidth, trim={0 1.2cm 0 5.5cm}, clip]{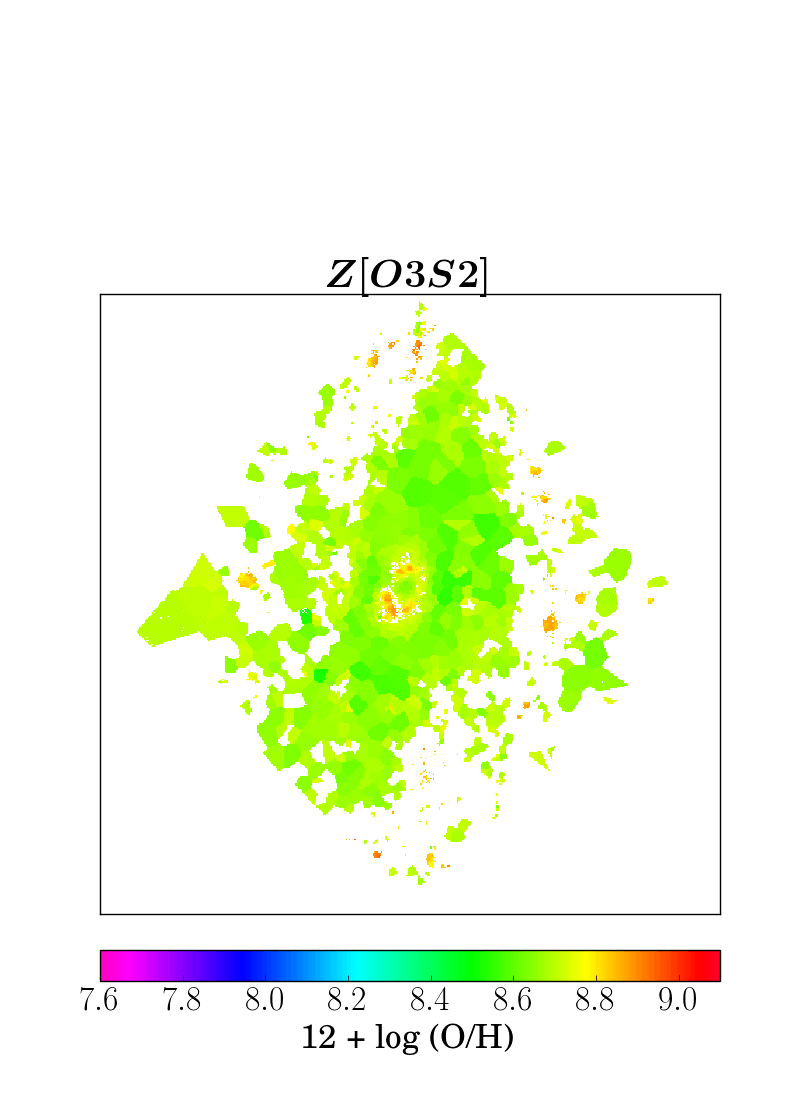}
	\includegraphics[width=0.28\textwidth, trim={2.8cm 0 2.8cm 0}, clip ]{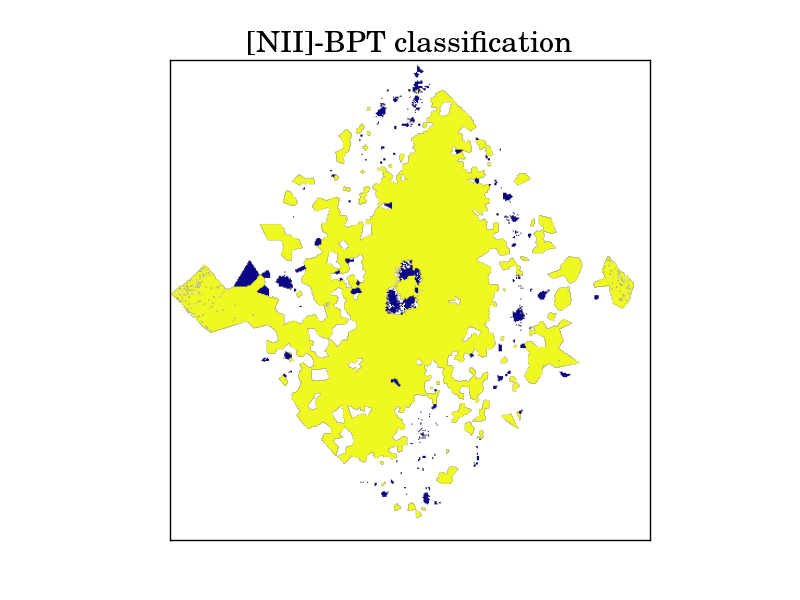}
	\caption{ Maps correspond to galaxy NGC5806, see caption of Figure \ref{fig:NGC1042} for details.}
	\label{fig:NGC5806}
\end{figure*}
\begin{figure*}
	\centering
	\includegraphics[width=0.28\textwidth, trim={0 1.2cm 0 5.5cm}, clip]{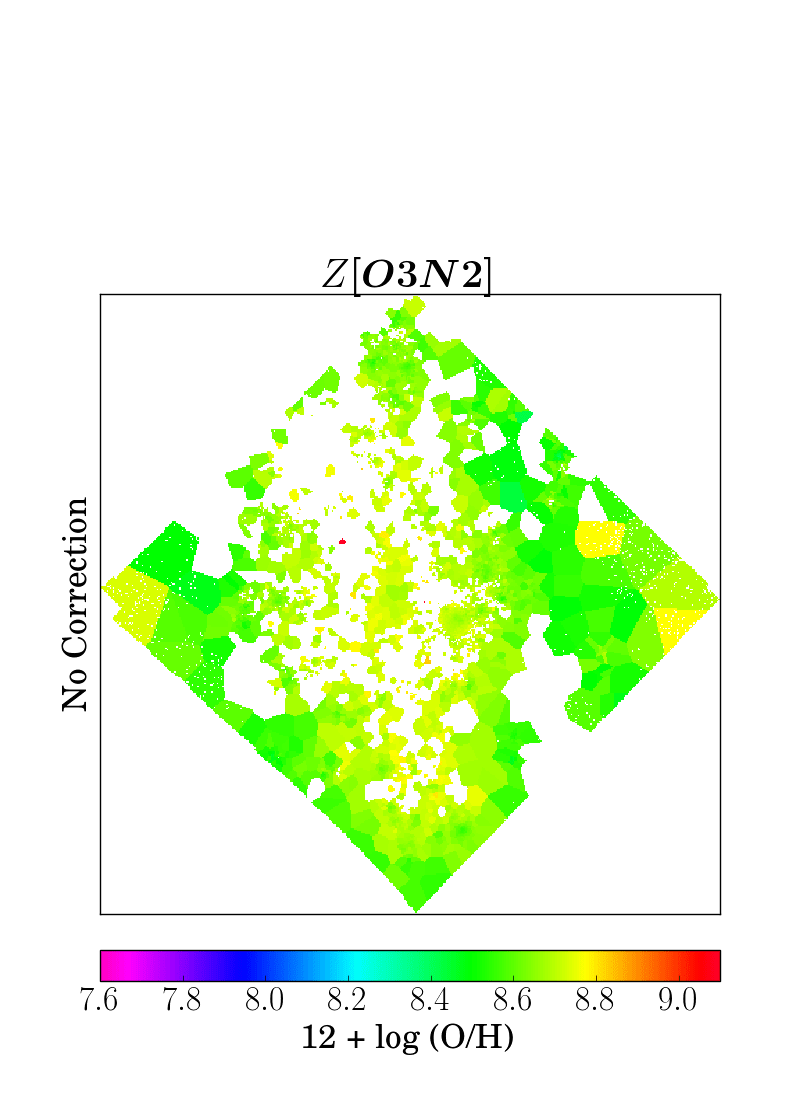}
	\includegraphics[width=0.28\textwidth, trim={0 1.2cm 0 5.5cm}, clip]{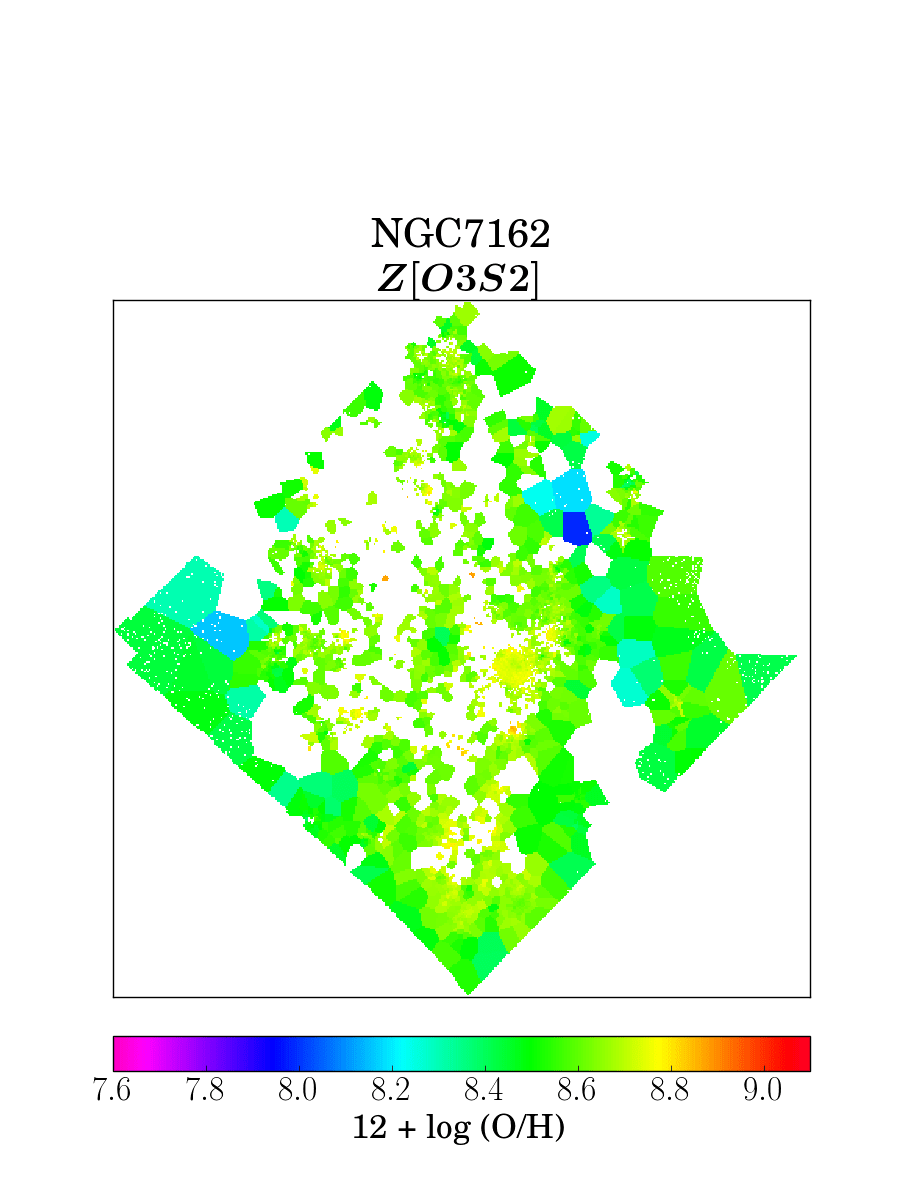}
	\includegraphics[width=0.28\textwidth, trim={0 1.2cm 0 5.5cm}, clip]{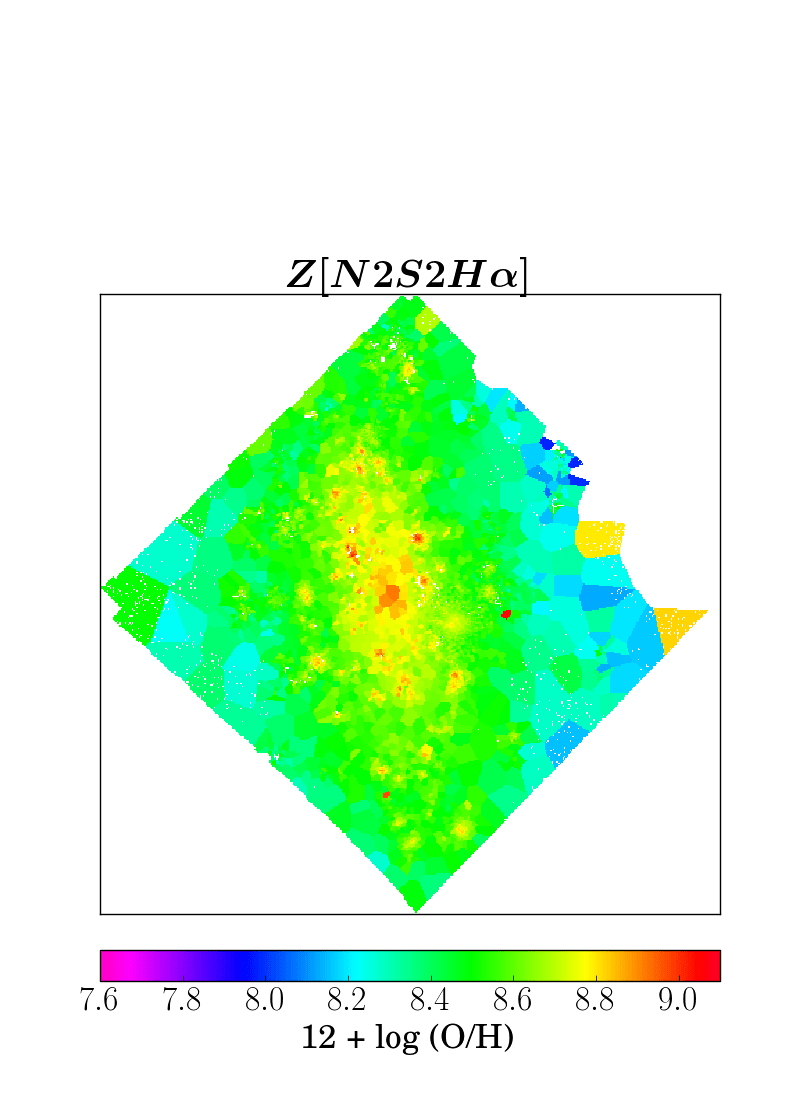}
	\includegraphics[width=0.28\textwidth, trim={0 1.2cm 0 5.5cm}, clip]{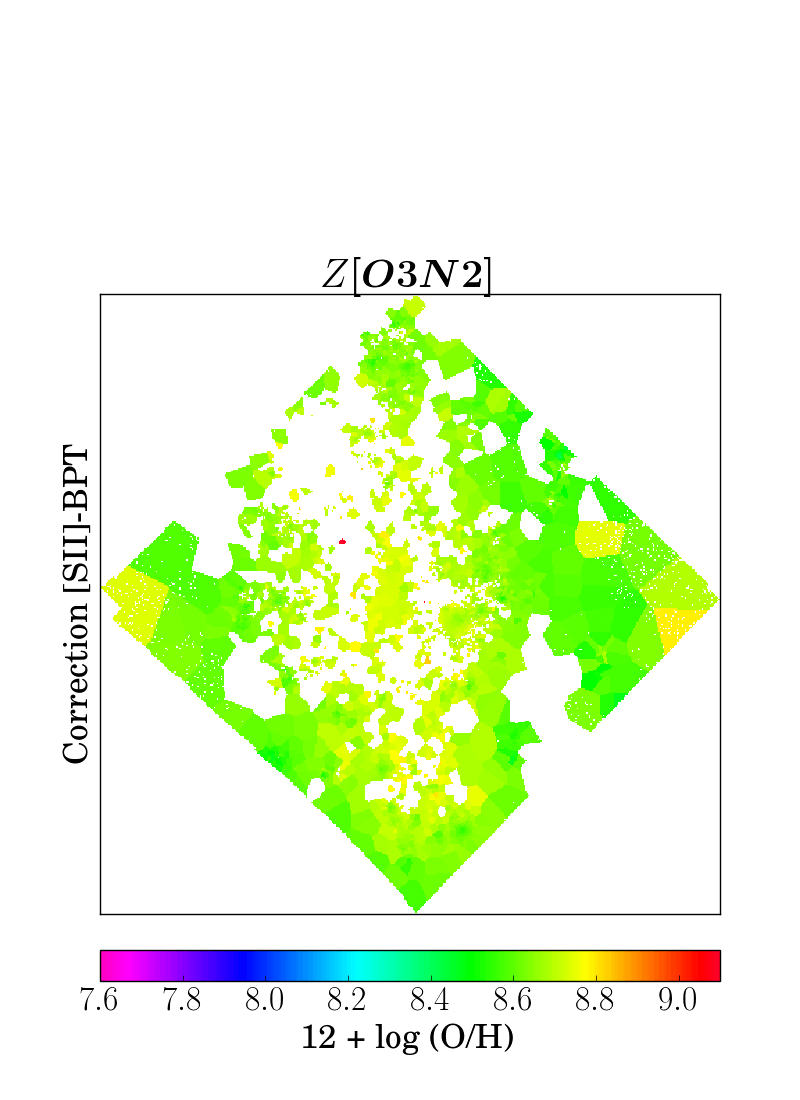}
	\includegraphics[width=0.28\textwidth, trim={0 1.2cm 0 5.5cm}, clip]{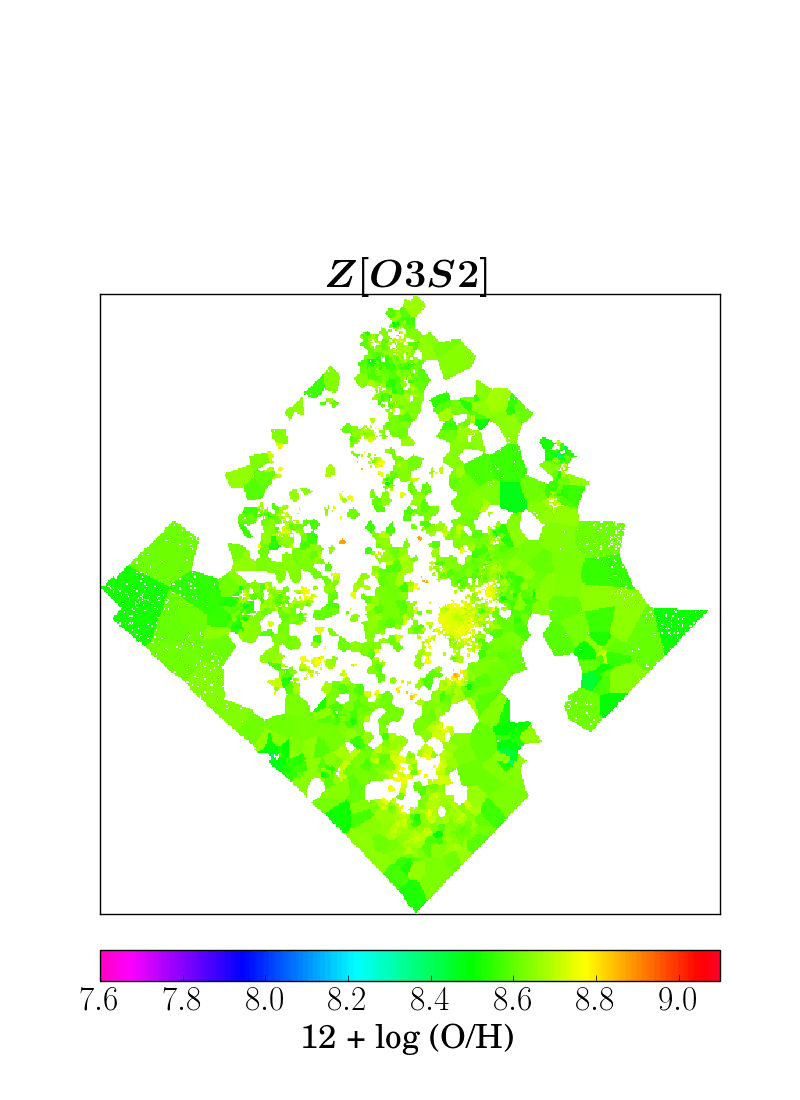}
	\includegraphics[width=0.28\textwidth, trim={2.8cm 0 2.8cm 0}, clip ]{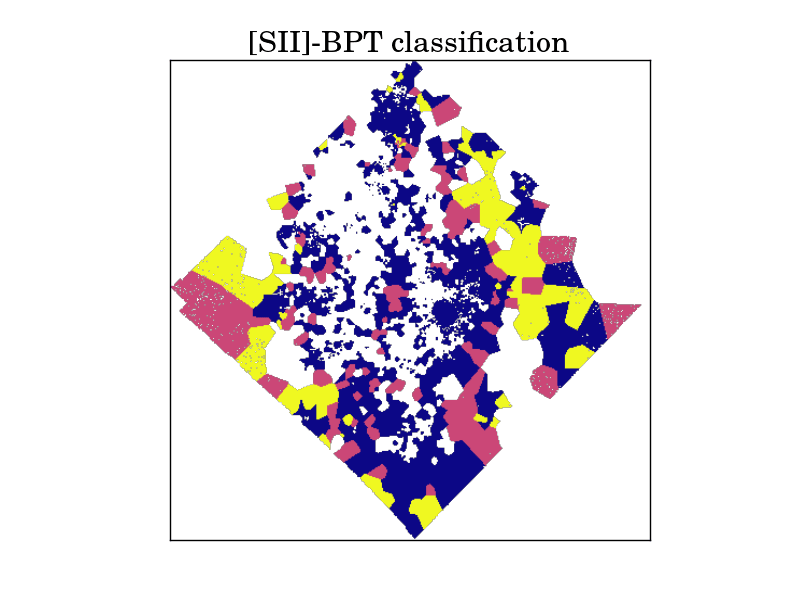}
	\includegraphics[width=0.28\textwidth, trim={0 1.2cm 0 5.5cm}, clip]{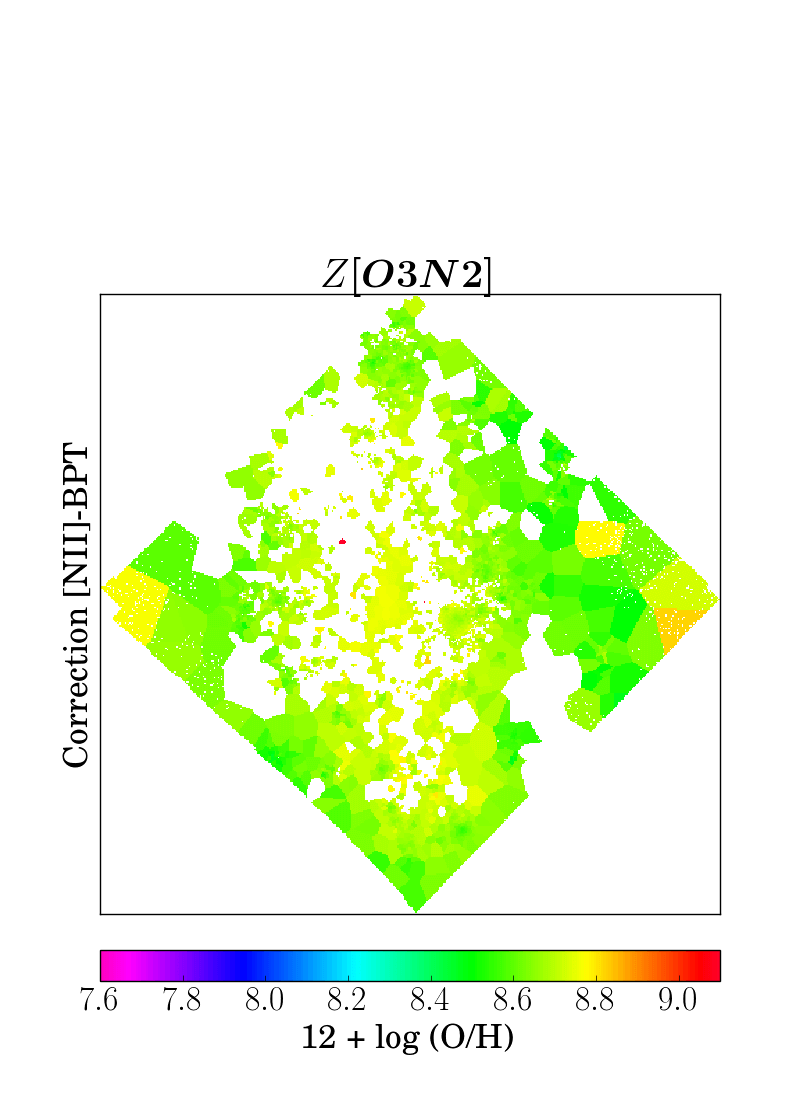}
	\includegraphics[width=0.28\textwidth, trim={0 1.2cm 0 5.5cm}, clip]{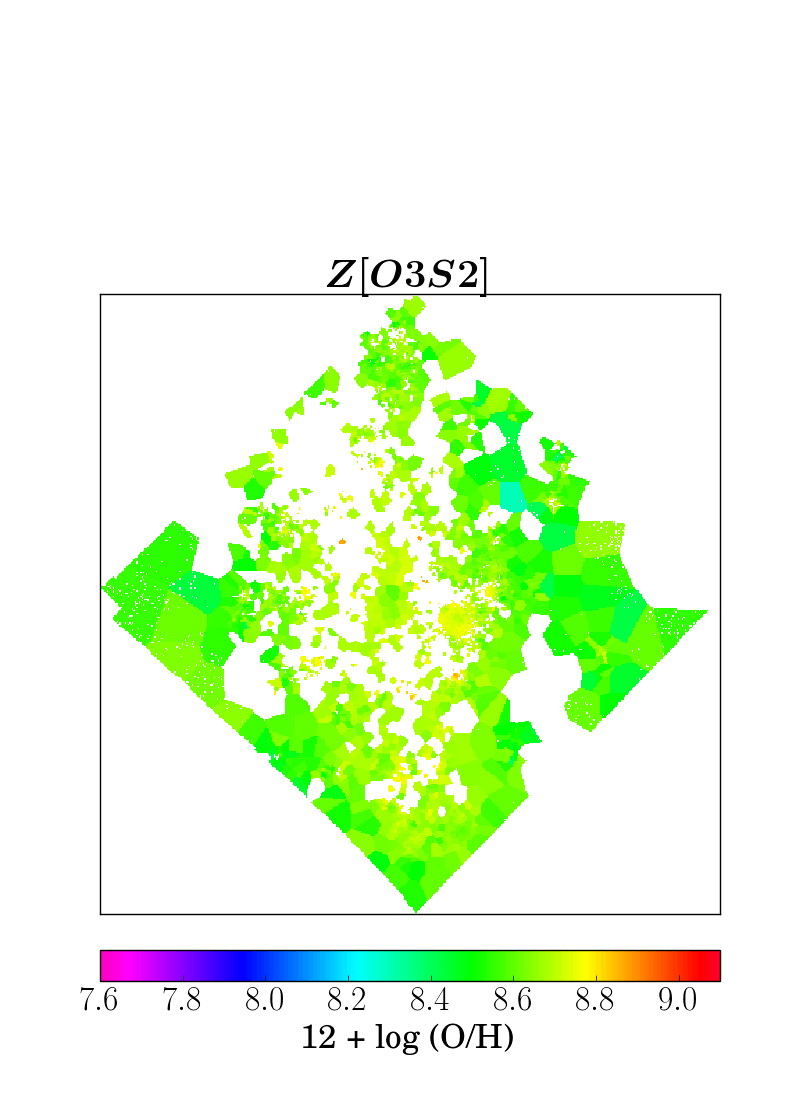}
	\includegraphics[width=0.28\textwidth, trim={2.8cm 0 2.8cm 0}, clip ]{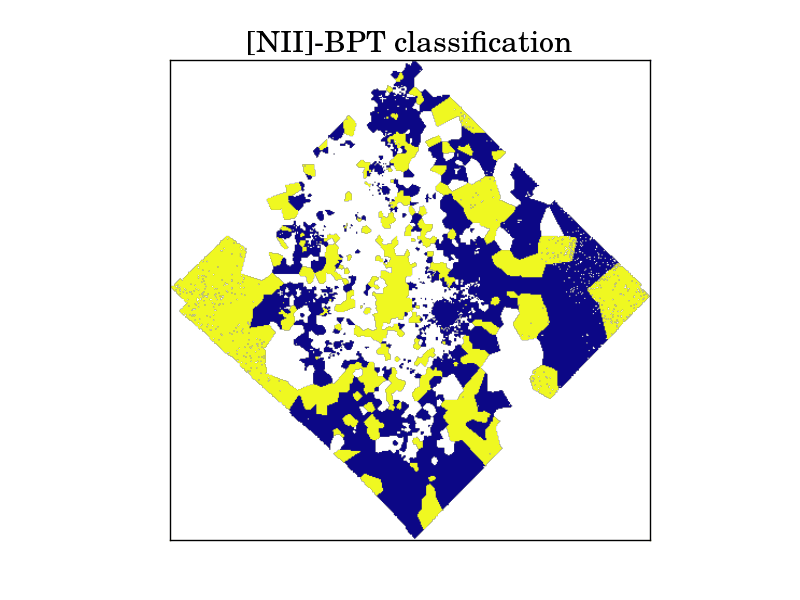}
	\caption{ Maps correspond to galaxy NGC7162, see caption of Figure \ref{fig:NGC1042} for details.}
	\label{fig:NGC7162}
\end{figure*}
\begin{figure*}
	\centering
	\includegraphics[width=0.28\textwidth, trim={0 1.2cm 0 5.5cm}, clip]{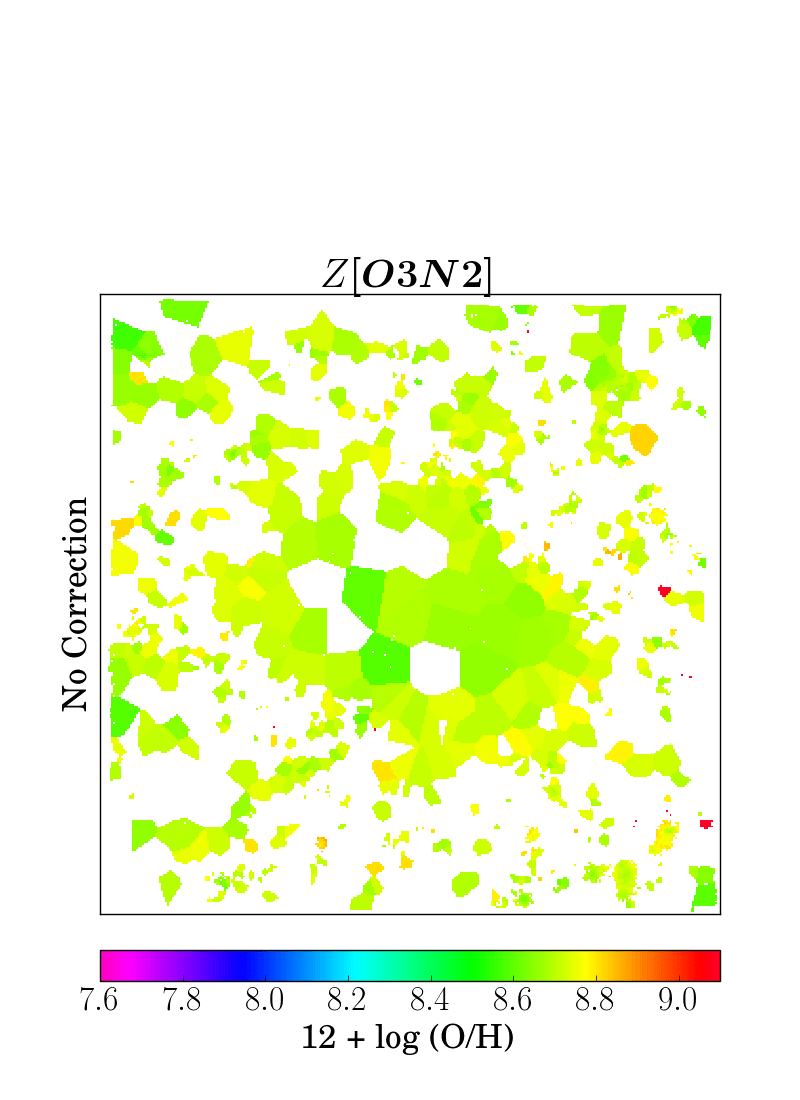}
	\includegraphics[width=0.28\textwidth, trim={0 1.2cm 0 5.5cm}, clip]{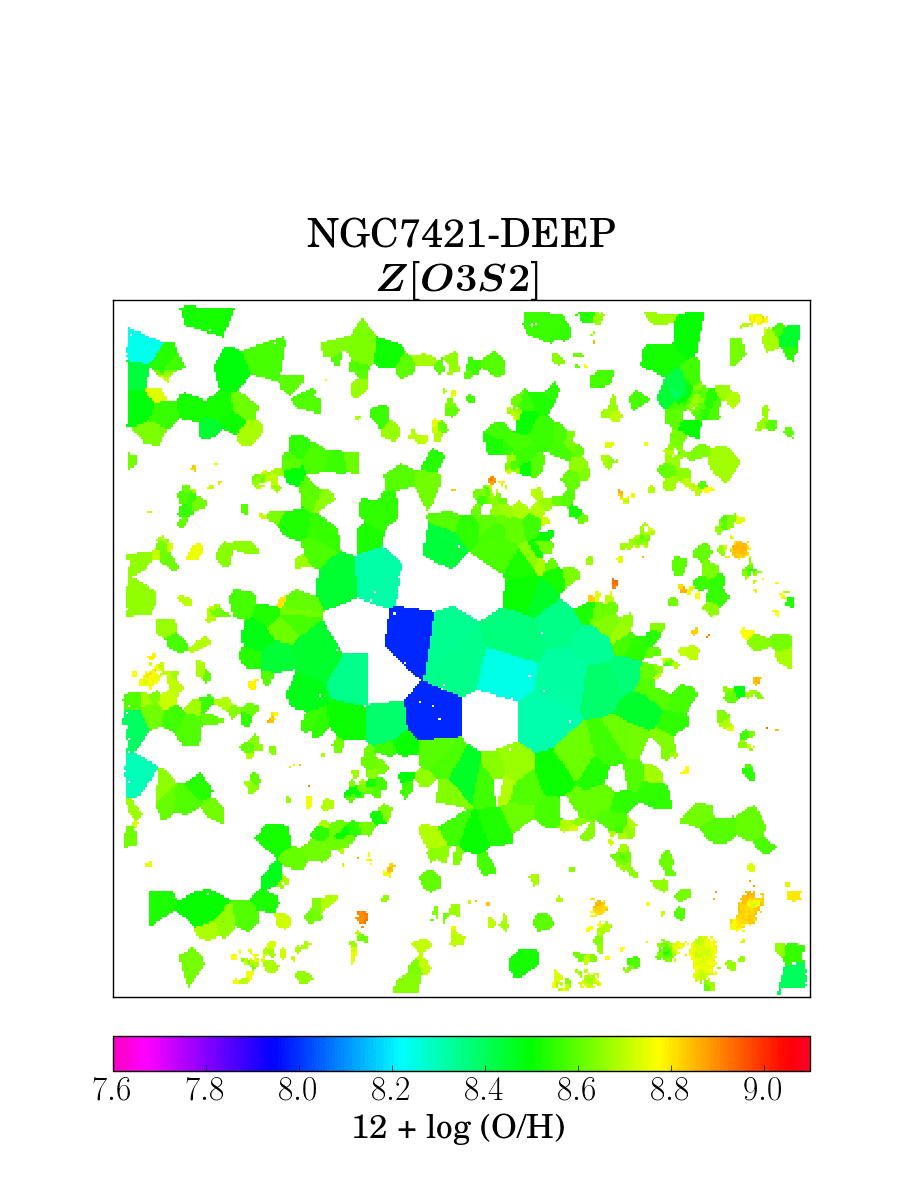}
	\includegraphics[width=0.28\textwidth, trim={0 1.2cm 0 5.5cm}, clip]{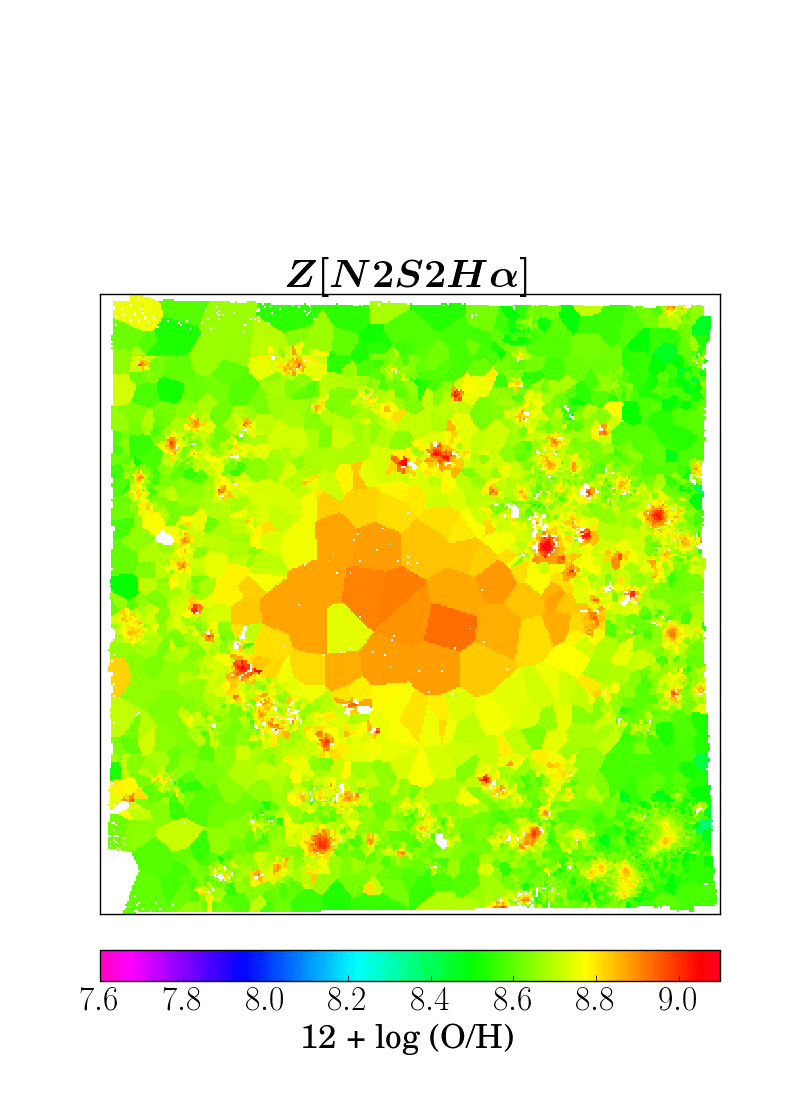}
	\includegraphics[width=0.28\textwidth, trim={0 1.2cm 0 5.5cm}, clip]{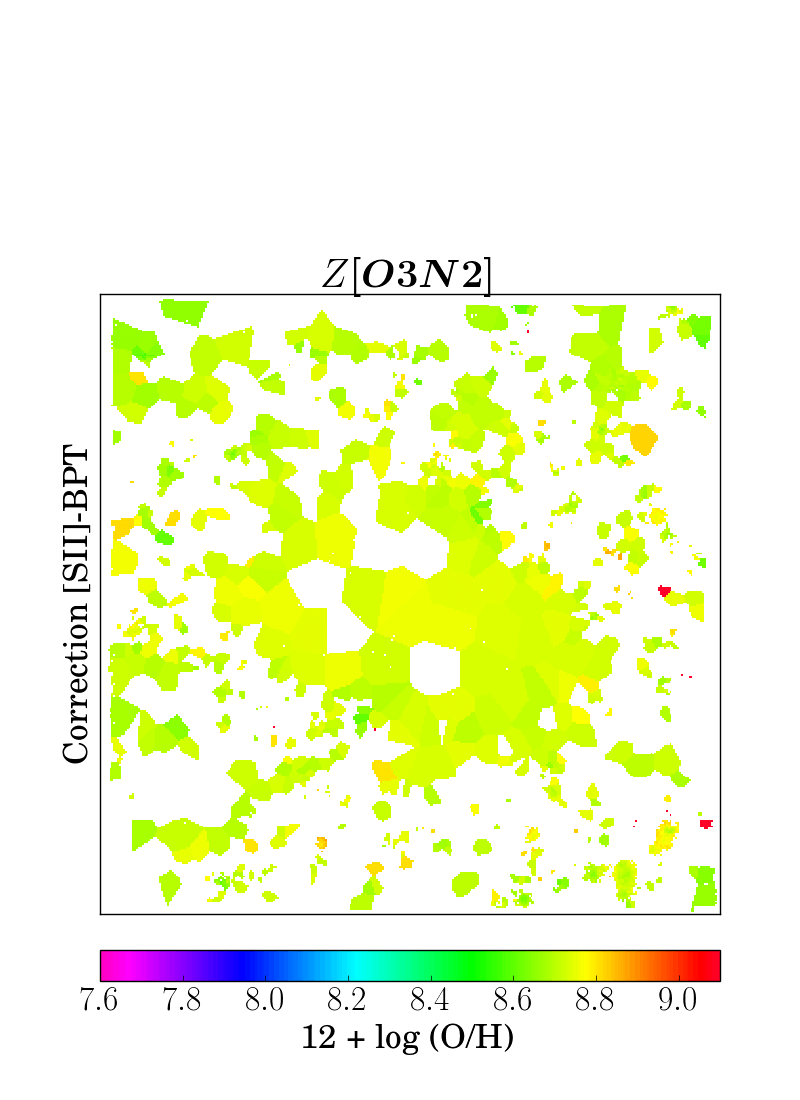}
	\includegraphics[width=0.28\textwidth, trim={0 1.2cm 0 5.5cm}, clip]{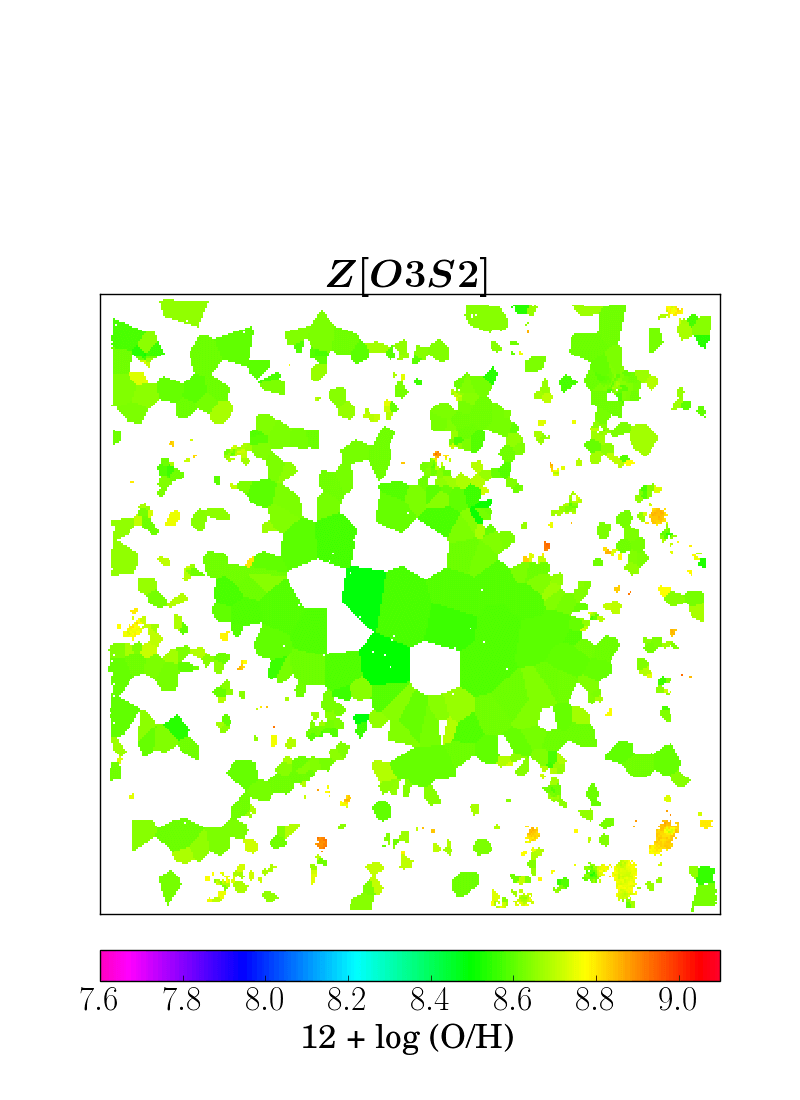}
	\includegraphics[width=0.28\textwidth, trim={2.8cm 0 2.8cm 0}, clip ]{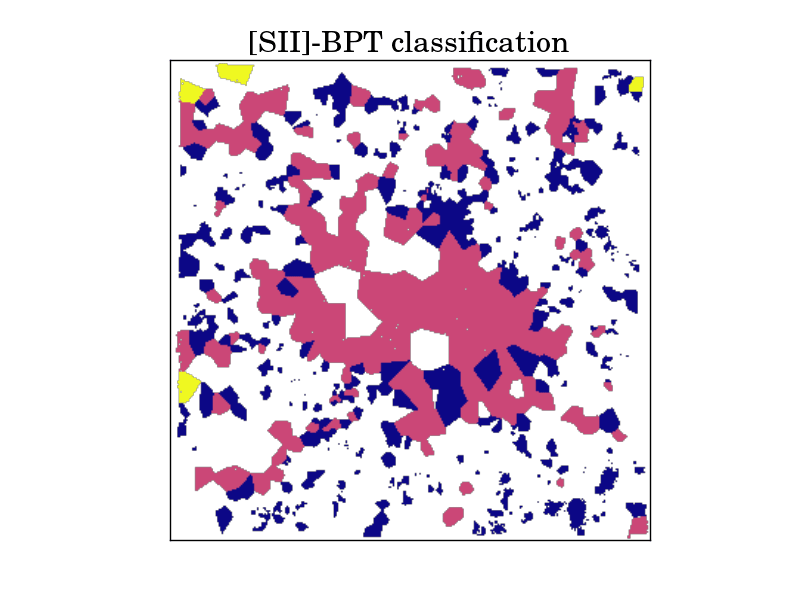}
	\includegraphics[width=0.28\textwidth, trim={0 1.2cm 0 5.5cm}, clip]{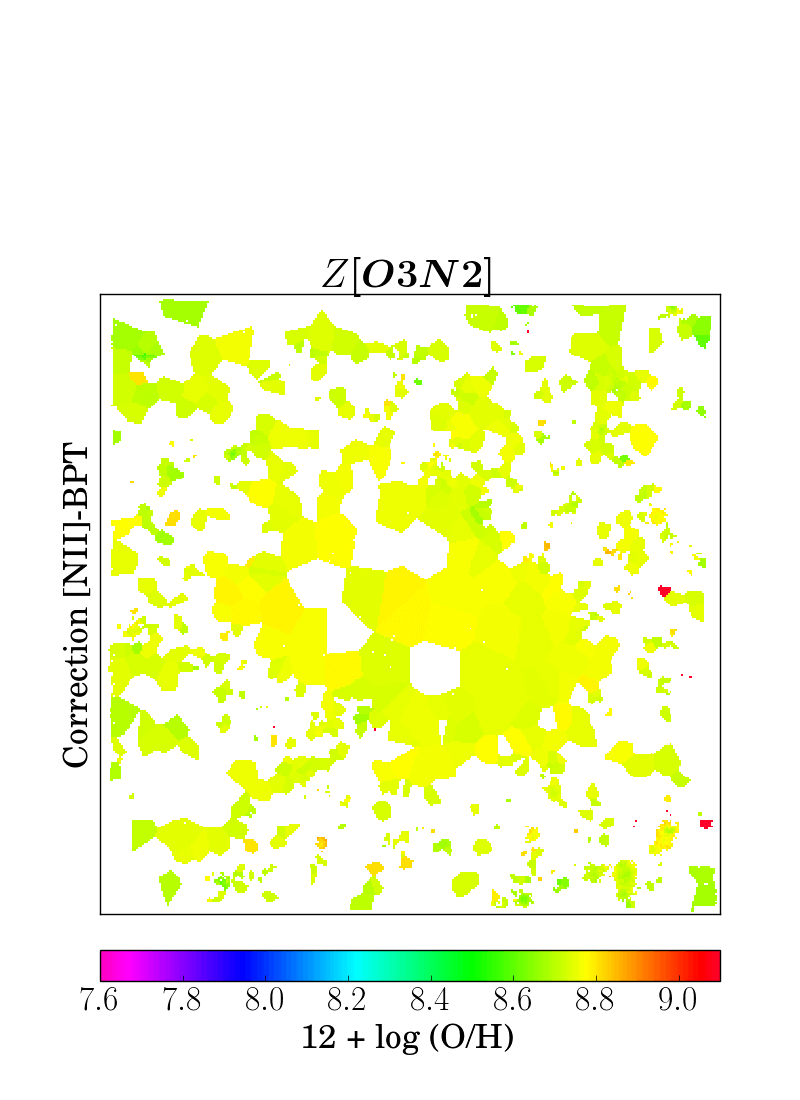}
	\includegraphics[width=0.28\textwidth, trim={0 1.2cm 0 5.5cm}, clip]{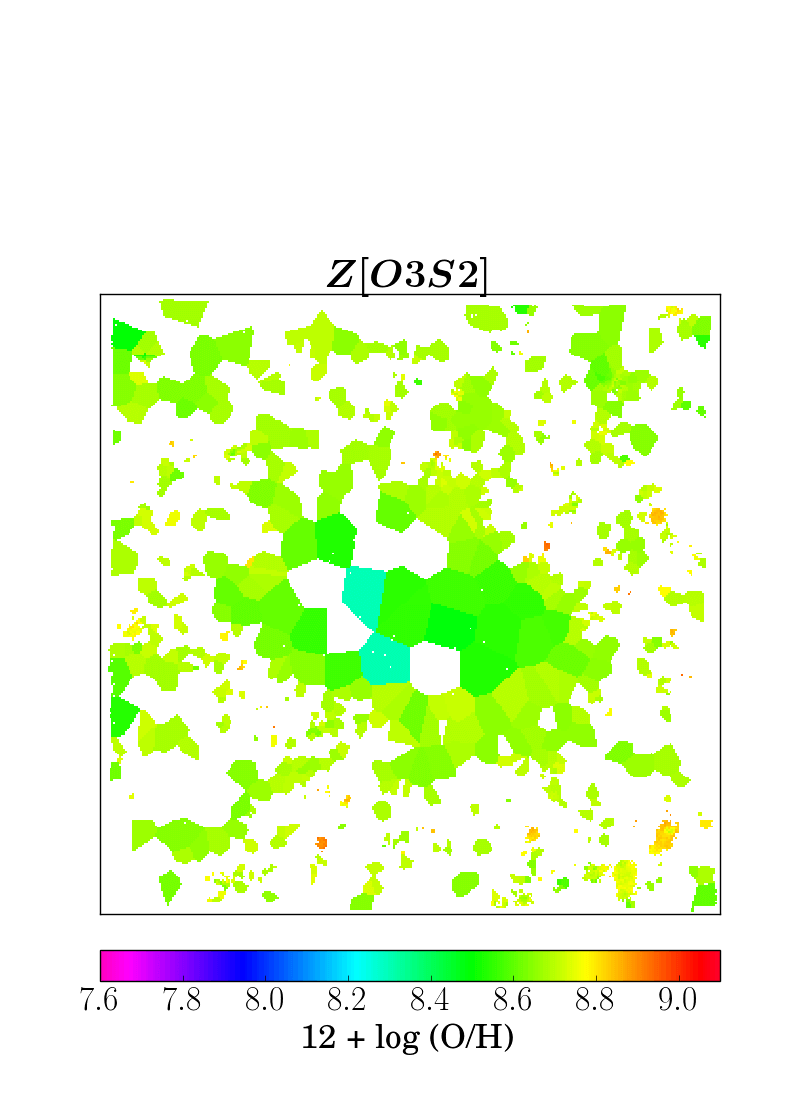}
	\includegraphics[width=0.28\textwidth, trim={2.8cm 0 2.8cm 0}, clip ]{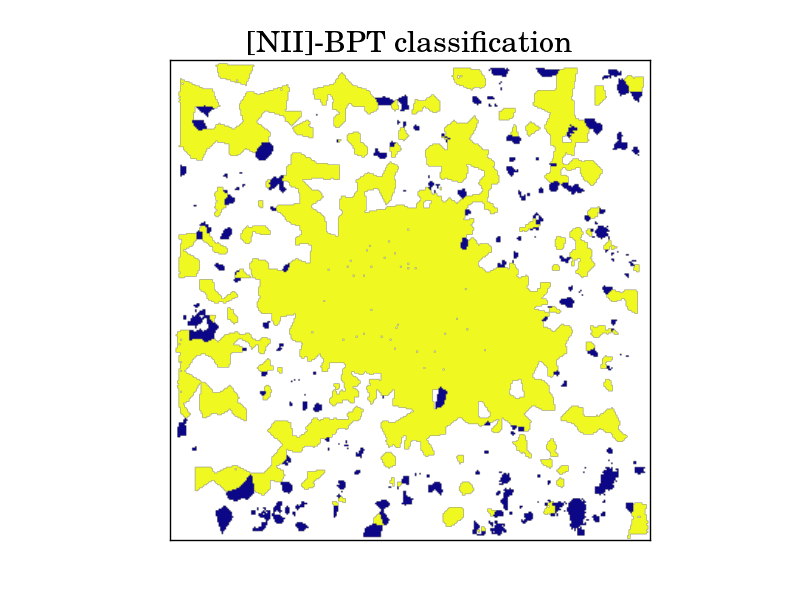}
	\caption{ Maps correspond to galaxy NGC7421-DEEP, see caption of Figure \ref{fig:NGC1042} for details.}
	\label{fig:NGC7421-DEEP}
\end{figure*}
\begin{figure*}
	\centering
	\includegraphics[width=0.28\textwidth, trim={0 1.2cm 0 5.5cm}, clip]{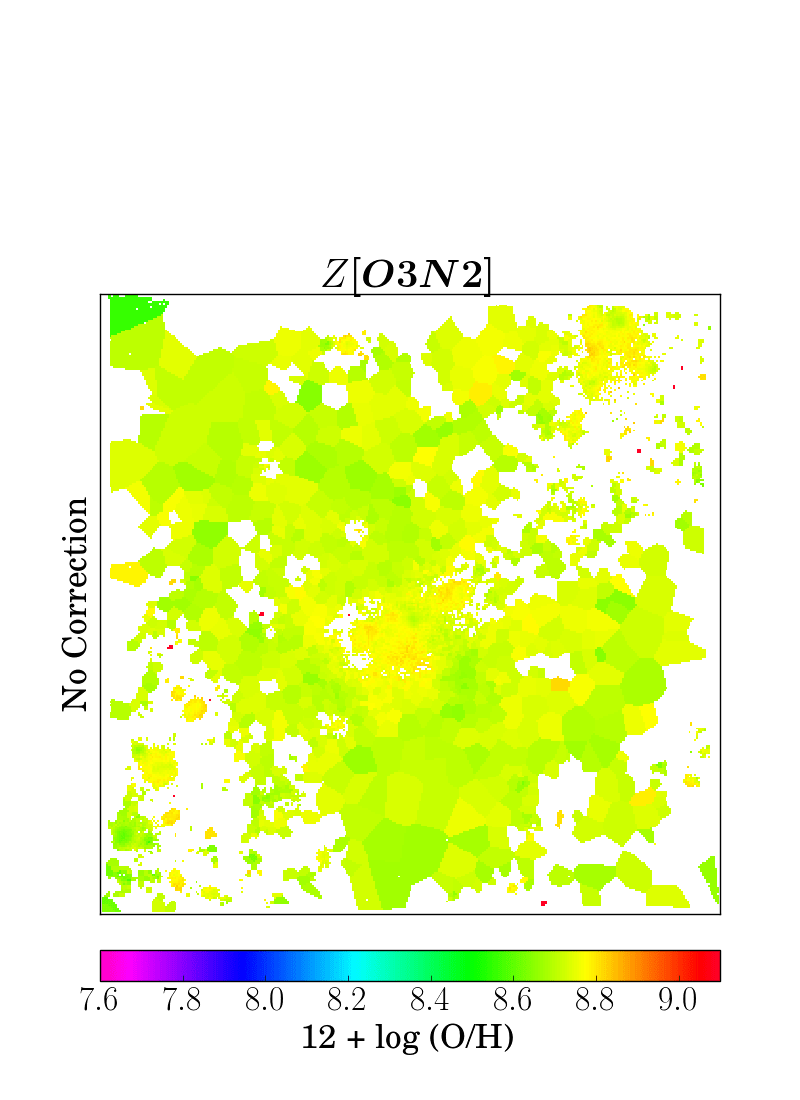}
	\includegraphics[width=0.28\textwidth, trim={0 1.2cm 0 5.5cm}, clip]{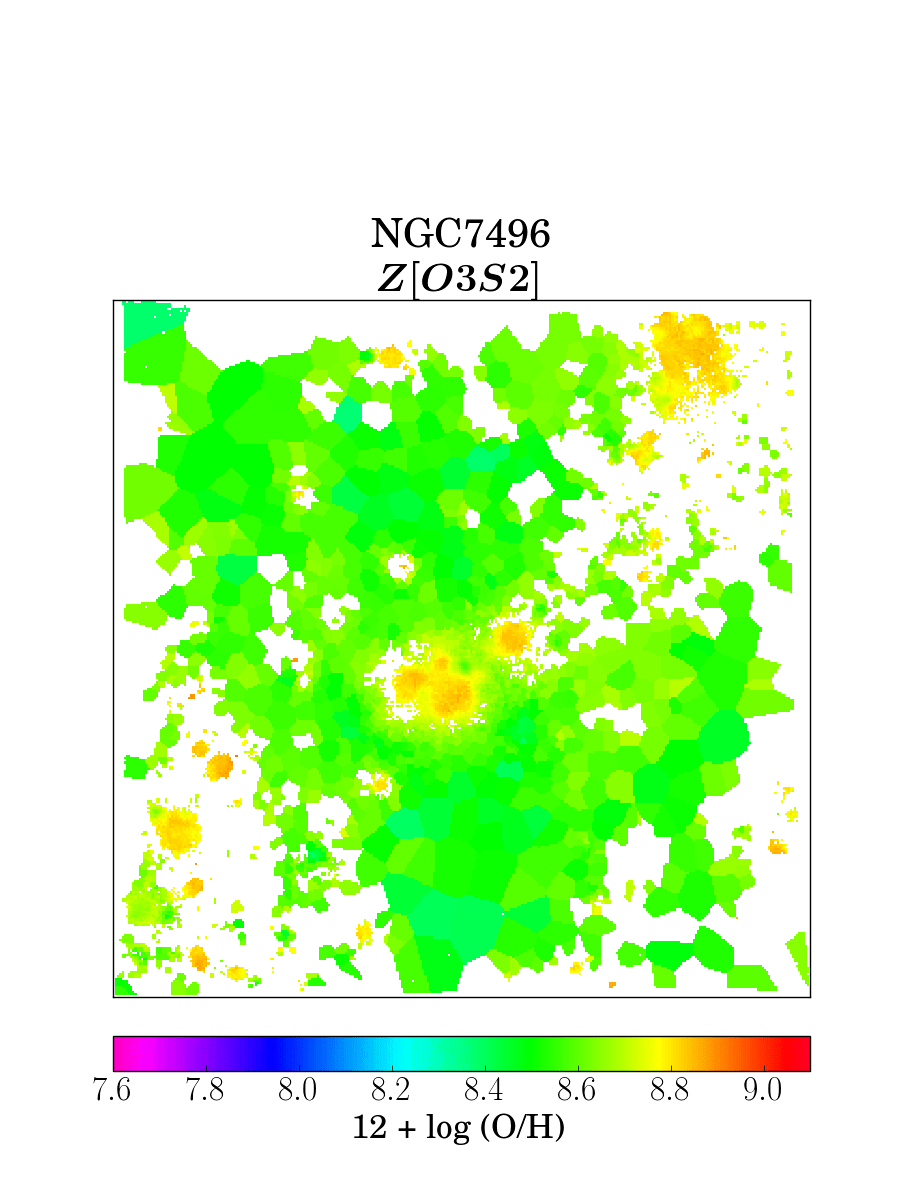}
	\includegraphics[width=0.28\textwidth, trim={0 1.2cm 0 5.5cm}, clip]{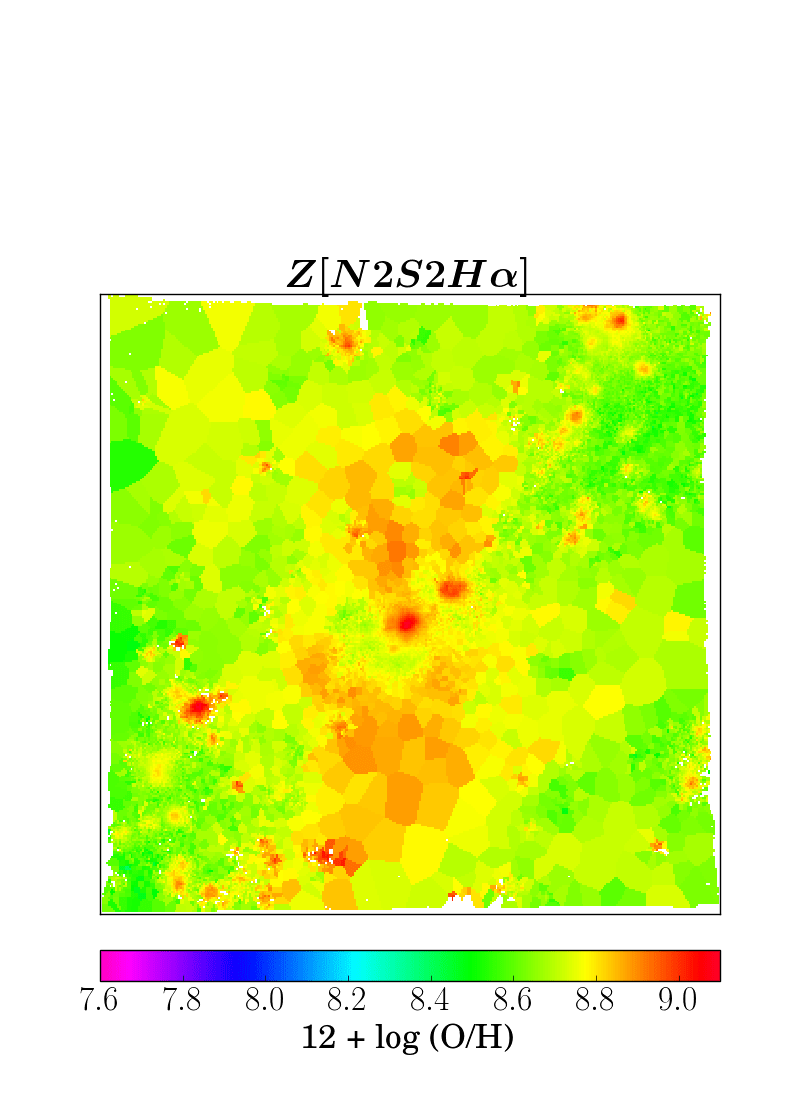}
	\includegraphics[width=0.28\textwidth, trim={0 1.2cm 0 5.5cm}, clip]{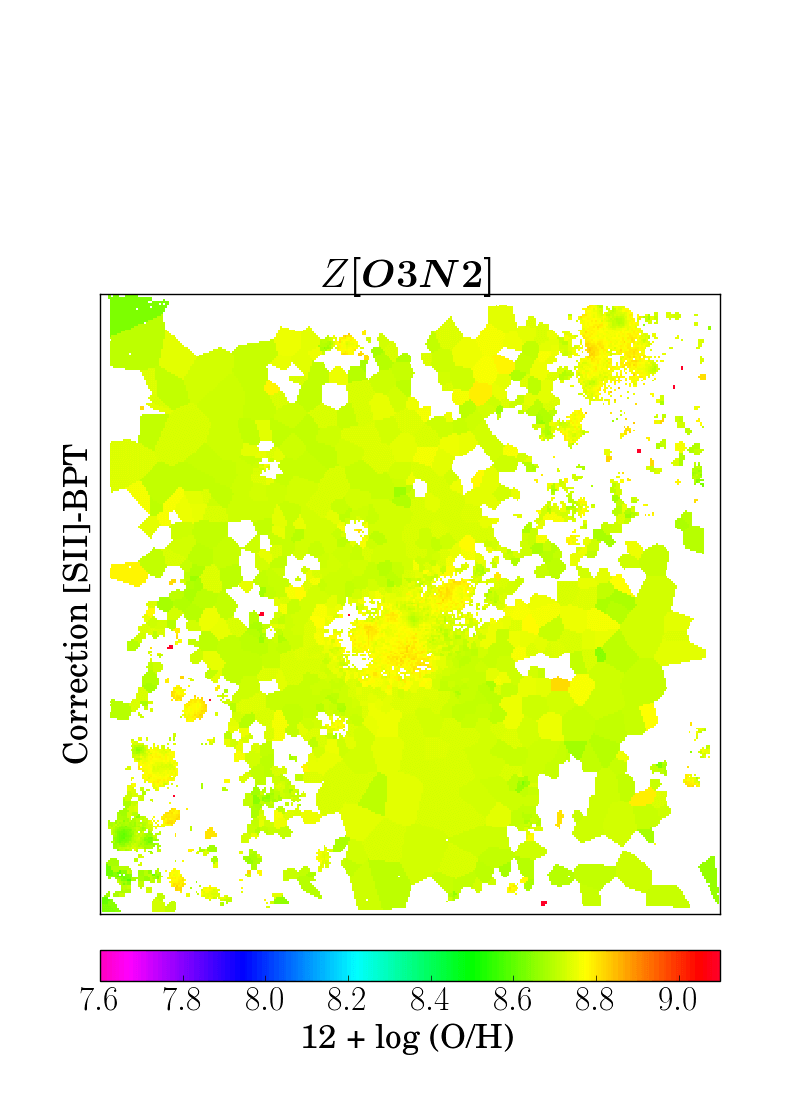}
	\includegraphics[width=0.28\textwidth, trim={0 1.2cm 0 5.5cm}, clip]{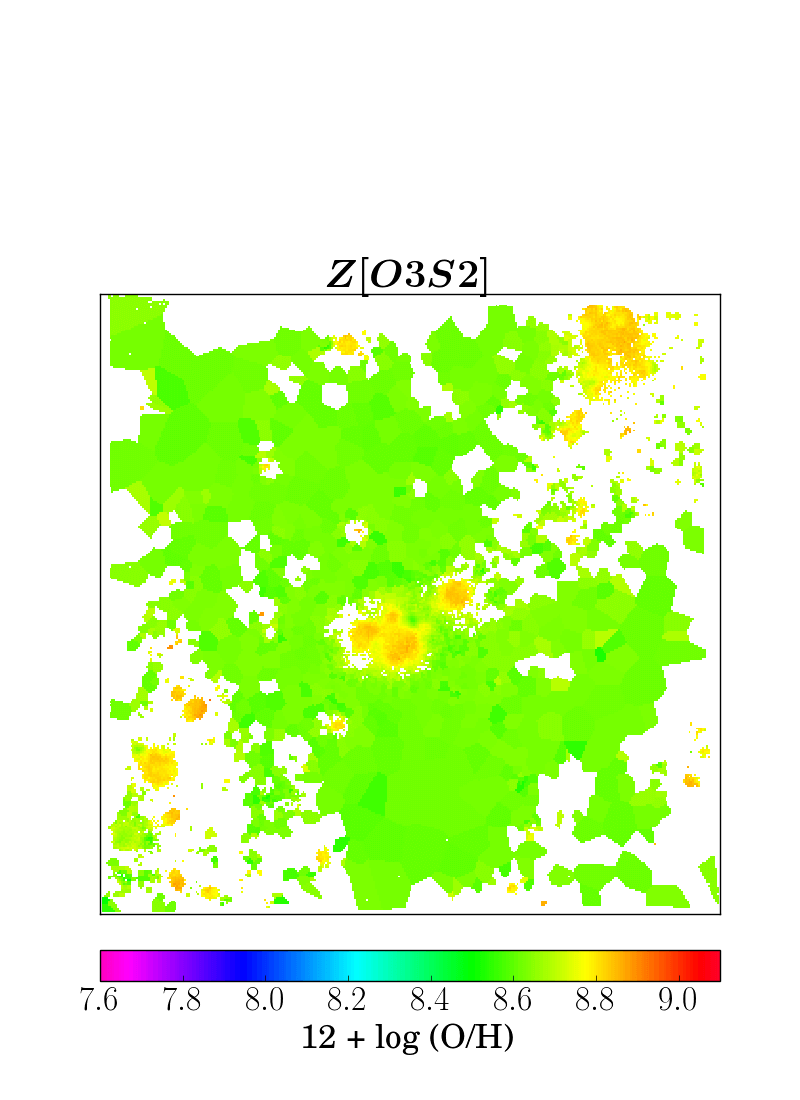}
	\includegraphics[width=0.28\textwidth, trim={2.8cm 0 2.8cm 0}, clip ]{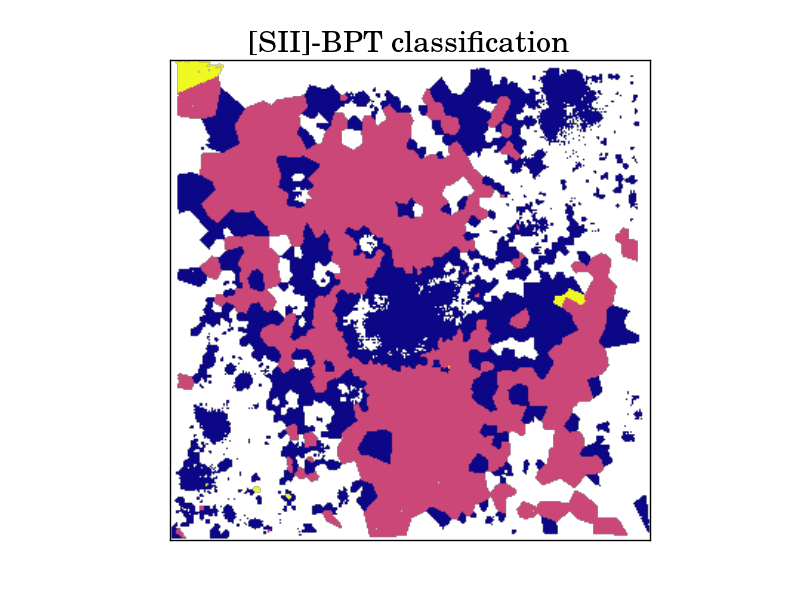}
	\includegraphics[width=0.28\textwidth, trim={0 1.2cm 0 5.5cm}, clip]{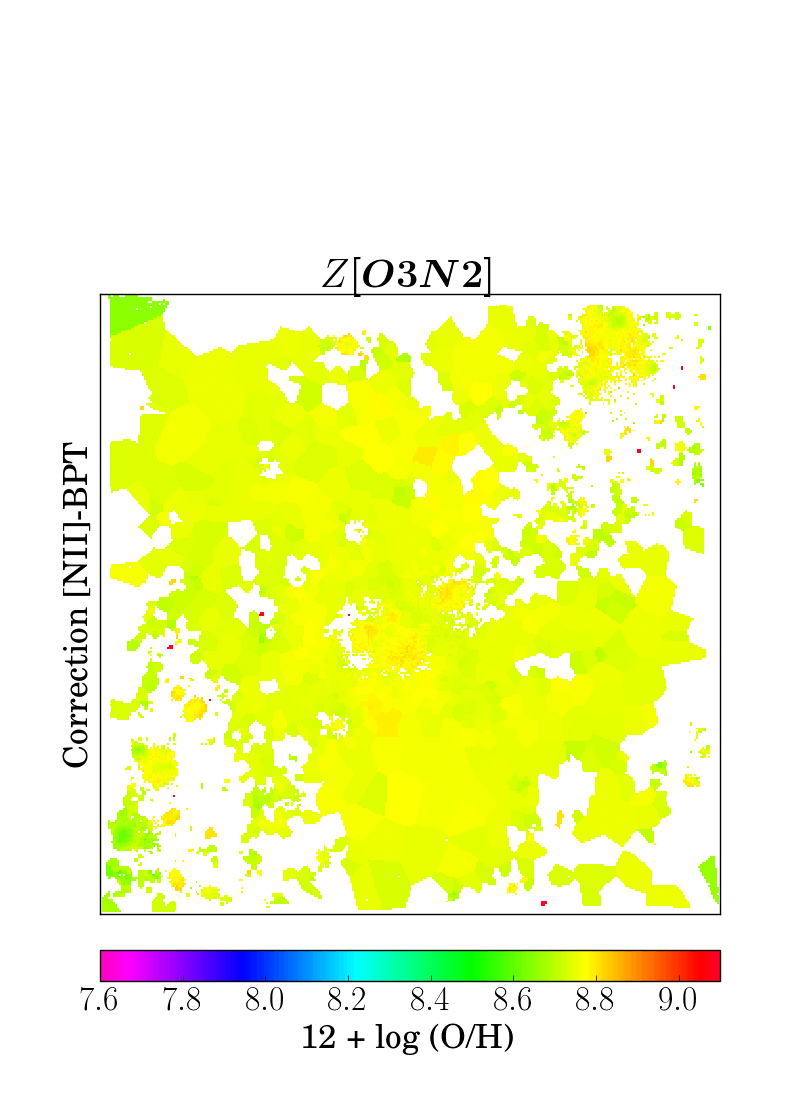}
	\includegraphics[width=0.28\textwidth, trim={0 1.2cm 0 5.5cm}, clip]{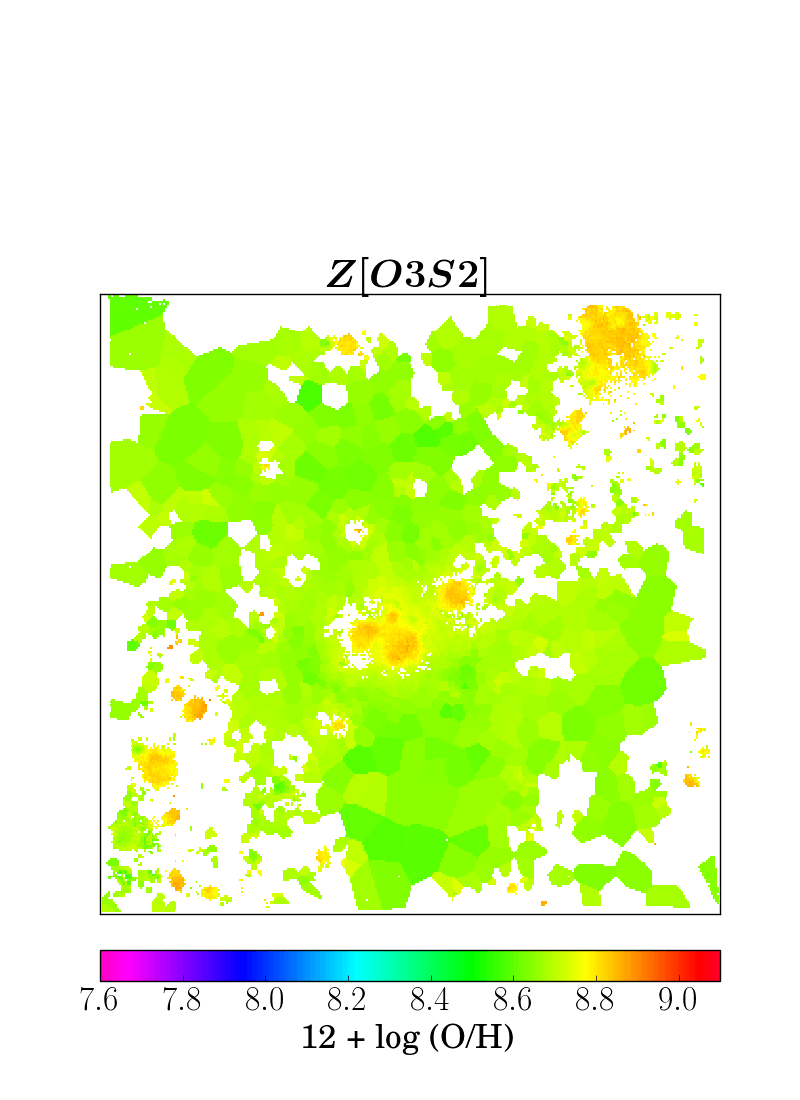}
	\includegraphics[width=0.28\textwidth, trim={2.8cm 0 2.8cm 0}, clip ]{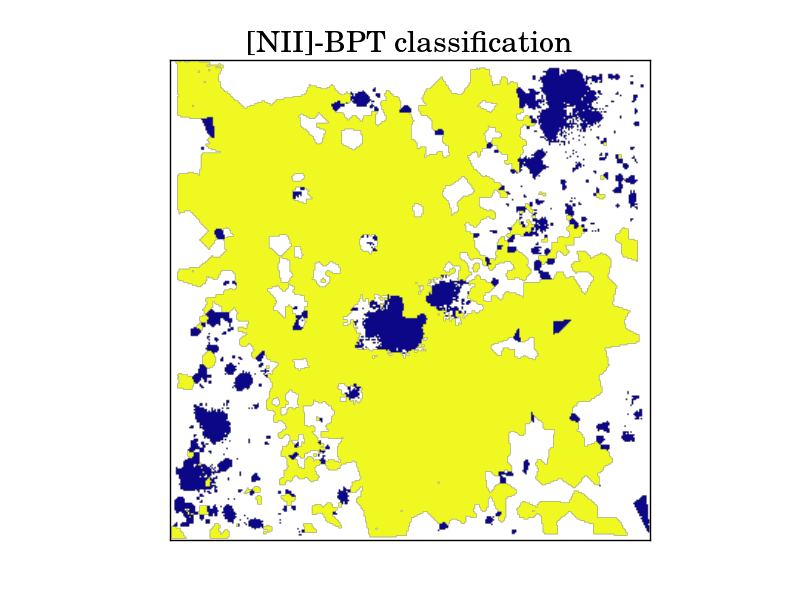}
	\caption{ Maps correspond to galaxy NGC7496, see caption of Figure \ref{fig:NGC1042} for details.}
	\label{fig:NGC7496}
\end{figure*}
\begin{figure*}
	\centering
	\includegraphics[width=0.28\textwidth, trim={0 1.2cm 0 5.5cm}, clip]{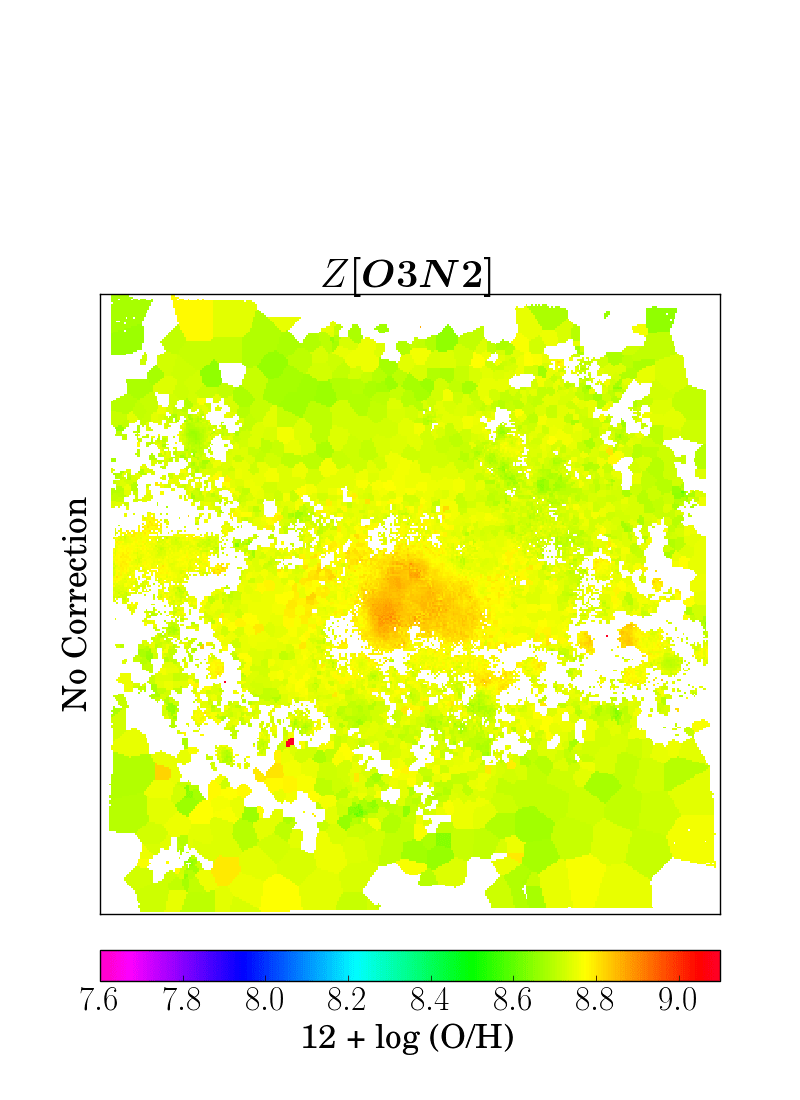}
	\includegraphics[width=0.28\textwidth, trim={0 1.2cm 0 5.5cm}, clip]{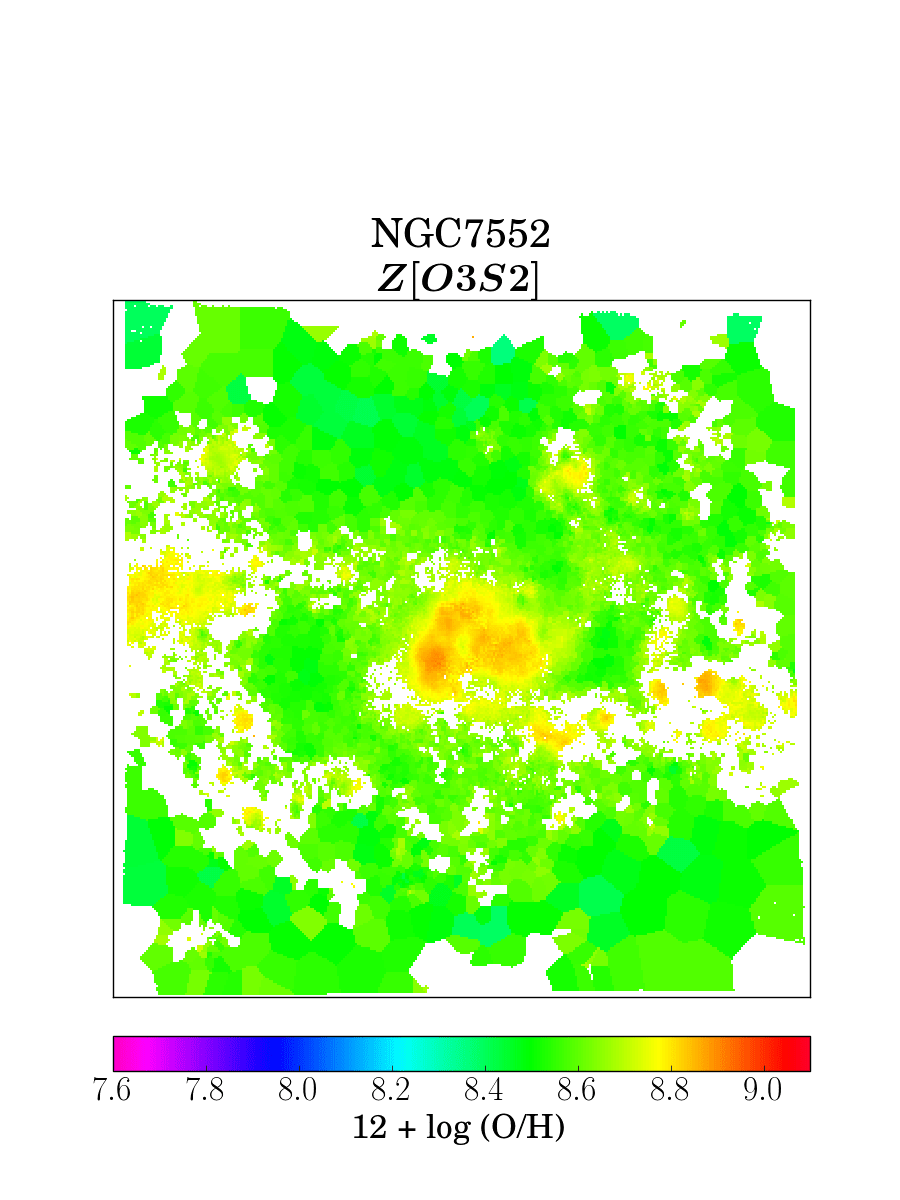}
	\includegraphics[width=0.28\textwidth, trim={0 1.2cm 0 5.5cm}, clip]{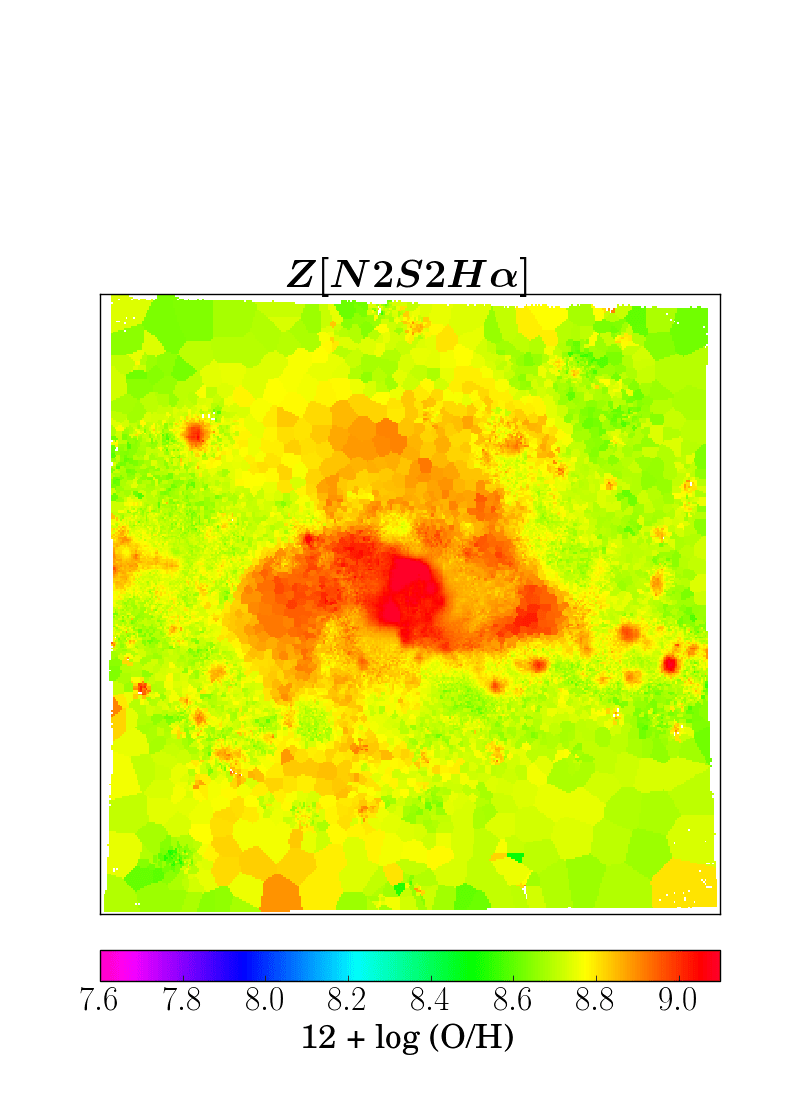}
	\includegraphics[width=0.28\textwidth, trim={0 1.2cm 0 5.5cm}, clip]{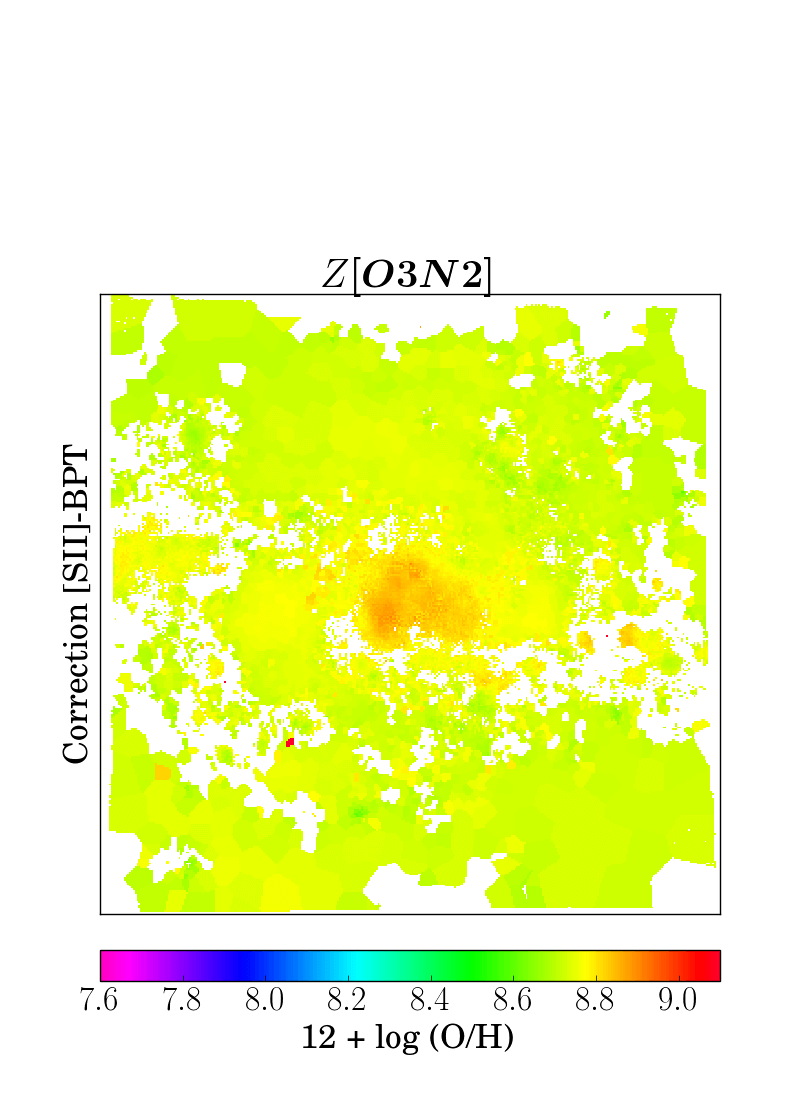}
	\includegraphics[width=0.28\textwidth, trim={0 1.2cm 0 5.5cm}, clip]{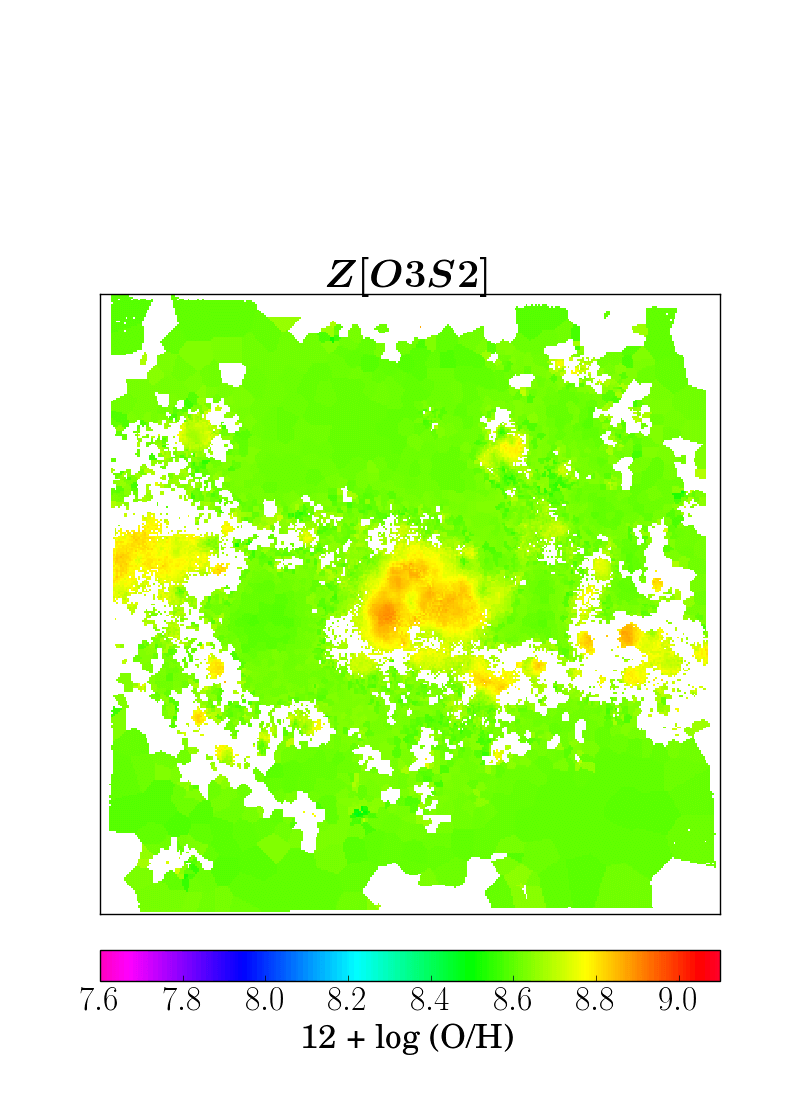}
	\includegraphics[width=0.28\textwidth, trim={2.8cm 0 2.8cm 0}, clip ]{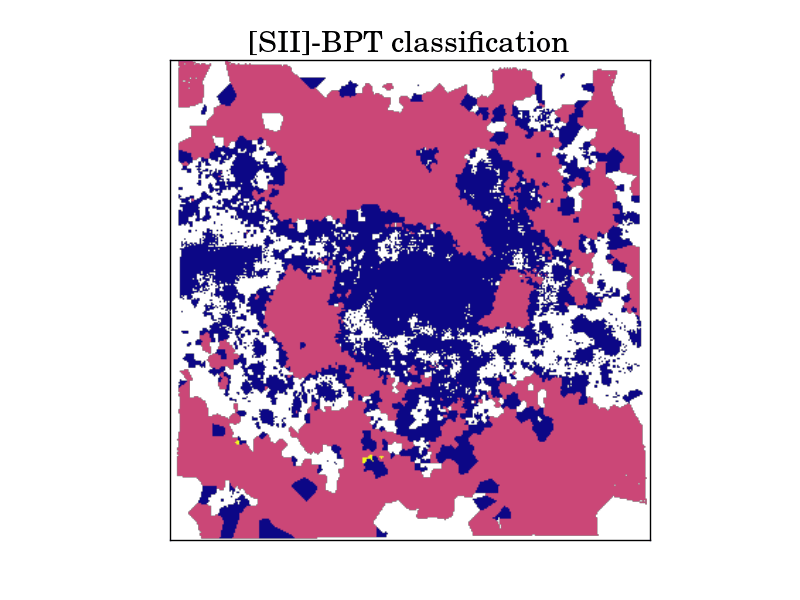}
	\includegraphics[width=0.28\textwidth, trim={0 1.2cm 0 5.5cm}, clip]{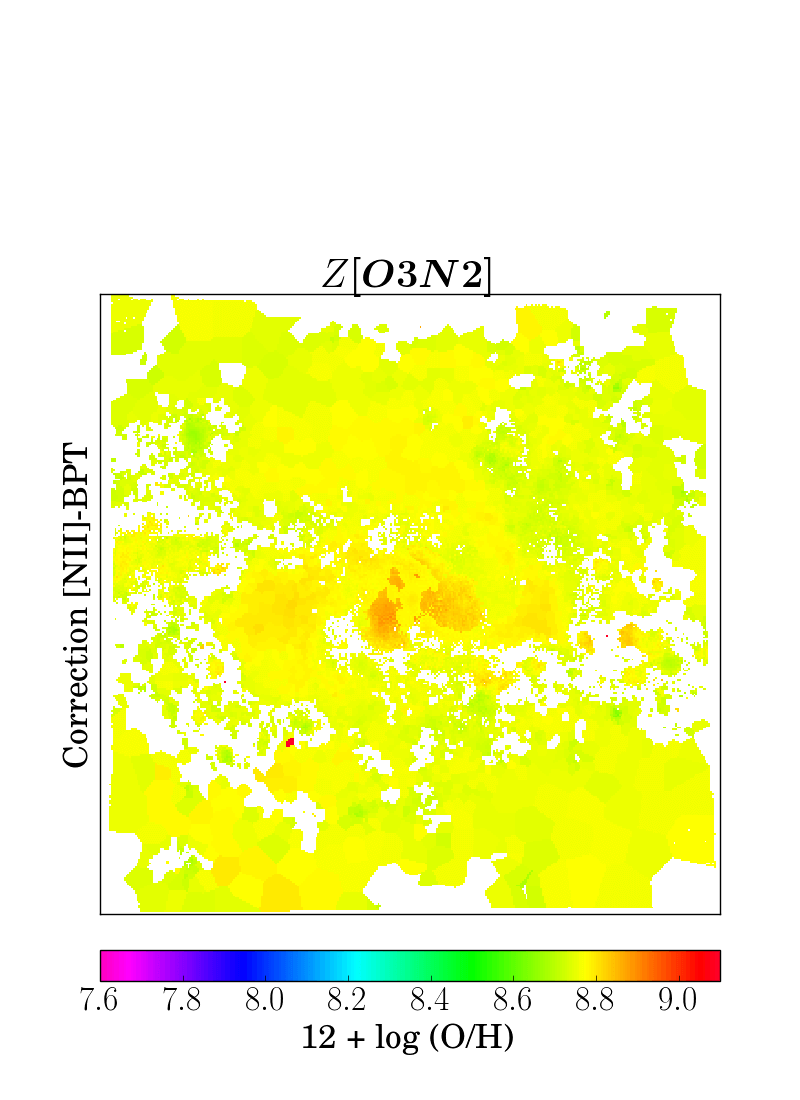}
	\includegraphics[width=0.28\textwidth, trim={0 1.2cm 0 5.5cm}, clip]{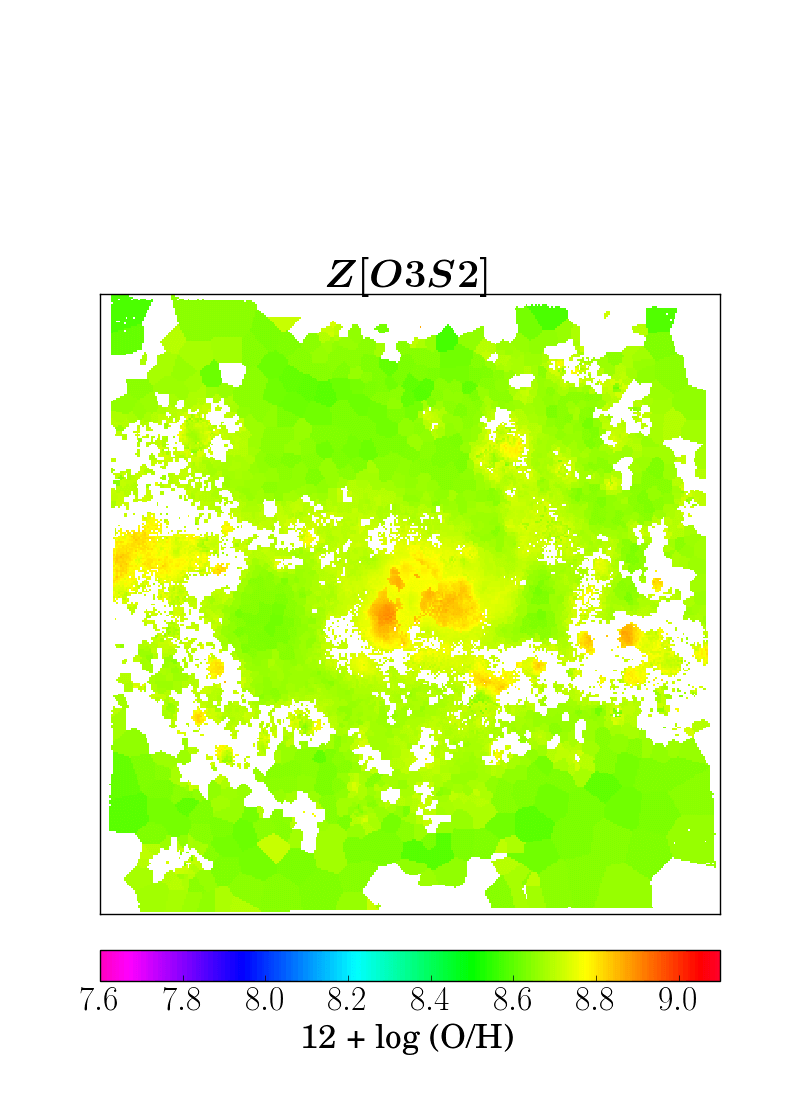}
	\includegraphics[width=0.28\textwidth, trim={2.8cm 0 2.8cm 0}, clip ]{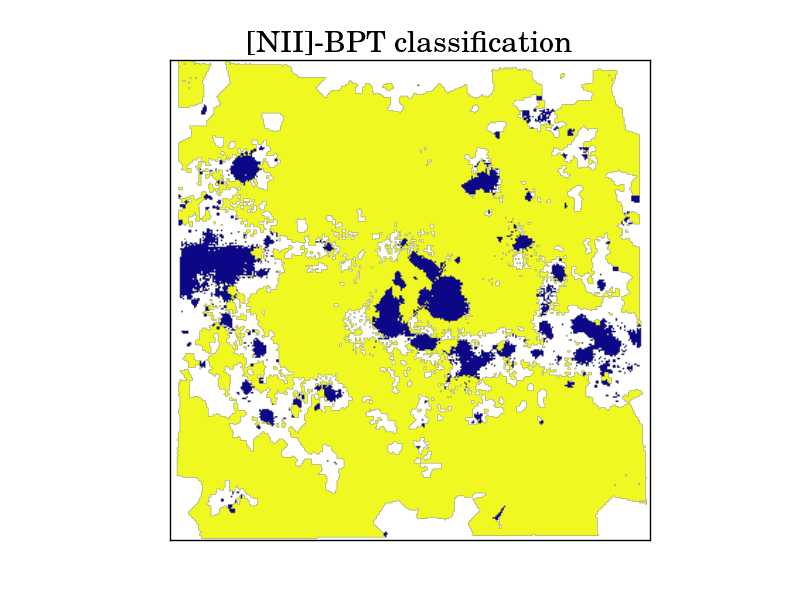}
	\caption{ Maps correspond to galaxy NGC7552, see caption of Figure \ref{fig:NGC1042} for details.}
	\label{fig:NGC7552}
\end{figure*}

\begin{figure*}
	\centering
	\includegraphics[width=0.28\textwidth, trim={0 1.2cm 0 5.5cm}, clip]{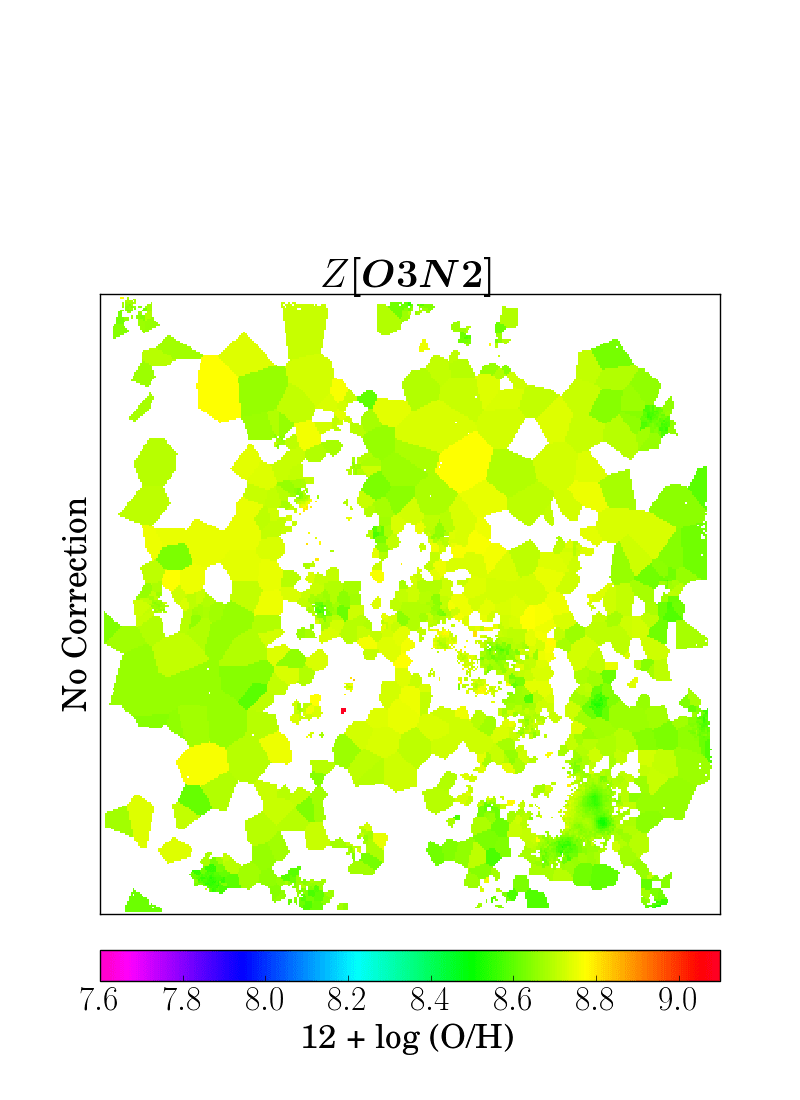}
	\includegraphics[width=0.28\textwidth, trim={0 1.2cm 0 5.5cm}, clip]{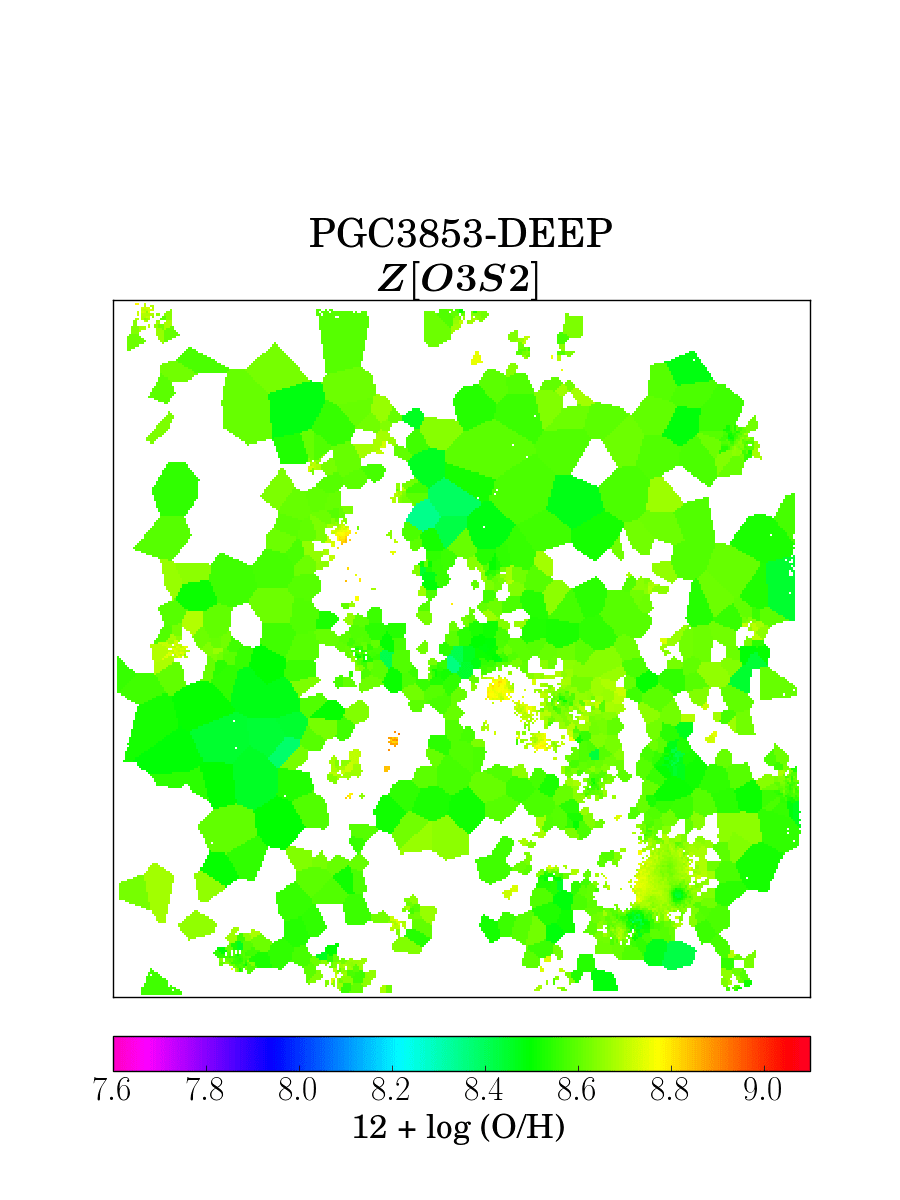}
	\includegraphics[width=0.28\textwidth, trim={0 1.2cm 0 5.5cm}, clip]{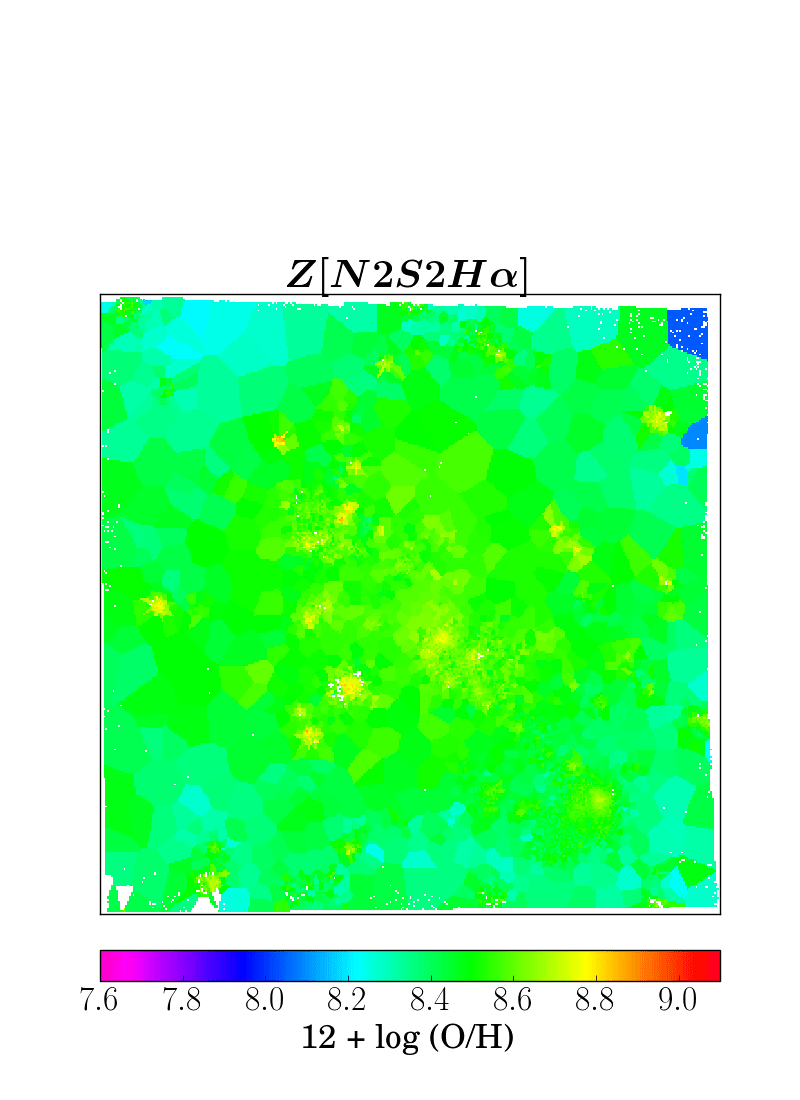}
	\includegraphics[width=0.28\textwidth, trim={0 1.2cm 0 5.5cm}, clip]{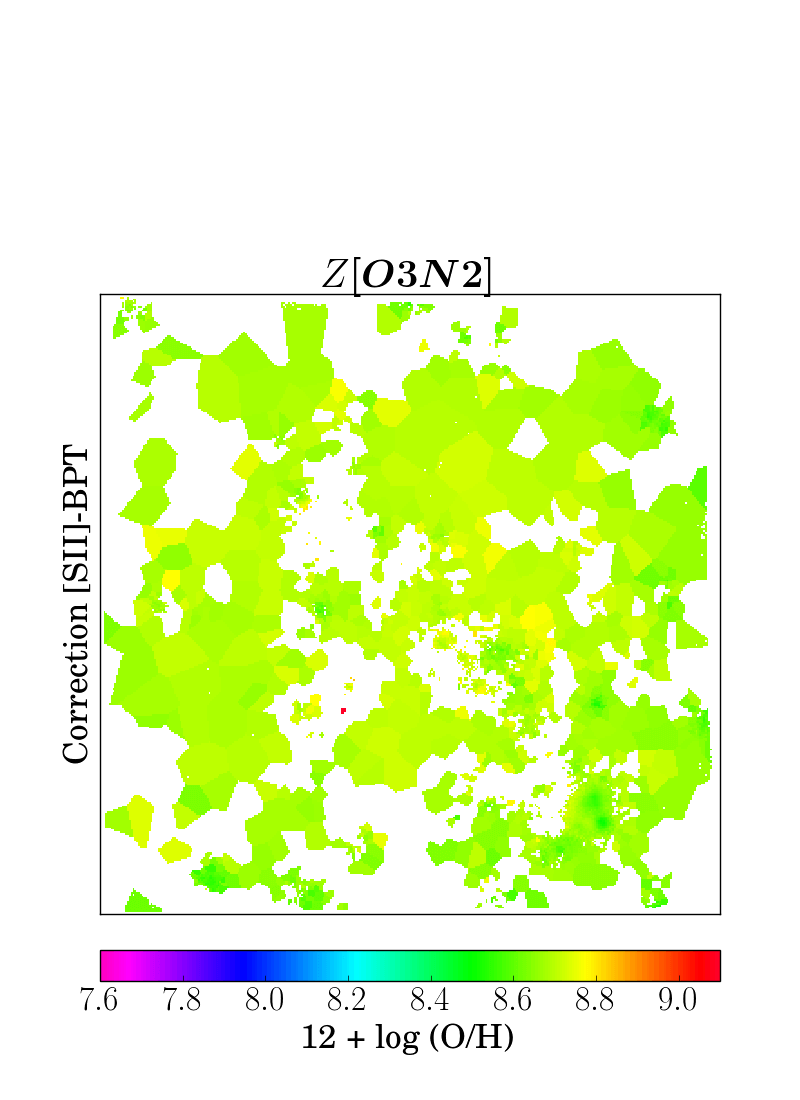}
	\includegraphics[width=0.28\textwidth, trim={0 1.2cm 0 5.5cm}, clip]{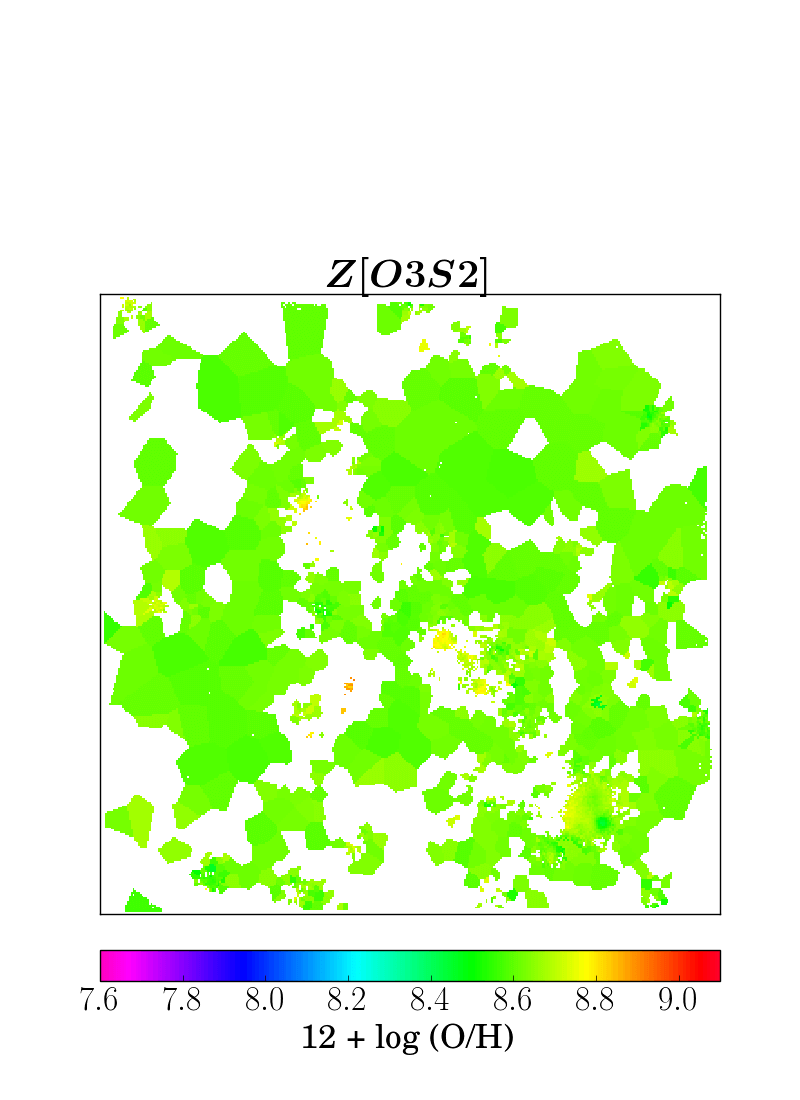}
	\includegraphics[width=0.28\textwidth, trim={2.8cm 0 2.8cm 0}, clip ]{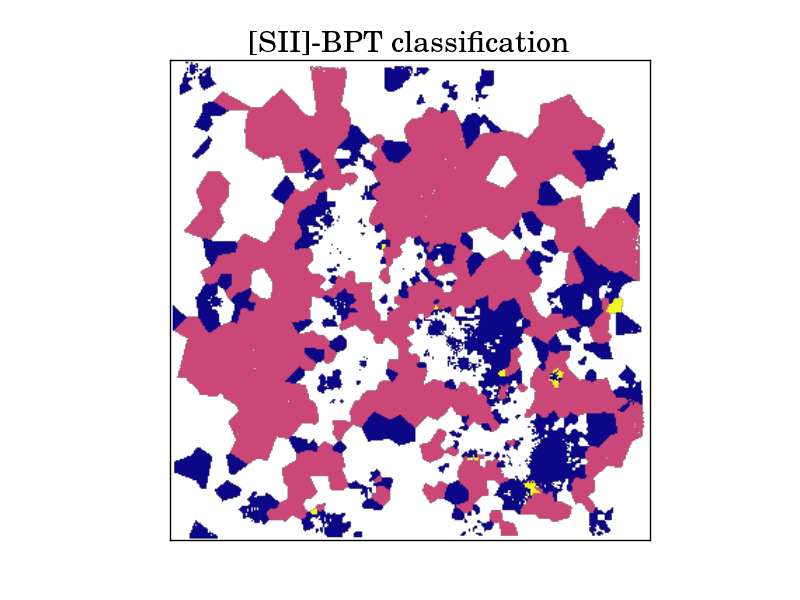}
	\includegraphics[width=0.28\textwidth, trim={0 1.2cm 0 5.5cm}, clip]{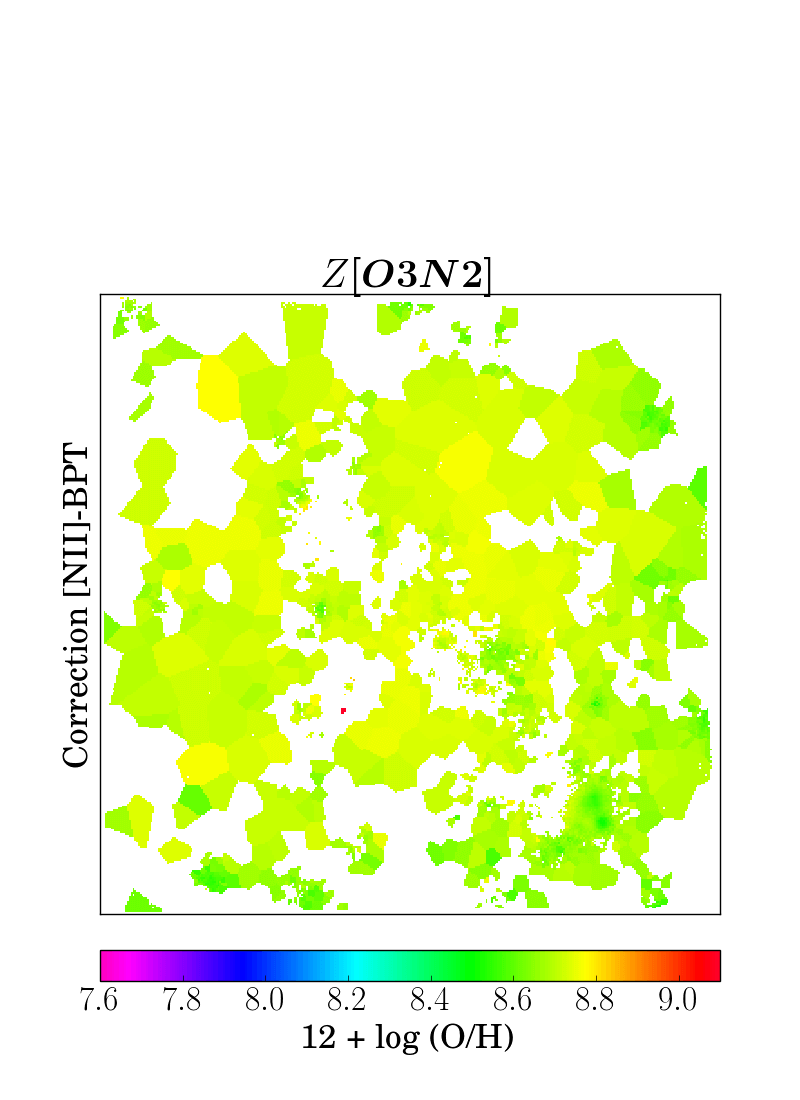}
	\includegraphics[width=0.28\textwidth, trim={0 1.2cm 0 5.5cm}, clip]{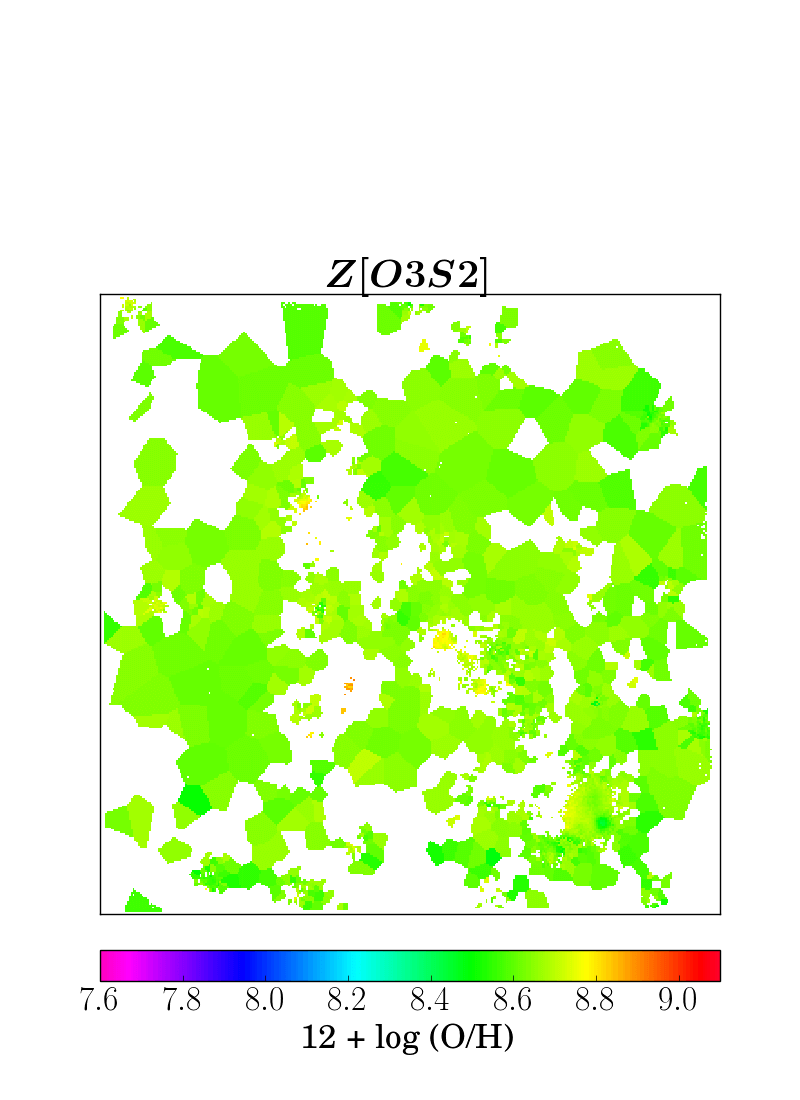}
	\includegraphics[width=0.28\textwidth, trim={2.8cm 0 2.8cm 0}, clip ]{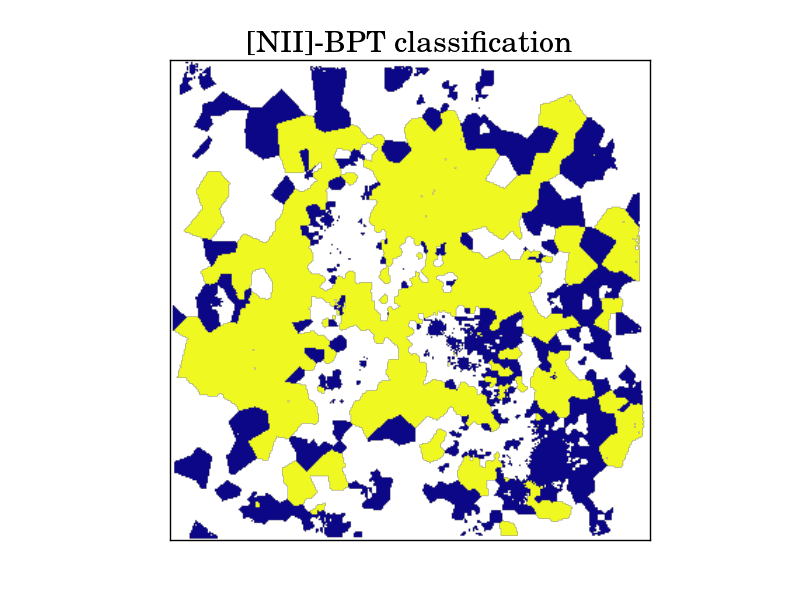}
	\caption{ Maps correspond to galaxy PGC3853-DEEP, see caption of Figure \ref{fig:NGC1042} for details.}
	\label{fig:PGC3853-DEEP}
\end{figure*}

\begin{figure*}
	\centering
	\includegraphics[width=0.28\textwidth, trim={0 1.2cm 0 5.5cm}, clip]{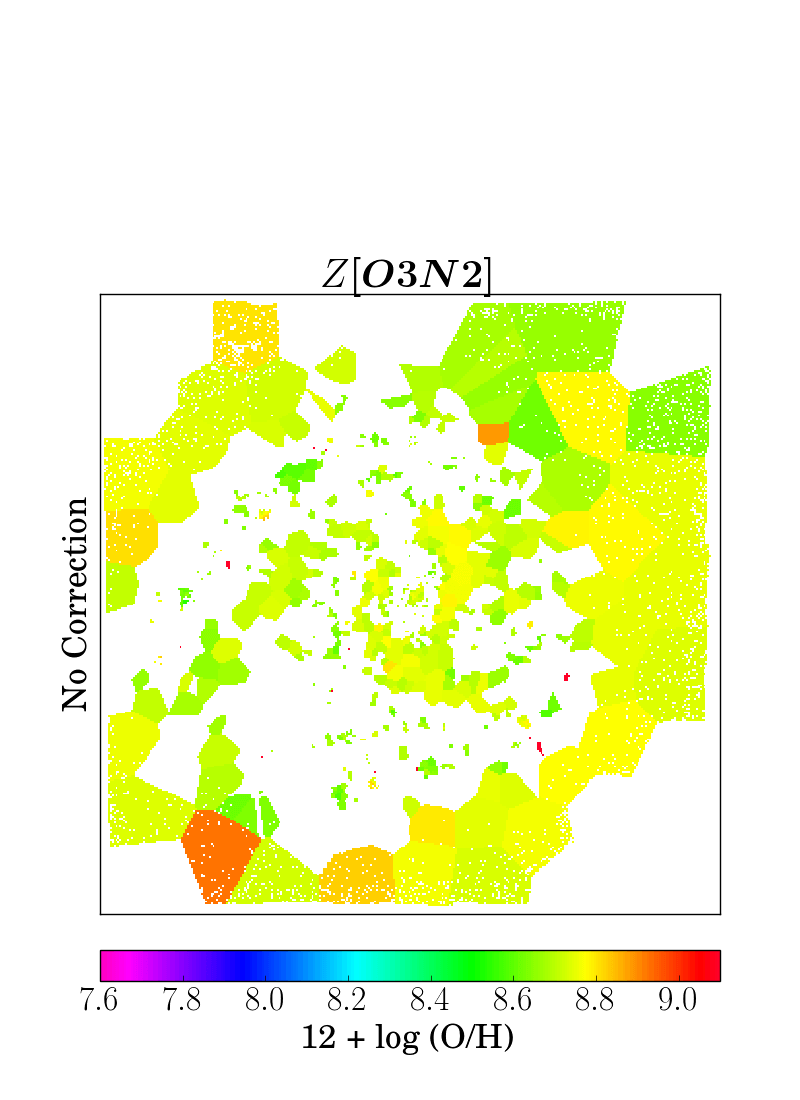}
	\includegraphics[width=0.28\textwidth, trim={0 1.2cm 0 5.5cm}, clip]{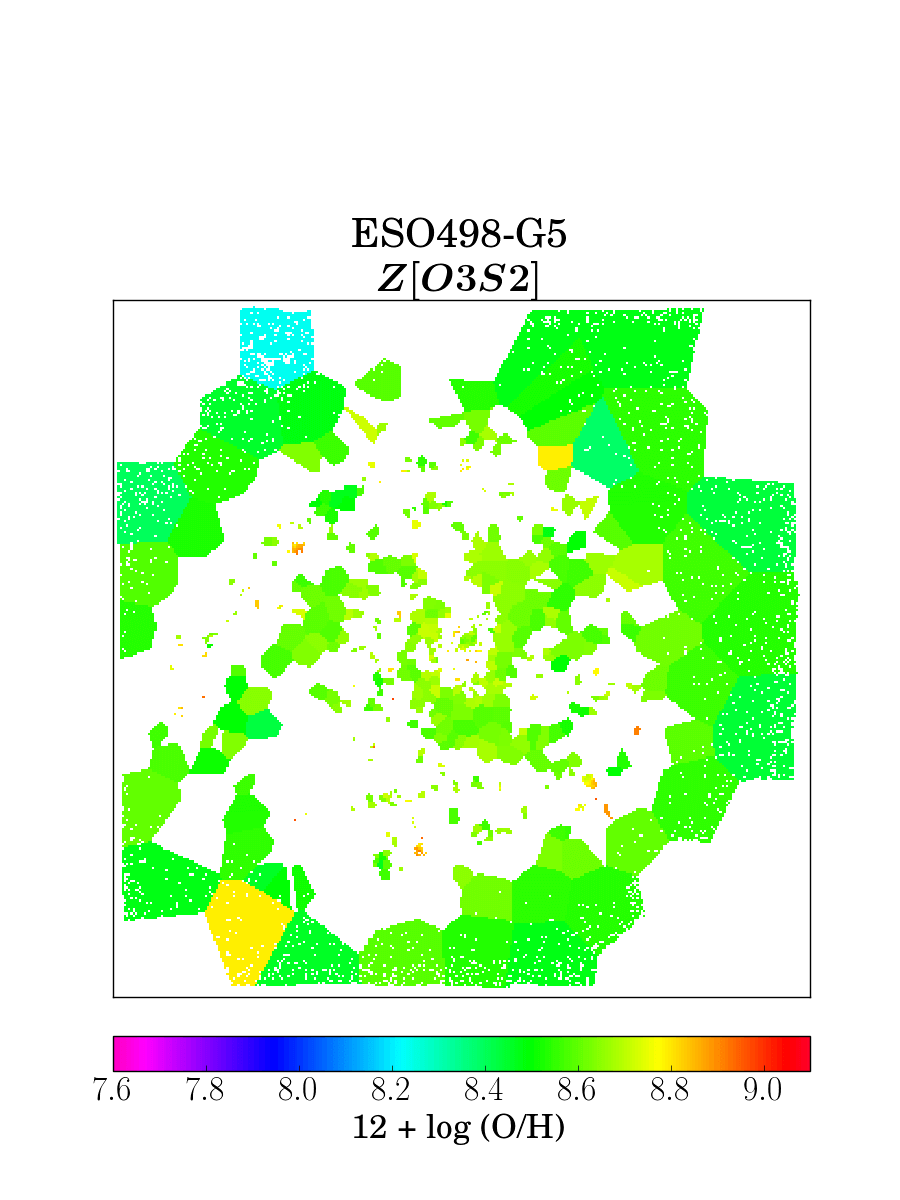}
	\includegraphics[width=0.28\textwidth, trim={0 1.2cm 0 5.5cm}, clip]{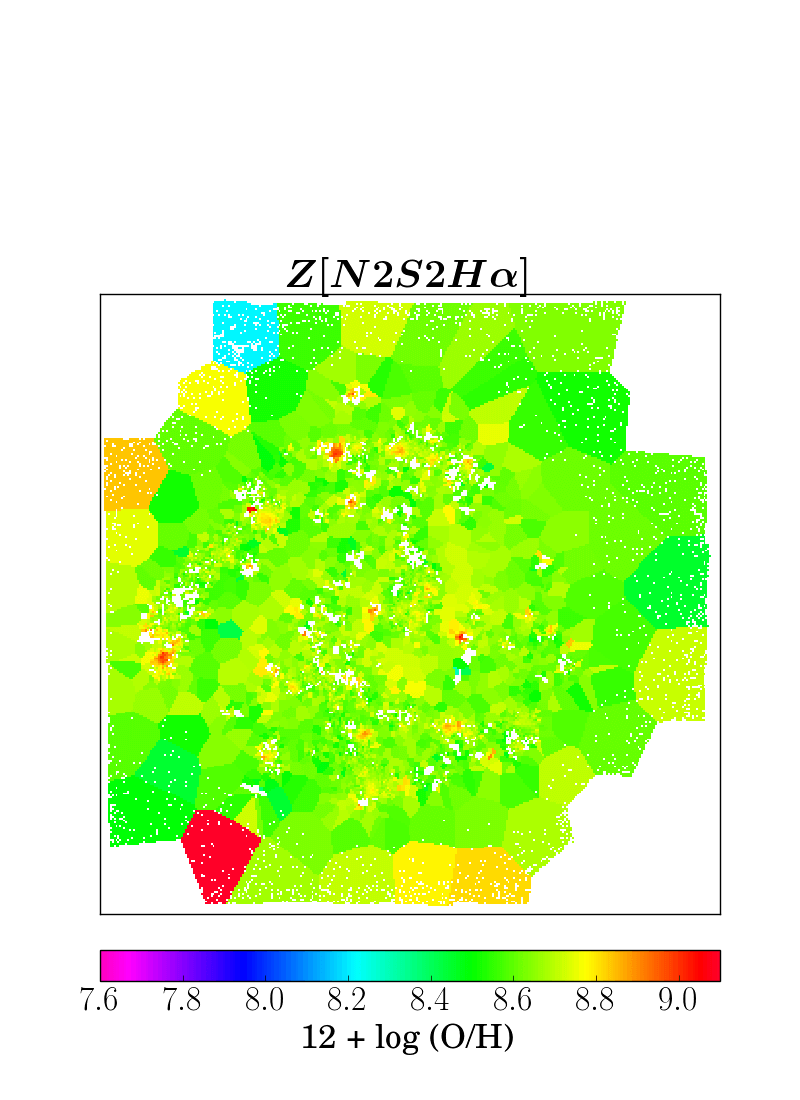}
	\includegraphics[width=0.28\textwidth, trim={0 1.2cm 0 5.5cm}, clip]{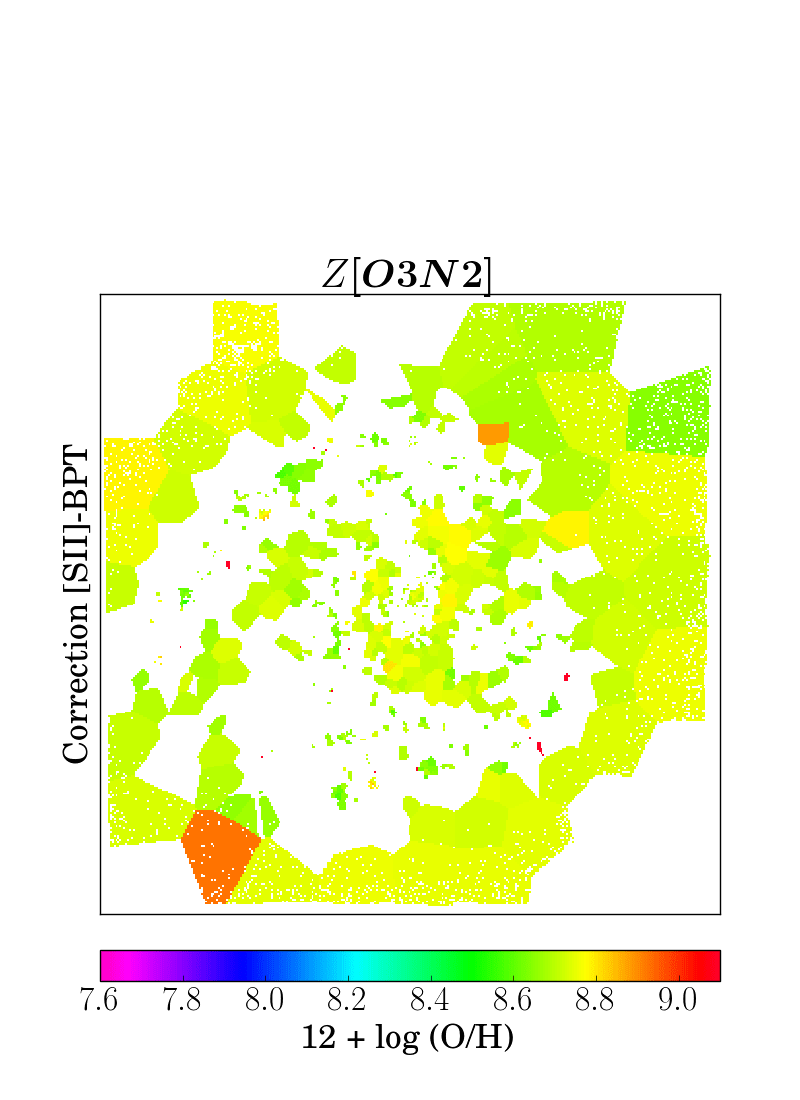}
	\includegraphics[width=0.28\textwidth, trim={0 1.2cm 0 5.5cm}, clip]{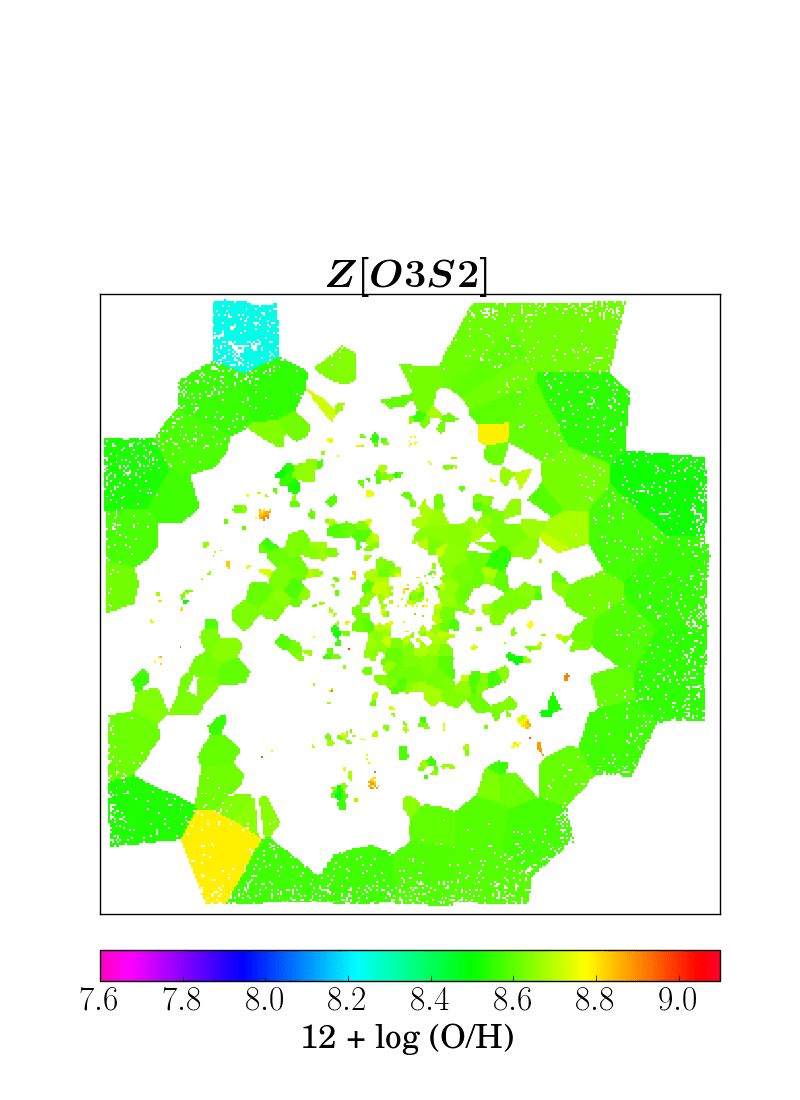}
	\includegraphics[width=0.28\textwidth, trim={2.8cm 0 2.8cm 0}, clip ]{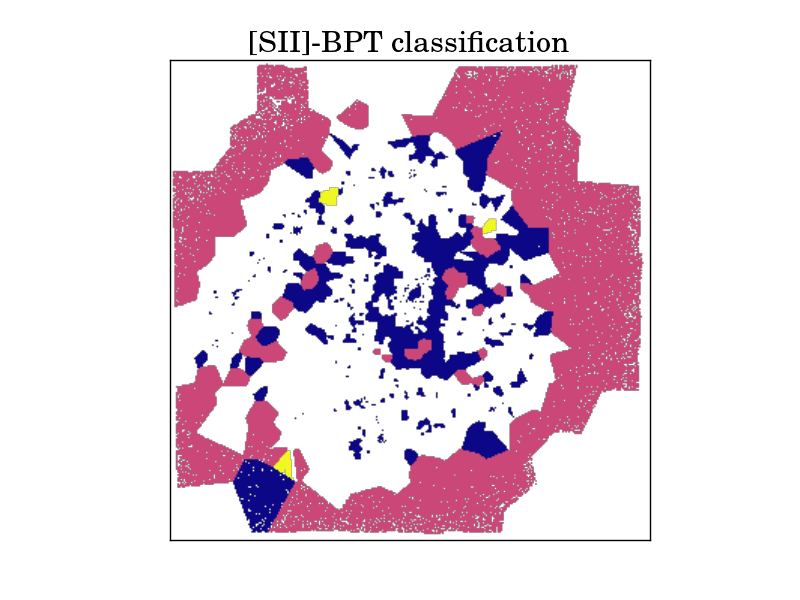}
	\includegraphics[width=0.28\textwidth, trim={0 1.2cm 0 5.5cm}, clip]{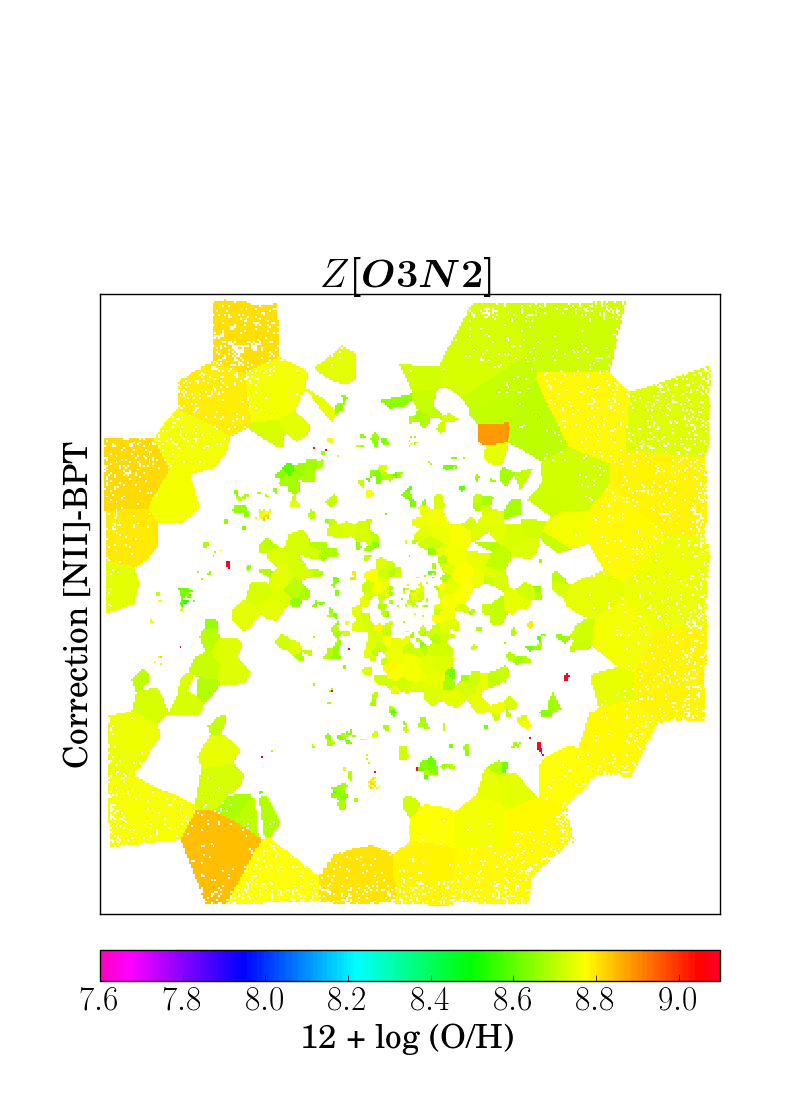}
	\includegraphics[width=0.28\textwidth, trim={0 1.2cm 0 5.5cm}, clip]{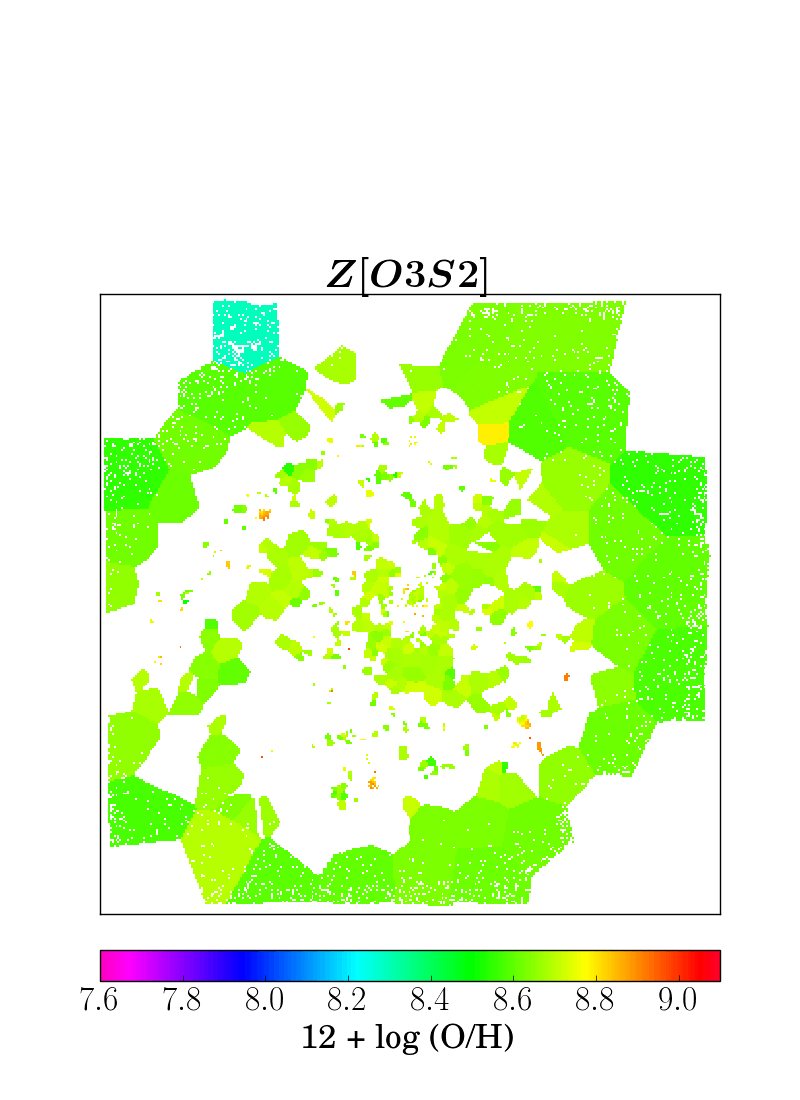}
	\includegraphics[width=0.28\textwidth, trim={2.8cm 0 2.8cm 0}, clip ]{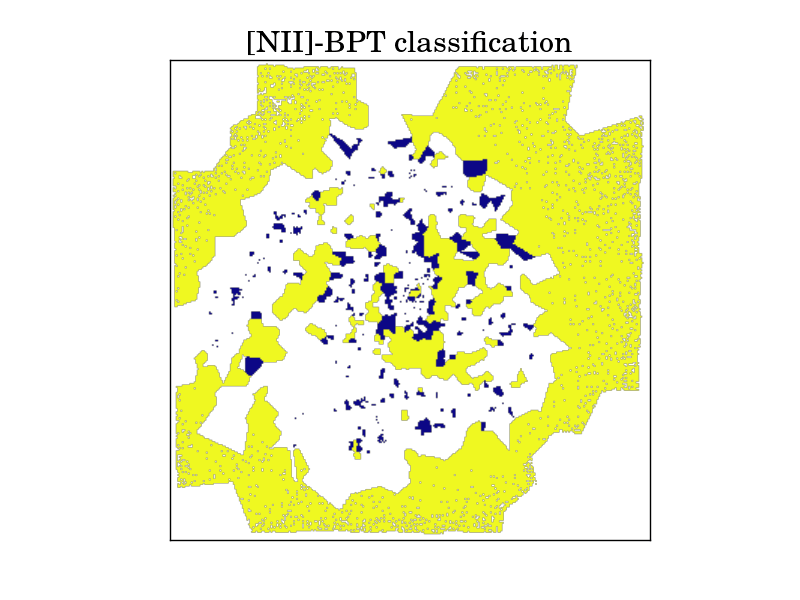}
	\caption{ Maps correspond to galaxy ESO498-G5, see caption of Figure \ref{fig:NGC1042} for details.}
	\label{fig:ESO498-G5}
\end{figure*}
\begin{figure*}
	\centering
	\includegraphics[width=0.28\textwidth, trim={0 1.2cm 0 5.5cm}, clip]{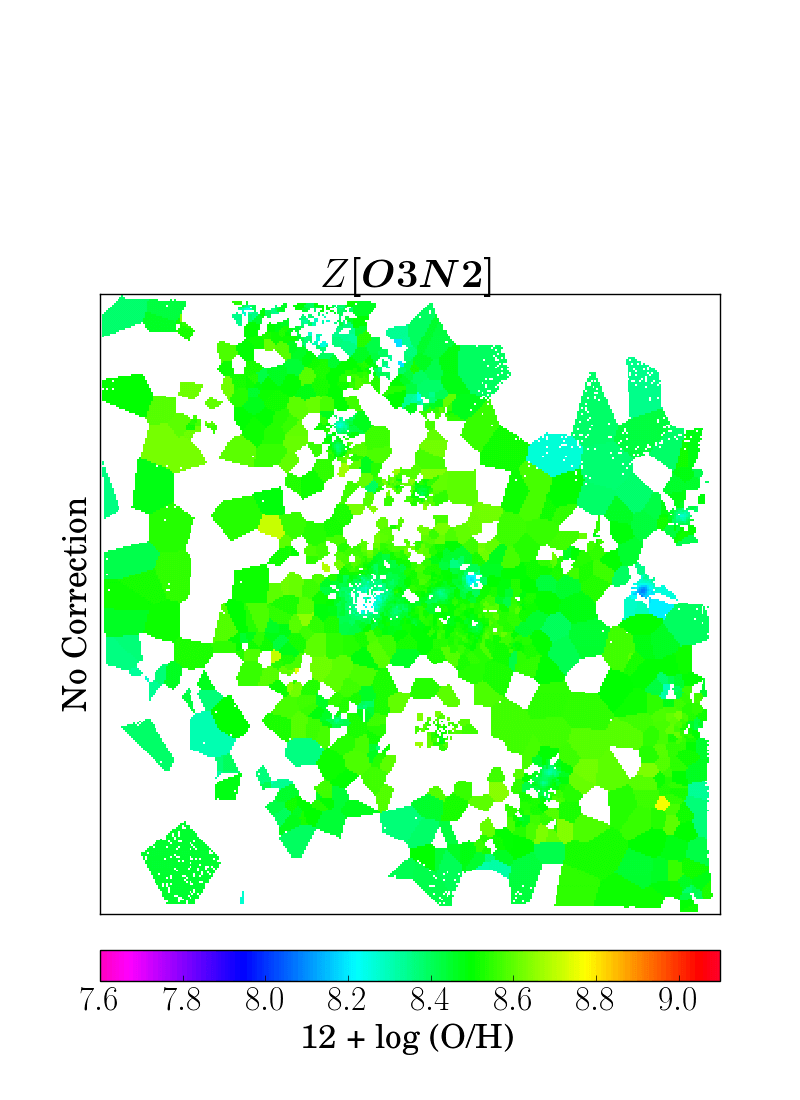}
	\includegraphics[width=0.28\textwidth, trim={0 1.2cm 0 5.5cm}, clip]{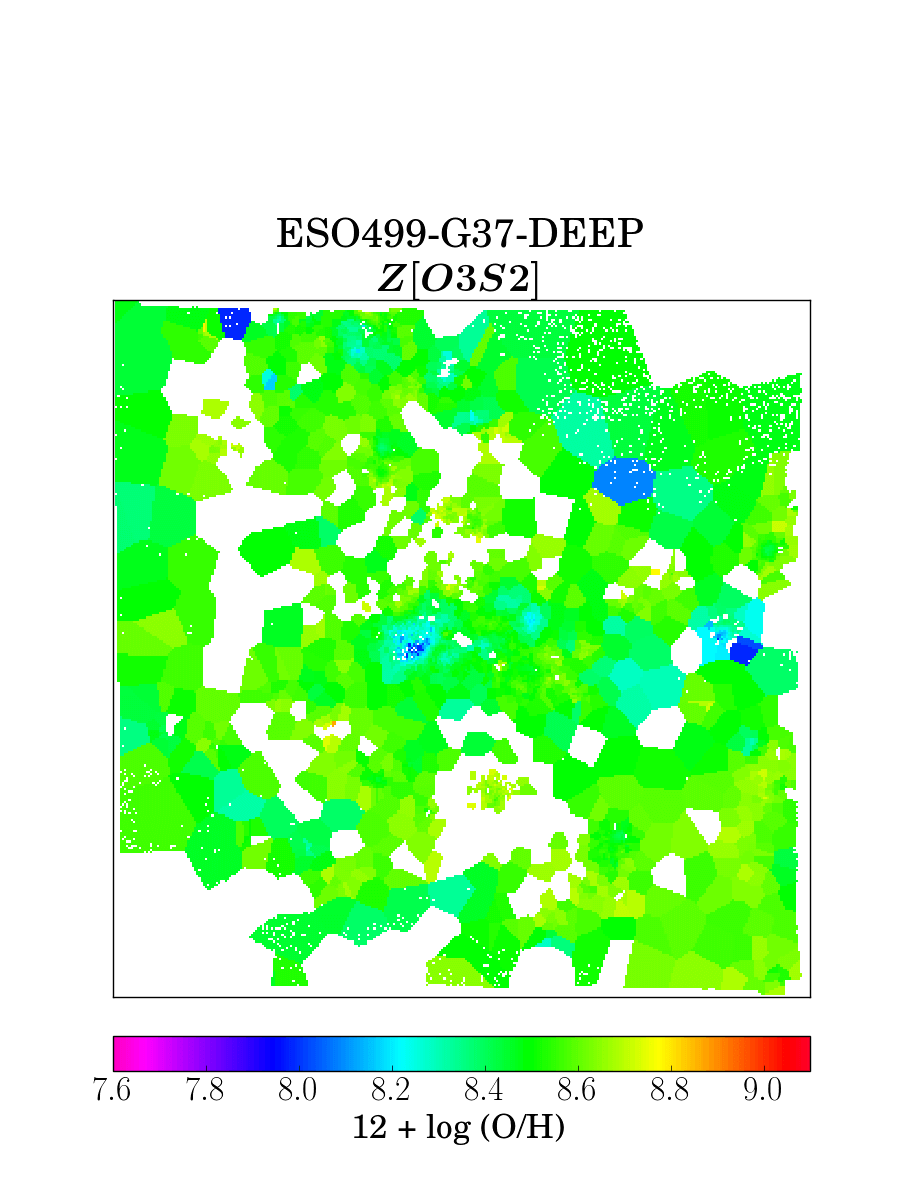}
	\includegraphics[width=0.28\textwidth, trim={0 1.2cm 0 5.5cm}, clip]{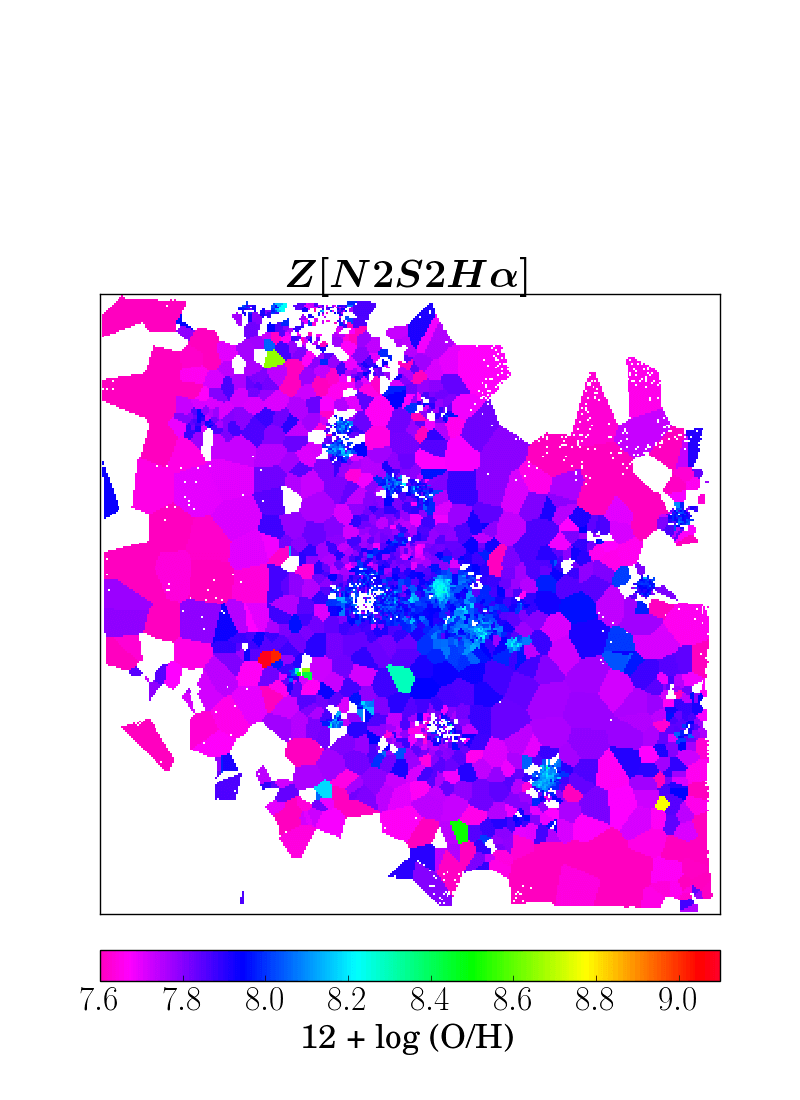}
	\includegraphics[width=0.28\textwidth, trim={0 1.2cm 0 5.5cm}, clip]{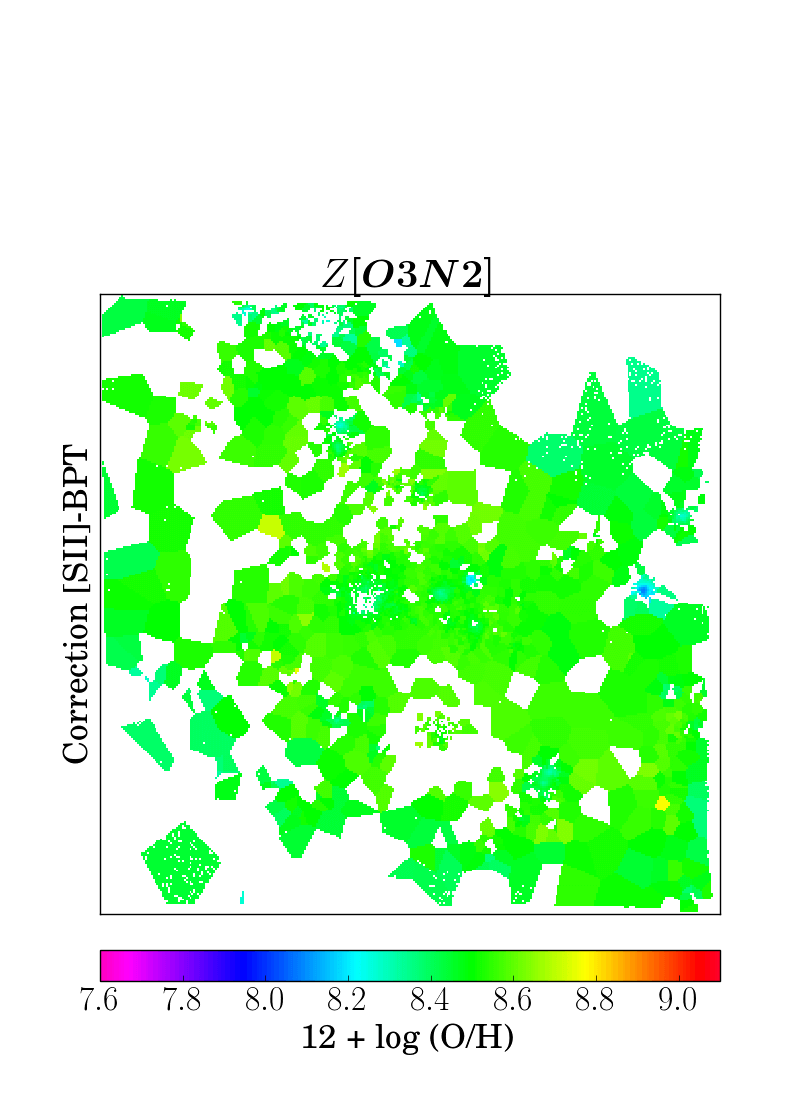}
	\includegraphics[width=0.28\textwidth, trim={0 1.2cm 0 5.5cm}, clip]{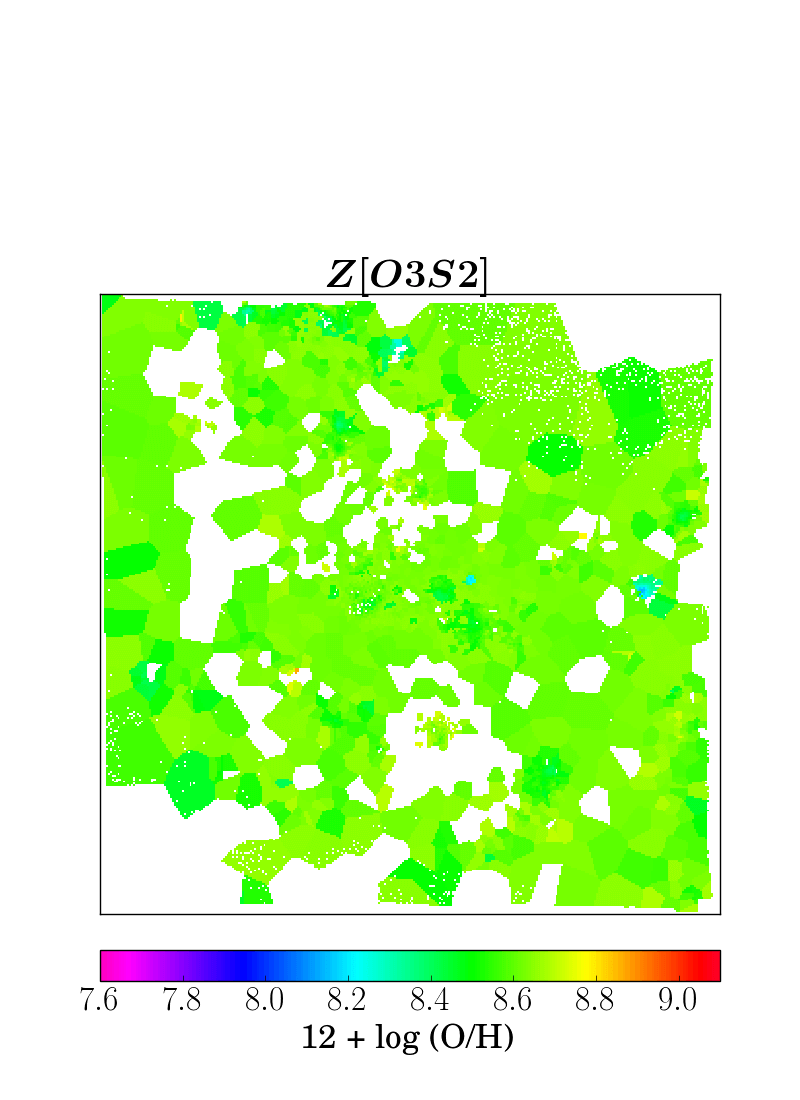}
	\includegraphics[width=0.28\textwidth, trim={2.8cm 0 2.8cm 0}, clip ]{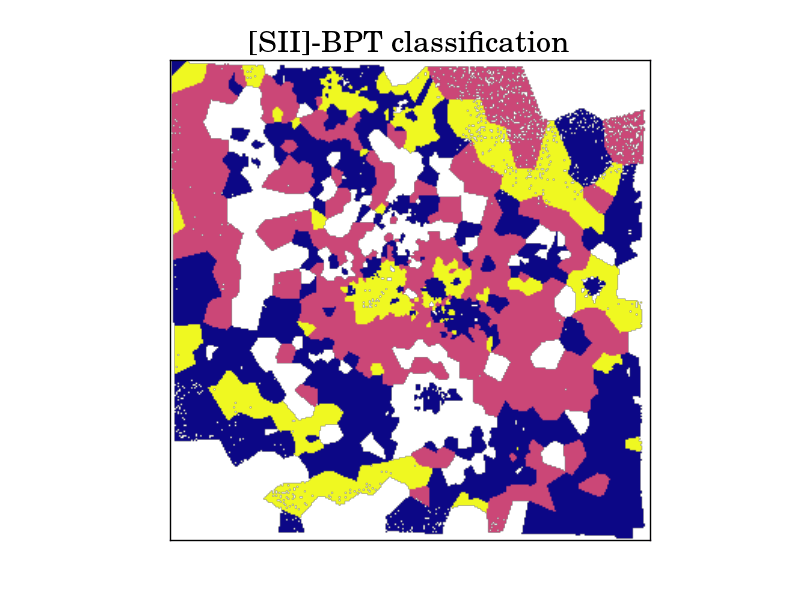}
	\includegraphics[width=0.28\textwidth, trim={0 1.2cm 0 5.5cm}, clip]{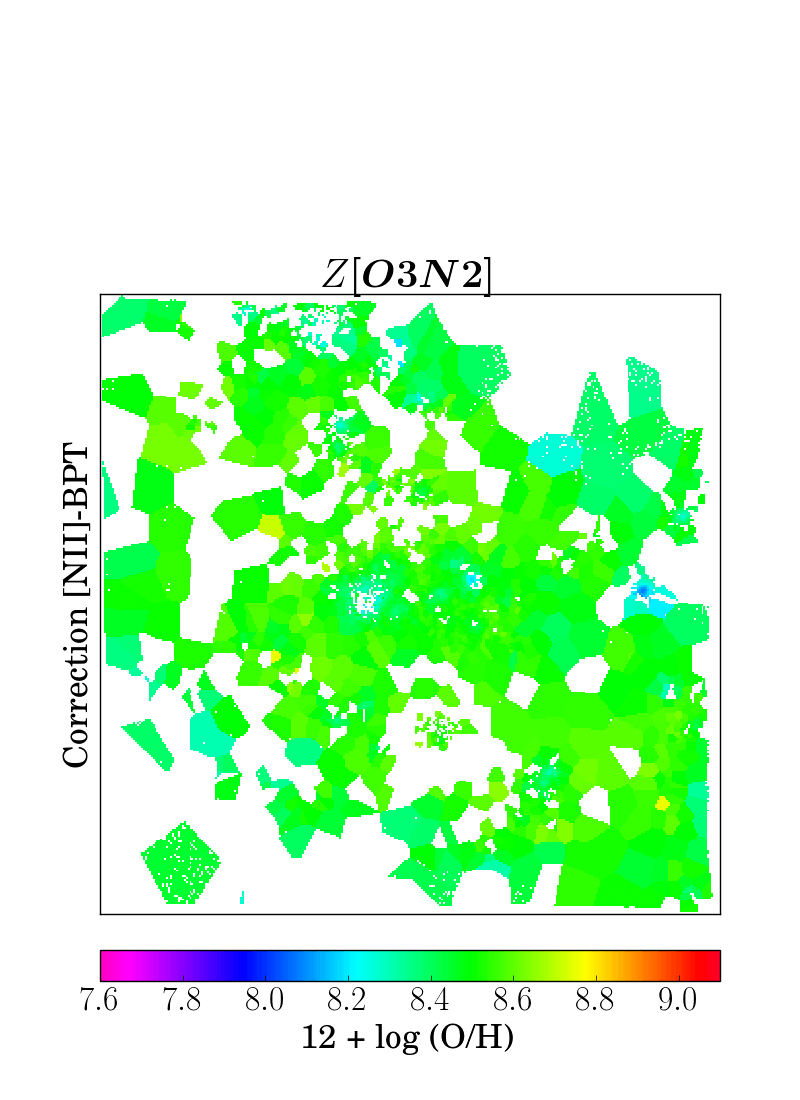}
	\includegraphics[width=0.28\textwidth, trim={0 1.2cm 0 5.5cm}, clip]{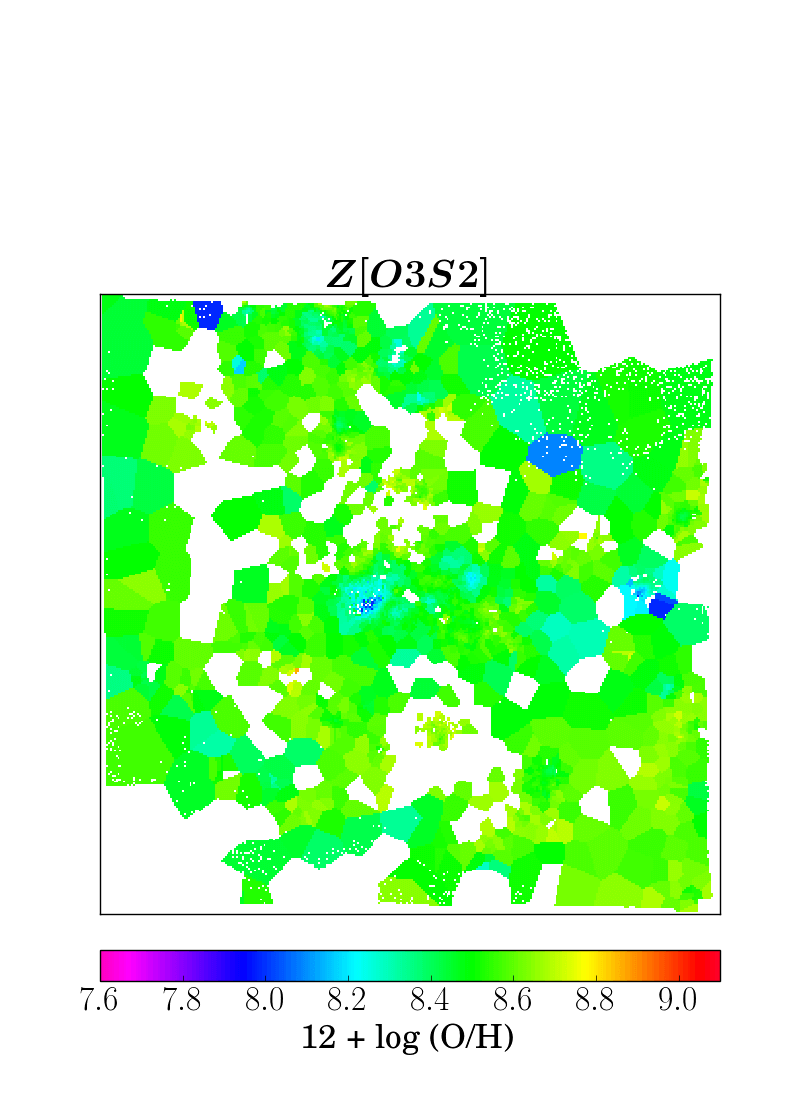}
	\includegraphics[width=0.28\textwidth, trim={2.8cm 0 2.8cm 0}, clip ]{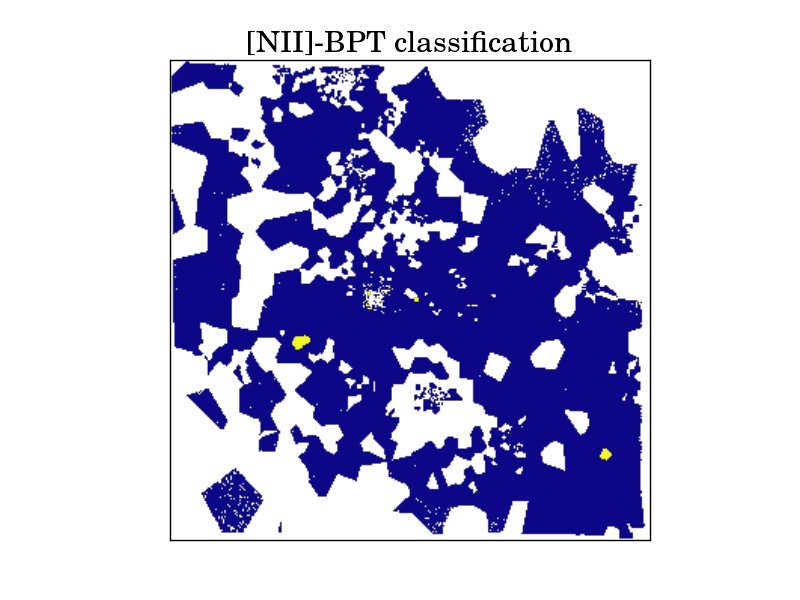}
	\caption{ Maps correspond to galaxy ESO499-G37-DEEP, see caption of Figure \ref{fig:NGC1042} for details.}
	\label{fig:ESO499-G37-DEEP}
\end{figure*}

\section{Comparison of new calibrations}
\indent Figure \textcolor{blue}{B1} shows a comparison of new O3N2 and O3S2 calibrations after
applying corrections derived in this work based on [S \textsc{ii}]-BPT (left
panel) and [N \textsc{ii}]-BPT (right panel) classifications. These data points
correspond to the \HII~(blue) and DIG/LIER/Seyfert counterparts (green
points) for all \HII-DIG/LIER/Seyfert pairs.
\begin{figure*}
	\centering
	\includegraphics[width=0.3\textwidth]{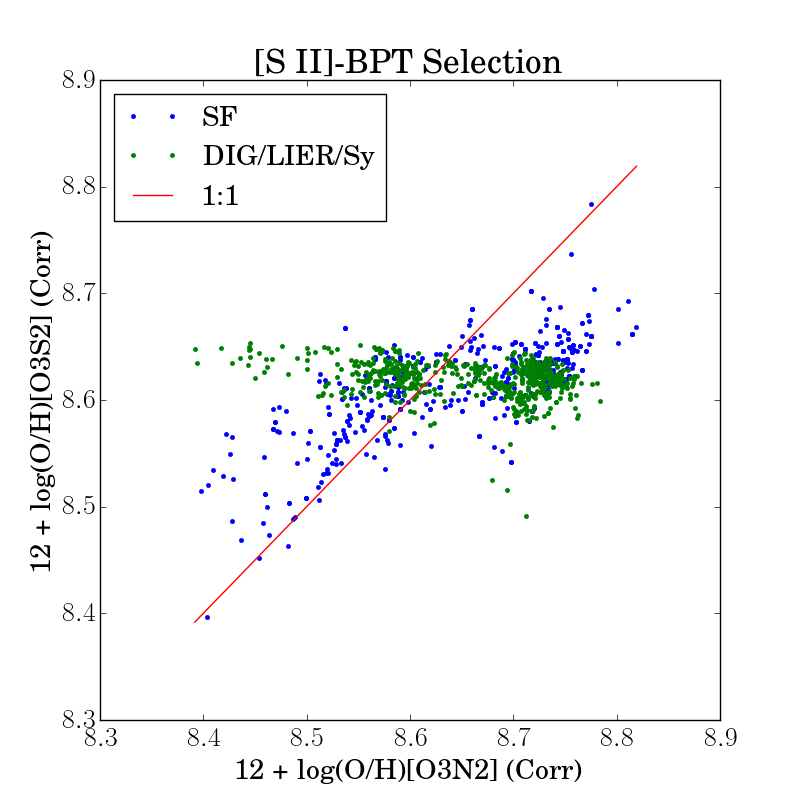}
	\includegraphics[width=0.3\textwidth]{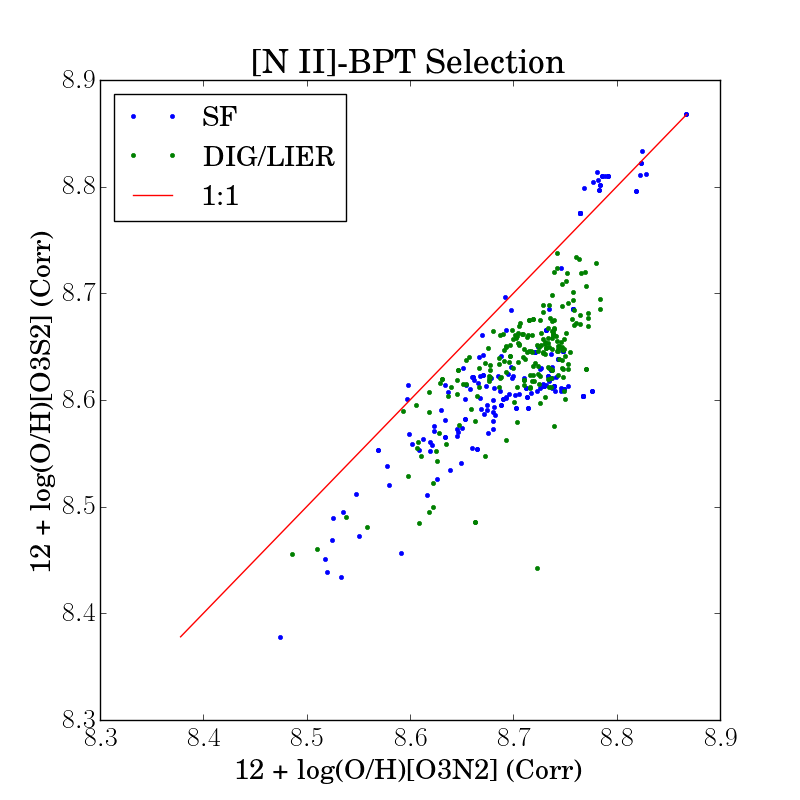}
	
	\caption{Comparison of metallicities using metallicity diagnositcs O3N2 and O3S2 after applying corrections derived in this work for [S \textsc{ii}]-BPT selection (left panel) and [N \textsc{ii}]-BPT selection (right panel). On each panel, blue points and green points represent metallicities of H \textsc{ii} (SF) and DIG/Sy/LIER, respectively, for all H \textsc{ii}-DIG/Sy/LIER pairs used for studying biases and deriving new calibrations.}
\end{figure*}

% Don't change these lines
\bsp	% typesetting comment
\label{lastpage}
\end{document}